\documentclass[12pt,a4paper,twoside,openright,titlepage,
               headinclude,,footinclude,BCOR=5mm,
               numbers=noenddot,cleardoublepage=empty,
               tablecaptionabove]{scrreprt}
               
\usepackage[T1]{fontenc}
\usepackage[utf8]{inputenc}
\usepackage[english]{babel}
\usepackage{amsmath,amssymb}
\usepackage[style=philosophy-modern,hyperref]{biblatex}
\usepackage{chngpage}
\usepackage{calc}
\usepackage{listings}
\usepackage{graphicx}
\usepackage{subfig}
\usepackage{lipsum}
\usepackage{shapepar}
\usepackage{pifont}
\usepackage[dottedtoc,eulerchapternumbers,subfig,beramono,eulermath,pdfspacing,listings]{classicthesis}
\usepackage{arsclassica}

\usepackage{rotating}

\usepackage{libertine} % NEW
\usepackage{libertinust1math} % NEW

\usepackage{float} % NEW

\usepackage{algorithmic,algorithm} % NEW
\newcommand{\Algo}[1]{Algorithm~\ref{algo:#1}} % NEW
\newcommand{\algolabel}[1]{\label{algo:#1}} % NEW

\usepackage{bm} % NEW

\usepackage{lscape} % NEW

\newcommand{\myName}{Minas Karamanis}
\newcommand{\myTitle}{Bayesian Computation in Astronomy}
\newcommand{\mySubTitle}{Novel methods for parallel and gradient--free inference}
\newcommand{\myUniversity}{Doctor of Philosophy\\The University of Edinburgh\\July 2022}

\graphicspath{{Graphics/}}

\definecolor{lightergray}{gray}{0.99}

\lstnewenvironment{code}% 
   {\setkeys{lst}{columns=fullflexible,keepspaces=true}%
   \lstset{basicstyle=\small\ttfamily}}{}

\bibliography{Bibliography, Ess_Bibliography, Zeus_Bibliography, PMC_Bibliography, PocoMC_Bibliography}

\defbibheading{bibliography}{%
\cleardoublepage
\manualmark
\phantomsection
\addcontentsline{toc}{chapter}{\tocEntry{\bibname}}
\chapter*{\bibname\markboth{\spacedlowsmallcaps{\bibname}}
{\spacedlowsmallcaps{\bibname}}}}

\begin{document}
\pagenumbering{roman}
\pagestyle{plain}

% !TEX TS-program = pdflatex
% !TEX root = ../ArsClassica.tex

%*******************************************************
% Titlepage
%*******************************************************
\begin{titlepage}
\pdfbookmark{Titlepage}{Titlepage}
\changetext{}{}{}{((\paperwidth  - \textwidth) / 2) - \oddsidemargin - \hoffset - 1in}{}
    \begin{center}
        {\LARGE  

        \hfill

        \vfill
        
        {\color{Maroon}\spacedallcaps{\myTitle}} \\ \bigskip
        
        {\mySubTitle} \\ \medskip
        
        \hfill
        
        {\spacedlowsmallcaps{\myName}}

        }

        \vfill

        \includegraphics[width=0.6\textwidth]{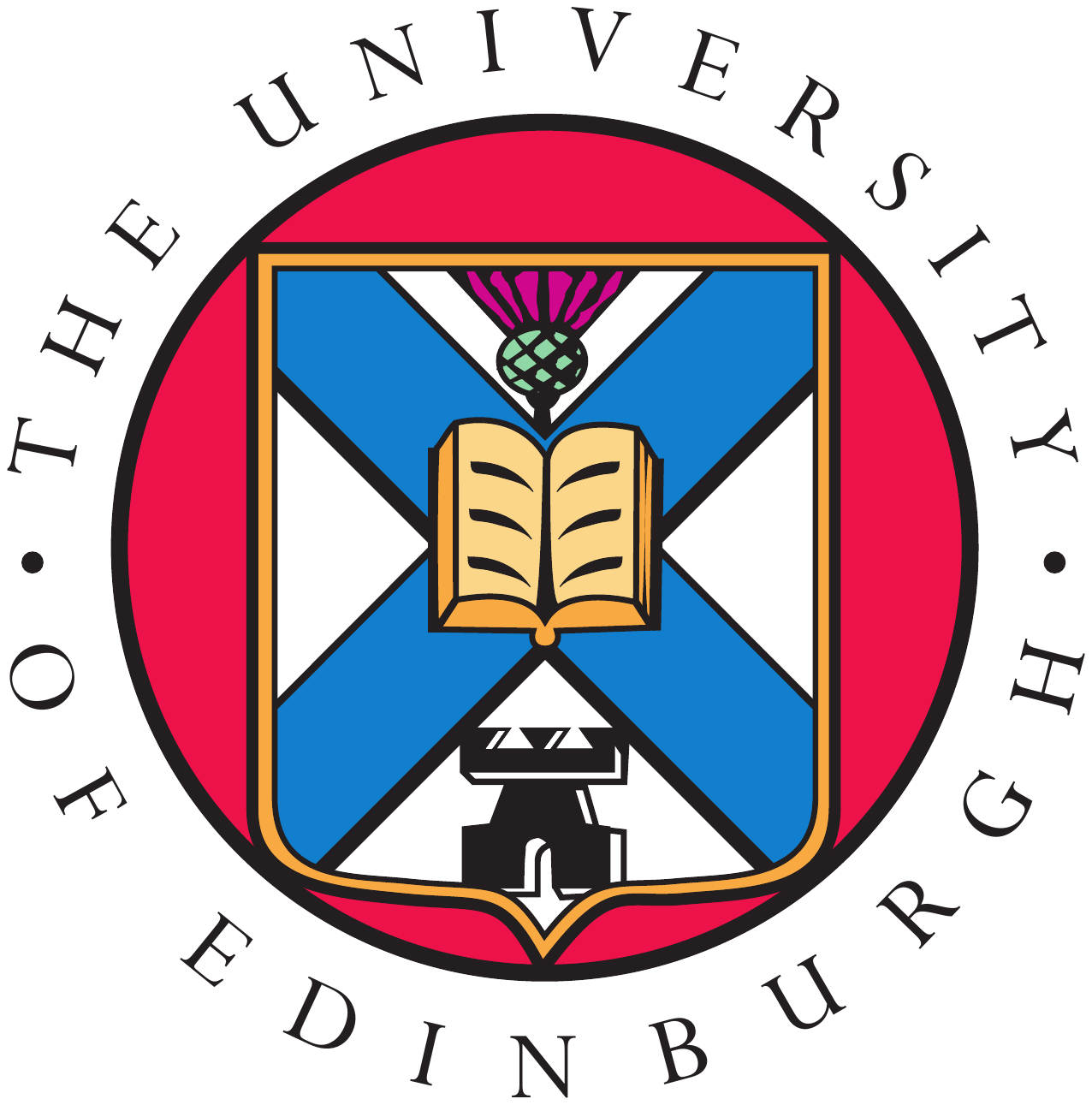} \\ \bigskip

        {\large

		\myUniversity
		
		}

        \vfill                      

    \end{center}        
\end{titlepage} 
\cleardoublepage

% !TEX TS-program = pdflatex
% !TEX root = ../ArsClassica.tex

%*******************************************************
% Dedication
%*******************************************************
\pdfbookmark{Dedication}{Dedication}
%\addcontentsline{toc}{chapter}{Dedication}

\begin{center}
    \thispagestyle{empty}
    \vspace*{\fill}
    %\textit{To ...}
    \begin{flushright}
    \itshape
    Heard melodies are sweet,\\
    but those unheard, are sweeter.\\
    \medskip
    --- John Keats
    \end{flushright}
    \vspace*{\fill}
\end{center}
\cleardoublepage

% !TEX TS-program = pdflatex
% !TEX root = ../ArsClassica.tex

%*******************************************************
% Lay Summary
%*******************************************************
\pdfbookmark{Lay summary}{Lay summary}
\addcontentsline{toc}{chapter}{Lay summary}

\chapter*{Lay summary}

\looseness=-1 Over the past few decades, the volume of astronomical and cosmological data has increased substantially. In response to that, a variety of astrophysical models have been proposed to explain the plethora of observations. As the information provided by the data is always incomplete and uncertain, inferring the properties of a model, including the values of its parameters, given the observed data, generally requires us to reason in the face of uncertainty. In the context of \textit{Bayesian inference}, uncertainty is represented by the notion of probability. One usually starts by quantifying their state of knowledge about the possible values of the model parameters \textit{prior} to seeing the data, in the form of a probability distribution called the \textit{prior}. The next step is to use the so--called \textit{Bayes' theorem} in order to update one's degree of belief about the model parameters given the available data. The outcome of this updating process is the \textit{posterior} probability distribution of the model parameters given the data which quantifies the plausibility of different parameter values.

Approximating the \textit{posterior} generally requires the use of probabilistic computational methods. Standard practice in astronomy often employs conventional computational tools (e.g. \textit{Markov chain Monte Carlo}) despite their specific theoretical limitations or narrow range of validity. The aim of this thesis is to first introduce the basic principles of Bayesian inference along with the basic methods used for Bayesian computation and then present two novel algorithms and their respective software implementations. A common element of these newly developed tools is their ability to exploit the available information about the geometry of the posterior in order to approximate it more quickly. Finally, both methods are able to benefit from the possible availability of multiple CPUs in order to accelerate their computation.
\cleardoublepage

% !TEX TS-program = pdflatex
% !TEX root = ../ArsClassica.tex

%*******************************************************
% Abstract
%*******************************************************
\addcontentsline{toc}{chapter}{\abstractname}
\pdfbookmark{Abstract}{Abstract}

\chapter*{Abstract}

The goal of this thesis is twofold; introduce the fundamentals of \textit{Bayesian inference and computation} focusing on astronomical and cosmological applications, and present recent advances in probabilistic computational methods developed by the author that aim to facilitate \textit{Bayesian data analysis} for the next generation of astronomical observations and theoretical models. 

The first part of this thesis familiarises the reader with the notion of probability and its relevance for science through the prism of \textit{Bayesian reasoning}, by introducing the key constituents of the theory and discussing its best practices. The second part includes a pedagogical introduction to the principles of \textit{Bayesian computation} motivated by the geometric characteristics of probability distributions and followed by a detailed exposition of various methods including \textit{Markov chain Monte Carlo (MCMC)}, \textit{Sequential Monte Carlo (SMC)} and \textit{Nested Sampling (NS)}. Finally, the third part presents two novel computational methods and their respective software implementations.

The first such development is \textit{Ensemble Slice Sampling (ESS)}, a new class of \textit{MCMC} algorithms that extend the applicability of the standard \textit{Slice Sampler} by adaptively tuning its only hyperparameter and utilising an ensemble of parallel walkers in order to efficiently handle strong correlations between parameters. The parallel, black--box and gradient--free nature of the method renders it ideal for use in combination with computationally expensive and \textit{non--differentiable} models often met in astronomy. \textit{ESS} is implemented in \textit{Python} in the well--tested and open-source software package called \textit{zeus} that is specifically designed to tackle the computational challenges posed by modern astronomical and cosmological analyses.  In particular, use of the code requires minimal, if any, hand--tuning of hyperparameters while its performance is insensitive to linear correlations and it can scale up to thousands of CPUs without any extra effort.

The next contribution includes the introduction of \textit{Preconditioned Monte Carlo (PMC)}, a novel \textit{Monte Carlo} method for \textit{Bayesian inference} that facilitates effective sampling of probability distributions with non--trivial geometry. \textit{PMC} utilises a \textit{Normalising Flow (NF)} in order to decorrelate the parameters of the distribution and then proceeds by sampling from the preconditioned target distribution using an adaptive \textit{SMC} scheme. \textit{PMC}, through its \textit{Python} implementation \textit{pocoMC}, achieves excellent sampling performance, including accurate estimation of the \textit{model evidence}, for highly correlated, non--Gaussian, and multimodal target distributions. Finally, the code is directly parallelisable, manifesting linear scaling up to thousands of CPUs. % New
\cleardoublepage % New

% !TEX TS-program = pdflatex
% !TEX root = ../ArsClassica.tex

%*******************************************************
% Acknowledgements
%*******************************************************
\pdfbookmark{Acknowledgements}{Acknowledgements}
\addcontentsline{toc}{chapter}{Acknowledgements}

\chapter*{Acknowledgements}

\looseness=-1 The completion of this journey would not have been possible without the support and guidance of a great number of people. This thesis is dedicated to all of them.

First and foremost, I would like to express my sincere gratitude to my supervisors Florian Beutler and John Peacock. Florian's constant encouragement and advice from day one of my PhD studies enabled me to explore freely and pursue whichever research avenues I found interesting. I only met John in the middle of my PhD studies but his enthusiasm and wisdom were enough to act as a huge source of inspiration ever since. My supervisors allowed me to become the researcher that I am today and also showed me the kind of researcher that I want to become. It was a pleasure and privilege to learn from and work with them.

I am also extremely grateful to Alan Heavens and Ross McLure, for being the two examiners for my viva. Their keen interest in the subject along with their insightful questions turned the examination into a fascinating and very enjoyable debate. 

This thesis is but the last stage of a long journey that started more than twenty years ago. Therefore, I would like to deeply thank my family for their continuous support during all these years, and especially my father and grandfather for cultivating my love for science when I was really young. My gratitude extends to my mother, aunt, and grandmother that enabled my involvement in science at an early age, always providing access to books, lectures, telescopes, microscopes, and spare parts that I required for my ``little science experiments'', and of course to my brother for always being my ``lab assistant''.

Special thanks goes to all my friends for making this process easier and more enjoyable and especially to Jamie, Mike, Tasos and anyone else who traveled this path along with me.

A big thanks goes to my girlfriend Denia for her unconditional support and for believing in me during the past four years. Finally, I would like to thank my two cats, Poco and Gatoulis, for keeping me company and lightening my mood during this stressful period.
\cleardoublepage % New

% !TEX TS-program = pdflatex
% !TEX root = ../ArsClassica.tex

%*******************************************************
% Declaration
%*******************************************************
\pdfbookmark{Declaration}{Declaration}
\addcontentsline{toc}{chapter}{Declaration}

\chapter*{Declaration}

\noindent I declare that this thesis was composed by myself, that the work contained herein is my own except where explicitly stated otherwise in the text, and that this work has not been submitted for any other degree or professional qualification except as specified.

\bigskip
\bigskip

\noindent Parts of this work have been published in \parencite{karamanis2021ensemble,karamanis2021zeus,karamanis2022pmc,karamanis2022pocomc}.

\bigskip

\begin{flushright}
    \itshape
    (M. Karamanis, July 2022)
\end{flushright} % New
\cleardoublepage % New

% !TEX TS-program = pdflatex
% !TEX root = ../ArsClassica.tex

%*******************************************************
% Contents
%*******************************************************
\phantomsection
\pdfbookmark{\contentsname}{tableofcontents}
\setcounter{tocdepth}{2}
\tableofcontents
\markboth{\spacedlowsmallcaps{\contentsname}}{\spacedlowsmallcaps{\contentsname}} 
\cleardoublepage

% !TEX TS-program = pdflatex
% !TEX root = ../ArsClassica.tex

%*******************************************************
% List of Figures
%*******************************************************
\addcontentsline{toc}{chapter}{\listfigurename}
\pdfbookmark{\contentsname}{listoffigures}
\listoffigures
%\markboth{\spacedlowsmallcaps{\contentsname}}{\spacedlowsmallcaps{\contentsname}}  % New
\cleardoublepage % New

% !TEX TS-program = pdflatex
% !TEX root = ../ArsClassica.tex

%*******************************************************
% List of Tables
%*******************************************************
\addcontentsline{toc}{chapter}{\listtablename}
\pdfbookmark{\contentsname}{listoftables}
\listoftables
%\markboth{\spacedlowsmallcaps{\contentsname}}{\spacedlowsmallcaps{\contentsname}}  % New
\cleardoublepage % New

% !TEX TS-program = pdflatex
% !TEX root = ../ArsClassica.tex

%*******************************************************
% List of Algorithms
%*******************************************************
\addcontentsline{toc}{chapter}{\listalgorithmname}
\pdfbookmark{\contentsname}{listofalgorithms}
\listofalgorithms
%\markboth{\spacedlowsmallcaps{\contentsname}}{\spacedlowsmallcaps{\contentsname}}  % New
\cleardoublepage % New

\pagenumbering{arabic}
% First Part : Bayesian Inference
% !TEX TS-program = pdflatex
% !TEX root = ../ArsClassica.tex

%************************************************
\part{Bayesian Inference}
\label{prt:fundamentals}
%************************************************

% !TEX TS-program = pdflatex
% !TEX root = ../ArsClassica.tex

%************************************************
\chapter{Probability Theory}
\label{chp:probability}
%************************************************

\begin{flushright}
\itshape
Science is more than a body of knowledge; it is a way of thinking. \\The method of science, as stodgy and grumpy as it may seem,\\ is far more important than the findings of science.\\
\medskip
--- Carl Sagan
\end{flushright}

This chapter introduces the basic principles of Bayesian inference and presents its fundamental ideas and distinctive features.

%************************************************

\section{The goal of science}

The key goal of science is to distil the patterns of nature into mathematical language and call them physical laws. To this end, science relies on a formal way of thinking and interrogating nature, asking the right questions, interpreting observations, and updating its beliefs and hypotheses in the light of new evidence. This way of thinking, inherent in all scientific pursuits seems to be deeply connected to the mathematical notion of probability.

Science proceeds towards this elusive target with careful steps following the scientific method. The latter is often illustrated as a loop. Hypotheses are proposed and models quantifying certain aspects of those hypotheses are developed. The hypotheses give rise to predictions, in a process called \textit{deductive inference}, to be tested against experimental data. Unfortunately, as the information that we extract from nature in the form of data is always incomplete and uncertain, testing our hypotheses by comparing our model predictions to the experimental data requires us to reason in the presence of uncertainty. We thus rely on \textit{plausible inference}, that is, the process of inferring the truth of our theories about the cosmos on the basis of incomplete and uncertain information. 

Scientific statements about the physical world are uncertain by necessity. No amount of new information will ever be enough to validate or disprove a hypothesis. Furthermore, our models, despite our best intentions, are often simpler than the natural processes which they attempt to capture. Our best hope is thus to accept the existence of this inherent and unavoidable uncertainty and instead try to quantify the plausibility of our statements about the cosmos. Assessing the plausibility of scientific theories is the subject of probability theory.

%************************************************

\section{The notion of probability}

There are few concepts in science and mathematics as controversial, with their meaning so contested during the centuries, as the notion of probability. Three centuries ago people started seriously thinking about how to best make decisions and reason in the face of uncertainty. Perhaps, the first to formally articulate this problem was \textit{Jacob Bernoulli} in his seminal work \textit{Ars Conjectandi} published in $1713$.

The answer to Bernoulli's question was provided by \textit{Reverend Thomas Bayes}, in an essay named \textit{An Essay towards solving a Problem in the Doctrine of Chances}, published posthumously by his friend \textit{Richard Price} in $1763$. The paper included theorems on \textit{conditional probability} which formed the basis of what we now call \textit{Bayes' theorem}. The discovery of the latter is actually due to \textit{Laplace}, who not only developed, extended and clarified probability theory, but also applied it successfully to a plethora of problems in astronomy, medicine, and economics. 
\begin{figure}[H]
    \centering
	\centerline{\includegraphics[scale=0.65]{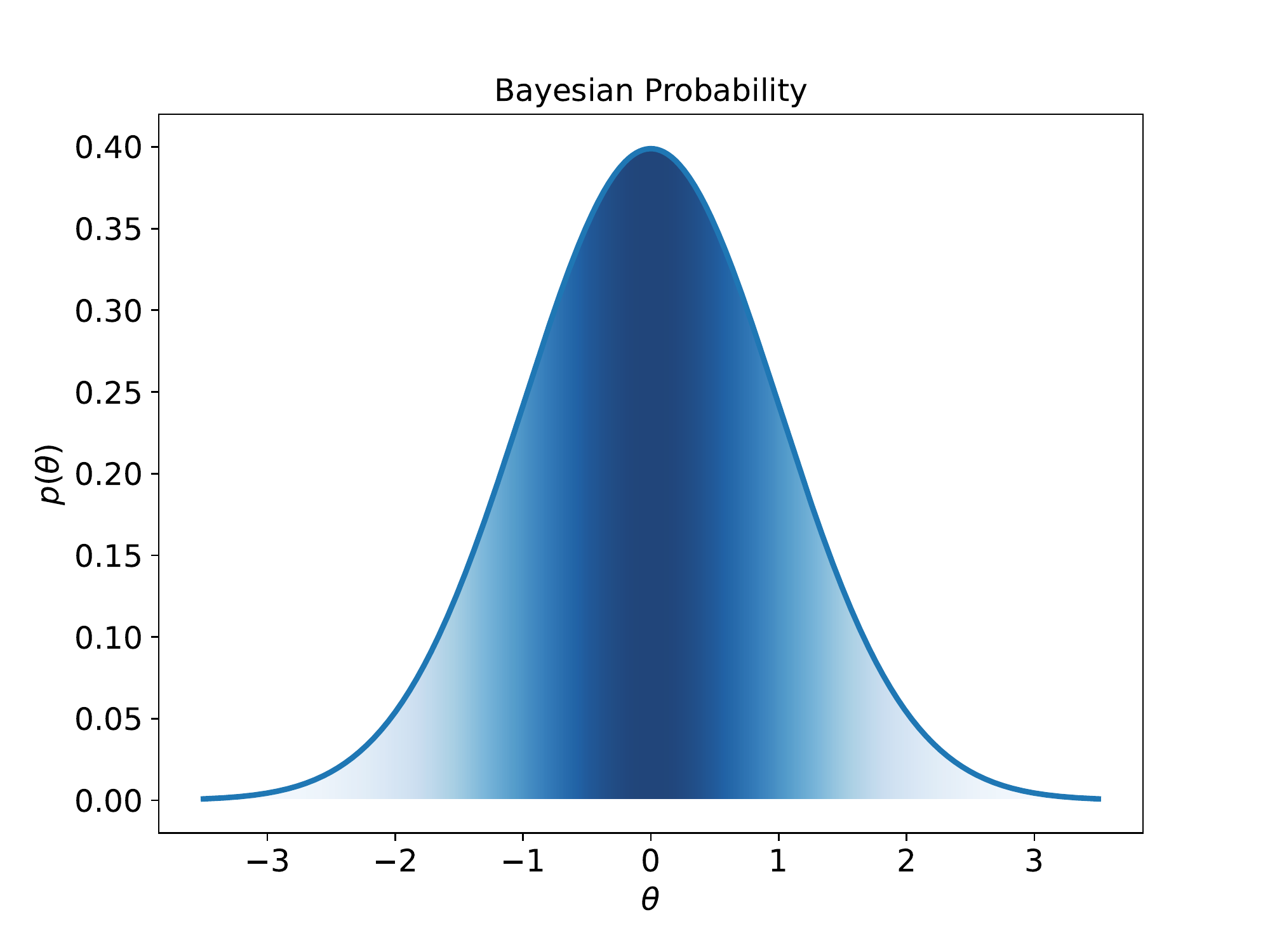}}
    \caption{In the Bayesian interpretation of probability, the degree of belief is distributed and the variable has a specific (unknown) fixed value.}
    \label{fig:probability_as_belief}
\end{figure}

Despite Laplace's indisputable empirical success, his theory was rejected by scholars soon after his death. Their problem with Laplace's probability theory was one of interpretation. For pioneers such as Bernoulli, Bayes, and Laplace, probability represented a degree--of--belief or plausibility of various hypotheses or statements based on the available evidence and prior knowledge. To $19$th century scholars though, this definition, or interpretation of probability, seemed too subjective and vague. For this reason, they redefined probability to mean the \textit{long--run relative frequency} with which an event occurs, given infinite trials. Since frequency can be measured experimentally, probability was then seen as an objective measure for dealing with \textit{randomness} and chance.
\begin{figure}[H]
    \centering
	\centerline{\includegraphics[scale=0.65]{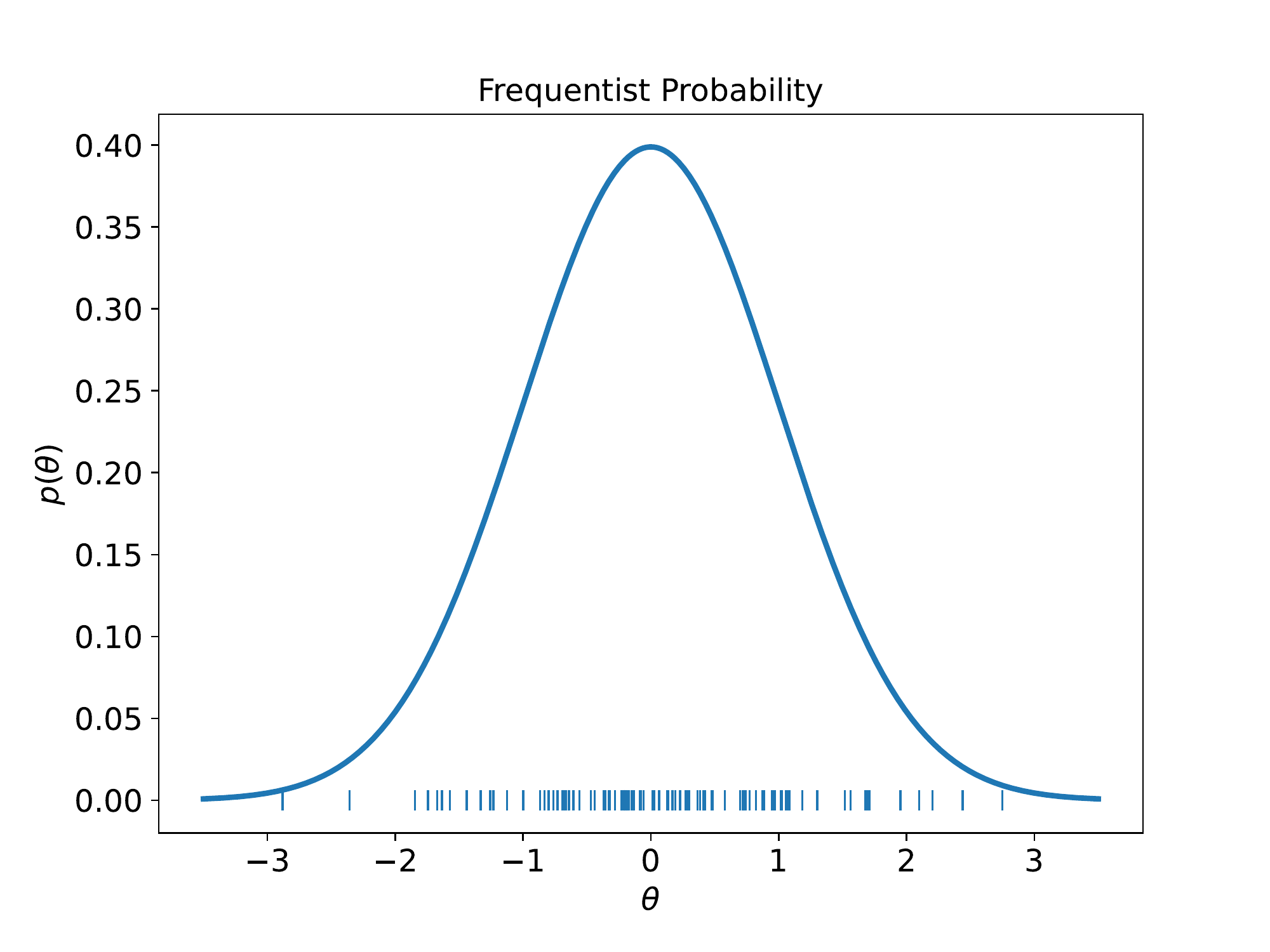}}
    \caption{In the frequentist interpretation of probability, the value of the variable itself is distributed along different experiments.}
    \label{fig:probability_as_frequency}
\end{figure}

\begin{figure}[H]
    \centering
	\centerline{\includegraphics[scale=0.65]{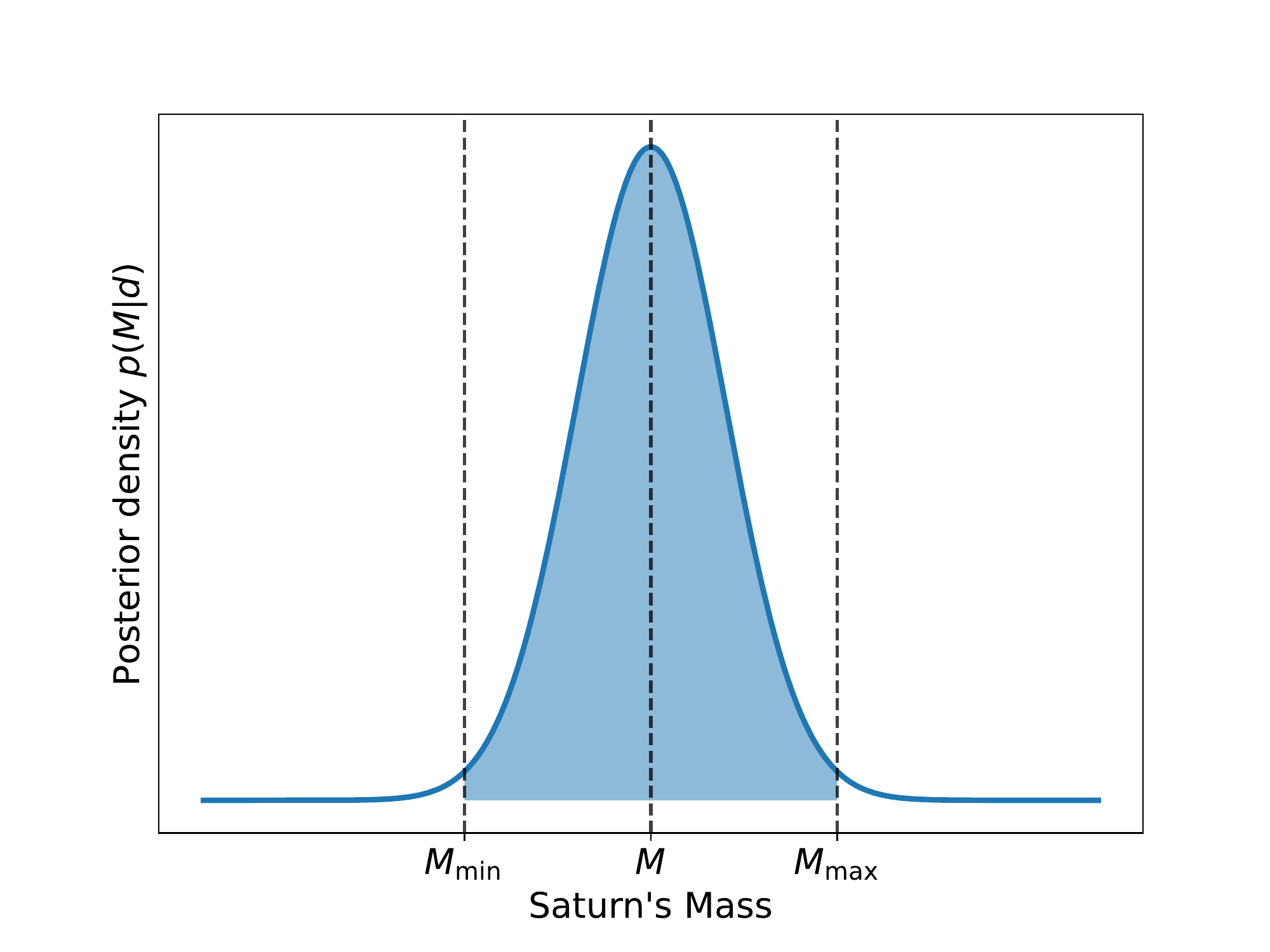}}
    \caption{Illustration of posterior probability density function of Saturn's mass. $M$ corresponds to the most probable value for the mass at the peak of the density. $M_{\min}$ and $M_{\max}$ denote the values of the mass that deviate by $1\%$ by the mean value $M$. The shaded area signifies the probability that the mass of Saturn is between the values of $M_{\min}$ and $M_{\max}$.}
    \label{fig:saturns_mass}
\end{figure}

Although the \textit{frequentist} interpretation of probability seems more objective, its range of applicability and validity is substantially more limited. For example, Laplace used \textit{Bayes' theorem} and his probability theory to estimate the mass of Saturn. He computed the \textit{posterior probability density function (pdf)} $p(M\vert d)$, that is, the probability that Saturn has a mass $M$ given the available data $d$ and model assumptions (e.g. validity of celestial mechanics). An illustration of this posterior pdf is shown in Figure \ref{fig:saturns_mass}, in which the value $M$ as the peak of the density corresponds to the most probable value for the mass of Saturn which also coincides with the mean value. $M_{\min}$ and $M_{\max}$ denote the values of the mass that deviate by $1\%$ from the mean value $M$, and the shaded area between them is the probability that the mass of Saturn is between the values of $M_{\min}$ and $M_{\max}$. Apart from the most probable value $M$, Laplace estimated that the probability (given by the area) that the real mass of Saturn is between these limits is $11327/11328=0.9999117$. In particular, he wrote ``applying to them my formulae of probability I find that it is a bet of 11,000 against one that the error of this result is not 1/100 of its value''. Today, almost two centuries after this statement was made, Laplace would have won this bet as the current best estimate of Saturn's mass differs only by $0.5\%$ from his.

However, according to the frequentist interpretation of probability one is not allowed to use probability theory to tackle this problem, as the mass of Saturn is a fixed constant and not a \textit{random variable} that follows a frequency distribution. If we were to interpret Laplace's results from a frequentist perspective we would have to imagine an infinitely large \textit{ensemble} of universes in which everything remains the same but the mass of a single planet. Although one has the liberty to make any kind of syllogisms in order to find a solution, having to seek a frequency interpretation for every problem can be cumbersome and at risk of detaching any notion of intuition from the physical problem.

Faced with the realisation that the frequentist interpretation of probability does not allow one to tackle most scientific questions, the new subject of \textit{statistics} was invented. For instance, in the problem of estimation of Saturn's mass since the mass is not a random variable, one has to create a function, called a statistic, that relates the data to the mass. Since the data are subject to random noise, so does the statistic. One is then free to apply the standard techniques to the statistic. However, the choice and construction of the statistic are often neither clear nor principled. There is no unifying principle relating the various techniques and practices used in order to choose which statistic is more appropriate for a given task. Historically, this lack of a common framework resulted in the creation of a large number of alternative schools of thought of frequentist statistics. Most notably, great statisticians such as \textit{Neyman}, \textit{Pearson} and \textit{Fisher} were responsible for promoting different approaches.

Early in the $20$th century something changed though, Sir \textit{Harold Jeffreys} rediscovered Laplace's Bayesian probability theory, and in $1930s$ he explained and presented it in greater detail and more clearly than Laplace ever did~\parencite{jeffreys1998theory}. Although apparently not enough to convince the most militant proponents of orthodox frequentist schools of the merits of probability theory, Jeffreys' work was the triggering event that acted as a catalyst for a change that lasted until the end of the $20$th and beginning of the $21$th century.

% Cox
In $1946$, \textit{Richard Cox} attempted to end the debate by approaching the problem of \textit{plausible inference} from a different perspective, that of its logical consistency~\parencite{cox1946probability}. Starting by the assumption that we can order different statements based on their plausibility, by assigning a real number to each statement representing how plausible it is, proved that for a calculus of plausible inference to be consistent (i.e. in the sense that if two different methods are permitted they should give the same results), it has to obey the rules of probability theory as defined by Laplace and Jeffreys. The work of Cox is of paramount importance as he effectively showed that any system of plausible inference that is logically consistent has to reduce to Bayesian inference.

% Jaynes

% 90s computer revolution and epilogue
By the last decade of the $20$th century, progress in computer technology and algorithms for probabilistic computation reached and surpassed the level of maturity required for the widespread application of Bayesian inference in most fields of physical science. Therefore, it is no surprise that the principles and methods of Bayesian probability theory have now become an integral and indispensable part of modern science. In the end, Laplace was right: ``It is remarkable that a science which began with the consideration of games of chance should have become the most important object of human knowledge''.

%************************************************

\section{Bayes' theorem}

\subsection{Rules of probability}

Any statement in probability theory can be derived by starting from the Laplace--Jeffreys sum and product rule given below. From these two formulas, expressions such as the ``or'' rule, the marginalisation rule and Bayes' theorem follow easily.

\subsubsection{The ``sum'' rule}
The ``sum'' rule expresses the relation between the probabilities of two mutually exclusive statements $A$ and $\Bar{A}$,
\begin{equation}
    \label{eq:sum_rule}
    p(A\vert B) + p(\Bar{A}\vert B) = 1\,,
\end{equation}
where $p(A\vert B)$ represents the plausibility (probability) of $A$ being true given that $B$ is true and $\Bar{A}$ simply means the opposite of $A$ or that $A$ is false.

\subsubsection{The ``product'' rule}
The ``product'' rule provides a way to compute the joint probability
\begin{equation}
    \label{eq:product_rule}
    p(A,B\vert C) = p(A\vert B,C)p(B\vert C) = p(B\vert A,C)p(A\vert C)\,.
\end{equation}
of both $A$ and $B$ being true given that $C$ is true.

\subsubsection{The ``or'' rule}
For instance, the ``or'' rule that expresses the probability that either $A$ or $B$ is true, given that $C$ is true, can be written as
\begin{equation}
    \label{eq:or_rule}
    p(A\cup B\vert C) = p(A\vert C) + p(B\vert C) - p(A\cap B\vert C)\,,
\end{equation}
follows easily, where $p(A\cap B\vert C)$ is simply another notation for the joint probability $p(A,B\vert C)$ for both events $A$ and $B$ being true given that $C$ is also.

\subsubsection{The marginalisation rule}

Another useful probability rule is the \textit{marginalisation rule} for discrete probability distributions,
\begin{equation}
    \label{eq:marginalisation_discrete}
    p(A\vert C) = \sum_{i} p(A,B_{i}\vert C) = \sum_{i} p(A\vert B_{i},C)p(B_{i}\vert C)\,,
\end{equation}
and for continuous probability distributions,
\begin{equation}
    \label{eq:marginalisation_continuous}
    p(A\vert C) = \int p(A,B\vert C)dB = \int p(A\vert B,C)p(B\vert C)dB\,.
\end{equation}
Equation \ref{eq:marginalisation_discrete} is straightforward to prove starting from the sum rule of equation \ref{eq:sum_rule}, extended to multiple mutually--exclusive events
\begin{equation}
    \label{eq:sum_rule_extended}
    \sum_{i} p(A_{i}\vert B) = 1\,.
\end{equation}
Therefore, starting from equation \ref{eq:sum_rule_extended} we have
\begin{equation}
    \label{eq:marginalisation_proof_1}
    \begin{split}
        \sum_{i} p(A,B_{i}\vert C) &= \sum_{i} p(B_{i}\vert A,C)p(A\vert C) \\
        &= p(A\vert C) \sum_{i} p(B_{i}\vert A,C) = p(A\vert C)\,.
    \end{split}
\end{equation}

\subsubsection{Bayes' theorem}
Arguably the most useful equation that can be derived is the so--called \textit{Bayes' theorem}~\parencite{bayes}
\begin{equation}
    \label{eq:bayes_theorem}
    p(A \vert B, C) = \frac{p(B\vert A, C)p(A\vert C)}{p(B\vert C)}\,,
\end{equation}
that follows directly from equation \ref{eq:product_rule}.

\subsection{Updating degrees of belief}

Although Bayes' theorem is a simple identity that holds for any statements $A$, $B$, and $C$, it also has a special role in the context of plausible inference. In particular, if we set $A\leftarrow \theta$ the parameters of a physical model, $B\leftarrow d$ the experimental data, and $C\leftarrow \mathcal{M}$ the physical model that also includes all assumptions made in an analysis, we get
\begin{equation}
    \label{eq:bayes_rule}
    p(\theta \vert d, \mathcal{M}) = \frac{p(d|\theta, \mathcal{M})p(\theta\vert \mathcal{M})}{p(d\vert \mathcal{M})}\,.
\end{equation}

The importance of equation \ref{eq:bayes_rule} for scientific inference is apparent if we examine each one of the constituent components individually.

\subsubsection{Posterior probability distribution -- $p(\theta\vert d, \mathcal{M})$}

This is the probability distribution of the parameters $\theta$, given the data $d$ and the modelling assumptions $\mathcal{M}$. The posterior is often what we are aspiring to approximate in a \textit{parameter estimation} analysis.

\subsubsection{Prior probability distribution -- $p(\theta\vert \mathcal{M})$}

This probability distribution quantifies any knowledge about the possible values of the parameters $\theta$ prior to seeing the data $d$. We have a whole chapter dedicated to the choice of the prior distribution.

\subsubsection{Likelihood function and sampling distribution -- $p(d\vert \theta, \mathcal{M})$}

The likelihood function is a key component of Bayes' theorem that plays a very important role, that of being the conduit that explains how the transition from prior to posterior takes place. Before we understand the role and properties of the likelihood function we first need to look into the so--called \textit{sampling distribution}.

The \textit{sampling distribution} $p(d\vert \theta )$ expresses the probability distribution of the data $d$ given the values of the model parameters $\theta$. In this picture, the parameters $\theta$ are known and fixed and $p(d\vert \theta )$ is a distribution over the data. If instead, we know the data $d$ and fix them to a specific value of set of values, and we let $\theta$ vary as a free parameter of a set of free parameters, then $p(d\vert \theta )$ is called the \textit{likelihood function}.

The likelihood function is often symbolised as $\mathcal{L}(\theta) = p(d\vert \theta)$ to denote that it is a function of parameters $\theta$ and not a probability distribution over the data $d$. This is very important as statements such as ``the likelihood of the data'' are meaningless and completely miss the point of the likelihood. The likelihood function shows how well the different sampling distributions $p(d\vert \theta)$, parameterised by $\theta$, predict the observed data.

\begin{figure}[H]
    \centering
	\centerline{\includegraphics[scale=0.45]{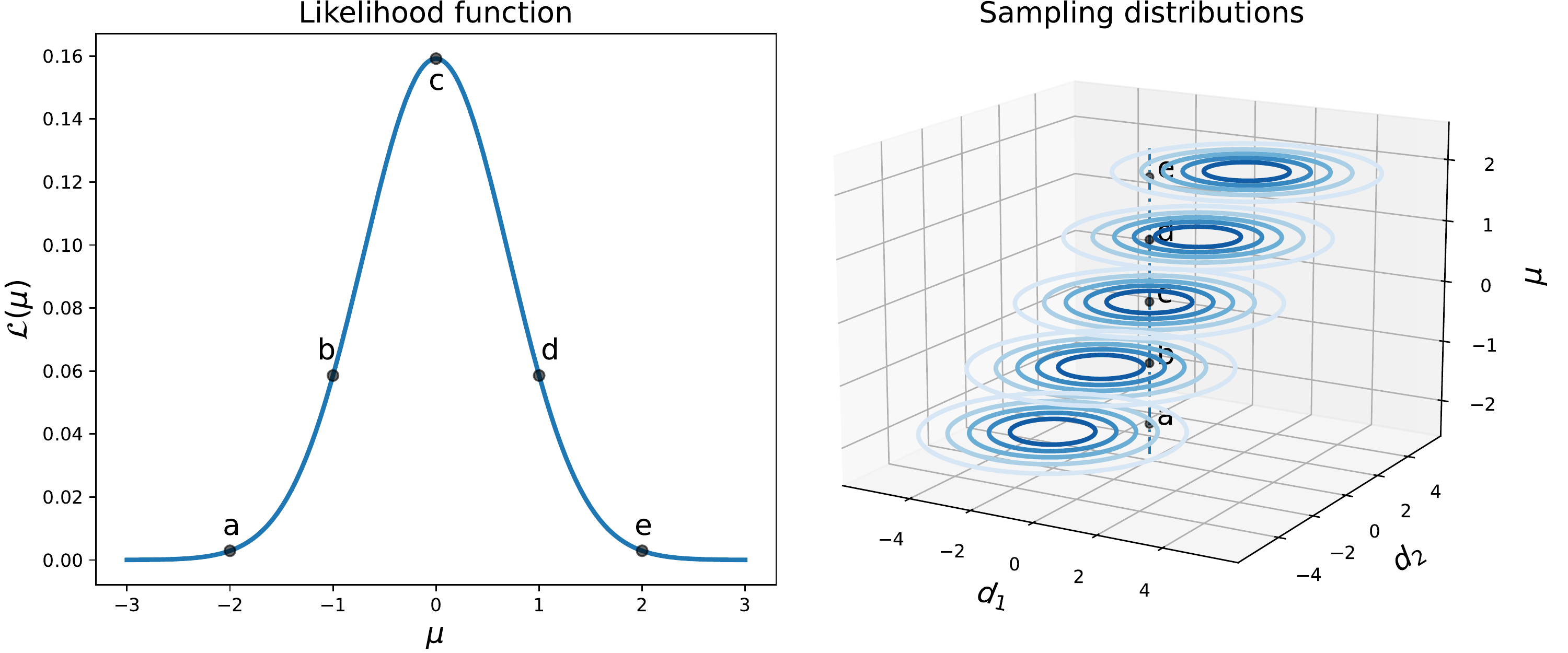}}
    \caption{Left: Likelihood function $\mathcal{L}(\mu)$ for the family of sampling distributions as shown on the right. The marked letters indicate the points in which the observed data intersect the sampling distributions. Right: The different groups of contours illustrate the sampling distribution $p(d_{1},d_{2}\vert\mu)$ for different values of $\mu$, ranging from $-2$ to $2$. The vertical line corresponds to the observed data $(d_{1},d_{2})=(0,0)$ and the marked letters indicate the points in which the observed data intersect the sampling distributions.}
    \label{fig:likelihood_vs_sampling}
\end{figure}

To make this more apparent, let us consider a simple example. Let us assume that we have a family of sampling distributions $p(d_{1},d_{2}\vert \mu)$ for the two dimensional data $d=(d_{1},d_{2})$, parameterised by a single parameter $\mu$. An example of such a family of sampling distributions for $\mu\in[-2,2]$ is shown in Figure \ref{fig:likelihood_vs_sampling} on the right. One can see that different values of $\mu$ correspond to different sampling distributions. In order to get a likelihood function $\mathcal{L}(\mu)$ from this family of sampling distributions, we need to specify some observed data for each member of the family. Without loss of generality, we choose the data to be simply $(d_{1}, d_{2}) = (0, 0)$. The data are indicated by a vertical line in the plot that intersects all members of the family of sampling distributions. The points of intersection, marked with letters \textit{a} to \textit{e} in the same plot, can either be in low or high probability regions of the respective sampling distributions. If we now monitor the value of the probability at the intersection points and plot this as a function of the parameter $\mu$ we get the likelihood function shown in the same figure on the left. 

In other words, although related, the notion of likelihood is really different from that of probability in the sense that it expresses the relative capacity of different sampling distributions, belonging however to the same family, to predict and explain the observed data. \textit{Sir Ronald Aylmer Fisher} wrote in 1922 about the difference between probability and likelihood, albeit in the frequentist tradition,
\begin{quote}
    \textit{If we need a word to characterise this relative property of different values of p, I suggest that we may speak without confusion of the likelihood of one value of p being thrice the likelihood of another, bearing always in mind that likelihood is not here used loosely as a synonym of probability, but simply to express the relative frequencies with which such values of the hypothetical quantity p would in fact yield the observed sample. [...] Likelihood also differs from probability in that it is a differential element, and is incapable of being integrated: it is assigned to a particular point of the range of variation, not to a particular element.}
\end{quote}

\subsubsection{Model evidence -- $p(d\vert\mathcal{M})$}

Finally, the \textit{model evidence} is a single real number that expresses the probability of observing the data $d$ given the model $\mathcal{M}$ and acts as a normalisation constant for the posterior
\begin{equation}
    \label{eq:model_evidence_integral}
    p(d\vert \mathcal{M}) = \int p(d\vert\theta,\mathcal{M})p(\theta\vert\mathcal{M})d\theta\,,
\end{equation}
such that $\int p(\theta\vert d, \mathcal{M})d\theta = 1$. The model evidence is often referred to as the \textit{marginal likelihood} due to the way it is represented as an integral. Its role in the task of \textit{model comparison} is great and it will be discussed in great length in the following chapters.

\begin{figure}[ht!]
    \centering
	\centerline{\includegraphics[scale=0.65]{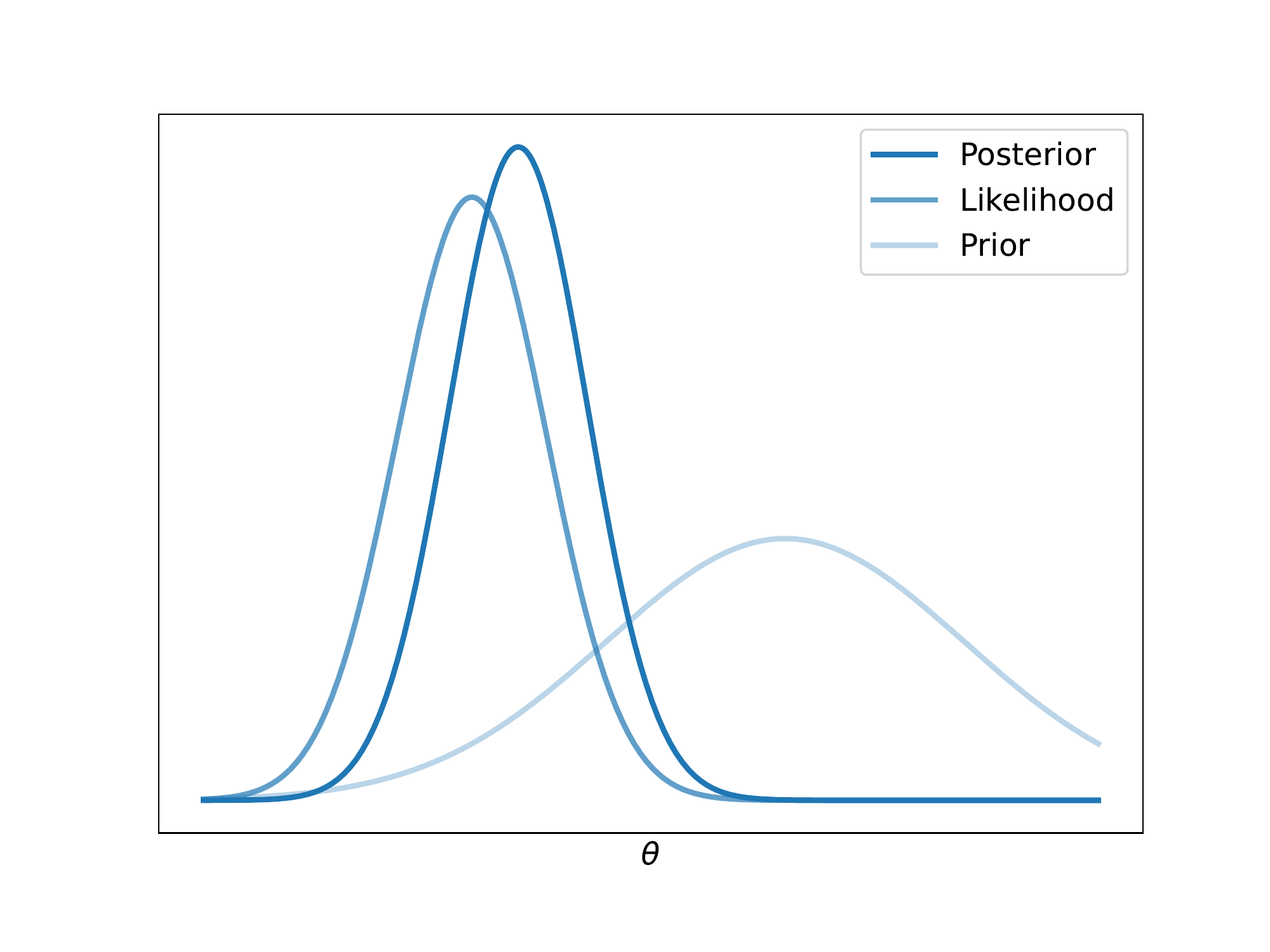}}
    \caption{Illustration of Bayes' theorem. The posterior probability is proportional to the product of the likelihood function and prior probability.}
    \label{fig:bayes_theorem}
\end{figure}

Schematically, we can summarise the above description of Bayes' theorem as
\begin{equation}
    \label{eq:bayes_rule_schematically}
    \text{posterior} = \frac{\text{likelihood}\times\text{prior}}{\text{evidence}}\,.
\end{equation}
In essence, Bayes' theorem in the form of equations \ref{eq:bayes_rule} and \ref{eq:bayes_rule_schematically} is a recipe for updating our degree of belief when new information, in the form of data, becomes available. The factor that upweights or downweights the prior $p(\theta\vert\mathcal{M})$ is the likelihood--to--evidence ratio $p(d\vert\theta,\mathcal{M})/p(d\vert\mathcal{M})$, also known as the \textit{predictive updating factor}. Keynes called this ratio the \textit{coefficient of influence} as it is this that determines how the prior is transformed into the posterior~\parencite{keynes1921treatise}. To better understand this, remember that the model evidence in the denominator is simply the expectation value of the likelihood over the prior probability distribution, so intuitively it expresses some sort of mean value of the likelihood. From this perspective, the predictive updating factor is simply the ratio of the likelihood to its mean value. This means that the prior will be upweighted for those values of $\theta$ that the likelihood is greater than its mean value and downweighted otherwise.

%************************************************

\section{Representing probability distributions}

Before we move on to any examples, it is important to explain how we represent probability distributions in practice. In general, there are two ways that we can do this, each one with different advantages and disadvantages.

\subsection{Function representation}

\subsubsection{Probability mass function}

When the parameter space $\Theta$ is discrete, then a probability distribution can be represented as a \textit{probability mass function (pmf)} $p(\theta)$ that assigns a probability value to each element of space, $p :\Theta\rightarrow [0,1]$. Any pmf has to obey the rule of total probability, that is
\begin{equation}
    \label{eq:total_probability_discrete}
    \sum_{\theta\in\Theta}p(\theta) = 1\,.
\end{equation}
The probability of any composite event $\Theta'\subset \Theta$ can be computed as
\begin{equation}
    \label{eq:composite_event_probability}
    p(\Theta') = \sum_{\theta\in\Theta'}p(\theta)\,.
\end{equation}
Finally, we can compute any expectation value
\begin{equation}
    \label{eq:expectation_value_discrete}
    \mathbb{E}_{p}[f] = \sum_{\theta\subset\Theta} f(\theta)p(\theta)
\end{equation}
for any function $f(\theta)$. Common examples of expectation values include the mean $\mu=\mathbb{E}_{p}[\theta]$ and the variance $\sigma^{2}=\mathbb{E}_{p}[(\theta-\mu)^{2}]$.

\subsubsection{Probability density function}

When the parameter space $\Theta\subseteq \mathbb{R}^{D}$ is continuous, then a probability distribution can be represented as a \textit{probability density function (pdf)} $p(\theta)$ that assigns a probability density value to each element of space, $p :\Theta\rightarrow \mathbb{R}$. Any pdf has to obey the rule of total probability, that is
\begin{equation}
    \label{eq:total_probability_continuous}
    \int_{\theta\in\Theta}p(\theta)d\theta = 1\,.
\end{equation}
Unlike a pmf, a pdf expresses probability density and thus it has to be integrated first to give probabilities. The probability of $\Theta'\subseteq\Theta$ is then
\begin{equation}
    \label{eq:probability_continuous}
    p(\Theta') = \int_{\Theta'\subseteq\Theta} p(\theta)d\theta \,.
\end{equation}
For instance, in 1--D we can compute the probability that $A\leq \theta \leq B$ as
\begin{equation}
    \label{eq:probability_of_region}
    p(A\leq \theta \leq B) = \int_{A}^{B}p(\theta)d\theta\,,
\end{equation}
as the area below the graph of $p(\theta)$ and between $A$ and $B$. Similarly, an expectation value can be computed as
\begin{equation}
    \label{eq:expectation_value_continuous}
    \mathbb{E}_{p}[f] = \int_{\Theta} f(\theta)p(\theta)d\theta\,.
\end{equation}

A crucial difference between probability mass functions and probability densities is that the latter do not transform quite as trivially under parameter transformations $g : \Theta \rightarrow \Phi$. The origin of this complication is that the differential volume $d\theta$ over which we integrate will generally change under such a transformation, and density functions have to change in the opposite way to compensate and ensure that probability is conserved. This change in volume is quantified by the absolute value of the determinant of the Jacobian matrix
\begin{equation}
    \label{eq:jacobian_matrix}
    J_{ij} = \frac{\partial g_{i}}{\partial \theta_{j}}\,,
\end{equation}
where $\phi = g(\theta)$ is the parameter transformation. Thus, the probability density $p(\theta)$ generally transforms as
\begin{equation}
    \label{eq:probability_transformation}
    p(\theta) = p(\phi) \vert \det J \vert\,.
\end{equation}

\subsection{Sample representation}

One of the inherent difficulties in the density function representation of probability distributions is that the computation of expectation values is often intractable as no closed--form solution exists for most applications. An alternative way of representing probability distributions is using a collection of points $S=\lbrace\theta_{1},\theta_{1},\dots,\theta_{n}\rbrace$ in the parameter space $\Theta$, called samples. The generation of samples for a given probability distribution will be the subject of discussion for most of this thesis. For now, it suffices to say that any probability distribution admits a sample representation
\begin{equation}
    \label{eq:sample_representation}
    \theta_{i} \sim p(\theta)\,,
\end{equation}
such that the empirical estimate
\begin{equation}
    \label{eq:empirical_expectation}
    \Hat{f}_{p} = \frac{1}{n}\sum_{i=1}^{n}f(\theta_{i})\,,
\end{equation}
asymptotically approaches the expectation value $\mathbb{E}_{p}[f]$ as $n\rightarrow \infty$.

\section{Asymptotic behaviour}

Let us now turn our attention to the question of the form of the posterior distribution in the limit of infinite data. Understanding the asymptotic behaviour of the posterior when the sample size is large is important for a number of reasons. First, there is practical utility as asymptotic results are often good first--order approximations. Second, as we will discuss in the following chapter, the asymptotic form of the posterior distribution can be utilised to automate the construction of prior distributions. Finally, the \textit{Bernstein--von Mises} theorem, which describes the asymptotic behaviour of the posterior in many cases, allows us to link Bayesian inference to frequentist results.

\subsection{Bernstein--von Mises theorem}

When the number $n$ of observations tends to infinity, the posterior distribution of a smooth finite--dimensional model approaches a normal distribution. In particular, if we denote $d^{(n)}=\lbrace d_{1}, \dots, d_{n}\rbrace$ the set of $n$ observations or data, then the posterior $p(\theta\vert d^{(n)})$ concentrates around the \textit{maximum likelihood estimate (MLE)}:
\begin{equation}
    \label{eq:maximum_likelihood_estimate}
    \Hat{\theta}_{n} = \underset{\theta}{\mathrm{arg\,max}} ~p(d^{(n)}\vert\theta)\,.
\end{equation}
Moreover, MLE is a consistent estimator which means that in the limit of infinite sample size (i.e. $n\to\infty$), $\Hat{\theta}_{n}$ converges to $\theta_{t}$, that is, the true value of the parameter vector. In other words, the asymptotic posterior is centred on the  true parameter value $\theta_{t}$. The precision matrix (i.e. inverse covariance matrix) is equal to $n\mathcal{I}(\Hat{\theta}_{n})$, where the \textit{Fisher information matrix} is defined as 
\begin{equation}
    \label{eq:fisher_information_matrix_definition}
    \begin{split}
        \mathcal{I}(\theta) &= \mathbb{E}_{d\sim p(d\vert \theta)}\left[- \frac{\partial^{2}\log p(d\vert \theta)}{\partial \theta_{i}\partial\theta_{j}}\right] \\
        &= -\int p(d\vert \theta)  \frac{\partial^{2}\log p(d\vert \theta)}{\partial \theta_{i}\partial\theta_{j}} d d \,.
    \end{split}
\end{equation}
In more mathematical terms, we can write down that
\begin{equation}
    \label{eq:posterior_converges_to_normal}
    p(\theta\vert d^{(n)}) \rightarrow \mathcal{N}(\theta\vert \Hat{\theta}_{n}, n^{-1}\mathcal{I}^{-1}(\Hat{\theta}_{n}))\,,
\end{equation}
as $n\to \infty$, where we use the notation $\mathcal{N}(\theta\vert \mu, \Sigma)$ to denote the Gaussian probability density function
\begin{equation}
    \label{eq:gaussian_probability_density_function}
    \mathcal{N}(\theta\vert \mu, \Sigma) = (2\pi )^{-D/2} \vert \Sigma \vert ^{-1/2} \exp \left\lbrace -\frac{1}{2}(\theta - \mu)^{T}\Sigma^{-1}(\theta - \mu)\right\rbrace \,,
\end{equation}
with mean $\mu$ and covariance matrix $\Sigma$, where $D$ is the number of components in the $\theta$ vector (i.e. dimensionality of parameter space). Although this result dates back to \textcite{laplace1810memoire}, today it is known as the \textit{Bernstein-von Mises} theorem~\parencite{van2000asymptotic}. 

One consequence of the above theorem, combined with the fact that the MLE asymptotically follows a normal distribution, allows us to interpret Bayesian \textit{credible intervals} as frequentist \textit{confidence intervals} in the limit of infinite data. 

\subsection{Heuristic argument}

We will now offer an intuitive heuristic argument, rather than a rigorous proof, of the \textit{Bernstein-von Mises} theorem. Let us begin by rewriting Bayes' theorem as
\begin{equation}
    \label{eq:rewrite_bayes_theorem}
    p(\theta\vert d^{(n)}) = \frac{\exp\left\lbrace \log p(\theta) + \log p(d^{(n)}\vert\theta)\right\rbrace}{P(d^{(n)})}\,,
\end{equation}
where $\log p(\theta)$ is the log--prior and
\begin{equation}
    \label{eq:log_likelihood_iid}
    \log p(d^{(n)}\vert\theta) = \sum_{i=1}^{n}\log p(d_{i}\vert \theta)\,,
\end{equation}
is the log--likelihood function of \textit{identically independently distributed (iid)} data, which readily follows from the fact that their sampling distributions are conditionally independent, meaning that
\begin{equation}
    \label{eq:likelihood_iid}
    p(d^{(n)}\vert\theta) = \prod_{i=1}^{n} p(d_{i}\vert \theta)\,.
\end{equation}

The next step is to Taylor--expand both the log--prior and the log--likelihood around their respective maxima. Starting with the log--prior, we can write
\begin{equation}
    \label{eq:log_prior_expansion}
    \log p(\theta) = \log p(\Hat{\theta}_{0}) - \frac{1}{2}(\theta - \Hat{\theta}_{0})^{T}\Lambda_{0}(\Hat{\theta}_{0})(\theta - \Hat{\theta}_{0}) + R_{0}\,,
\end{equation}
where
\begin{equation}
    \label{eq:prior_precision_matrix}
    \Lambda_{0}(\Hat{\theta}_{0}) = \left( - \frac{\partial^{2}\log p(\theta)}{\partial \theta_{i}\partial \theta_{j}}\right) \bigg\vert_{\theta = \Hat{\theta}_{0}}
\end{equation}
and $R_{0}$ denotes any higher--order terms. Notice that since the expansion takes place around the prior maximum $\Hat{\theta}_{0}$, there is no first--order term (i.e. the first derivative is equal to zero). Similarly, we can expand the log--likelihood around the MLE as follows:
\begin{equation}
    \label{eq:log_likelihood_expansion}
    \log p(d^{(n)}\vert\theta) = \log p(d^{(n)}\vert\Hat{\theta}_{n}) - \frac{1}{2}(\theta - \Hat{\theta}_{n})^{T}\Lambda_{n}(\Hat{\theta}_{n})(\theta - \Hat{\theta}_{n}) + R_{n}\,,
\end{equation}
where
\begin{equation}
    \label{eq:likelihood_precision_matrix}
    \Lambda_{n}(\Hat{\theta}_{n}) = \left( - \frac{\partial^{2}\log p(d^{(n)}\vert\theta)}{\partial \theta_{i}\partial \theta_{j}}\right) \bigg\vert_{\theta = \Hat{\theta}_{n}} = \left( - \sum_{\ell=1}^{n}\frac{\partial^{2}\log p(d_{\ell}\vert\theta)}{\partial \theta_{i}\partial \theta_{j}}\right) \bigg\vert_{\theta = \Hat{\theta}_{n}}
\end{equation}
and $R_n$ denotes any terms beyond the second order.

Assuming that the prior and likelihood are sufficiently smooth such that $R_{0}$ and $R_{n}$ can be safely ignored we can write equation \ref{eq:rewrite_bayes_theorem} as
\begin{equation}
    \label{eq:rewrite_bayes_theorem2}
    \begin{split}
        p(\theta\vert d^{(n)}) 
        &\propto \exp \bigg\lbrace  - \frac{1}{2}(\theta - \Hat{\theta}_{0})^{T}\Lambda_{0}(\Hat{\theta}_{0})(\theta - \Hat{\theta}_{0}) \\ 
        &\quad\quad\quad\;\; - \frac{1}{2}(\theta - \Hat{\theta}_{n})^{T}\Lambda_{n}(\Hat{\theta}_{n})(\theta - \Hat{\theta}_{n})\bigg\rbrace \\
        &\propto \exp \left\lbrace  - \frac{1}{2}(\theta - \Tilde{\theta}_{n})^{T}\Tilde{\Lambda}_{n}(\Tilde{\theta}_{n})(\theta - \Tilde{\theta}_{n}) \right \rbrace\,,
    \end{split}
\end{equation}
where $\Tilde{\Lambda}_{n} = \Lambda_{n} + \Lambda_{0}$ and $\Tilde{\theta}_{n}= \Tilde{\Lambda}_{n}^{-1}(\Lambda_{n}\Hat{\theta}_{n}+\Lambda_{0}\Hat{\theta}_{0})$. Comparing the above expression to equation \ref{eq:gaussian_probability_density_function}, we find that the posterior has a Gaussian probability density function
\begin{equation}
    \label{eq:assymptotic_posterior_is_normal}
    p(\theta\vert d^{(n)}) = \mathcal{N}(\theta\vert \Tilde{\theta}_{n}, \Tilde{\Lambda}_{n}^{-1})\,. 
\end{equation}

In the limit that $n\to\infty$, the sum in equation \ref{eq:likelihood_precision_matrix} completely dominates the calculation leading to $\Tilde{\Lambda}_{n} \to \Lambda_{n}$ and $\Tilde{\theta}_{n}\to \Hat{\theta}_{n}$. This means that asymptotically
\begin{equation}
    \label{eq:assymptotic_posterior_is_normal2}
    p(\theta\vert d^{(n)}) \to \mathcal{N}(\theta\vert \Hat{\theta}_{n}, \Lambda_{n}^{-1})\,. 
\end{equation}
Furthermore, according to the \textit{law of large numbers}, which states that ``the average of a large number of trials approaches the expectation value``,  $\Lambda_{n}$ as given by the sum in equation \ref{eq:likelihood_precision_matrix} is asymptotically equal to $n \mathcal{I}(\Hat{\theta}_{n})$. Therefore, we can write down that
\begin{equation}
    \label{eq:assymptotic_posterior_is_normal3}
    p(\theta\vert d^{(n)}) \to \mathcal{N}(\theta\vert \Hat{\theta}_{n}, n^{-1} \mathcal{I}^{-1}(\Hat{\theta}_{n}))\,,
\end{equation}
which concludes our heuristic derivation.

\section{Estimating parameters}

\subsection{Coin--tossing experiment}

Let us now consider a simple example of Bayesian parameter estimation. Suppose that we have a coin and we want to determine whether the coin is fair or not. A simple way to quantify the fairness of a coin is to introduce a \textit{bias} parameter $F$ such that $F=1/2$ means that the coin is fair, whereas any other value in the range $0\leq F \leq 1$ denotes that the coin is biased. $F=0$ corresponds to a coin which always lands on \textit{tails} and $F=1$ to a one that always lands on \textit{heads}. We can then divide the continuous range of $F$ into a discrete number of propositions (e.g. $ 0\leq F \leq 0.01$, $0.01\leq F \leq 0.02$, etc.). Our state of knowledge about the fairness of the coin is summarised by our degree of belief, quantified as a probability, of each one of those intervals (e.g. $p(0\leq F \leq 0.01)$, $p(0.01\leq F \leq 0.02)$, etc.).

In order to collect some data we just have to toss the coin a few times and monitor the number of times $H$ the coin lands on heads as well as the total number of trials $N$. The number of times that the coin lands on tails is simply $N-H$. Furthermore, to better understand the iterative nature of Bayes' theorem for updating our degree of belief, we will keep not only the final outcome of the experiment (i.e. the total number $H$ that the coin landed on heads in $N$ trials) but also all the intermediate values.

\begin{figure}[ht!]
    \centering
	\centerline{\includegraphics[scale=0.65]{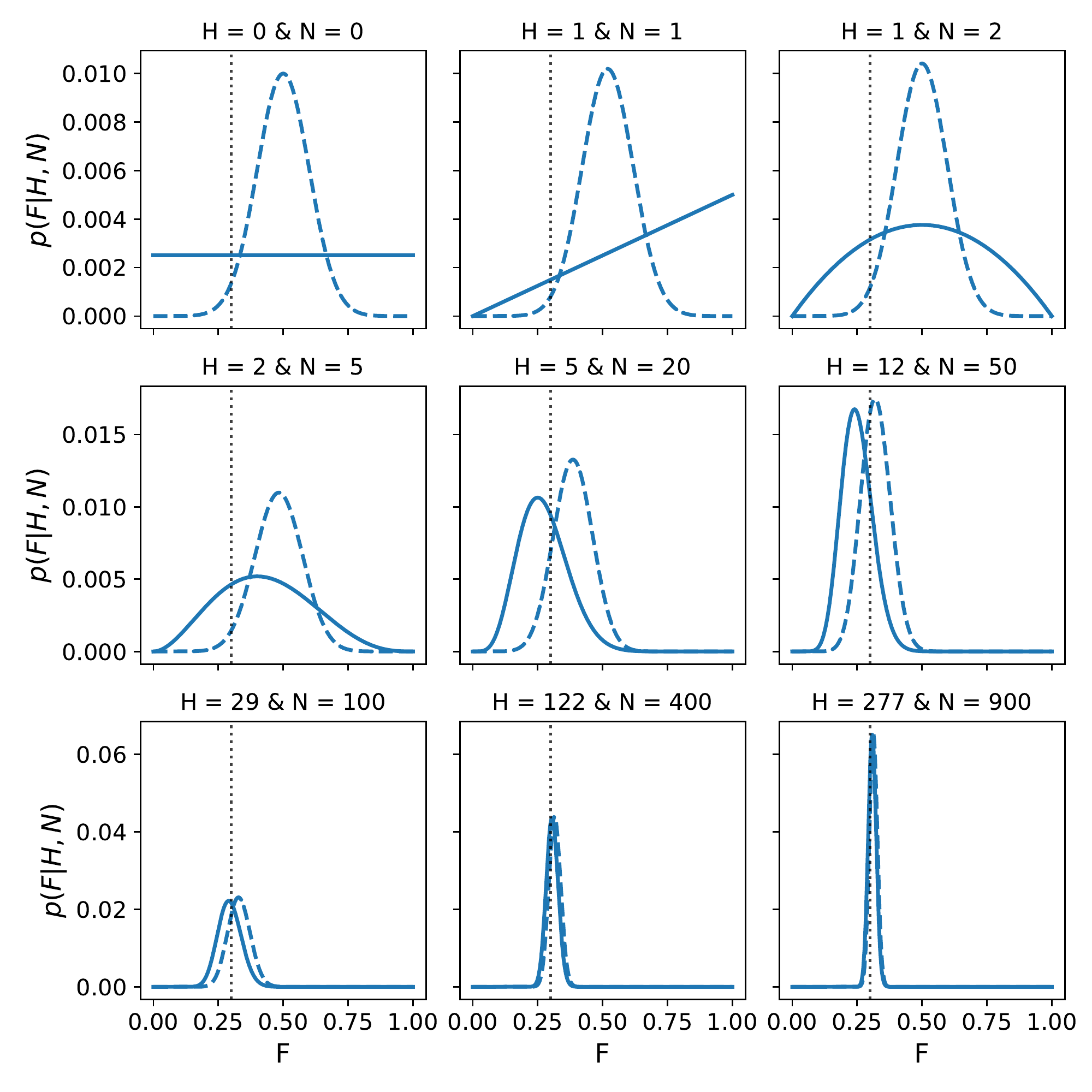}}
    \caption{The evolution of the posterior probability distribution of a coin--tossing experiment for increasing number of trials. $F$ is the bias parameter that we want to estimate. $H$ is the number of times the coin landed on \textit{heads} and $N$ is the total number of trials. The continuous line corresponds to the case of using a uniform prior whereas the dashed line to a normal prior. The dotted line shows the true (unknown) value of $F$.}
    \label{fig:coins}
\end{figure}

Since our aim is to estimate the posterior distribution $p(F\vert H,N)$, that is, the probability distribution of $F$ given the observed data in terms of the number of heads $H$ and the number of trials $N$, we need to define all the components that enter Bayes's theorem. Starting with the prior probability distribution $p(F)$ we will use two choices in order to demonstrate their effect on the posterior. The first choice of prior is to be agnostic, before seeing the data, about the fairness of the coin and thus assume that intervals of the same size in the range $0\leq F \leq 1$ are equally probable. This is quantified by the uniform probability density function
\begin{equation}
    \label{eq:uniform_prior}
    p(F) = 
    \begin{cases}
    1, & \text{if } 0 \leq F \leq 1 \\
    0, & \text{otherwise}\,.
    \end{cases}
\end{equation}
The other prior that we will test is more informative than the first and assumes that it is more probable that the coin is fair, or at least close to it. To this end, we will use a normal prior with a Gaussian probability density
\begin{equation}
    \label{eq:gaussian_prior}
    p(F\vert \mu, \sigma) = \frac{1}{\sqrt{2\pi}\sigma}\exp\left(-\frac{(F-\mu)^{2}}{2\sigma^{2}}\right)\,,
\end{equation}
centred around the mean value $\mu=1/2$ with standard deviation $\sigma=0.1$. This kind of prior assigns most prior probability to values of $F$ close to that of $1/2$ that correspond to a fair coin. Both priors can be seen in the top--left panel of Figure \ref{fig:coins} where the uniform prior corresponds to the continuous line and the normal prior to the dashed line.

To get the \textit{likelihood function}, we start by choosing the sampling distribution $p(H, N\vert F)$, that is, the probability distribution of the data $H$ and $N$ given the value of $F$. For this task, we choose the \textit{binomial} probability distribution with probability density given by
\begin{equation}
    \label{eq:binomial_likelihood_coins}
    p(H, N \vert F) = \binom{N}{H} F^{H} (1-F)^{N-H}\,.
\end{equation}
The above formula can be understood as follows: $R$ heads occur with probability $F^{H}$ and $N-H$ tails with probability $(1-F)^{N-H}$. The combinatorial factor that $H$ heads can occur anywhere among the $N$ trials, and there are $\binom{F}{H}$ of distributing $H$ heads between $N$ trials. If we fix $H$ and $N$ to their observed values then $\mathcal{L}(F)=p(H, N \vert F)$ is simply the likelihood function $F$.

According to Bayes' theorem then, the posterior distribution can be written as
\begin{equation}
    \label{eq:posterior_coins}
    p(F\vert H, N) = \frac{p(H, N\vert F)p(F)}{p(H,N)}\,,
\end{equation}
where the model evidence is simply the normalisation factor
\begin{equation}
    \label{eq:model_evidence_coins}
    p(H,N) = \int_{0}^{1}p(H, N\vert F)p(F)d F\,.
\end{equation}
In the case of the uniform prior of equation \ref{eq:uniform_prior} the above integral can be computed analytically. This is not however true for the case of the normal prior of equation \ref{eq:gaussian_prior}, for which numerical integration is necessary.

Figure \ref{fig:coins} shows the evolution of the posterior distribution of equation \ref{eq:posterior_coins}, starting from the prior distribution in the top--left panel, as we gradually increase the number of data points that are included in the analysis. The posteriors with both prior choices are illustrated, also highlighting the effect of the prior choice on the posterior. As we can see from the same figure, while the number of trials $N$ remains small (e.g. $N\leq 5$), the posterior corresponding to the informative normal prior remains unaffected. On the other hand, the posterior corresponding to the more agnostic uniform prior responds rapidly to the new data and concentrates close to the lower half of the $F$ range. The reason for this difference is the fact that the few initial data points do not carry sufficient information compared to the normal prior, but they do so compared to the less informative uniform prior. For a higher number of trials $N$ the behaviour
is changing though. Both posteriors rapidly concentrate around the same value of $F$. This indicates that the prior, while important in the low--data regime, does not affect the posterior when the amount of data is substantial. This behaviour is a direct consequence of the \textit{Bernstein--von Mises theorem}~\parencite{van2000asymptotic} which, under quite general conditions, states that ``for sufficiently nice prior probabilities, in the limit of infinite data the posterior converges to a Gaussian distribution independently of the initial prior''. This also explains the symmetric form of the posterior in Figure \ref{fig:coins} when the number of trials is large, as well as its reduced width.

\subsection{Fitting a model to data}

A general problem that scientists are often called to solve is that of fitting a mathematical model $m(t\vert\theta)$ to the data $d$, where the pairs $(t,d)$ constitute the measured data points. A simple example of a model is the straight line $m(t\vert \alpha, \beta) = \alpha + \beta t$. $t=\lbrace t_{i}\rbrace$ could be a sequence of time instances, positions or any other physical quantity in which the measurements $d=\lbrace d_{i} \rbrace$ are collected. The task of \textit{model fitting} lies within the context of Bayesian parameter estimation as the main goal is to approximate the posterior probability distribution $p(\theta\vert d,\mathcal{M})$, that is, the probability distribution of the parameters $\theta$, given the data $d$ and the model $\mathcal{M}$. The latter consists of the actual mathematical model $m(\theta)$ plus all the assumptions made during the analysis.
\begin{figure}[ht!]
    \centering
	\centerline{\includegraphics[scale=0.65]{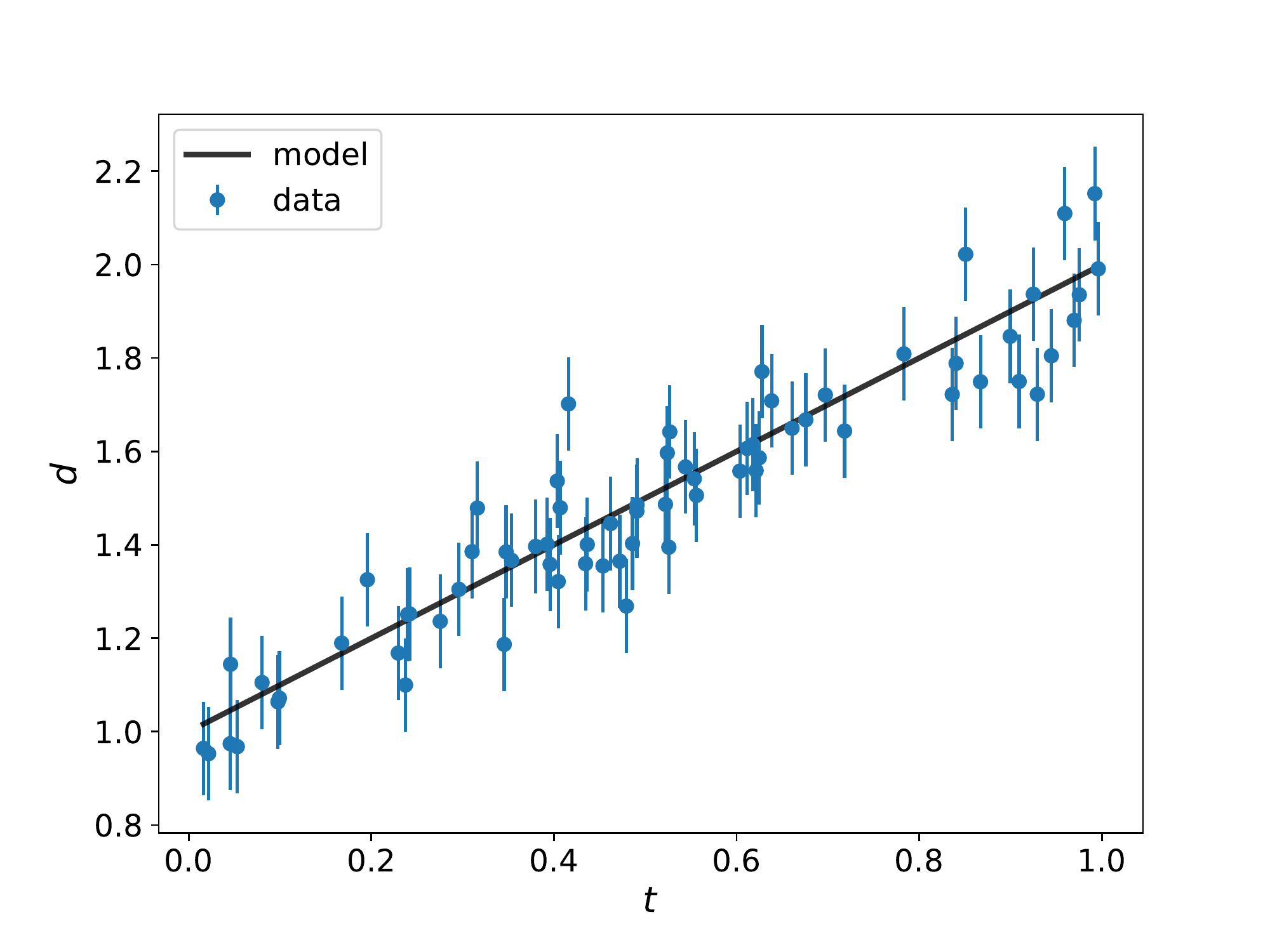}}
    \caption{Example of data $d$ and the straight line model $m(t\vert,\alpha, \beta)=\alpha + \beta t$ that was used to generate them assuming true values $\theta^{*}=(\alpha^{*},\beta^{*})=(1,1)$ and $\epsilon=0.1$.}
    \label{fig:fitting_data}
\end{figure}
Usually, the data $d$ are assumed to be a noise--corrupted realisation of the model, meaning
\begin{equation}
    \label{eq:forward_model}
    d = m(t\vert\theta^{*}) + \epsilon\,,
\end{equation}
where $\theta^{*}$ are the true values of the parameters that we, as scientists, are aspiring to approximate, and $\epsilon$ is the noise or uncertainty added to the model realisation $m(t\vert\theta^{*})$ in order to generate the data $d$. In the absence of any noise (i.e. $\epsilon = 0$), the data are no longer corrupted and the value of $\theta^{*}$ can be estimated with certainty. As we have discussed already, this is an idealised scenario and in real life, our incomplete knowledge about the physical mechanism which produced the data introduces a non--zero noise contribution $\epsilon$.

As the assumed physical model $m(\theta)$ is often \textit{deterministic}, it follows that the sampling probability of the noise $\epsilon$ is identical to that of the data $d$, or in other words that
\begin{equation}
    \label{eq:sampling_distribution_data_noise}
    p(d\vert \theta, \mathcal{M}) = p(\epsilon\vert \theta, \mathcal{M})\,.
\end{equation}
Furthermore, as the underlying physical mechanisms that give rise to the noise, often consist of a plethora of contributing factors one usually employs the \textit{central limit theorem (CLT)} in order to justify the use of a zero--mean normal sampling distribution
\begin{equation}
    \label{eq:sampling_distribution_noise}
    \epsilon \sim \mathcal{N}(0, \Sigma)\,,
\end{equation}
where $\Sigma$ is the $D\times D$ positive--definite symmetric covariance matrix of the noise. The likelihood function is thus assumed to be Gaussian
\begin{equation}
    \label{eq:gaussian_likelihood}
    p(d\vert \theta, \mathcal{M}) = \det(2\pi\Sigma)^{-\frac{1}{2}}\exp\left\lbrace-\frac{1}{2}\left[ d - m(t\vert\theta)\right]^{T}\Sigma^{-1}\left[ d - m(t\vert\theta)\right]\right\rbrace\,.
\end{equation}
Contrary to popular opinion, and as we will discover in the next chapter where the principle of maximum entropy is discussed, a Gaussian function, or equivalently a normal sampling distribution, is quite often a very good choice. There are of course applications in which other sampling distributions will be more appropriate (e.g. \textit{Poisson} for number counts). However, when only the (co--)variance of the noise is known, the normal distribution is the most conservative choice one can make~\parencite{gregory2005bayesian,  hogg2010data,sivia2006data}. Of course, the accurate estimation of the covariance is on its own a difficult problem. Furthermore, if the covariance matrix $\Sigma$ is estimated using simulated data $d_{i}\sim p(d)$, for instance
\begin{equation}
    \label{eq:covariance_estimate}
    \Hat{\Sigma} = \frac{1}{n-1}\sum_{i=1}^{n}(d_{i}-\Bar{d})(d_{i}-\Bar{d})^{T}\,,
\end{equation}
where $\Bar{d}=n^{-1}\sum_{i=1}^{n}d_{i}$, the Gaussian likelihood function must to be modified to account for the uncertainty of the covariance estimate~\parencite{sellentin2015parameter}.

\begin{figure}[ht!]
    \centering
	\centerline{\includegraphics[scale=0.85]{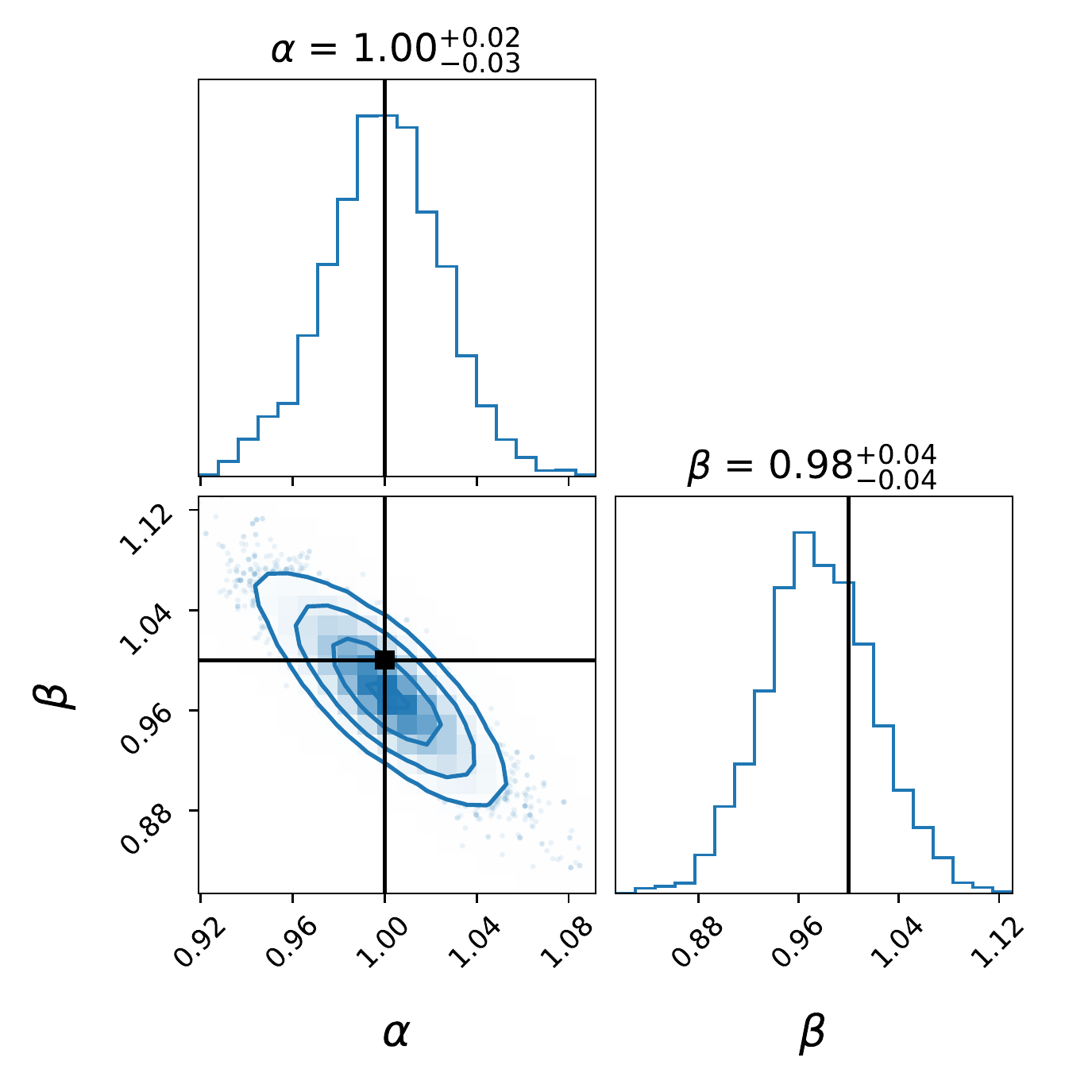}}
    \caption{1--D and 2--D marginal posterior contours of fitting the straight line model $m(t\vert,\alpha, \beta)=\alpha + \beta t$ to the data of Figure \ref{fig:fitting_data} assuming flat/uniform priors $\alpha, \beta \sim \mathcal{U}(-5,5)$. The black lines show the true values of the parameters $\theta^{*}=(\alpha^{*},\beta^{*})=(1,1)$ which were used to generate the data.}
    \label{fig:fitting_posterior}
\end{figure}

Given the model, the data, and the likelihood, the final requirement in order to conduct Bayesian inference is the prior distribution $p(\theta\vert\mathcal{M})$. This will of course depend on the specific application and we will discuss the choice of prior in more detail in the next chapter. The task of approximating the posterior $p(\theta\vert d, \mathcal{M})$, that we have discussed so far, is in general analytically intractable for all but the simplest models and prior choices. In the rest of this thesis, we will present various methods and computational tools that will allow us to tackle problems such as this one. As an illustration, we offer Figure \ref{fig:fitting_posterior} which shows the 1--D and 2--D marginal posteriors of fitting the straight line model $m(t\vert,\alpha, \beta)=\alpha + \beta t$ to the data of Figure \ref{fig:fitting_data} assuming flat/uniform priors $\alpha, \beta \sim \mathcal{U}(-5,5)$. Although this is a relatively simple model, the same principles and techniques that were used to estimate its posterior also extend to more complicated applications.

%************************************************
% !TEX TS-program = pdflatex
% !TEX root = ../ArsClassica.tex

%************************************************
\chapter{Quantifying prior knowledge}
\label{chp:priors}
%************************************************

\begin{flushright}
\itshape
Only entropy comes easy. \\
\medskip
--- Anton Chekhov
\end{flushright}

The discussion about Bayes' theorem so far explains how one can update one's prior knowledge in the light of new data. The question that naturally arises is how does one quantify their prior knowledge in the form of a probability distribution in the first place? In this chapter, we will attempt to provide a series of methods and practices that aim to do exactly that.

Ever since its initial development, many have criticised Bayesian inference for its dependence on prior knowledge~\parencite{efron1986isn, gelman2008objections}. Arguments against it mostly focus on the alleged subjectivity of its derived results. We maintain however that those claims are unfounded as all statistical analyses, Bayesian or not, employ prior information in some form or another. The difference with Bayesian inference is that this is explicitly done and taken into account. Indeed, anytime one has to perform a statistical analysis they have to assume a specific model (or a collection of them), often a specific set of parameters, a procedure of collecting data and a set of assumptions about the process that generated the data. In terms of the objectivity of its results, Bayesian inference is objective in the sense that any researcher possessing the same model assumptions, data, and prior knowledge will naturally reach exactly the same conclusions. Finally, the use of prior information can be understood as a great strength of Bayesian inference as it allows for the numerous scientific analyses which employ posterior distributions from old experiments as the priors for new ones, thus updating our knowledge of the world in a sequential and accumulative manner without discarding previous results. In this chapter, we will present both methods which employ this philosophy and those which attempt to provide a systematic procedure for generating prior distributions.

\section{Conjugate priors}

A prior distribution is said to be conjugate to the likelihood function if it belongs to the same family of distributions as the posterior~\parencite{gelman2013bayesian}. For instance, if the prior is a Gamma distribution and the likelihood is described by a Poisson probability mass function then the posterior is also Gamma.

From a mathematical point of view, conjugate priors are the most convenient choice as they allow us to compute the posterior analytically without the requirement of any computational method. From a scientific point of view however, conjugate priors are not well justified as they exist solely for the merit of algebraic convenience and they are not designed in order to encode the actual prior information. They are however a useful pedagogical and illustrative example of a method for choosing prior distributions.

\subsection{Binomial likelihood function with Beta prior}

Let us now consider the case of a binomial distribution
\begin{equation}
    \label{eq:binomial_distribution}
    p(s,n\vert \theta ) = \binom{n}{s}\theta^{s}(1-\theta)^{n-s}\,,
\end{equation}
which is the sampling distribution for the number of successes $s$ in $n$ \textit{Bernoulli trials} with probability of success equal to $\theta$. Fixing the number of successes $s$ and trials $n$ and letting $\theta$ vary as a free parameter, the above probability mass function will be the likelihood function for this example. It is also more convenient to express it in terms of the number of failures $f=n-s$ instead of the number of trials $n$ as
\begin{equation}
    \label{eq:binomial_distribution_in_terms_of_sf}
    p(s,f\vert \theta ) = \binom{s+f}{s}\theta^{s}(1-\theta)^{f}\,.
\end{equation}
The prior distribution that is conjugate to this likelihood function turns out to be the \textit{Beta} distribution
\begin{equation}
    \label{eq:beta_distribution}
    p(\theta ) = \frac{\theta^{\alpha-1}(1-\theta)^{\beta - 1}}{B(\alpha, \beta)}\,
\end{equation}
where $B(\alpha, \beta)$ is the \textit{Beta} function
\begin{equation}
    \label{eq:beta_function}
    B(\alpha, \beta) = \int_{0}^{1}\theta^{\alpha-1}(1-\theta)^{\beta - 1}d\theta \,,
\end{equation}
that acts as a normalisation factor for the distribution and $\alpha$ and $\beta$ are hyperparameters of the distribution. In the Bayesian context, a hyperparameter is a parameter of a prior distribution; the term is used to distinguish them from parameters of the model. For $\alpha=1$ and $\beta=1$ the \textit{Beta} distribution reduces to the uniform distribution. We can now apply \textit{Bayes' theorem} to produce the posterior distribution
\begin{equation}
    \label{eq:posterior_beta_sf}
    p(\theta\vert s, f) = \frac{p(s,f\vert\theta)p(\theta)}{p(s,f)}\,,
\end{equation}
where
\begin{equation}
    \label{eq:evidence_beta_sf}
    p(s, f) = \int_{0}^{1}p(s,f\vert\theta)p(\theta)d\theta\,,
\end{equation}
is the \textit{evidence}. Substituting equations \ref{eq:binomial_distribution_in_terms_of_sf} and \ref{eq:beta_distribution} into \ref{eq:posterior_beta_sf} and \ref{eq:evidence_beta_sf} we have
\begin{equation}
    \label{eq:posterior_beta_calculation}
    \begin{split}
        p(\theta\vert s, f) &= \frac{\binom{s+f}{s}\theta^{s}(1-\theta)^{f}\theta^{\alpha-1}(1-\theta)^{\beta - 1}/B(\alpha, \beta)}{\int_{0}^{1} \binom{s+f}{s}\theta^{s}(1-\theta)^{f}\theta^{\alpha-1}(1-\theta)^{\beta - 1}/B(\alpha, \beta) d\theta} \\
        &= \frac{\theta^{s+\alpha-1}(1-\theta)^{f+\beta -1}}{B(s+\alpha,f+\beta)}\,,
    \end{split}
\end{equation}
which is another \textit{Beta} distribution with hyperparameters $\alpha' = s+\alpha$ and $\beta ' = f+\beta$.

\section{Jeffreys priors}

There is often the need for priors that are invariant under reparameterisation, meaning that two different parameterisations $\theta$ and $\phi$ of the same model $\mathcal{M}$ yield consistent results. This type of prior was named after Sir Harold Jeffreys and it has the key feature that it is invariant under reparameterisations \parencite{robert2007bayesian}. One natural consequence of this approach is that a Jeffreys prior is fully determined by the choice of parameters, model and likelihood function. In that sense, it is often categorised as an objective prior as the preferences of the researcher affect it only indirectly through the choice of model and likelihood function. Although it is often characterised as an uninformative prior, this is actually far from true as all priors encode prior information. Perhaps a more appropriate name would be the \textit{reparametersation invariant prior}.

\subsection{One--dimensional case}

Let us assume that $\theta$ and $\phi$ are two possible parameterisations of the same model $\mathcal{M}$, and $\theta$ is a continuously differentiable function of $\phi$, then we say that the prior density $p_{\theta}(\theta)$ is invariant under the reparameterisation $\theta = \theta(\phi)$ if it is related to the prior density $p_{\phi}(\phi)$ by the usual change--of--variables theorem
\begin{equation}
    \label{eq:change_of_variables_1d}
    p_{\phi}(\phi) = p_{\theta}(\theta) \left| \frac{d\theta }{d\phi} \right|\,.
\end{equation}
Furthermore, the expected Fisher information is defined as
\begin{equation}
    \label{eq:fisher_information_theta}
    \mathcal{I}_{\theta}(\theta) = - \mathbb{E}_{d} \left[ \frac{d^{2}}{d\theta^{2}}\log p(d|\theta)\right]\,,
\end{equation}
and similarly for the $\phi$ parameterisation, where $\log p(d|\theta)$ is the logarithm of likelihood function, is transformed as
\begin{equation}
    \label{eq:fisher_information_transform}
    \mathcal{I}_{\phi}(\phi) = \mathcal{I}_{\theta}(\theta) \left( \frac{d\theta}{d\phi}\right)^{2}\,,
\end{equation}
under the reparametrisation $\theta = \theta(\phi)$.

Comparing the equations \ref{eq:change_of_variables_1d} and \ref{eq:fisher_information_transform} one can see that defining the priors as
\begin{equation}
    \label{eq:jeffreys_prior_theta_1d}
    p_{\theta}(\theta ) \propto \sqrt{\mathcal{I}_{\theta}(\theta)} \,,
\end{equation}
and
\begin{equation}
    \label{eq:jeffreys_prior_phi_1d}
    p_{\phi}(\phi ) \propto \sqrt{\mathcal{I}_{\phi}(\phi)} \,,
\end{equation}
yields the desired invariance under reparameterisation.

\subsection{Multi--dimensional case}

The generalisation to multiple dimensions is straightforward. The change--of--variables formula has the general form
\begin{equation}
    \label{eq:change_of_variables_nd}
    p_{\phi}(\phi) = p_{\theta}(\theta) \vert \det J\vert \,,
\end{equation}
where $\theta$ and $\phi$ are now sets of parameters (i.e. vectors), and $J$ is the Jacobian matrix of the transformation with components given by
\begin{equation}
    \label{eq:jacobian_matrix_components}
    J_{ij} = \frac{\partial \theta_{i}}{\partial \phi_{i}}\,,
\end{equation}
where the indices $i$ and $j$ point to the $i$--th and $j$--th component of the parameter vectors $\theta$ and $\phi$ respectively.
Similarly, the expected Fisher information matrix, defined as
\begin{equation}
    \label{eq:fisher_information_matrix_theta}
    \mathcal{I}_{\theta}(\theta)_{ij} = -\mathbb{E}_{d}\left[ \frac{\partial^{2}}{\partial \theta_{i} \partial \theta_{j}}\log p(d|\theta )\right]\,,
\end{equation}
is transformed as
\begin{equation}
    \label{eq:fisher_information_matrix_transform}
    \mathcal{I}_{\phi}(\phi) = J^{T} \mathcal{I}_{\theta}(\theta) J\,.
\end{equation}
Computing the determinant of both parts of equation \ref{eq:fisher_information_matrix_transform} leads to
\begin{equation}
    \label{eq:fisher_information_matrix_transform_determinant}
    \det \mathcal{I}_{\phi}(\phi) = \det \mathcal{I}_{\theta}(\theta) (\det J )^{2}\,.
\end{equation}
Comparing equations \ref{eq:change_of_variables_nd} and \ref{eq:fisher_information_matrix_transform_determinant} one can see that defining the priors as
\begin{equation}
    \label{eq:jeffreys_prior_theta_nd}
    p_{\theta}(\theta ) \propto \sqrt{\det \mathcal{I}_{\theta}(\theta)} \,,
\end{equation}
and
\begin{equation}
    \label{eq:jeffreys_prior_phi_nd}
    p_{\phi}(\phi ) \propto \sqrt{\det \mathcal{I}_{\phi}(\phi)} \,,
\end{equation}
once again yields the desired invariance under reparameterisation.

\subsection{Gaussian distribution with mean parameter}

Assuming that the data $d$ are Gaussian--distributed with unknown mean $\mu$ and known standard deviation $\sigma$, the probability density function of $d$ given $\mu$ can be written as
\begin{equation}
    \label{eq:gaussian_distribution_unknown_mean}
    p(d|\mu)=\frac{1}{\sqrt{2\pi}\sigma} e^{-\frac{(d-\mu)^2}{2\sigma^{2}}}\,,
\end{equation}
where $\sigma$ is fixed. Applying equation \ref{eq:jeffreys_prior_theta_1d} using equation \ref{eq:fisher_information_theta} in this case, the Jeffreys prior for parameter $\theta \equiv \mu$ is simply
\begin{equation}
    \label{eq:jeffreys_prior_gaussian_mean}
    \begin{split}
        p(\mu) &\propto \sqrt{I(\mu)} = \sqrt{- \mathbb{E}_{d} \left[ \frac{d^{2}}{d\mu^{2}}\log p(d|\mu)\right]} = \sqrt{\mathbb{E}_{d}\left[ \left( \frac{d-\mu}{\sigma} \right)^{2}\right]} \\
        &= \sqrt{\int_{-\infty}^{\infty} p(d|\mu) \left( \frac{d-\mu}{\sigma} \right)^{2} dd} = \frac{1}{\sigma} \propto 1\,.
    \end{split}
\end{equation}
The prior of $\mu$ in this case is independent of $\mu$ which means that is an improper (i.e. unnormalised) uniform prior.

\subsection{Gaussian distribution with scale parameter}

Assuming now that we know the mean parameter $\mu$ (i.e. $\mu$ is fixed) and the standard deviation $\sigma$ is unknown, the Gaussian probability density function of $d$ given $\sigma$ is simply
\begin{equation}
    \label{eq:gaussian_distribution_unknown_scale}
    p(d|\sigma)=\frac{1}{\sqrt{2\pi}\sigma} e^{-\frac{(d-\mu)^2}{2\sigma^{2}}}\,.
\end{equation}
Applying equation \ref{eq:jeffreys_prior_theta_1d} using equation \ref{eq:fisher_information_theta} in this case, the Jeffreys prior for parameter $\theta \equiv \sigma$ is
\begin{equation}
    \label{eq:jeffreys_prior_gaussian_scale}
    \begin{split}
        p(\sigma) &\propto \sqrt{I(\sigma)} = \sqrt{- \mathbb{E}_{d} \left[ \frac{d^{2}}{d\sigma^{2}}\log p(d|\sigma )\right]} = \sqrt{\mathbb{E}_{d}\left[ \left( \frac{(d-\mu)^{2}-\sigma^{2}}{\sigma^{3}} \right)^{2}\right]} \\
        &= \sqrt{\int_{-\infty}^{\infty} p(d|\sigma) \left( \frac{(d-\mu)^{2}-\sigma^{2}}{\sigma^{3}} \right)^{2} dd} = \frac{\sqrt{2}}{\sigma} \propto \frac{1}{\sigma}\,.
    \end{split}
\end{equation}

\section{Maximum entropy priors}

% Principle of indifference

In the absence of any information, one should distribute their degree of belief equally between all possible outcomes. This simple rule for assigning probabilities to discrete outcomes was considered so apparent to the fathers of probability theory, \textit{Jacob Bernoulli} and \textit{Pierre Simon Laplace}, that they did not even bother to give it a name. However, its importance in the context of probability theory was clear to both of them. In particular, Laplace wrote:
\begin{quote}
    \textit{The theory of chance consists in reducing all the events of the same kind to a certain number of cases equally possible, that is to say, to such as we may be equally undecided about in regard to their existence, and in determining the number of cases favourable to the event whose probability is sought. The ratio of this number to that of all the cases possible is the measure of this probability, which is thus simply a fraction whose numerator is the number of favourable cases and whose denominator is the number of all the cases possible.}
\end{quote}
This rule was later named the \textit{principle of insufficient reason}, possibly as a play on Leibniz's \textit{principle of sufficient reason} \parencite{brading2003symmetries}. Finally, it was renamed to the \textit{principle of indifference} by economist \textit{John Maynard Keynes} that noted that it can only be applied when one has no additional information~\parencite{keynes1921treatise}.

\looseness=-1 But what if we have some additional information, perhaps in the form of expectation values? Can we somehow incorporate that information and minimally extend the \textit{principle of indifference}? The answer to this question was provided by Jaynes in the form of the \textit{principle of maximum entropy (MaxEnt)}~\parencite{jaynes1982rationale}.

Using the notion of Shannon's ``entropy'' that quantifies the uncertainty of a probability distribution, MaxEnt is a mathematical procedure for the derivation of the maximally agnostic (i.e. least informative) probability distribution subject to a collection of known constraints. The MaxEnt principle can be applied in the assignment of prior probabilities in cases where we know some constraints about the parameters \textit{a priori} in the form of expectation values (e.g. mean, variance, lower or upper bounds, etc.) and we seek to find the least informative distribution that respects those constraints and still complies as much as possible to the \textit{principle of indifference}.

\looseness=-1 The MaxEnt principle turns the problem of defining a prior distribution into a task of optimisation. In particular, one seeks the probability distribution with the maximum entropy, that is, the least informative, subject to a collection of algebraic constraints in the form of expectation values. Before we move on to discuss some explicit examples that demonstrate the application of the MaxEnt principle, let us first present a couple of definitions for the ``entropy''.

\subsection{Shannon's entropy}

In 1948, \textit{Claude Shannon}'s pioneering work on \textit{information theory}~\parencite{shannon1948mathematical} introduced a measure of the uncertainty of a discrete probability distribution which he termed ``entropy'' and defined as
\begin{equation}
    \label{eq:shannon_entropy}
    S(p) = - \sum_{i=1}^{n} p_{i} \log p_{i}\,.
\end{equation}
The entropy of a probability distribution quantifies the amount of missing information, uncertainty or ``surprisal'' inherent in the distribution~\parencite{mackay2003information}. In other words, entropy is the expectation value $\mathbb{E}_{p}[I]$ of the information
\begin{equation}
    \label{eq:self_information}
    I_{i} = -\log p_{i}\,,
\end{equation}
which quantifies the information content of an event $i$ with probability $p_{i}$. For example, a fair coin landing ``heads--tails--heads`` with probability $1/2^{3}$ provides information of $-\log(1/2^{3})=3\log 2$ or $3\,\mathrm{bits}$. Information is measured in ``bits'' if the logarithm has base $2$ or in ``nats'' if it has base $e$. Information has a series of desired properties, namely
\begin{enumerate}
    \item An event $i$ with probability $p_{i}=1$ is certain and offers no information (i.e. $I_{i}=0$),
    \item The lower the probability of an event, the more surprising it is and thus the higher its information contribution,
    \item Information is additive, meaning that the total amount of information is the sum of the information of the individual events.
\end{enumerate}
It turns out the form of equation \ref{eq:self_information} for information is the only option if we want less probable events to have more information, and information to add for independent events. 

Let us consider a simple example in order to make the notions of information and entropy better understood. Suppose that according to the weather forecast there is a $p=1/2$ chance that it rains $1\,cm$, $p=1/4$ chance that it rains $2\,cm$ and $p=1/4$ chance that it does not rain at all. The expected amount of rain is simply $1/2\times 1\,cm+1/4\times2\,cm+1/4\times 0 = 1\,cm$. The expected amount of information that you gain when you find out how much it rains is $-1/2\log(1/2)-1/4\log(1/4)-1/4\log(1/4)=3/2 \log 2$ or $3/2\,\mathrm{bits}$, this is the Shannon entropy of the weather report.

Shannon's entropy naturally assumes that the uniform distribution $p_{i}=1/n$ where $n$ is the number of discrete events holds a very special role. In the absence of any other constraints, assigning equal probability to all outcomes (i.e. $p_{i}$ is constant) corresponds to the state of complete ignorance. In other words, the distribution that maximises Shannon's entropy is the uniform distribution in accordance with the \textit{principle of indifference}.

\subsection{Relative entropy}

Another useful, entropy--like quantity is the following
\begin{equation}
    \label{eq:kl_divergence_discrete}
    D_{KL}(p|q) = \sum_{i=1}^{n} p_{i} \log \left( \frac{p_{i}}{q_{i}}\right)\,,
\end{equation}
that has been given many names, including \textit{relative entropy}, \textit{cross--entropy}, and \textit{Kullback--Leibler (KL) divergence} as \parencite{kullback1951information} were the first ones to demonstrate its potential for statistical applications.

The latter is a measure of information gained when one updates their beliefs, initially quantified by a distribution $q$ to an updated distribution $p$. In that sense, relative entropy is a measure of statistical distance between the two distributions and it is defined as
\begin{equation}
    \label{eq:relative_entropy_discrete}
    D_{KL}(p|q) = \sum_{i=1}^{n} p_{i} \log \left( \frac{p_{i}}{q_{i}}\right)\,,
\end{equation}
for the discrete case, and
\begin{equation}
    \label{eq:relative_entropy_continuous}
    D_{KL}(p|q) = \int p(\theta) \log \left( \frac{p(\theta)}{q(\theta)}\right) d\theta\,,
\end{equation}
for the continuous case.

Relative entropy has a collection of desired properties too, including
\begin{enumerate}
    \item It is always non--negative,
        \begin{equation}
            \label{eq:Gibbs_inequality}
            D_{KL}(p|q) \ge 0\,,
        \end{equation}
        a result commonly known as \textit{Gibbs' inequality}. Relative entropy is zero if and only if $p=q$ almost everywhere.
        
    \item Relative entropy, unlike Shannon's entropy, is well defined for continuous distributions.
    
    \item Given a transformation $\theta = \theta(\phi)$ such that $p(\theta ) d\theta = p(\phi ) d\phi$ and $q(\theta ) d\theta = q(\phi ) d\phi$ the relative entropy is parameterisation invariant, meaning
        \begin{equation}
            \label{eq:relative_entropy_transform}
            \begin{split}
                D_{KL}(p|q) &= \int p(\theta) \log \left( \frac{p(\theta)}{q(\theta)}\right) d\theta \\
                &= \int p(\phi) \log \left( \frac{p(\phi)\frac{d\phi}{d\theta}}{q(\phi)\frac{d\phi}{d\theta}}\right) d\phi \\
                &= \int p(\phi) \log \left( \frac{p(\phi)}{q(\phi)}\right) d\phi\,.
            \end{split}
        \end{equation}
        
    \item Relative entropy reduces to the well known Shannon's entropy, up to a sign, for the case of a uniform distribution $q$ in the discrete case.
    
\end{enumerate}

As we mentioned in the previous sub--section, using \textit{Shannon}'s definition of entropy places the uniform distribution into a very special place, that of the maximum entropy distribution in the absence of any other constraints that provide additional information. Although this is in accordance with the \textit{principle of indifference}, there are cases in practice in which one requires a different prior distribution $q$, before taking into account any constraints. For instance, one may seek to find a distribution that minimally deviates from a Jeffreys prior subject to some constraints. In those cases, instead of maximising \textit{Shannon}'s entropy, one can minimise the relative entropy. For simplicity, we will refer to the principle of minimum relative entropy as MaxEnt too \textcite{shore1980axiomatic} proved that minimising the relative entropy is the uniquely correct way of updating probability distributions in the face of new information in the form of expectation values for both discrete and continuous cases. Furthermore, the \textit{relative entropy}, unlike Shannon's entropy, is easily generalisable to the continuous case too. 

\subsection{Lagrange multipliers}

The method of Lagrange multipliers~\parencite{riley1999mathematical} offers a powerful way of finding the local extrema (i.e. maxima and minima) of a function $f(p )$ subject to a constraint $g(p )=0$. If no constraint is available then the extrema of $f$ can be found by solving
\begin{equation}
    \label{eq:df_equals_zero}
    df = \frac{\partial f}{\partial p_{1}}dp_{1} + \dots + \frac{\partial f}{\partial p_{n}}dp_{n} = 0\,.
\end{equation}
Here the $dp_{i}$ are independent so one concludes that the extrema are simply given by $\partial f/ \partial p_{i} =0$. However, the existence of the constraint means that $dp_{i}$ are not actually independent since
\begin{equation}
    \label{eq:dg_equals_zero}
    dg = \frac{\partial g}{\partial p_{1}}dp_{1} + \dots + \frac{\partial g}{\partial p_{n}}dp_{n} = 0\,,
\end{equation}
because $g(p )$ is constant. We can combine equations \ref{eq:df_equals_zero} and \ref{eq:dg_equals_zero} by first multiplying the second by an unknown factor $\lambda$ called \textit{Lagrange multiplier}, thus yielding
\begin{equation}
    \label{eq:lagrage_multiplier}
    d(f-\lambda g) = \left(\frac{\partial f}{\partial p_{1}}-\lambda \frac{\partial g}{\partial p_{1}}\right)dp_{1} + \dots + \left(\frac{\partial f}{\partial p_{n}}-\lambda \frac{\partial g}{\partial p_{n}}\right)dp_{n} = 0\,.
\end{equation}
We can now choose $\lambda$ such that
\begin{equation}
    \label{eq:lagrange_factors}
    \frac{\partial f}{\partial p_{i}}-\lambda \frac{\partial g}{\partial p_{i}} = 0\,,
\end{equation}
for all $i\in\lbrace1,\dots,n\rbrace$. Equations \ref{eq:lagrange_factors} along with the constraint equation $g(p)$ are sufficient to determine the value of $\lambda$ and coordinates $p_{i}$ of the stationary point.

\subsection{Uniform distribution}

Assuming that the only constraint or form of information is that the sum of all probabilities is equal to one, meaning
\begin{equation}
    \label{eq:probabilities_sum_to_one}
    \sum_{i=1}^{n}p_{i} = 1\,,
\end{equation}
then in order to find the maximum entropy distribution, we have to solve
\begin{equation}
    \label{eq:lagrange_uniform}
    d\left[ -\sum_{i=1}^{n}p_{i}\log\frac{p_{i}}{q_{i}} - \lambda \left( \sum_{i=1}^{n}p_{i} - 1\right) \right] = 0\,,
\end{equation}
where the first term is the relative entropy and the second term is the constraint multiplied with the unknown Lagrange multiplier $\lambda$. Doing some simple calculus on the expression of \ref{eq:lagrange_uniform} we get
\begin{equation}
    \label{eq:lagrange_uniform_2}
    \begin{split}
        d\left[ -\sum_{i=1}^{n}p_{i}\log p_{i}+\sum_{i=1}^{n}p_{i}\log q_{i} - \lambda \left( \sum_{i=1}^{n}p_{i} - 1\right) \right] = 0 \Rightarrow \\
        -\sum_{i=1}^n dp_i\,\log p_i - \sum_{i=1}^n p_i\,d(\log p_i) \\
        + \sum_{i=1}^n dp_i\,\log q_i - \lambda\sum_{i=1}^n dp_i = 0 \Rightarrow \\
        -\sum_{i=1}^n dp_i\,\log p_i - \sum_{i=1}^n p_i\,\left( \sum_{j=1}^n \frac{\partial \log p_i}{\partial p_j}\,dp_j \right) \\
        + \sum_{i=1}^n dp_i\,\log q_i - \lambda\sum_{i=1}^n dp_i = 0 \Rightarrow \\
        -\sum_{i=1}^n dp_i\,\log p_i - \sum_{i=1}^n p_i\,\left( \sum_{j=1}^n \delta_{ij} \frac{1}{p_j}\,dp_j \right) \\
        + \sum_{i=1}^n dp_i\,\log q_i - \lambda\sum_{i=1}^n dp_i = 0 \Rightarrow \\
        -\sum_{i=1}^n dp_i\,\log p_i - \sum_{i=1}^n p_i\, \frac{1}{p_i}\,dp_i + \sum_{i=1}^n dp_i\,\log q_i - \lambda\sum_{i=1}^n dp_i = 0
        \Rightarrow \\
        \sum_{i=1}^{n}\left( -\log \frac{p_{i}}{q_{i}} - 1 - \lambda  \right)d p_{i} = 0
    \end{split}
\end{equation}
According to our previous discussion on Lagrange multipliers, for equation \ref{eq:lagrange_uniform_2} to hold, the terms in the parentheses need to vanish for every $i=1,\dots,n$, therefore we get
\begin{equation}
    \label{eq:lagrange_uniform_with_lambda}
    p_{i} = q_{i} e^{-(1+\lambda )}\,.
\end{equation}
We can determine the value of $\lambda$ using the constraint equation \ref{eq:probabilities_sum_to_one}
\begin{equation}
    \label{eq:lagrange_uniform_get_lambda}
    \sum_{i=1}^{n}q_{i} e^{-(1+\lambda )} = 1\,,
\end{equation}
$\sum_{i=1}^{n}q_{i}=1$ so $\lambda = -1$. Therefore, the distribution in the discrete case is
\begin{equation}
    \label{eq:lagrange_uniform_discrete}
    p_{i} = q_{i}\,,
\end{equation}
and in the continuous case is
\begin{equation}
    \label{eq:lagrange_uniform_continuous}
    p(\theta) = q(\theta)\,,
\end{equation}
Furthermore, assuming that $q$ is a uniform (i.e. $q_{i}=1/n$) in accordance with the \textit{principle of indifference}, then $p$ is also uniform. This means that the distribution of maximum entropy under the minimal constraint that the total probability needs to sum up to one is the uniform distribution. Let us now move on to a few more intriguing examples.

\subsection{Exponential distribution}

Suppose now that we have an additional constraint apart from equation \ref{eq:probabilities_sum_to_one} for the sum of probabilities,
\begin{equation}
    \label{eq:constraint_mean}
    \sum_{i=1}^{n}p_{i}\theta_{i} = \mu\,,
\end{equation}
indicating that the mean value of $\theta$ is known and equal to $\mu$. Having two constraints requires us to introduce two Lagrange multipliers, $\lambda$ and $\Tilde{\lambda}$, and solve
\begin{equation}
    \label{eq:lagrange_exponential}
    d\left[ -\sum_{i=1}^{n}p_{i}\log\frac{p_{i}}{q_{i}} - \lambda \left( \sum_{i=1}^{n}p_{i} - 1\right) - \Tilde{\lambda} \left( \sum_{i=1}^{n}p_{i}\theta_{i} - \mu\right) \right] = 0\,,
\end{equation}
in order to find the appropriate maximum entropy distribution. Similarly to before, after some calculus, we get
\begin{equation}
    \label{eq:lagrange_exponential_2}
    \sum_{i=1}^{n} \left( -\log\frac{p_{i}}{q_{i}} - 1 - \lambda - \theta_{i}\Tilde{\lambda} \right) d p_{i} = 0\,.
\end{equation}
Again, the term in the parentheses needs to vanish for any value of $i$, thus
\begin{equation}
    \label{eq:lagrange_exponential_with_lambda}
    p_{i} = q_{i} e^{-(1+\lambda )}e^{-\Tilde{\lambda}\theta_{i}}\,.
\end{equation}
We can now apply the two constraints to determine the values of the Lagrange multipliers. From equation \ref{eq:probabilities_sum_to_one} for the first constraint, we have
\begin{equation}
    \label{eq:lagrange_exponential_get_lambda}
    e^{-(1+\lambda )} = \frac{1}{\sum_{i=1}^{n} q_{i} e^{-\Tilde{\lambda}\theta_{i}}}\,.
\end{equation}
Similarly, from equation \ref{eq:constraint_mean} for the second constraint, we have
\begin{equation}
    \label{eq:lagrange_exponential_get_lambda_tilde}
    \sum_{i=1}^{n} q_{i} \theta_{i} e^{-\Tilde{\lambda}\theta_{i}} - \mu \sum_{i=1}^{n} q_{i} e^{-\Tilde{\lambda}\theta_{i}} = 0\,,
\end{equation}
which can only be solved numerically. Equation \ref{eq:lagrange_exponential_with_lambda} can also be written for the continuous case as
\begin{equation}
    \label{eq:lagrange_exponential_with_lambda_continuous}
    p(\theta) = q(\theta) e^{-(1+\lambda )}e^{-\Tilde{\lambda}\theta}\,.
\end{equation}

Equation \ref{eq:lagrange_exponential_with_lambda_continuous} is the general MaxEnt prior for an arbitrary pseudo--prior $q(\theta)$. However, the expression can be simplified more if we assume that our state of knowledge about $\theta$ prior to the information provided by the constraints \ref{eq:probabilities_sum_to_one} and \ref{eq:constraint_mean}, is that it is positive (i.e. $\theta > 0$). This means, that $q(\theta)$ is a uniform distribution, in agreement with the principle of indifference. Furthermore, to avoid issues with ``infinities'' and render $q$ a proper pseudo--prior, we can set an upper limit $L$ on the possible values of $\theta$, therefore
\begin{equation}
    \label{eq:bounded_uniform}
    q(\theta) = \mathcal{U}(\theta\vert 0, L) = 
    \begin{cases}
      1/L & \text{if $0<\theta\leq L$}\\
      0 & \text{otherwise}
    \end{cases} \,.
\end{equation}
Once we have derived the form of the MaxEnt prior $p(\theta)$ we can then take the limit of $L\to\+\infty$ to allow $\theta$ to be any positive real number. Including this particular choice of $q(\theta)$, equation \ref{eq:lagrange_exponential_with_lambda_continuous} takes the form
\begin{equation}
    \label{eq:exponential_useless_form}
    p(\theta) = \frac{1}{L} e^{-(1+\lambda )}e^{-\Tilde{\lambda}\theta}\,,\quad \theta > 0\,.
\end{equation}

Using the pseudo-prior of equation \ref{eq:bounded_uniform}, the first constraint, for the total probability, given by equation \ref{eq:probabilities_sum_to_one}, can be expressed as
\begin{equation}
    \label{eq:first_constraint_expression}
    \begin{split}
        \int_{-\infty}^{\infty}p(\theta)d\theta = 1 \Rightarrow \\
        \frac{1}{L}e^{-(1+\lambda)}\int_{0}^{L}e^{-\Tilde{\lambda}\theta}d\theta = 1 \Rightarrow \\
        \frac{1}{L}e^{-(1+\lambda)}\left[ -\frac{e^{-\Tilde{\lambda}\theta}}{\Tilde{\lambda}} \right]_{\theta=0}^{\theta=L} = 1 \Rightarrow \\
        \frac{1}{L}e^{-(1+\lambda)} = \frac{\Tilde{\lambda}}{1 - e^{-\Tilde{\lambda}L} }\,.
    \end{split}
\end{equation}

Similarly, the second constraint, for the expected or mean value of $\theta$, given by equation \ref{eq:constraint_mean}, can be expressed as
\begin{equation}
    \label{eq:second_constraint_expression}
    \begin{split}
        \int_{-\infty}^{\infty}\theta p(\theta)d\theta = \mu \Rightarrow \\
        \frac{1}{L}e^{-(1+\lambda)}\int_{0}^{L}\theta e^{-\Tilde{\lambda}\theta}d\theta = \mu \Rightarrow \\
        \frac{1}{L}e^{-(1+\lambda)} \left[ - \frac{e^{-\Tilde{\lambda}\theta}(\Tilde{\lambda}\theta + 1)}{\Tilde{\lambda}^{2}}\right]_{\theta=0}^{\theta=L} = \mu \Rightarrow \\
        \frac{1}{L}e^{-(1+\lambda)} \frac{1-e^{-\Tilde{\lambda}L}(\Tilde{\lambda}L + 1)}{\Tilde{\lambda}^{2}} = \mu \Rightarrow \\
        \frac{\Tilde{\lambda}}{1 - e^{-\Tilde{\lambda}L} } \times \frac{1-e^{-\Tilde{\lambda}L}(\Tilde{\lambda}L + 1)}{\Tilde{\lambda}^{2}} = \mu \Rightarrow \\
        \mu = \frac{1}{\Tilde{\lambda}} - \frac{L  e^{-\Tilde{\lambda}L}}{1-e^{-\Tilde{\lambda}L}}\,,
    \end{split}
\end{equation}
where we used equation \ref{eq:first_constraint_expression} to simplify the result.

Taking the limit $L\to +\infty$ and using equation \ref{eq:second_constraint_expression}, we find that
\begin{equation}
    \label{eq:second_constraint_limit}
    \mu \to \frac{1}{\Tilde{\lambda}}\,,
\end{equation}
as the second term vanishes. Similarly, using equation \ref{eq:first_constraint_expression}, we find that
\begin{equation}
    \label{eq:second_constraint_limit}
    \frac{1}{L}e^{-(1+\lambda)} \to \Tilde{\lambda}\,.
\end{equation}
Substituting these results into equation \ref{eq:exponential_useless_form}, we are lead to
\begin{equation}
    \label{eq:lagrange_exponential_continuous}
    p(\theta|\mu) = \frac{1}{\mu}e^{-\frac{\theta}{\mu}}\,,
\end{equation}
the well known exponential distribution. What this paragraph taught us is crucial, if we only know the mean of a non--negative parameter and nothing else, then the exponential distribution is the one that best represents the current state of knowledge, by making the fewest assumptions.

\subsection{Normal distribution}

Suppose that we also know the the standard deviation $\sigma$ given by
\begin{equation}
    \label{eq:standard_deviation_discrete}
    \sum_{i=0}^{n} p_{i} (\theta_{i} - \mu )^{2} = \sigma^{2}\,,
\end{equation}
as an additional constraint. We now have to solve
\begin{equation}
    \label{eq:lagrange_gaussian}
    d\left[ -\sum_{i=1}^{n}p_{i}\log\frac{p_{i}}{q_{i}} - \lambda \left( \sum_{i=1}^{n}p_{i} - 1\right) - \Tilde{\lambda} \left( \sum_{i=1}^{n}p_{i}(\theta_{i} - \mu)^{2}-\sigma^{2}\right) \right] = 0\,,
\end{equation}
where the first term corresponds to the entropy, the second to the constraint that the sum of all probabilities needs to add up to one, and the last term to the standard deviation constraint that also includes that about the mean $\mu$. Thus we have two \textit{Lagrange} multipliers and we follow the same procedure as before, solving equation \ref{eq:lagrange_gaussian} we have
\begin{equation}
    \label{eq:lagrange_gaussian_2}
    \sum_{i=1}^{n} \left( -\log\frac{p_{i}}{q_{i}} - 1 - \lambda - \Tilde{\lambda} (\theta_{i}-\mu)^{2} \right) d p_{i} = 0\,.
\end{equation}
The terms in the parentheses need to vanish for all values of $i$, thus
\begin{equation}
    \label{eq:lagrange_gaussian_with_lambda}
    p_{i} = q_{i} e^{-(1+\lambda )}e^{-\Tilde{\lambda}(\theta_{i}-\mu)^{2}}\,.
\end{equation}
The corresponding continuous probability density function is simply
\begin{equation}
    \label{eq:lagrange_gaussian_with_lambda_continuous}
    p(\theta) = q(\theta) e^{-(1+\lambda )}e^{-\Tilde{\lambda}(\theta-\mu)^{2}}\,.
\end{equation}
Furthermore, assuming a uniform prior $q(\theta)\propto 1$ in accordance with the \textit{principle of indifference}, equation \ref{eq:lagrange_exponential_with_lambda_continuous} reduces to
\begin{equation}
    \label{eq:lagrange_gaussian_with_lambda_continuous_2}
    p(\theta) = e^{-(1+\lambda )}e^{-\Tilde{\lambda}(\theta-\mu)^{2}}\,.
\end{equation}
We can now apply the constraint equations \ref{eq:probabilities_sum_to_one} and \ref{eq:standard_deviation_discrete} in order to uniquely determine the values of the two \textit{Lagrange} multipliers. In the continuous limit, the first constraint given by equation \ref{eq:probabilities_sum_to_one}, is written as
\begin{equation}
    \label{eq:probabilities_integrate_to_one}
    \int_{-\infty}^{\infty}p(\theta)d\theta = 1\,.
\end{equation}
Substituting equation \ref{eq:lagrange_gaussian_with_lambda_continuous_2} into equation \ref{eq:probabilities_integrate_to_one} yields
\begin{equation}
    \label{eq:lagrange_gaussian_get_lambda}
    e^{-(1+\lambda)}\int_{-\infty}^{\infty}e^{-\Tilde{\lambda}(\theta-\mu)^{2}}d\theta = 1\,.
\end{equation}
Doing the change of variables $z=\sqrt{\Tilde{\lambda}}(\theta - \mu)$ brings equation \ref{eq:lagrange_gaussian_get_lambda} into the simpler form
\begin{equation}
    \label{eq:lagrange_gaussian_get_lambda_2}
    \frac{e^{-(1+\lambda)}}{\sqrt{\Tilde{\lambda}}}\int_{-\infty}^{\infty}e^{-z^{2}}dz = 1\,,
\end{equation}
where the integral is the so called \textit{Gaussian integral} with value equal to $\sqrt{\pi}$. Therefore, 
\begin{equation}
    \label{eq:lagrange_gaussian_lambda}
    e^{-(1+\lambda)} = \sqrt{\frac{\Tilde{\lambda}}{\pi}}\,,
\end{equation}
and equation \ref{eq:lagrange_gaussian_with_lambda_continuous_2} reduces to
\begin{equation}
    \label{eq:lagrange_gaussian_with_lambda_continuous_3}
    p(\theta ) = \sqrt{\frac{\Tilde{\lambda}}{\pi}} e^{-\Tilde{\lambda}(\theta-\mu)^{2}} \,.
\end{equation}
We can now move on to determine the second \textit{Lagrange} multiplier $\Tilde{\lambda}$ by substituting equation \ref{eq:lagrange_gaussian_with_lambda_continuous_3} into \ref{eq:standard_deviation_discrete}, thus
\begin{equation}
    \label{eq:lagrange_gaussian_get_lambda_3}
    \sqrt{\frac{\Tilde{\lambda}}{\pi}} \int_{-\infty}^{\infty}e^{-\Tilde{\lambda}(\theta-\mu)^{2}}(\theta-\mu)^{2}d\theta = \sigma^{2}\,.
\end{equation}
Applying the same change of variables as before, $z=\sqrt{\Tilde{\lambda}}(\theta - \mu)$, we have
\begin{equation}
    \label{eq:lagrange_gaussian_get_lambda_4}
    \frac{1}{\Tilde{\lambda}\sqrt{\pi}} \int_{-\infty}^{\infty}e^{-z^{2}}z^{2}dz = \sigma^{2}\,.
\end{equation}
The integral can be computed using integration by parts
\begin{equation}
    \label{eq:lagrange_gaussian_get_lambda_5}
    \frac{1}{\Tilde{\lambda}\sqrt{\pi}} \left\lbrace\left[z\left(-\frac{1}{2}e^{-z^{2}}\right) \right]_{-\infty}^{\infty}-\int_{-\infty}^{\infty}-\frac{1}{2}e^{-z^{2}}dz\right\rbrace = \sigma^{2}\,,
\end{equation}
in which the first term in the braces vanishes and the second is equal to $\sqrt{\pi}/2$, thus
\begin{equation}
    \label{eq:lagrange_gaussian_lambda_2}
    \Tilde{\lambda} = \frac{1}{2\sigma^{2}}\,.
\end{equation}
Finally, substituting equation \ref{eq:lagrange_gaussian_lambda_2} into \ref{eq:lagrange_gaussian_with_lambda_continuous_3} leads to the usual Gaussian function
\begin{equation}
    \label{eq:gaussian_pdf_1d}
    p(\theta) = \frac{1}{\sqrt{2\pi}\sigma}e^{-\frac{(\theta-\mu)^{2}}{2\sigma^{2}}}\,,
\end{equation}
as the maximum entropy probability density function. In other words, the maximum entropy probability distribution subject to the constraints of known mean and standard deviation is the normal distribution.

\section{Reference priors}

The method of \textit{reference priors}, originally proposed by~\textcite{bernardo1979reference} and later expanded by others~\parencite{berger2009formal, bernardo2005reference, kass1996selection, bernardo2009bayesian}, is another approach that utilises information--theoretic ideas. The main idea behind reference priors is to choose the  prior $p(\theta)$ to maximise some notion of discrepancy between the prior $p(\theta)$ and the posterior $p(\theta\vert d)$. One reason to do this is that such a prior would allow the data $d$ to be maximally informative and have the greatest effect on the posterior distribution. In one--dimensional cases it turns out that reference priors and Jeffreys priors are equivalent. In higher dimensional cases, however, they are generally different. The research field of reference priors has expanded substantially during the past decades. For this reason, we will cover the fundamentals in this section and direct the reader to the aforementioned references for more information.

As we discussed in the previous section regarding maximum entropy priors, a common measure of the discrepancy between two distributions is the relative entropy or KL divergence, given by equation \ref{eq:relative_entropy_continuous}. In the case of the prior and posterior distribution, this can be written as
\begin{equation}
    \label{eq:relative_entropy_prior_posterior}
    D_{KL}\left[p(\theta\vert d)\vert p(\theta)\right] = \int p(\theta\vert d)\log\frac{p(\theta\vert d)}{p(\theta)}d\theta\,.
\end{equation}
One might then wonder how can we maximise the above discrepancy measure, in order to find the prior $p(\theta)$, without knowing the posterior distribution $p(\theta\vert d)$. Reference priors address this point by maximising the \textit{expectation value} of the relative entropy of equation \ref{eq:relative_entropy_prior_posterior} over the distribution of the data $p(d^{(n)})$, where $d^{(n)}=\lbrace d_{1},\dots, d_{n}\rbrace$ are $n$ conditionally independent instances of the data. At first, this appears to be a frequentist procedure as one will base the choice of the prior on unseen fictional data, such as infinite repetitions of the same experiment (e.g. in the limit that $n\rightarrow \infty$). However, unlike frequentist approaches, once the prior is determined, the analysis proceeds in the usual Bayesian manner. Furthermore, \textit{Bernardo} argued that taking the limit of $n$ to infinity does not just lead to a convenient mathematical procedure but it is also, philosophically, the right thing to do. His argument is that when choosing a prior we should consider many future experiments than just a single one. In this sense, the reference prior procedure aims to maximise the \textit{missing information} about the parameters $\theta$ which can be obtained by repeated experiments.

\subsection{Mutual information}

The expected relative entropy between the prior and posterior, also known as the \textit{mutual information}, quantifies the missing information and can be derived as follows:
\begin{equation}
    \label{eq:mutual_information}
    \begin{split}
        I(\theta,d^{(n)})&=\mathbb{E}_{d^{(n)}}\left[ D_{KL}\left[p(\theta\vert d^{(n)})\vert p(\theta)\right]\right] = \int p(d^{(n)}) D_{KL}\left[p(\theta\vert d^{(n)})\vert p(\theta)\right] d d^{(n)} \\
        &= \int p(d^{(n)}) \int p(\theta\vert d^{(n)})\log\frac{p(\theta\vert d^{(n)})}{p(\theta)}d\theta dd^{(n)} \\
        &= \int \int p(\theta\vert d^{(n)}) p(d^{(n)})\log\frac{p(\theta\vert d^{(n)})}{p(\theta)}d\theta d  d^{(n)} \\
        &= \int \int p(\theta, d^{(n)}) \log\frac{p(\theta, d^{(n)})}{p(\theta) p(d^{(n)})}d\theta dd^{(n)} \,,
    \end{split}
\end{equation}
where we used Bayes' theorem to introduce the joint probability $p(\theta, d^{(n)})$. In order to understand the meaning and significance of the above expression, let us consider the simple case in which the parameters $\theta$ and the data $d^{(n)}$ are independent. In this case the joint probability of the two is separable, or $p(\theta, d^{(n)})=p(\theta) p(d^{(n)})$, and the mutual information $I(\theta,d^{(n)})$ is zero. In other words, the data $d^{(n)}$ have no effect on the parameters $\theta$. However, those two quantities are not generally independent, and the mutual information quantifies the influence or effect of the data $d^{(n)}$ on the parameters $\theta$. Finally, defining the reference priors in terms of the mutual information, has the advantage of sharing its reparameterisation invariance.

\subsection{Maximising the mutual information}

The reference prior $p^{*}(\theta)$ is simply the prior which maximises the mutual information in the limit that $n\rightarrow \infty$, or:
\begin{equation}
    \label{eq:maxmise_mutual_information}
    p^{*}(\theta) = \lim_{n\rightarrow \infty}p_{n}^{*}(\theta)\,, \text{ where }\,\,p_{n}^{*}(\theta)=\underset{p(\theta)}{\mathrm{arg\,max}}~I(\theta, d^{(n)})\,.
\end{equation}
The mutual information of equation \ref{eq:mutual_information} can be written as:
\begin{equation}
    \label{eq:mutual_information_short}
    I(\theta,d^{(n)})=\int p(\theta) \log \frac{f(\theta)}{p(\theta)}d\theta\,,
\end{equation}
where we have introduced the function
\begin{equation}
    \label{eq:reference_f}
    f_{n}(\theta)=\exp\left\lbrace \int p(d^{(n)}\vert \theta)\log p(d^{(n)}\vert \theta) dd^{(n)}\right\rbrace\,,
\end{equation}
where the product sampling distribution is defined as
\begin{equation}
    \label{eq:product_sampling_distribution}
    p(d^{(n)}\vert \theta) = \prod_{i=1}^{n}p(d_{i}\vert \theta)\,.
\end{equation} 
Finding the prior distribution $p_{n}(\theta)$, which maximises the mutual information $I(\theta,d^{(n)})$ subject to the constraint $\int p(\theta)d\theta = 1$, is essentially a problem that can be solved via the methods of calculus of variations. It may be simpler to derive the result by working in the discrete case. This means that we have to solve
\begin{equation}
    \label{eq:lagrange_mutual_info}
    d\left[ \sum_{i}p_{i}\log \left( \frac{f_{i}}{p_{i}}\right)+\lambda \left( \sum_{i}p_{i}-1\right)\right] = 0
\end{equation}
where $\lambda$ is a Lagrange multiplier, in order to find the prior $p_{i}$. The derivation is as follows:
\begin{equation}
    \label{eq:lagrange_mutual_info2}
    \begin{split}
       \sum_{i}dp_{i}\log \left( \frac{f_{i}}{p_{i}}\right)+\sum_{i}p_{i}\left[ \sum_{j}\frac{\partial \log (f_{i}/p_{i})}{\partial p_{j}}dp_{j}\right] +\lambda \sum_{i}dp_{i} &= 0 \Rightarrow \\
       \sum_{i}dp_{i}\log \left( \frac{f_{i}}{p_{i}}\right)+\sum_{i}p_{i}\left[ \sum_{j}\delta_{ij} \left( -\frac{1}{p_{j}}\right)dp_{j}\right] +\lambda \sum_{i}dp_{i} &= 0 \Rightarrow \\
       \sum_{i}\left[ \log \left( \frac{f_{i}}{p_{i}}\right) - 1 +\lambda\right] dp_{i} &= 0
    \end{split}
\end{equation}
For the above equation to be true, all terms in the sum must be zero, in other words we get that $p_{i}=f_{i}\exp(\lambda - 1)$ or simply $p_{i}\propto f_{i}$. Rewriting this in the continuous case, in the limit that $n\rightarrow \infty$ we have:
\begin{equation}
    \label{eq:reference_prior_expression}
    p(\theta) = \lim_{n\rightarrow \infty} \frac{f_{n}(\theta)}{f_{n}(\theta_{0})}\,,
\end{equation}
where $\theta_{0}$ is an internal point in parameter space and $f_{n}(\theta)$ is given by equation \ref{eq:reference_f}. Alternatively, $f_{n}(\theta)$ can be defined as
\begin{equation}
    \label{eq:reference_fh}
    f_{n}(\theta)=\exp\left\lbrace \int p(d^{(n)}\vert \theta)\log\left[ \frac{p(d^{(n)}\vert \theta) h(\theta)}{\int p(d^{(n)}\vert \theta) h(\theta) d\theta}  \right] dd^{(n)}\right\rbrace\,,
\end{equation}
where we have included an arbitrary pseudo--prior $h(\theta)$. Carefully selecting the functional form of $h(\theta)$ (e.g. conjugate prior) can significantly simplify the calculations.

Intuitively, equations \ref{eq:reference_prior_expression} and \ref{eq:reference_fh} state that the reference prior $p(\theta)$ depends only on the asymptotic behaviour of the posterior, and schematically can be written in the form
\begin{equation}
    \label{eq:reference_prior_schematic}
    \begin{split}
        p(\theta) &\propto \exp\left\lbrace \int p(d^{(n)}\vert \theta)\log p^{*}(\theta\vert d^{(n)}) dd^{(n)}\right\rbrace \\
        &\propto \exp\left\lbrace\mathbb{E}_{p(d^{(n)}\vert \theta)}\left[ \log p^{*}(\theta\vert d^{(n)})\right]\right\rbrace \,,
    \end{split}
\end{equation}
where $p^{*}(\theta\vert d^{(n)})$ is the asymptotic form of the posterior.

\subsection{Asymptotic solution}

Finding the reference prior is now reduced to computing $f_{n}(\theta)$ using equation \ref{eq:reference_f} or \ref{eq:reference_fh}. However, this can be quite challenging in practice. The problem can be simplified by using the \textit{Bernstein--von Mises} theorem, which, as we discussed in Chapter \ref{chp:probability}, states that under certain conditions, as the sample size approaches infinity (i.e. $n\rightarrow \infty$), the posterior distribution converges to a normal distribution centred on the \textit{maximum likelihood estimate (MLE)} $\theta_{n}$ with variance equal to $n^{-1}I^{-1}(\theta_{n})$, where $I(\theta)$ is the \textit{Fisher information} given by
\begin{equation}
    \label{eq:fisher_information}
    \mathcal{I}(\theta) = - \mathbb{E}_{p(d\vert \theta)}\left[ \frac{\partial^{2}\log p(d\vert \theta)}{\partial \theta^{2}} \right]\,.
\end{equation}

We can use the fact that MLE is a \textit{consistent} and \textit{asymptotically} sufficient estimator, meaning that
\begin{equation}
    \label{eq:mle_consistent}
    \lim_{n\to\infty}\Hat{\theta}_{n} = \theta \,,
\end{equation}
and
\begin{equation}
    \label{eq:mle_sufficient}
    \lim_{n\to\infty} \int p(d^{(n)}\vert\theta) \log \frac{p^{*}(\theta\vert d^{(n)})}{p^{*}(\theta\vert \Hat{\theta}_{n})} d d^{(n)} = 0 \,,
\end{equation}
respectively, in order to simplify the form of the reference prior. Starting with equation \ref{eq:reference_fh}, we can write
\begin{equation}
    \label{eq:reference_mle}
    \begin{split}
        f_{n}^{*}(\theta) &= \exp \left\lbrace \int p(d^{(n)}\vert\theta)\log p^{*}(\theta\vert d^{(n)})d d^{(n)}\right\rbrace \\
        &= \exp \left\lbrace \int p(d^{(n)}\vert\theta)\log p^{*}(\theta\vert \Hat{\theta}_{n})d d^{(n)}\right\rbrace \\
        &= \exp \left\lbrace \int p(\Hat{\theta}_{n}\vert\theta)\log p^{*}(\theta\vert \Hat{\theta}_{n}) d \Hat{\theta}_{n}\right\rbrace \\
        &= \exp \left\lbrace \log p^{*}(\theta\vert \Hat{\theta}_{n})\Big\vert_{\Hat{\theta}_{n}=\theta} \right\rbrace \\
        &= p^{*}(\theta\vert \Hat{\theta}_{n})\Big\vert_{\Hat{\theta}_{n}=\theta}\,.
    \end{split}
\end{equation}
Therefore, the asymptotically normal form of the posterior with mean $\Hat{\theta}_{n}$ and variance $n^{-1}\mathcal{I}^{-1}(\theta_{n})$ can be written as
\begin{equation}
    \label{eq:asymptotic_posterior}
    p_{n}^{*}(\theta\vert \Hat{\theta}_{n}) = (2\pi)^{-1/2}n^{1/2}I^{1/2}(\theta_{n}) \exp \left[ -\frac{1}{2}n \mathcal{I}(\theta_{n})(\theta - \theta_{n})^{2}\right]\bigg\vert_{\Hat{\theta}_{n}=\theta}
\end{equation}
Substituting this into equation \ref{eq:reference_mle} we find that
\begin{equation}
    \label{eq:fn_normal}
    f_{n}^{*}(\theta) = (2\pi)^{-1/2}n^{1/2}\mathcal{I}^{1/2}(\theta)\,,
\end{equation}
and using equation \ref{eq:reference_prior_expression} we get
\begin{equation}
    \label{eq:reference_is_jeffreys}
    p(\theta) \propto \mathcal{I}^{1/2}(\theta)\,.
\end{equation}
This means that the reference prior, in asymptotically normal models described by one parameter, is equivalent to the Jeffreys prior. As we will discuss shortly, this is not the case for models with many parameters where the two approaches generally produce different results.

\subsection{Numerical solution}

In many cases, equation \ref{eq:reference_fh} cannot be computed analytically and a numerical solution is required to derive the reference prior. This approach can be applied to one--parameter models and results in a numerical representation of the reference prior in the form of pairs $\lbrace \theta , p(\theta) \rbrace$ of values which can be interpolated and used to define the prior's pdf. The numerical procedure, described below, generally requires that it is computationally possible to simulate data from the sampling distribution (i.e. $d\sim p(d\vert \theta)$) in order to approximate the outer integral of equation \ref{eq:reference_fh}, and use numerical integration (e.g. quadrature) in order to compute the inner integral in the normalisation of the asymptotic posterior.

\begin{algorithm}[ht!]
\caption{Numerical reference prior} \algolabel{reference}
\begin{algorithmic}[1]
\REQUIRE{Values of $\theta_{t}\in\lbrace \theta_{1}, \dots, \theta_{T}\rbrace$ for which to compute the reference prior, a moderate value of $n$ to simulate the asymptotic posterior, number of samples $m$, an arbitrary pseudo--prior (e.g. $h(\theta)=1$)}
\ENSURE{Pairs $\lbrace \theta_{t}, p(\theta_{t})\rbrace $}
\FOR{$t=1$ \TO $T$}
    \FOR{$j=1$ \TO $m$}
        \STATE{Simulate a data set $\lbrace d_{1j}\,\dots, d_{nj}\rbrace \sim p(d\vert \theta_{t})$,}
        \STATE{Compute the integral $c_{j}=\int \prod_{i=1}^{n}p(d_{ij}\vert \theta )h(\theta)d\theta$ numerically, where the integration takes place in the prior domain of $\theta$,}
        \STATE{Evaluate $r_{j}=\log\left[c_{j}^{-1}\prod_{i=1}^{n}p(d_{ij}\vert \theta_{t})h(\theta_{t}) \right]$,}
    \ENDFOR
    \STATE{Compute and store $p(\theta_{t}) = m^{-1}\sum_{j=1}^{m}r_{j}(\theta_{t})$.}
\ENDFOR

\end{algorithmic}
\end{algorithm}

\subsection{Many parameters}

So far, we have only discussed cases where the model has a single parameter $\theta$, in which case the reference prior is identical to the Jeffreys prior under the assumption of asymptotic normality. However, the reference prior procedure can be extended to models with more than one parameter where it generally differs from the Jeffreys prior. 

In the multivariate case, the reference prior can be decomposed as
\begin{equation}
    \label{eq:decomposed_reference_prior}
    p(\theta_{1}, \dots, \theta_{D}) = p(\theta_{D}\vert \theta_{1}, \dots , \theta_{D-1})p(\theta_{D-1}\vert \theta_{1}, \dots , \theta_{D-2})\dots p(\theta_{2}\vert \theta_{1}) p(\theta_{1})\,,
\end{equation}
where $D$ is the number of dimensions and we assumed that the parameters are $\lbrace \theta_{1}, \dots, \theta_{D}\rbrace$, in decreasing degree of ``importance'' or ``relevance''. The specific ordering of the parameters in terms of their importance matters as different arrangements can result in different reference priors. Given the aforementioned parameter arrangement, the reference prior procedure works by sequentially deriving the aforementioned conditional priors in reverse order, starting with $p(\theta_{D}\vert \theta_{1}, \dots , \theta_{D-1})$ and ending with $p(\theta_{1})$. Intuitively, this means that we are seeking the reference prior that maximises the missing information about $\theta_{1}$, then $\theta_{2}\vert\theta_{1}$, then $\theta_{3}\vert \theta_{1},\theta_{2}$, and so on.

In practice, we first fix all parameters but $\theta_{D}$ and we estimate $p(\theta_{D}\vert \theta_{1}, \dots , \theta_{D-1})$ by treating the problem as one--dimensional. Assuming that the prior is proper, then $\theta_{D}$ can be marginalised, and the sampling distribution becomes
\begin{equation}
    \label{eq:marginalised_sampling_distribution}
    p(d^{(n)}\vert \theta_{1}, \dots, \theta_{D-1}) = \int p(d^{(n)}\vert \theta_{1}, \dots, \theta_{D}) p(\theta_{D}\vert \theta_{1}, \dots , \theta_{D-1}) d\theta \,,
\end{equation}
The process is then repeated for the next conditional prior $p(\theta_{D-1}\vert \theta_{1}, \dots , \theta_{D-2})$ using $p(d^{(n)}\vert \theta_{1}, \dots, \theta_{D-1})$ as the sampling distribution. After $D$ iterations of the above procedure, all conditional priors are known and the reference prior can be computed as their product according to equation \ref{eq:decomposed_reference_prior}. Although it is possible to use numerical methods in more than one dimension, it often simpler to derive results by employing the asymptotic normality of the posterior distribution when this assumption holds.

\subsubsection{Multivariate reference prior under asymptotic normality}

To derive reference priors, using the asymptotic normality of the posterior distribution, we first need to understand its conditional structure. In particular, we want to know how we can express the variance and precision of each conditional posterior distribution in terms of the components of the covariance and precision matrices of the unconditional posterior distribution. 

Let us assume that the asymptotic posterior distribution can be described as a normal distribution with covariance matrix $\Sigma$ or precision matrix $P=\Sigma^{-1}$. When the conditions of the Bernstein--von Mises theorem are met, the precision matrix can be written as $P=n \mathcal{I}(\Hat{\theta}_{n})$, where $\mathcal{I}$ is the Fisher information matrix, $n$ is the sample size, and $\Hat{\theta}_{n}$ is the MLE. Following the usual conventions, we can identify the elements of those matrices using two indices, that is, $\Sigma_{ij}$ is the element in the intersection of the $i$--th row and $j$--th column.

One way to decompose the asymptotic posterior into its conditionals is
\begin{equation}
    \label{eq:asymptotic_posterior_conditionals}
    \begin{split}
        p^{*}(\theta_{1},\dots,\theta_{D}\vert d^{(n)}) = &p^{*}(\theta_{D}\vert \theta_{1},\dots,\theta_{D-1}, d^{(n)})\\ 
        &\times p^{*}(\theta_{D-1}\vert \theta_{1},\dots,\theta_{D-2}, d^{(n)}) \\
        &\dots p^{*}(\theta_{2}\vert \theta_{1}, d^{(n)}) p^{*}(\theta_{1}\vert d^{(n)})
    \end{split}
\end{equation}

The steps that we need to follow to compute the precision of a conditional $p^{*}(\theta_{j}\vert \theta_{1}, \dots,\theta_{j-1}, d^{(n)})$ are the following:
\begin{enumerate}
    \item Construct the matrix $\Sigma_{j}$ from the upper $j\times j$ sub--matrix of $\Sigma$,
    \item Compute the inverse matrix $P_{j}=\Sigma_{j}^{-1}$,
    \item Drop the rows and columns that correspond to the conditional parameters $\theta_{1},\dots, \theta_{j-1}$. For 1--D conditionals of the form $p^{*}(\theta_{j}\vert \theta_{1}, \dots,\theta_{j-1}, d^{(n)})$, this leaves only the lower right element of $P_{j}$ that we denote as $P_{j*}$ and is equal to the precision of the conditional posterior.
\end{enumerate}

Using the above formula, the reference prior which corresponds to the ordered parameterisation $\lbrace \theta_{1}\,, \dots\,, \theta_{D}$, in terms of importance or relevance, is
\begin{equation}
    \label{eq:ordered_reference_prior}
    p(\theta_{1},\dots,\theta_{D})=p(\theta_{D}\vert \theta_{1},\dots,\theta_{D-1})\dots p(\theta_{2}\vert\theta_{1})p(\theta_{1})\,,
\end{equation}
where
\begin{equation}
    \label{eq:last_parameter_prior}
    p(\theta_{D}\vert \theta_{1},\dots,\theta_{D-1}) \propto P_{D*}^{1/2}(\theta)\,,
\end{equation}
following equation \ref{eq:reference_is_jeffreys}, and for $j=1,\dots,D-1$
\begin{equation}
    \label{eq:all_other_parameters_prior}
    \begin{split}
        p(\theta_{j}\vert \theta_{1},\dots,\theta_{j-1}) \propto \exp\bigg\lbrace &\int \prod_{\ell=j+1}^{D}p(\theta_{\ell}\vert \theta_{1},\dots,\theta_{\ell -1}) \\
        &\times\log P_{j*}^{1/2}(\theta) d\theta_{j+1}\dots d\theta_{D}\bigg\rbrace\,,
    \end{split}
\end{equation}
where we used equations \ref{eq:reference_prior_schematic} and \ref{eq:marginalised_sampling_distribution} to derive the above expression. 

In the special case that the functions $P_{j*}^{1/2}(\theta)$ factorise in the form
\begin{equation}
    \label{eq:p_factorise}
    P_{j*}^{1/2}(\theta) \propto f_{j}(\theta_{j})g_{j}(\theta_{1},\dots,\theta_{j-1},\theta_{j+1},\dots\theta_{D})\,,
\end{equation}
the reference prior is simply
\begin{equation}
    \label{eq:reference_prior_factorise}
    p(\theta_{1},\dots,\theta_{D}) = \prod_{j=1}^{D}f_{j}(\theta_{j})\,.
\end{equation}

\subsubsection{2--D example}

In this example, the joint posterior distribution $p(\theta_{1}, \theta_{2}\vert d^{(n)})$ is asymptotically normal with precision matrix $P=n\mathcal{I}(\theta_{n})$ and covariance matrix $\Sigma = P^{-1}$. Without loss of generality, we can order the parameters in increasing importance or relevance as $\lbrace \theta_{1}, \theta_{2} \rbrace$ and seek to find the reference prior $p(\theta_{1}, \theta_{2})=p(\theta_{2}\vert\theta_{1})p(\theta_{1})$. According to equation \ref{eq:last_parameter_prior}, the conditional prior $p(\theta_{2}\vert\theta_{1})$ is given by
\begin{equation}
    \label{eq:conditional_prior_2d}
    p(\theta_{2}\vert\theta_{1}) \propto P_{2*}^{1/2}(\theta_{1},\theta_{2}) \propto \mathcal{I}_{22}^{1/2}(\theta_{1}, \theta_{2})\,.
\end{equation}
The marginal prior $p(\theta_{1})$ can be derived using equation \ref{eq:all_other_parameters_prior}, and it is given by
\begin{equation}
    \label{eq:marginal_prior_2d}
    p(\theta_{1}) \propto \exp \left\lbrace \int p(\theta_{2}\vert\theta_{1})\log P_{1*}^{1/2}(\theta_{1},\theta_{2})d\theta_{2}\right\rbrace\,,
\end{equation}
where $P_{1*}^{1/2}(\theta_{1},\theta_{2}) = P_{11}-P_{12}P_{22}^{-1}P_{21} \propto \mathcal{I}_{11}-\mathcal{I}_{12}\mathcal{I}_{22}^{-1}\mathcal{I}_{21}$.

So far we have not specified any particular model for this example. In other words, the aforementioned equations hold for any 2--D likelihood function $p(d^{(n)}\vert \theta_{1}, \theta_{2})$. To make the example more specific, we choose the sampling distribution to be normal with the likelihood function parameterised by the mean $\theta_{1}=\mu$ and standard deviation $\theta_{2}=\sigma$, that is,
\begin{equation}
    \label{eq:likelihood_function_gaussian_2d}
    p(d\vert\mu,\sigma) = \mathcal{N}(d\vert\mu,\sigma)\,.
\end{equation}
Substituting the above equation into the definition of the Fisher information matrix given by
\begin{equation}
    \label{eq:fisher_matrix_once_more}
    \mathcal{I}_{ij}(\mu, \theta) = -\int p(d\vert\mu,\sigma) \frac{\partial^{2}\log p(d\vert\mu,\sigma)}{\partial \theta_{i} \partial \theta_{j}} dd\,,
\end{equation}
leads to
\begin{equation}
    \label{eq:fisher_matrix_mu_sigma}
    \mathcal{I}_{ij}(\mu, \theta) = 
    \begin{pmatrix}
        \sigma^{-2} &  0 \\
        0 & 2\sigma^{-2}
    \end{pmatrix}\,.
\end{equation}
It follows directly that the terms $P_{j*}^{1/2}$ are given by
\begin{equation}
    \label{eq:p_j_terms}
    P_{1*}^{1/2}(\mu,\theta) = \sigma^{-1}\,,\quad P_{2*}^{1/2}(\mu,\theta) = \sqrt{2}\sigma^{-1}\,.
\end{equation}
We notice that the above terms factorise into the form of equation \ref{eq:p_factorise}, thus the reference prior is simply
\begin{equation}
    \label{eq:rererence_prior_2d}
    p(\mu,\sigma) = p(\sigma\vert \mu) p(\mu) \propto \sigma^{-1} \times 1 \propto \sigma^{-1}\,.
\end{equation}

It is worth noting that, the alternative ordering of the parameters (i.e. $\theta_{1}=\sigma$ and $\theta_{2}=\mu$), which prioritises $\sigma$ over $\mu$, results in the same reference prior in this example. Furthermore, in this case, the bivariate reference prior $p_{R}(\mu,\sigma)=\sigma^{-1}$ differs markedly from the corresponding Jeffreys prior $p_{J}(\mu,\sigma)=\sigma^{-2}$. Indeed, even Jeffreys himself criticised his multivariate method, which is known to lead to marginalisation paradoxes~\parencite{dawid1973marginalization}.

\section{Weakly informative and regularisation priors}

All options that were discussed so far consist of automated methods of generating prior distributions. There is however another class of priors that is distinctly different in purpose than the ones presented above. Those are the weakly informative priors. 

In most analyses, we have some limited prior information about the range and possible values that a parameter can take based on domain expertise and the model assumptions. For instance, when constraining the mass of an elementary particle we know that it must be smaller than the mass of macroscopic objects and at the same time it has to be greater than or equal to zero. This sort of weakly informative knowledge, although not as well quantified as that in the case of \textit{Jeffreys} and \textit{maximum entropy} priors, can still be included in a Bayesian analysis with the hope of guiding the computation by providing regularisation without significantly affecting the outcome. Weakly informative and regularisation priors are used very often in practice, mostly in cases where the data are very informative and the posterior concentrates to a distribution approaching a multivariate normal in accordance with the \textit{Bernstein--von Mises} theorem~\parencite{van2000asymptotic}.

\section{Informative priors}

Finally, the last class of priors are the \textit{informative priors} the purpose of which is, unlike \textit{Jeffreys} and \textit{maximum entropy} priors which attempt to minimise the amount of prior information, to include and take into account useful information for an analysis. They are often highly concentrated in parameter space and might have been the outcome (i.e. in the form of a posterior distribution) of a previous experiment or analysis of older data. Their aim is clearly to inform the analysis and often no attempt is made to restrict the amount of information provided.

%************************************************
% !TEX TS-program = pdflatex
% !TEX root = ../ArsClassica.tex

%************************************************
\chapter{Making predictions and evaluating models}
\label{chp:predictions}
%************************************************

\begin{flushright}
\itshape
Tomorrow belongs to those who can hear it coming. \\
\medskip
--- David Bowie
\end{flushright}

\section{Making predictions}

Making predictions is a paramount task for most scientific analyses. Often the parameters of a model are not observable quantities and we have to rely on simulated data to assess the validity of our models. In this section, we will discuss how different kinds of predictive checks can help us avoid various common pitfalls in Bayesian analyses.

\subsection{Prior predictive checks}

A very useful practice, that is always recommended, is to check the predictions of the prior distribution under the specified model~\parencite{gelman2013bayesian}. Prior predictive checks constitute an elegant way of finding out what kind of data are compatible (i.e. can be described or explained) by our choice of prior and model. The main benefits of this approach are two. First of all, this can help diagnose priors that are either too restrictive or too wide. Furthermore, assuming that the choice of prior distribution is justified, prior predictive checks can help shield against severe cases of \textit{model misspecification} in which no specific set of parameters corresponds to a model that describes the observed data sufficiently well.

In order to assess whether a particular choice of prior distribution is appropriate we need a way to produce simulated data that are consistent with the prior. The Bayesian way of doing this is by sampling the simulated data
\begin{equation}
    \label{eq:samples_from_prior_predictive}
    d_{sim} \sim p(d)\,,
\end{equation}
from the \textit{prior predictive distribution} 
\begin{equation}
    \label{eq:prior_predictive_distribution}
    p(d)=\int p(d\vert\theta)p(\theta)d\theta\,.
\end{equation}
Generating simulated data using the prior predictive distribution in practice can be done easily by first simulating parameters from the prior distribution
\begin{equation}
    \label{eq:sample_from_prior}
    \theta_{sim} \sim p(\theta)\,,
\end{equation}
and then simulating the data according to the sampling distribution
\begin{equation}
    \label{eq:sample_from_sampling_distribution}
    d_{sim} \sim p(d\vert\theta_{sim})\,,
\end{equation}
given the simulated parameters. The simulated pairs $(d_{sim},\theta_{sim})$ constitute samples from the joint distribution
\begin{equation}
    \label{eq:sample_from_joint_distribution}
    (d_{sim},\theta_{sim}) \sim p(d,\theta)\,,
\end{equation}
and thus
\begin{equation}
    \label{eq:samples_from_prior_predictive2}
    d_{sim} \sim p(d)\,,
\end{equation}
are simulated from the prior predictive distribution.

\subsection{Posterior predictive checks}

Similarly to the prior predictive checks, but this time conditioned on the observed data $d_{obs}$, one can perform \textit{posterior predictive checks} \parencite{gelman2013bayesian}. The latter offer a way of measuring whether a model is able to capture aspects of the data sufficiently well. Just like prior predictive checks that simulate data consistent with the prior, posterior predictive checks on the other hand simulate data that are consistent with the posterior distribution. 

In practice, the process of generating simulated data
\begin{equation}
    \label{eq:sample_from_posterior_predictive}
    d_{sim}\sim p(d\vert d_{obs})\,,
\end{equation}
from the posterior predictive distribution
\begin{equation}
    \label{eq:posterior_preditive_distribution}
    p(d \vert d_{obs}) = \int p(d\vert\theta)p(\theta \vert d_{obs})d\theta\,,
\end{equation}
starts by simulating parameters from the posterior distribution
\begin{equation}
    \label{eq:sample_from_posterior}
    \theta_{sim} \sim p(\theta\vert d_{obs})\,.
\end{equation}
It is important to remind the reader that this last step, unless conjugate priors are used, is highly non--trivial and often requires advanced computational algorithms that are the subject of the next chapter. For now, it suffices to understand that the simulation of parameters as described by the relation \ref{eq:sample_from_posterior} is possible although generally difficult, requiring careful steps. The last step is to generate the simulated data according to sampling distribution
\begin{equation}
    \label{eq:sample_from_sampling_distribution2}
    d_{sim} \sim p(d\vert\theta_{sim})\,,
\end{equation}
given the simulated parameters. $d_{sim}$ then constitute samples from the posterior predictive distribution.

%************************************************

\section{Evaluating and comparing models}

A key goal of science is to determine which model under consideration better accounts for the observed data. As we will discover shortly, this is generally done by assessing the predictive power of different models. From a Bayesian perspective, there are two ways one can approach this subject. The first uses \textit{Bayes factors} and compares models based on their \textit{prior predictive performance}, meaning their capacity to explain the observed data using only the information encoded in the prior distribution. On the other hand, the second approach uses the notion of \textit{cross--validation} in order to compare models based on their \textit{posterior predictive performance}, meaning their ability to make predictions for out--of--sample data, meaning, future or unseen data, using what we learned from the observed data.

In the \textit{prior predictive} approach, the main quantity that goes into the calculation of the \textit{Bayes factor} is the \textit{prior predictive probability} $p(d\vert \mathcal{M}_{i})$ of the observed data $d$ given a model $\mathcal{M}_{i}$, also known as the \textit{marginal likelihood} or the model \textit{evidence} \parencite{jaynes2003probability, gregory2005bayesian}. Naturally, the \textit{prior predictive} approach is sensitive to the choice of priors. On the other hand, in the \textit{posterior predictive} approach, we compute the \textit{posterior predictive probability} of some subset of the observed data given the rest of the data. Typically, \textit{cross--validation} means that this process is repeated several times, trying to predict different subsets of data, until the entire data set is assessed as held--out data.

\subsection{Bayes factors}

The probability of a model $\mathcal{M}_{i}$ given the data $d$ can be computed using Bayes' theorem
\begin{equation}
    \label{eq:probability_of_model_given_data}
    p(\mathcal{M}_{i}\vert d) = \frac{p(d\vert\mathcal{M}_{i})p(\mathcal{M}_{i}) }{p(d)}\,,
\end{equation}
where $p(d\vert\mathcal{M}_{i})$ is the probability of the data given the model, $p(\mathcal{M}_{i})$ is the prior probability of the model, and $p(d)$ is the prior predictive probability of the data. We can compare two models, $\mathcal{M}_{i}$ and $\mathcal{M}_{j}$, by computing their \textit{odds ratio}
\begin{equation}
    \label{eq:model_odds_ratio}
    \frac{p(\mathcal{M}_{i}\vert d)}{p(\mathcal{M}_{j}\vert d)} = \frac{p(d\vert\mathcal{M}_{i})p(\mathcal{M}_{i})}{p(d\vert\mathcal{M}_{j})p(\mathcal{M}_{j})}
\end{equation}
The ratio $BF_{ij}=p(d\vert\mathcal{M}_{i})/p(d\vert\mathcal{M}_{j})$ in the above expression is called the \textit{Bayes factor}. Once the model priors $p(\mathcal{M}_{i})$ and $p(\mathcal{M}_{j})$ are specified, model comparison using Bayes factors amounts to the calculation of the model evidences $p(d\vert\mathcal{M}_{i})$ and $p(d\vert\mathcal{M}_{j})$.

\begin{figure}[ht!]
    \centering
	\centerline{\includegraphics[scale=0.65]{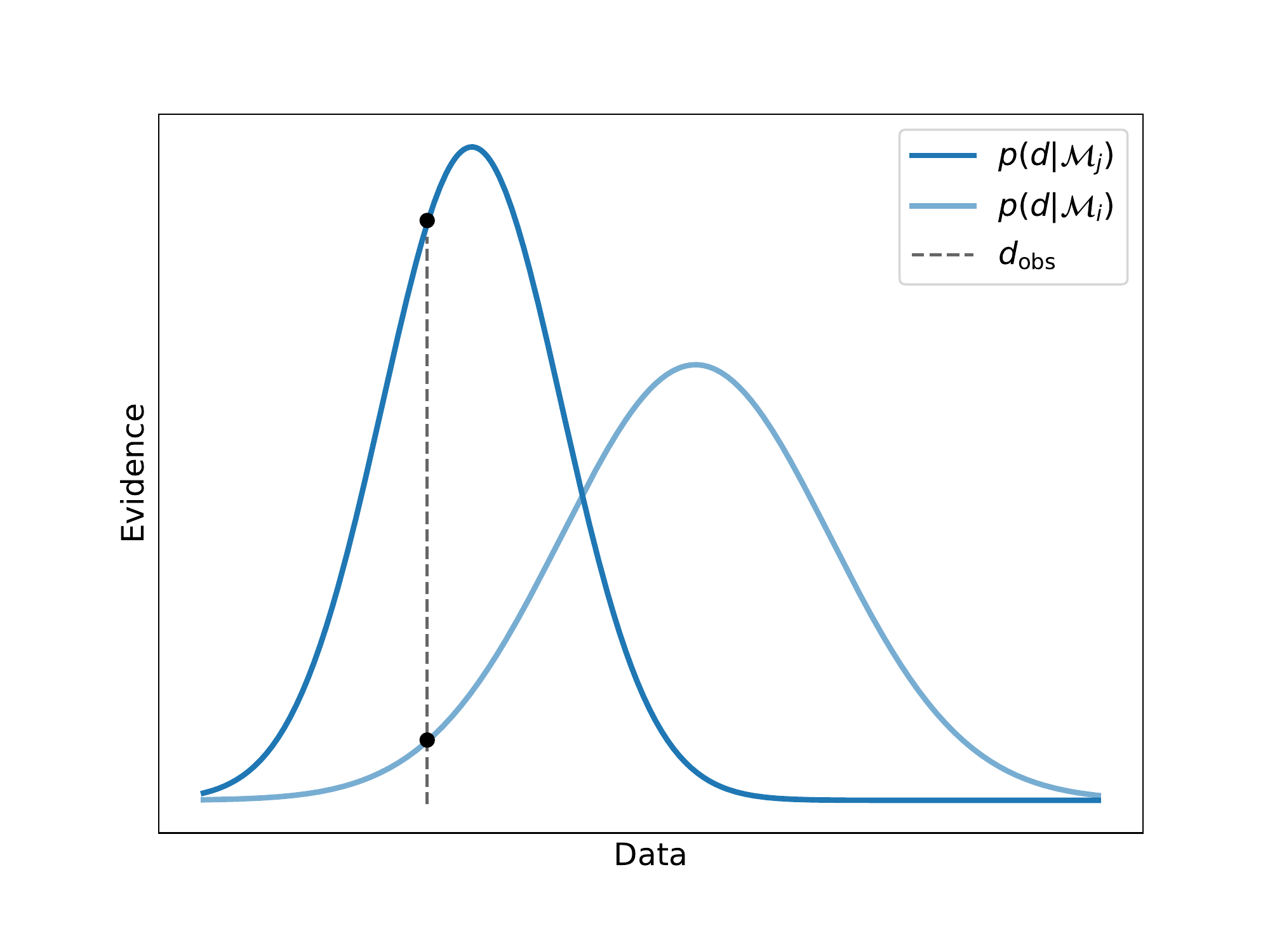}}
    \caption{\textit{Prior predictive distributions} $p(d\vert\mathcal{M}_{i})$ and $p(d\vert\mathcal{M}_{j})$ for models ${M}_{i}$ and ${M}_{j}$ respectively. The dashed line, that intersects both distributions, corresponds to the actual observed data. The Bayes factor is simply the ration between the values at the two points of intersection, in this case favouring ${M}_{j}$ over ${M}_{i}$. It is clear that for other realisations of the actual observed data (e.g. on the right part of the data vector) the other model would be favoured.}
    \label{fig:bayes_factor}
\end{figure}

Despite the apparent simplicity of model comparison using Bayes factors, caution must be exercised when applying the method in real analyses. There are three main reasons for this warning, all of which are sometimes overlooked in practice leading to catastrophic results.

The first reason has to do with the computational difficulty of estimating the model evidence $p(d\vert \mathcal{M})$, particularly in problems with many parameters. In fact, as we will see in detail in the next part of this thesis, a large collection of methods have been designed with the sole purpose of estimating the model evidence. Therefore, the practitioner has to be familiar with the range of applicability of each method as well as their intrinsic limitations when deciding which technique to use.

The second reason, equally important with the first, is the sensitivity of the model evidence to the choice of prior distribution. This sensitivity is apparent if we just notice that the model evidence is simply the \textit{prior predictive distribution},
\begin{equation}
    \label{eq:prior_predictive_distribution_2}
    p(d\vert\mathcal{M}) = \int p(d\vert\theta,\mathcal{M})p(\theta\vert \mathcal{M})d\theta\,,
\end{equation}
evaluated at the observed data $d$. However, we argue that this sensitivity is not a weakness of the method as it is often portrayed, but a strength that needs to be properly understood.

\begin{figure}[H]
    \centering
	\centerline{\includegraphics[scale=0.65]{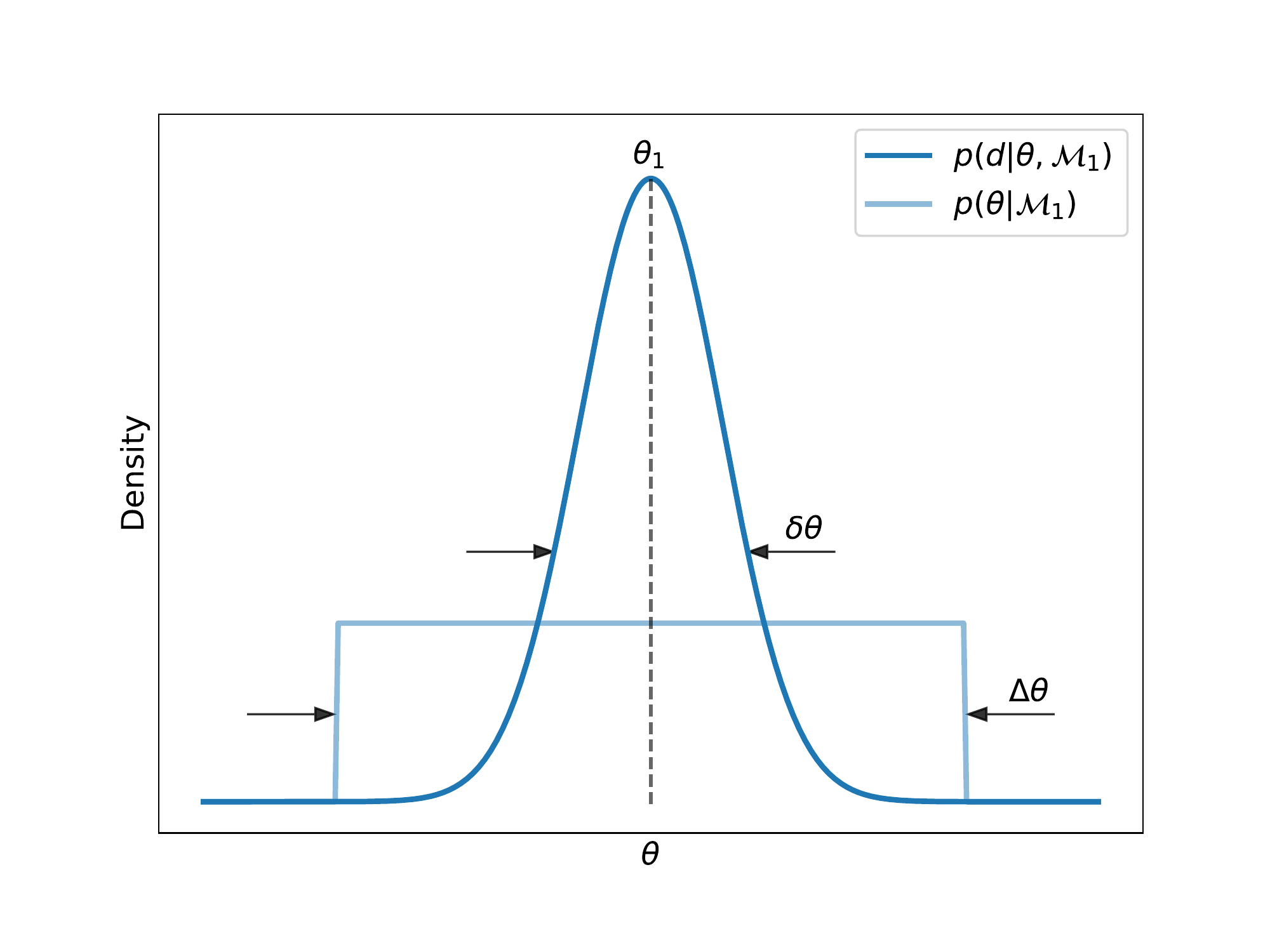}}
    \caption{The characteristic width $\delta \theta$ of the likelihood function $p(d\vert\theta,\mathcal{M}_{1})$ and $\Delta \theta$ of the prior distribution.}
    \label{fig:model_evidence}
\end{figure}

In order to understand the effects of the choice of priors on the Bayes factor, let us consider a simple example. Imagine that we have to compare two models, $\mathcal{M}_{1}$ with a single scalar parameter $\theta$ and $\mathcal{M}_{0}$ with no free parameters. Furthermore, let us assume that $\mathcal{M}_{0}$ is nested in $\mathcal{M}_{1}$, meaning that the more complex model, $\mathcal{M}_{1}$, reduces to the simpler one, $\mathcal{M}_{0}$, for a specific parameter value, $\theta=\theta_{0}$.

Let us now assume that the prior on parameter $\theta$ is flat or uniform, such that,
\begin{equation}
    \label{eq:flat_prior_theta}
    p(\theta\vert \mathcal{M}_{1}) = \frac{1}{\Delta \theta}\,,
\end{equation}
and that the likelihood function is sharply peaked around a value $\theta_{1}$ such that,
\begin{equation}
    \label{eq:sharply_peaked_likelihood}
    \int p(d\vert \theta, \mathcal{M}_{1}) d\theta = p(d\vert \theta_{1}, \mathcal{M}_{1})\times \delta \theta\,,
\end{equation}
where $\delta \theta$ is its characteristic width. It is easy to show that for the case of Gaussian likelihood, centred around $\theta_{1}$, the characteristic width is simply $\delta\theta = \sqrt{2\pi}\sigma$, where $\sigma$ is the standard deviation.

The model evidence of $\mathcal{M}_{1}$ is then simply,
\begin{equation}
    \label{eq:model_evidence_m1}
    \begin{split}
        p(d\vert \mathcal{M}_{1}) &= \int p(d\vert \theta, \mathcal{M}_{1}) p(\theta\vert \mathcal{M}_{1}) d\theta \\
        &= p(d\vert \theta_{1}, \mathcal{M}_{1})\times \frac{\delta\theta}{\Delta \theta}\,.
    \end{split}
\end{equation}
Since the model $\mathcal{M}_{0}$ has no free parameters, no integration is required and its model evidence $p(d\vert\mathcal{M}_{0})$ is simply the likelihood function of $\mathcal{M}_{1}$ evaluated at $\theta=\theta_{0}$, or,
\begin{equation}
    \label{eq:model_evidence_m0}
    p(d\vert \mathcal{M}_{0}) = p(d\vert\theta_{0},\mathcal{M}_{1})\,.
\end{equation}
Therefore, the Bayes factor is,
\begin{equation}
    \label{eq:bayes_factor_approximate}
    B_{10} = \frac{p(d\vert \mathcal{M}_{1})}{p(d\vert \mathcal{M}_{0})} = \frac{p(d\vert \theta_{1}, \mathcal{M}_{1})}{p(d\vert\theta_{0},\mathcal{M}_{1})} \times \frac{\delta\theta}{\Delta \theta}\,.
\end{equation}
The first term in equation \ref{eq:bayes_factor_approximate} is the likelihood ratio that always favours the most complex model $\mathcal{M}_{1}$ since it contains $\mathcal{M}_{0}$ as a special case. In other words, the first term is always greater than one as the most complex model can always fit the data better than the simpler one.

On the other hand, the second term in equation \ref{eq:bayes_factor_approximate} that consists of the ratio of the likelihood width $\delta\theta$ to the prior width $\Delta\theta$ penalises the most complex model $\mathcal{M}_{1}$, since $\delta\theta < \Delta \theta$, for any ``wasted'' regions of parameter space that are ruled out by the data. This term quantifies the so--called \textit{principle of parsimony} or \textit{Occam's razor} as it most commonly known. The principle, often attributed to \textit{William of Ockham}, states that ``entities should not be multiplied beyond necessity'', meaning that between competing models or hypotheses the simplest one is often preferred. Therefore, the Bayes factor will only favour the most complex model (i.e. $\mathcal{M}_{1}$) only if the likelihood ratio is large enough to overcome the penalty introduced by \textit{Occam's razor}. This intrinsic property of Bayes factors to prefer simpler models, that originates directly from the reliance to the prior distributions, is what makes them so useful in practice.

Now that we understand how sensitive the Bayes factor is to the choice of priors we can discuss some ways that we can shield our analyses against potential problems. First and foremost, Bayesian model comparison can be performed only when \textit{proper} priors are used. By that we mean that \textit{improper} priors such as uniform/flat priors ranging from $-\infty$ to $+\infty$ are not acceptable. Only prior distributions that can be integrated and normalised to unity are \textit{proper} in this sense. However, the use of \textit{proper} priors is not enough, the choice of priors needs to be well--justified too. Priors that are not defined using a principled process (e.g. MaxEnt, Jeffreys, etc.), and sometimes even those that do, can lead to significant deviations in the value of a Bayes factor. For this reason, we urge caution not to over--emphasise the significance of, and instead mostly neglect Bayes factors of $\mathcal{O}(1)$. 

The third, and final in our list of reasons, has to do with the open--ended nature of the task model comparison. In particular, model comparison often takes place in the context of a finite set of possible models under investigation with no guarantee whatsoever that one of those models captures perfectly, or even sufficiently, the true data generating process. In that sense, in almost all cases, inference takes place under conditions of \textit{model misspecification}. This brings to mind the saying by \textit{Box}, that ``all models are wrong, but some are useful''. The fact that the value of a Bayes factor might seem to favour one model over another does not mean that the first model is ``correct'', only that it is better than the second. Both models might be far from the true data generating process and the Bayes factor will offer generally no indication of that.

\subsection{Cross--validation}

For a model to be useful in practice it must be able to make accurate predictions regarding unseen data. The generalisation uncertainty of a model is often quantified using some measure of the out--of--sample predictive accuracy. A commonly used scoring rule for the out--of--sample predictive accuracy for $n$ data points is the \textit{expected log--pointwise predictive density},
\begin{equation}
    \label{eq:elpd}
    \mathrm{ELPD} = \sum_{i=1}^{n}\int p_{t}(d_{i})\log p(d_{i}\vert d_{obs}) d d_{i}\,,
\end{equation}
where $p_{t}(d_{i})$ is the probability density of the true data generative process which is in general unknown and $p(d_{i}\vert d_{obs})$ is the posterior predictive density.

Another useful quantity is the \textit{log--pointwise predictive density},
\begin{equation}
    \label{eq:lpd}
    \mathrm{LPD} = \sum_{i=1}^{n}\log p(d_{i}\vert d_{obs}) = \sum_{i=1}^{n}\log \int p(d_{i}\vert \theta) p(\theta\vert d_{obs})d\theta\,.
\end{equation}
LPD of the observed data $d_{obs}$ is an overestimate of ELPD. We can compute LPD in practice as,
\begin{equation}
    \label{eq:lpd_estimate}
    \Hat{\mathrm{LPD}} = \sum_{i=1}^{n}\log \left( \frac{1}{J}\sum_{j=1}^{J}p(d_{i}\vert\theta_{j})\right)\,,
\end{equation}
where $\theta_{j}\sim p(\theta\vert d_{obs})$ are samples from the posterior distribution.

\subsubsection{Leave--one--out cross--validation}

The term cross--validation refers to the practice of estimating the out--of--sample predictive accuracy of a model. In general, the method requires running the analysis multiple times, each time excluding a different portion of the data. The excluded part of the data is then used in order to assess the predictive accuracy of the model. Once the whole dataset is covered, the total accuracy is computed as the average accuracy over all runs,
\begin{equation}
    \label{eq:elpd_loo}
    \mathrm{ELPD}_{\mathrm{LOO}} = \sum_{i=1}^{n}\log p(d_{i}\vert d_{-i})\,,
\end{equation}
where,
\begin{equation}
    \label{eq:loo_predictive_density}
    p(d_{i}\vert d_{-i}) = \int p(d_{i}\vert \theta) p(\theta\vert d_{-i})d\theta \,,
\end{equation}
is the leave--one--out predictive density given the data without the $i$--th datapoint~\parencite{geisser1979predictive, bernardo2009bayesian, gneiting2007strictly}.

Assuming that the $n$ datapoints are conditionally independent in the data generative model, then we can approximate equation \ref{eq:loo_predictive_density} using draws from the posterior $\theta_{j}\sim p(\theta\vert d_{obs})$ and importance weights~\parencite{gelfand1992model},
\begin{equation}
    \label{eq:loo_importance_weights}
    w_{ij} = \frac{1}{p(d_{i}\vert\theta_{j})} \propto \frac{p(\theta_{j}\vert d_{-i})}{p(\theta_{j}\vert d_{obs})}\,,
\end{equation}
leading to the importance sampling leave--one--out predictive density,
\begin{equation}
    \label{eq:isloo_predictive_density}
    p(d_{i}\vert d_{-i}) = \frac{\sum_{j=1}^{J}w_{ij}p(d_{i}\vert\theta_{j})}{\sum_{j=1}^{J}w_{ij}} = \frac{1}{\frac{1}{J}\sum_{j=1}^{J}[p(d_{i}\vert\theta_{j})]^{-1}}\,.
\end{equation}
However the posterior $p(\theta\vert d_{obs})$ is likely to have a smaller variance than then $p(\theta\vert d_{-i})$ distributions leading to insufficient overlap between their typical sets and high--variance importance weights. \textcite{ionides2008truncated} showed that truncating the importance weights,
\begin{equation}
    \label{eq:truncated_importance_weights}
    \Tilde{w}_{ij} = \min\left( w_{ij}, \sqrt{J} \,\Bar{w}_{i} \right)\,,
\end{equation}
where
\begin{equation}
    \label{eq:mean_importance_weight}
    \Bar{w}_{i} = \frac{1}{J} \sum_{j=1}^{J}w_{ij}\,,
\end{equation}
leads to provable finite--variance weights at the cost of introducing bias. \textcite{vehtari2017practical} proposed instead to fit a generalised Pareto distribution to the upper tail of the importance weights, in order to smooth the weights, leading to improved estimates.

\subsubsection{WAIC}

The \textit{Watanabe--Akaike} or \textit{widely applicable information criterion (WAIC)}~\parencite{watanabe2010asymptotic} offers a different way to approximate ELPD and is defined as,
\begin{equation}
    \label{eq:waic}
    \Hat{\mathrm{ELPD}}_{\mathrm{WAIC}} = \Hat{\mathrm{LPD}} - \Hat{p}_{\mathrm{WAIC}}\,,
\end{equation}
where,
\begin{equation}
    \label{eq:p_waic}
    \Hat{p}_{\mathrm{WAIC}} = \sum_{i=1}^{n} \mathrm{Var}_{\theta\sim p(\theta\vert d_{obs})}[\log p(d_{i}\vert \theta)]\,,
\end{equation}
is the estimated effective number of parameters expressed as the posterior variance of the log predictive density of each datapoint. Equation \ref{eq:p_waic} can be computed using posterior samples.

\subsection{Model averaging}

Standard practice ignores model uncertainty and instead focuses on the most probable models as deduced by their Bayes factors. This approach leads to over--confident estimates and ignores the fact that often more than one model can describe the data sufficiently. There is, however, a different approach that we can follow in order to deal with the model uncertainty and properly account for the plethora of plausible models, called \textit{Bayesian model averaging} \parencite{madigan1996bayesian}.

Let $\mathcal{M}_{1}, \mathcal{M}_{2}, \dots, \mathcal{M}_{M}$ be a set of $M$ models with posterior model probabilities $p(\mathcal{M}_{1}\vert d),\, p(\mathcal{M}_{2}\vert d),\, \dots,\, p(\mathcal{M}_{M}\vert d)$ and posterior distributions $p(\theta\vert d, \mathcal{M}_{1}),\,\allowbreak p(\theta\vert d, \mathcal{M}_{2}),\,\allowbreak \dots,\,\allowbreak p(\theta\vert d, \mathcal{M}_{M})$ respectively. Then \textit{Bayesian model averaging} relies on the \textit{marginal} posterior density,
\begin{equation}
    \label{eq:marginal_posterior_density}
    p(\theta\vert d) = \sum_{i=1}^{M} p(\theta\vert d, \mathcal{M}_{i}) p(\mathcal{M}_{i}\vert d)\,,
\end{equation}
which is no longer conditioned on a model.

Moreover, predictions can be made by averaging over all models, weighted proportional to their posterior model probabilities, thereby incorporating model uncertainty using the marginal posterior predictive density,
\begin{equation}
    \label{eq:marginal_posterior_predictive_density}
    p(d\vert d_{\mathrm{obs}} ) = \sum_{i=1}^{M} p(d\vert d_{\mathrm{obs}}, \mathcal{M}_{i}) p(\mathcal{M}_{i}\vert d_{\mathrm{obs}})\,,
\end{equation}
where $d_{\mathrm{obs}}$ are the available observed data and $d$ are the new predicted data. \textcite{madigan1994model} note that averaging over all models in this fashion leads to higher predictive accuracy than using any single model individually.
% Second Part : Bayesian Computation
\cleardoublepage
% !TEX TS-program = pdflatex
% !TEX root = ../ArsClassica.tex

%************************************************
\part{Bayesian Computation}
\label{prt:computation}
%************************************************
% !TEX TS-program = pdflatex
% !TEX root = ../ArsClassica.tex

%************************************************
\chapter{Principles of Bayesian Computation}
\label{chp:computation}
%************************************************

\begin{flushright}
\itshape
Anyone who considers arithmetical methods of producing\\
random digits is, of course, in a state of sin. \\
\medskip
--- John von Neumann
\end{flushright}

This chapter introduces the various methods that are used in practice in order to tackle the computational challenges of Bayesian analyses. We begin this journey by discussing some fundamental ideas about the geometry of high--dimensional probability distributions, while gradually introducing the concepts and algorithms that constitute the modern mathematical machinery of Bayesian computation.

\section{Expectation values}

Probability theory teaches us the only well defined way to extract information from probability distributions is through expectation values. By this term, we mean high--dimensional integrals of the form
\begin{equation}
    \label{eq:expectation_value_integral_formula}
    \mathbb{E}_{p}[f] = \int f(\theta) p(\theta) d\theta\,,
\end{equation}
where $p(\theta)$ is the probability density function that often corresponds to the posterior density for problems of scientific inference, $\theta$ signifies the parameters of the distribution, and $f(\theta)$ is the function that we aim to integrate. In this sense, an expectation value of a function $f(\theta)$ over a probability distribution $p(\theta$) is technically a functional of the product of the function and the probability density.

To see why expectation values hold such a central role in scientific parameter inference, let us discuss a few characteristic and common examples that a scientist often has to compute.

\begin{itemize}
    \item \textbf{Mean value} -- Perhaps the most commonly computed expectation value is the mean value. This can be calculated by choosing the function to be $f(\theta)=\theta$, the expectation value then reduces to
    \begin{equation}
        \label{eq:expectation_value_mean}
        \mu \equiv \mathbb{E}_{p}[\theta] = \int \theta\, p(\theta) d\theta\,.
    \end{equation}
    
    \item \textbf{Variance} -- One might also want to compute higher moments of the probability distribution. The first moment is the variance that corresponds to the following expectation value
    \begin{equation}
        \label{eq:expectation_value_variance}
        \sigma^{2} \equiv \mathbb{E}_{p}\left[(\theta - \mu)^{2}\right] = \int (\theta - \mu)^{2}\, p(\theta) d\theta\,.
    \end{equation}
    
    \item \textbf{Marginal distributions} -- Even marginal distribution can be thought of as expectation values. In this case, the function $f$ corresponds to a conditional distribution, for instance
    \begin{equation}
        \label{eq:expectation_value_marginal}
        p(\phi) \equiv \mathbb{E}_{p}\left[p(\phi |\theta)\right] = \int p(\phi |\theta) p(\theta) d\theta \,.
    \end{equation}
\end{itemize}

\section{Quadrature and uniform grids}

\begin{figure}[ht!]
    \centering
	\centerline{\includegraphics[scale=0.65]{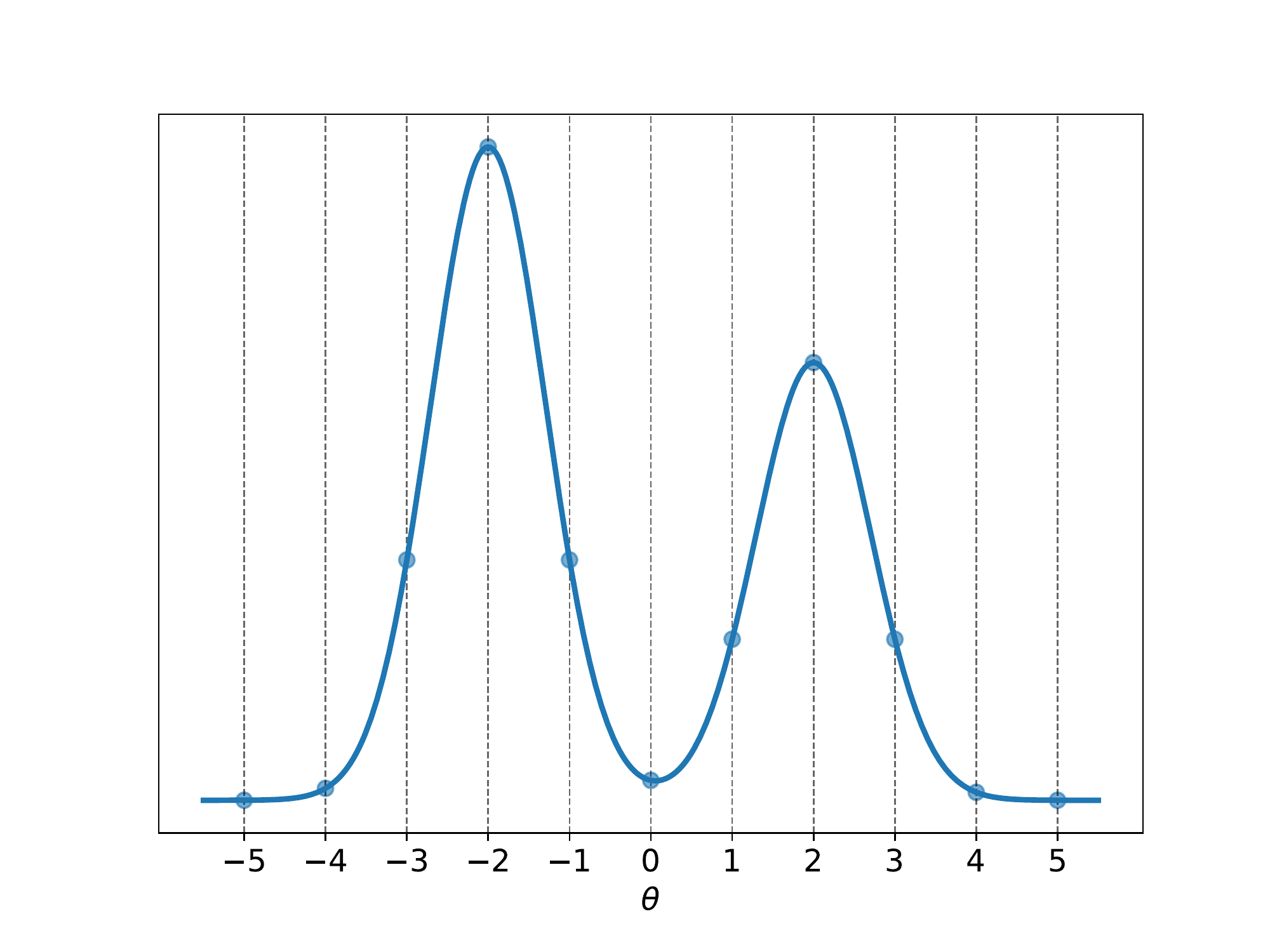}}
    \caption{Uniform grid approximation of 1--dimensional probability distribution.}
    \label{fig:grid_1d}
\end{figure}
Having discussed a number of examples of commonly used expectation values, we can now turn our attention to the methods that are used for their computation. As we mentioned before, these expectation values are defined as high--dimensional integrals. As those integrals are not generally tractable analytically, one might attempt to approximate their value by means of a Riemann sum over a discreet grid of $n$ points:
\begin{equation}
    \label{eq:riemann_sum}
    \mathbb{E}_{p}[f] = \int f(\theta) p(\theta) d\theta \approx \sum_{i=1}^{n}f(\theta_{i})p(\theta_{i})\Delta\theta_{i}\,,
\end{equation}
where
\begin{equation}
    \label{eq:riemann_interval}
    \Delta \theta_{i} = \theta_{j+1} - \theta_{j}\,,
\end{equation}
is simply the interval between two subsequent points, $\theta_{j}$ and $\theta_{j+1}$ on the underlying grid, and
\begin{equation}
    \label{eq:riemann_midpoint}
    \theta_{i} = \frac{\theta_{j+1} + \theta_{j}}{2}\,,
\end{equation}
is just the mid--point between $\theta_{j}$ and $\theta_{j+1}$.

\begin{figure}[ht!]
    \centering
	\centerline{\includegraphics[scale=0.65]{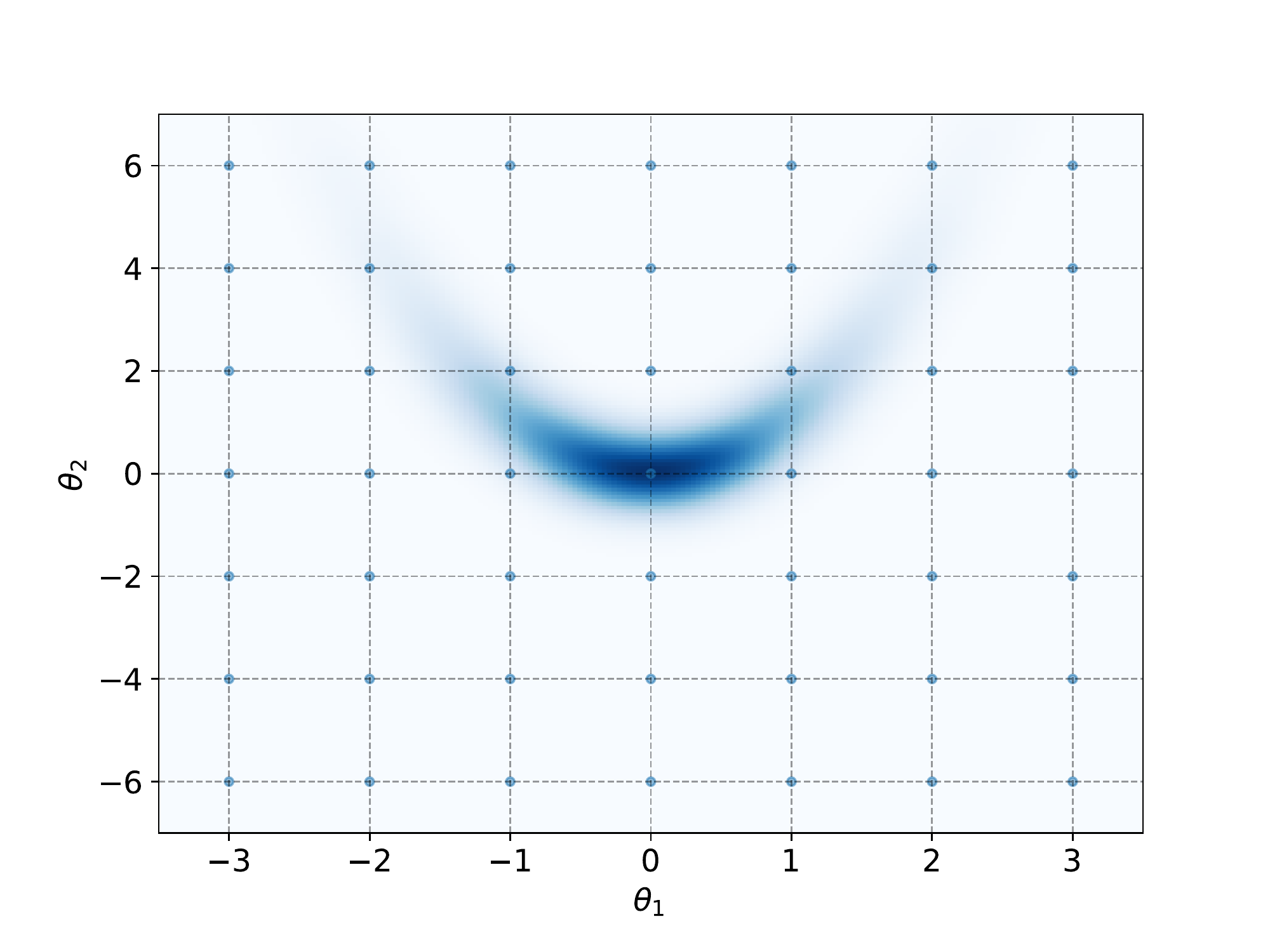}}
    \caption{Uniform grid approximation of 2--dimensional probability distribution.}
    \label{fig:grid_2d}
\end{figure}

In principle, this idea can be extended to high dimensions by replacing the 1--dimensional intervals $\Delta \theta_{i}$ with D--dimensional hypercubes. Figure \ref{fig:grid_2d} shows one such example for a 2--dimensional probability distribution. However, as the number of dimensions increases, one immediately has to face a significant difficulty, the \emph{curse of dimensionality}~\parencite{bellman1966dynamic}. Already in 2 dimensions we require $n^{2}$ grid points to approximate the distribution. As it turns out, the number of grid points required for the evaluation of the Riemann sum increases exponentially with the number of dimensions, rendering this method of computing expectation values unusable for $D>3$. Overcoming the difficulties imposed by the \emph{curse of dimensionality} is one of the key goals of \textit{probabilistic computing}.

\begin{figure}[ht!]
    \centering
	\centerline{\includegraphics[scale=0.65]{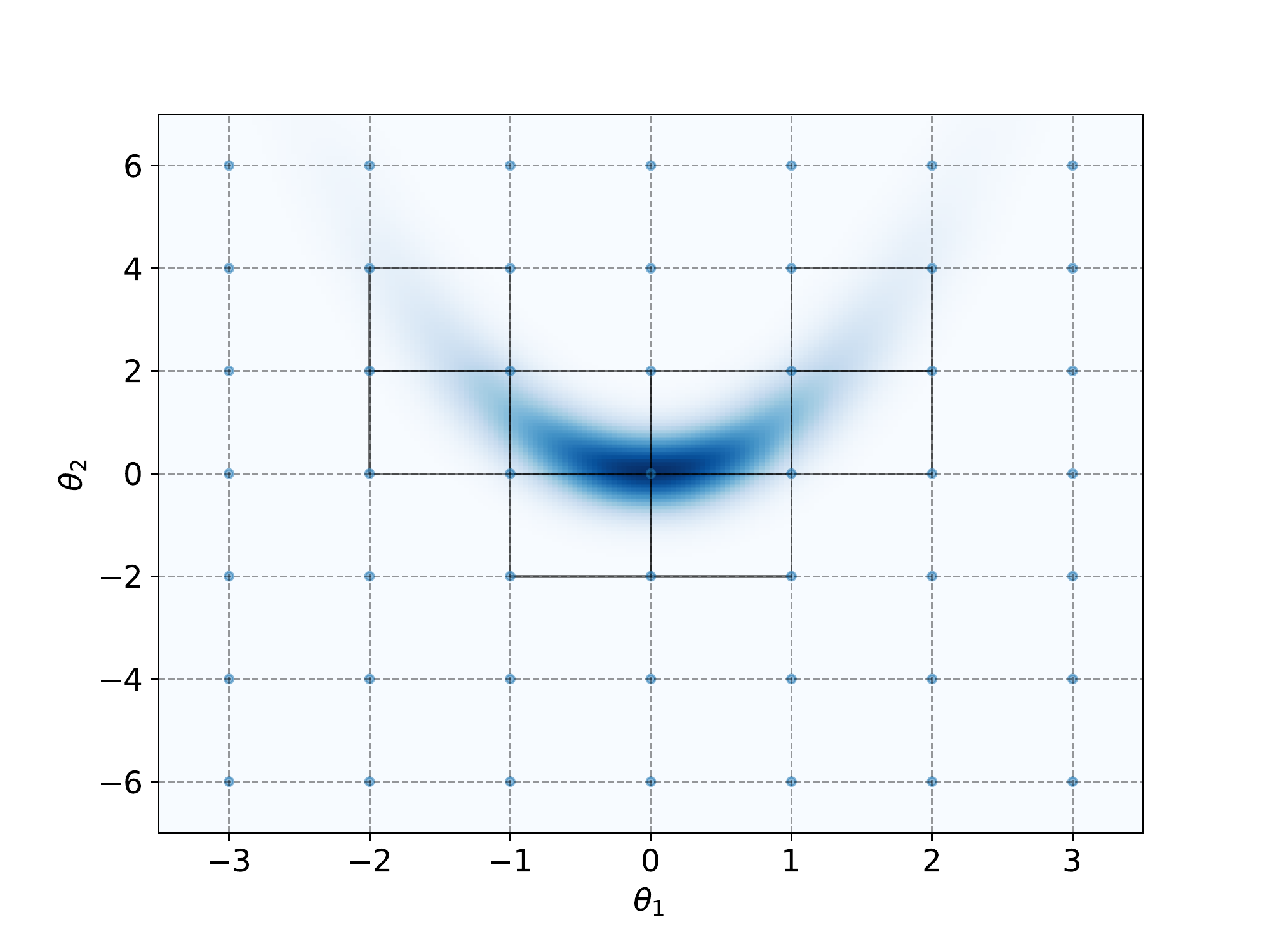}}
    \caption{Uniform grid approximation of 2--dimensional probability distribution with highlighted the grid--cells that actually contribute to the calculation of an expectation value.}
    \label{fig:grid_2d_highlighted}
\end{figure}

In order to reduce the computational cost of estimating expectation values in high dimensions, we need to find a way to focus our effort and computation only on those regions of parameter space that are relevant for the integral that we aim to evaluate. One simple idea would be to remove any points of the grid that the integrand $f(\theta)p(\theta)$ is very close to zero. Applying this technique would certainly reduce the total computational cost since only a few grid--cells have a non--negligible value of $f(\theta)p(\theta)$ as shown in Figure \ref{fig:grid_2d_highlighted}. The problem that we face however is that by focusing our attention on $f(\theta)p(\theta)$ we ignore a key factor in the estimation of any expectation value, the \textit{volume}.

\section{Geometry of high--dimensional spaces}

The concepts of volume and distance in high--dimensional spaces defy our everyday intuition in ways that matter for the computation of expectation values. To understand this, we will go through an example that illustrates these peculiar effects.

Let us assume that we inscribe a circle of radius $R$ inside a square of side $2 R$. We are interested in computing the area of the circle as a fraction of that of the square. We can get to the result easily using basic geometry, in particular, the ratio of the two areas is
\begin{equation}
    \label{eq:area_ratio_sphere_2d}
    \frac{A_{\textrm{circle}}}{A_{\textrm{square}}} = \frac{\pi R^{2}}{(2 R)^{2}} = \frac{\pi}{4}\,.
\end{equation}

We can now extend the same problem into three dimensions, in which we have a sphere of radius $R$ inscribed in a cube of side $2 R$. The ratio of the volume of the sphere to the volume of the cube is simply
\begin{equation}
    \label{eq:volume_ratio_sphere_3d}
    \frac{V_{\textrm{sphere}}}{V_{\textrm{cube}}} = \frac{\frac{4}{3}\pi R^{3}}{(2 R)^{3}} = \frac{\pi}{6}\,.
\end{equation}

By comparing equations \ref{eq:area_ratio_sphere_2d} and \ref{eq:volume_ratio_sphere_3d} one realises that the volume ratio has decreased going from $2$ dimensions to $3$. We will now show that this result in fact holds for any number of dimensions $D$. In $D$ dimensions, the volume of a hyper--sphere is given by
\begin{equation}
    \label{eq:volume_sphere_nd}
    V_{\textrm{sphere}} = \frac{\pi^{D/2}}{\Gamma \left( \frac{D}{2} + 1 \right)} R^{D}\,,
\end{equation}
where $\Gamma$ is Euler's gamma function which extends the factorial operation to non--integer arguments and satisfies
\begin{equation}
    \label{eq:gamma_function_integer}
    \Gamma (D) = (D-1)!\,,
\end{equation}
for positive integer $D$, and
\begin{equation}
    \label{eq:gamma_function_halfinteger}
    \Gamma \left(D+\frac{1}{2}\right) = \left(D-\frac{1}{2}\right)\times\left(D-\frac{3}{2}\right)\times\dots\times\frac{1}{2}\times\pi^{1/2}\,,
\end{equation}
for non--negative integer $D$. 

The volume of a hyper--cube in $D$ dimensions is simply
\begin{equation}
    \label{eq:volume_cube_nd}
    V_{\textrm{cube}} =(2 R)^{D}\,.
\end{equation}
Taking the ratio of the terms of equations \ref{eq:volume_sphere_nd} and \ref{eq:volume_cube_nd} yields
\begin{equation}
    \label{eq:volume_ratio_sphere_nd}
    \frac{V_{\textrm{sphere}}}{V_{\textrm{cube}}} = \frac{\pi^{D/2}}{2^{D}\Gamma \left( \frac{D}{2} + 1 \right)}\,.
\end{equation}

\begin{figure}[ht!]
    \centering
	\centerline{\includegraphics[scale=0.65]{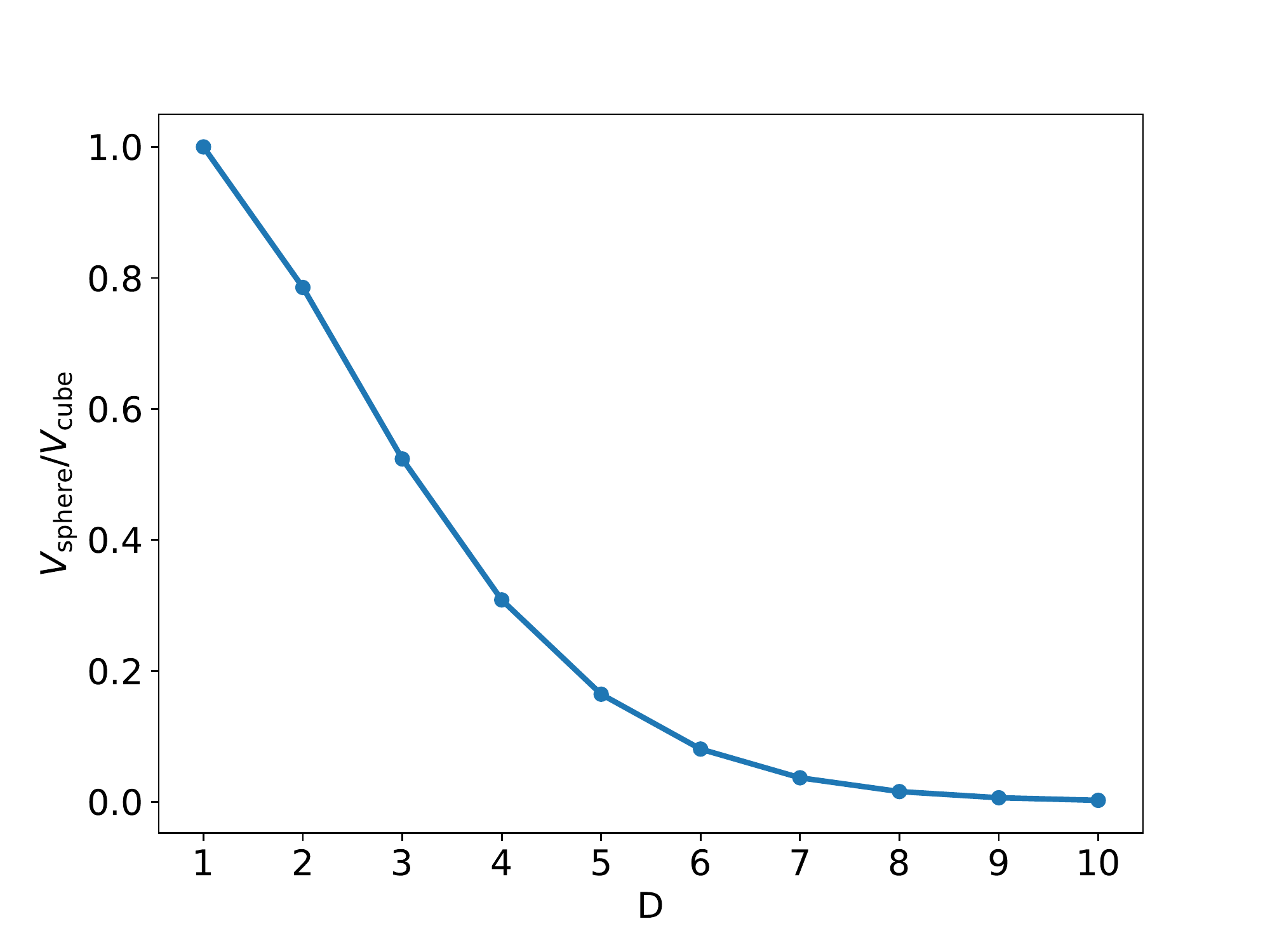}}
    \caption{The ratio of the volume of a hyper--sphere of radius $R$ to the volume of a hyper--cube of edge size $2 R$ as function of the number of dimensions $D$.}
    \label{fig:volume_ratio_sphere_nd}
\end{figure}

Figure \ref{fig:volume_ratio_sphere_nd} shows the ratio of the volume of a hypersphere to a hypercube as a function of the number of dimensions $D$. As the number of dimensions $D$ increases, the volume ratio of equation \ref{eq:volume_ratio_sphere_nd} asymptotically approaches $0$. This means that in high dimensions, almost all of the volume of a hypercube is concentrated in the corners.

\section{Concentration of measure}

As we will see shortly, the strange behaviour of volume is of paramount importance in the calculation of expectation values over probability distributions. To understand this one need to think not about the probability density but instead about the probability mass. When evaluating an expectation value, not all regions of parameter space are contributing equally to the value of the integral. In fact, the contribution from some regions of parameter space dominates the calculation. We only need to take a look into the form of the expectation value integral to notice that is essentially the product of three terms that contributes. These terms are the function $f(\theta )$, the probability density function $p(\theta )$, as well as the differential volume element $d\theta$. In other words, it is the product of these three terms and their dependence on $\theta$ that determines the value of the integral. Assuming that the function $f(\theta )$ is well behaved, we can ignore its presence for a while.

For the sake of simplicity let us assume that the probability distribution is characterised by an D--dimensional Gaussian probability density function
\begin{equation}
    \label{eq:gaussian_pdf}
    p(\theta ) = \det(2\pi\Sigma)^{-\frac{1}{2}} \exp{\left[-\frac{1}{2}(\theta-\mu)^{T}\Sigma^{-1}(\theta-\mu)\right]}\,,
\end{equation}
where $\mu$ is the mean and $\Sigma$ is the covariance matrix of the distribution. Without loss of generality let us also assume that the density is centred at zero (i.e. $\mu = 0$) and the covariance matrix diagonal with the elements of its diagonal equal to $\sigma^{2}$, meaning that equation \ref{eq:gaussian_pdf} simplifies into
\begin{equation}
    \label{eq:gaussian_pdf_unit_variance}
    p(\theta ) = \frac{1}{\sqrt{(2\pi)^{D}}\sigma^{D}}\exp{\left(-\frac{|\theta|^{2}}{2\sigma^{2}}\right)}\,.
\end{equation}
Assuming further spherical coordinates, the density only depends on the magnitude $r$ of the $\theta$ parameter vector
\begin{equation}
    \label{eq:gaussian_pdf_unit_variance_spherical}
    p(r) = \frac{1}{(2\pi)^{\frac{D}{2}}\sigma^{D}}\exp{\left(-\frac{r^{2}}{2\sigma^{2}}\right)}\,.
\end{equation}
Keep in mind that $p(r)$ is not a probability density function of the magnitude $r=\vert\theta \vert$, but a D--dimensional density of $\theta$.

Let us now turn our attention to the differential volume element $dV$. Differentiating equation \ref{eq:volume_sphere_nd} that provides the volume of the hyper--sphere we get
\begin{equation}
    \label{eq:differential_volume_sphere_nd}
    dV = \frac{D \pi^{D/2}}{\Gamma \left( \frac{D}{2} + 1 \right)} r^{D-1} dr\,.
\end{equation}
\begin{figure}[H]
    \centering
	\centerline{\includegraphics[scale=0.65]{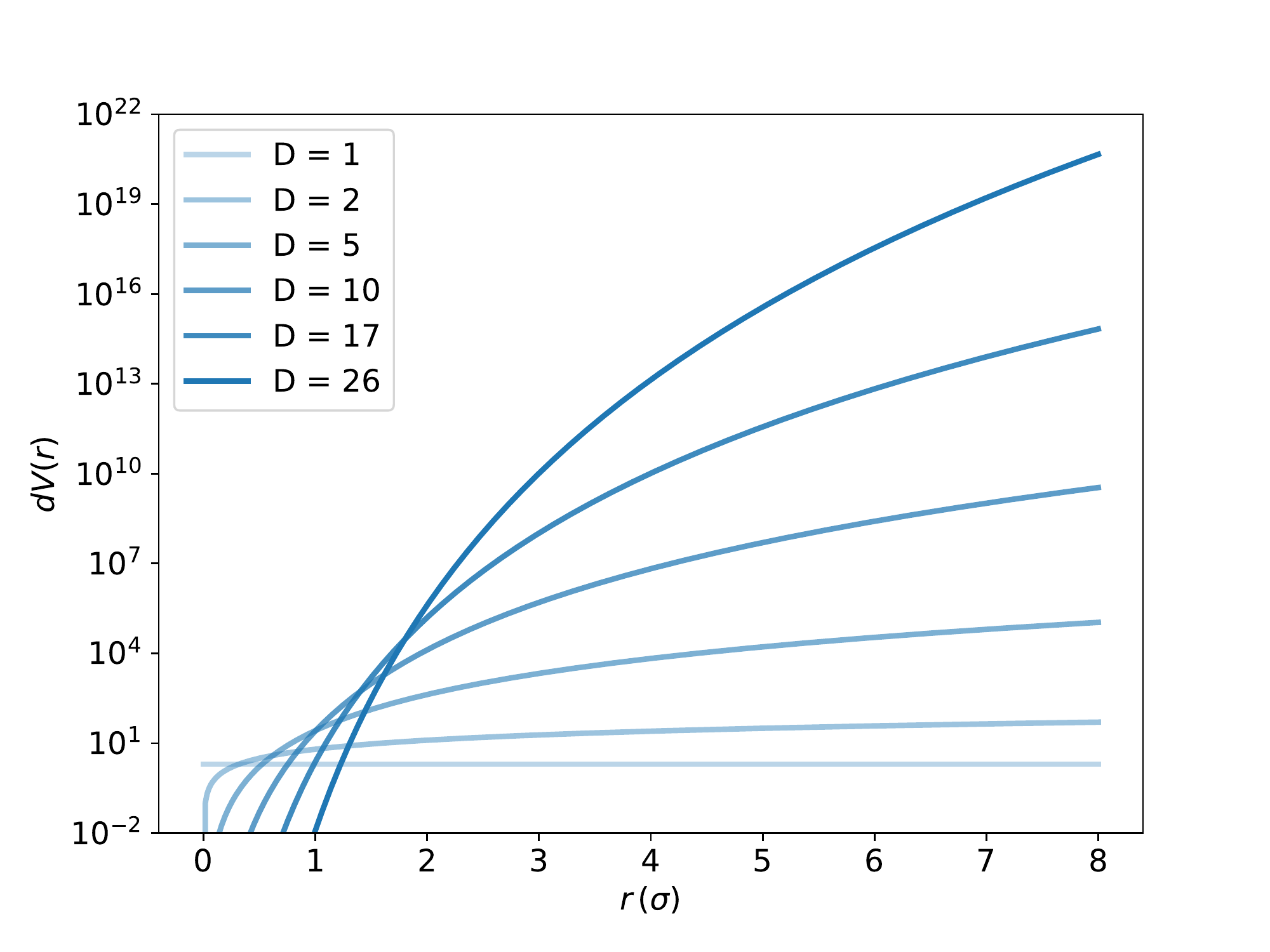}}
    \caption{Scaling of differential volume with the number of dimensions as a function of distance.}
    \label{fig:differential_volume}
\end{figure}

The differential probability mass $dm(r)$ is then just the product of $p(r)$ and $dV$ given by equations \ref{eq:gaussian_pdf_unit_variance_spherical} and \ref{eq:differential_volume_sphere_nd} respectively
\begin{equation}
    \label{eq:differential_mass_sphere_nd}
    dm(r) = \frac{D r^{D-1}}{\Gamma \left( \frac{D}{2} + 1 \right)2^{\frac{D}{2}}\sigma^{D}} \exp{\left(-\frac{r^{2}}{2\sigma^{2}}\right)} dr\,.
\end{equation}
The differential mass $dm(r)$ has a clear physical meaning, that of the probability mass enclosed in a hyper--spherical shell of radius $r$ and width $dr$. The probability mass differential $dm(r)$ peaks (i.e. is maximised) at the typical radius
\begin{equation}
    \label{eq:typical_radius}
    r_{\textrm{peak}} = \sqrt{D-1}\sigma\,.
\end{equation}
Equation \ref{eq:typical_radius} indicates that while in 1--D the probability mass peaks at $r_{\textrm{peak}}=0$, in higher dimensions this is not the case. As the number of dimensions increases the radius in which the probability mass peaks moves to greater distances. Table \ref{tab:typical_radius} shows the typical radius of the probability mass for a different number of dimensions for our problem. This is a direct consequence of the rapid increase of the differential volume for large $r$ values.
\begin{figure}[H]
    \centering
	\centerline{\includegraphics[scale=0.65]{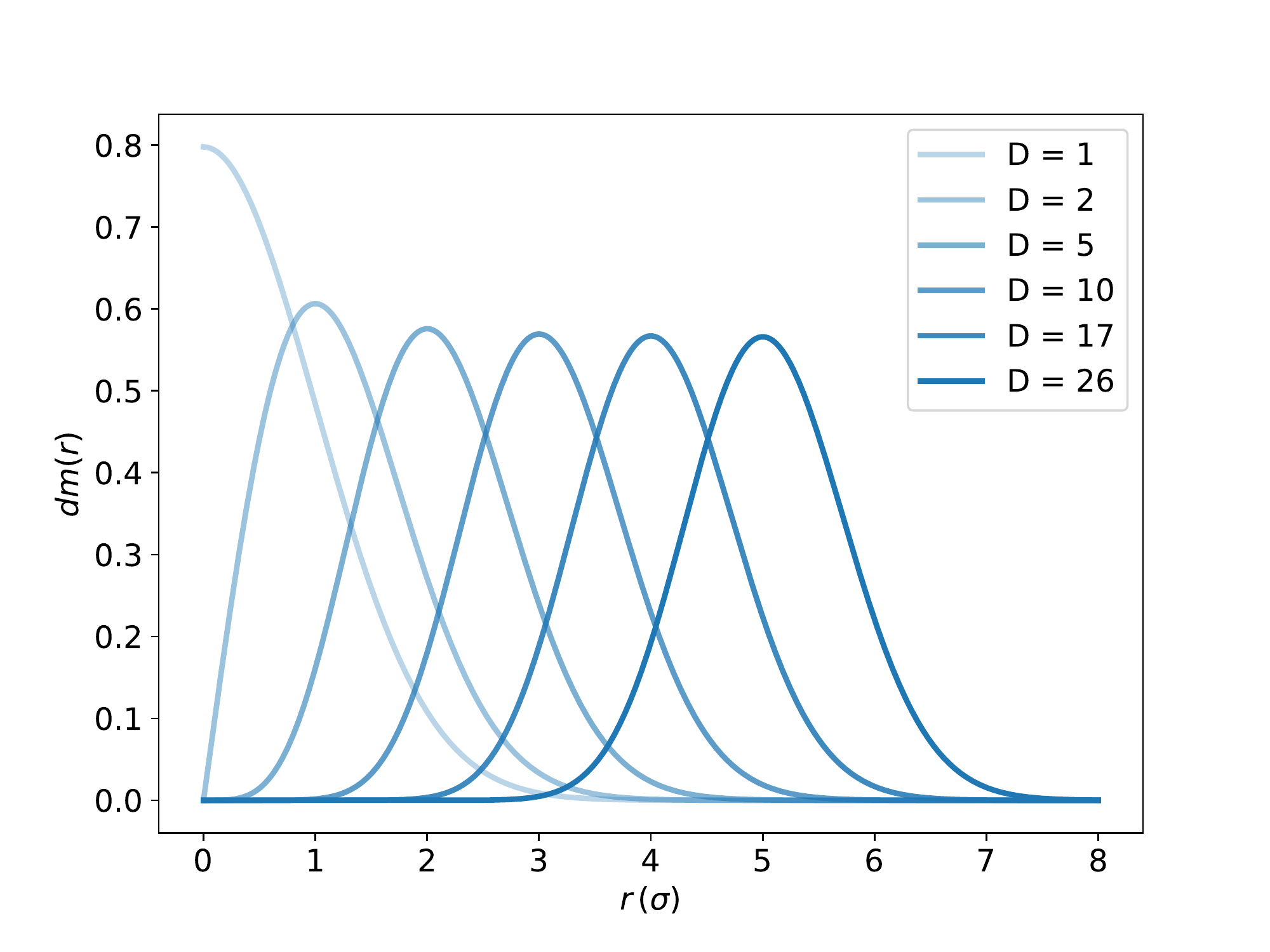}}
    \caption{Scaling of differential probability mass with the number of dimensions as a function of distance.}
    \label{fig:typical_set_unscaled}
\end{figure}

\begin{table}[h!]
    \centering
    \caption{The typical radius $r_{\textrm{peak}}$ as function of the number of dimensions $D$.}
    \def\arraystretch{1.1}
    \begin{tabular}{cc}
        \toprule[0.75pt]
        Number of dimensions $D$ & Typical radius $r_{\textrm{peak}}$ \\
        \midrule[0.5pt]
        $1$ & $0$  \\
        $2$ & $1\sigma$  \\
        $5$ & $2\sigma$  \\
        $10$ & $3\sigma$  \\
        $17$ & $4\sigma$  \\
        $26$ & $5\sigma$  \\
        \bottomrule[0.75pt]
    \end{tabular}
    \label{tab:typical_radius}
\end{table}

In general, we expect the probability mass to form a hyper--shell of mean radius 
\begin{equation}
    \label{eq:shell_mean_radius}
    r_{\textrm{mean}} \equiv \mathbb{E}_{p}[r] = \int_{0}^{+\infty}r dm(r)
\end{equation}
and width (i.e. standard deviation) 
\begin{equation}
    \label{eq:shell_width}
    \Delta r \equiv  \sqrt{\mathbb{E}_{p}[(r-r_{\textrm{mean}})^{2}]} = \sqrt{\int_{0}^{+\infty}(r-r_{\textrm{mean}})^{2} dm(r)}
\end{equation}

\begin{figure}[H]
    \centering
	\centerline{\includegraphics[scale=0.65]{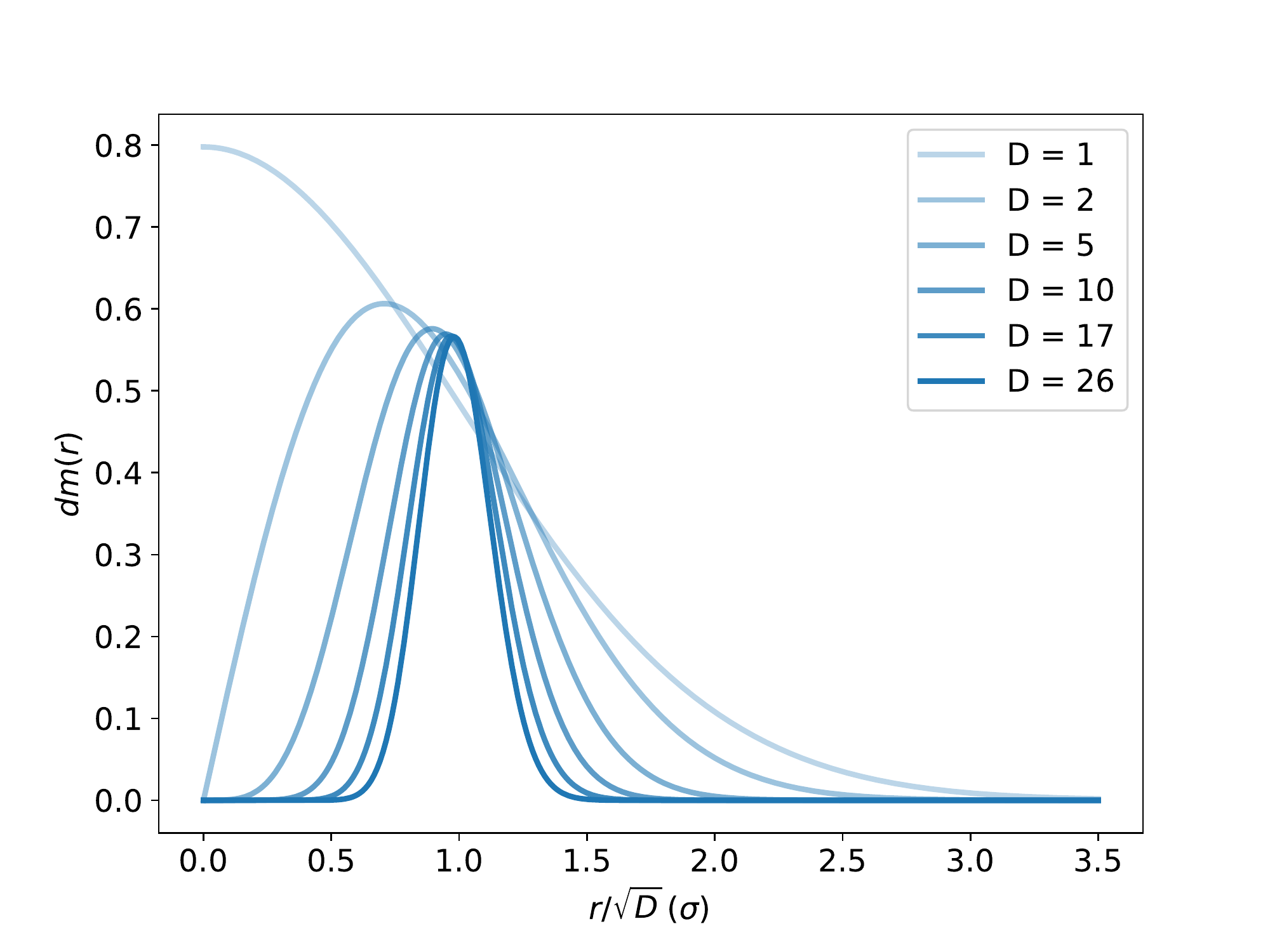}}
    \caption{Scaling of differential probability mass with the number of dimensions as a function of distance normalised by the square root of the number of dimensions.}
    \label{fig:typical_set_scaled}
\end{figure}

\section{Typical set}

\begin{figure}[ht!]
    \centering
	\centerline{\includegraphics[scale=0.65]{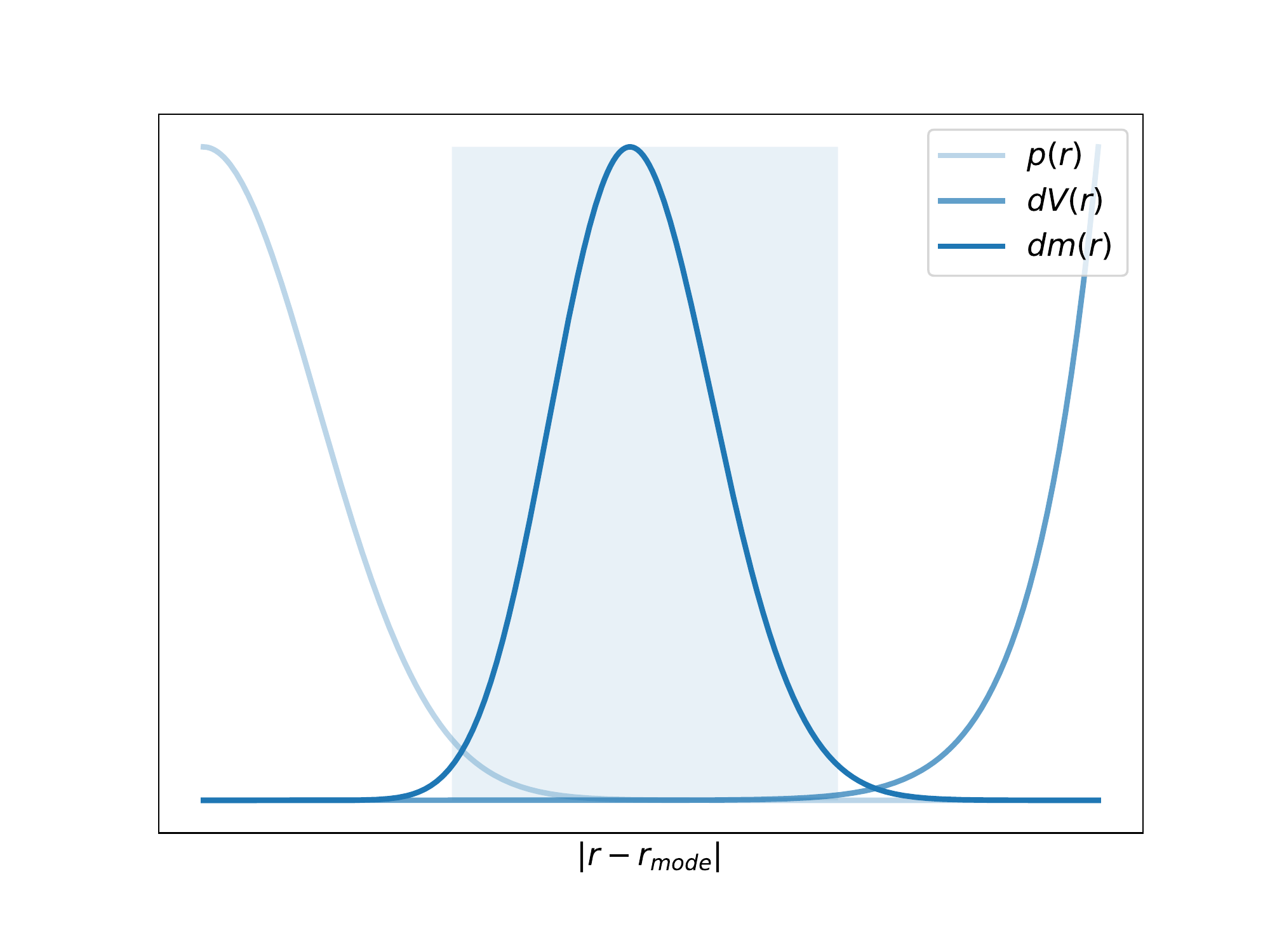}}
    \caption{Illustration of the typical set as the region in parameter space that the product of probability density and differential volume is non--negligible.}
    \label{fig:typical_set}
\end{figure}
\begin{figure}[ht!]
    \centering
	\centerline{\includegraphics[scale=0.65]{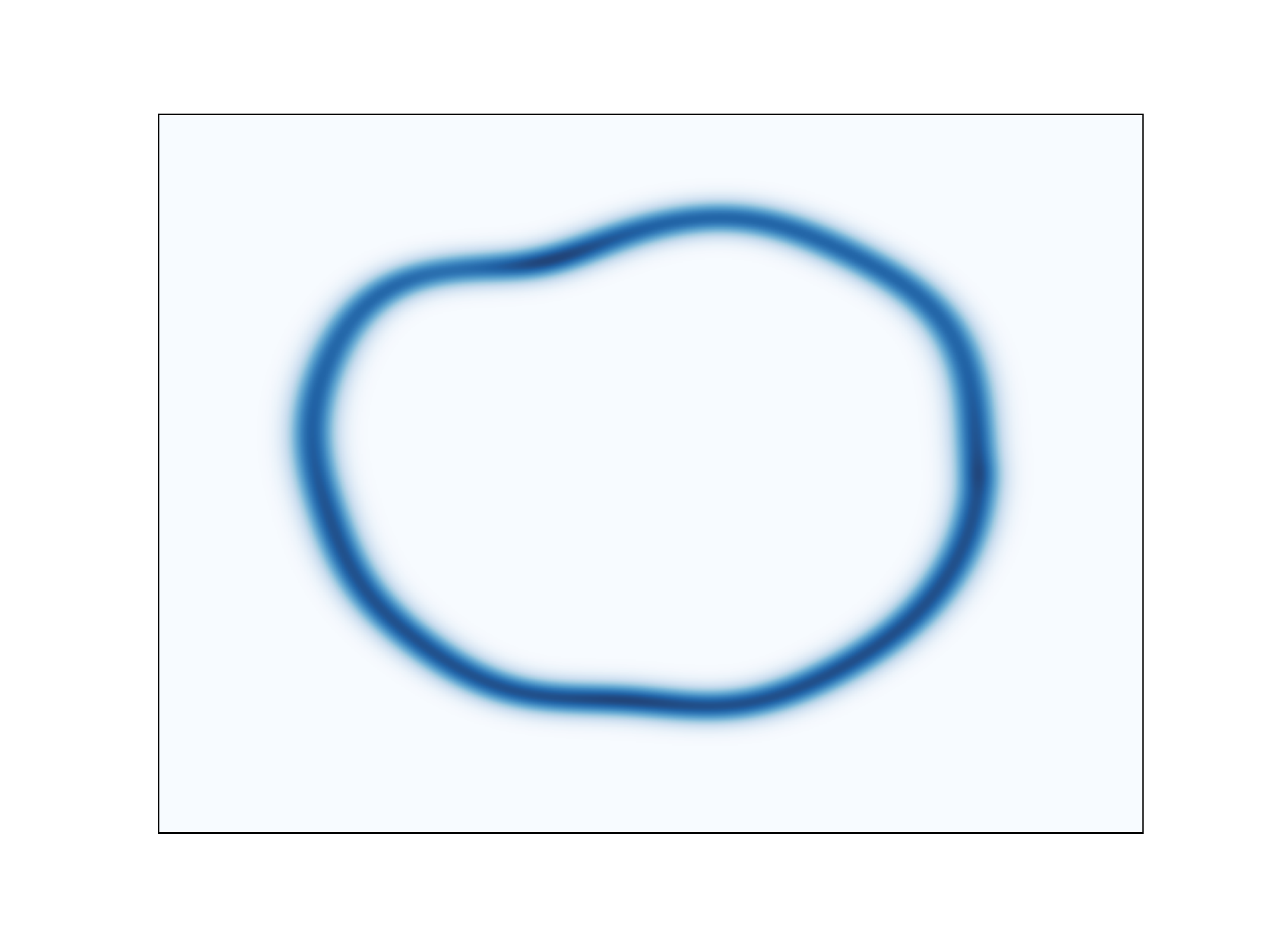}}
    \caption{Illustration of the typical set as a thin hyper--shell surrounding the mode of the probability distribution.}
    \label{fig:shell}
\end{figure}

The qualitative conclusions of the previous section are general and hold for any continuous probability distribution. The probability mass does not concentrate close to the mode where the probability density is high as there is not sufficient volume there. On the other hand, it does not concentrate on large distances because the density vanishes. Instead, it compromises on some region of intermediate distance surrounding the mode, as shown in Figure \ref{fig:typical_set}. This region is called the \textit{typical set}, and has the form of a high--dimensional thin hyper--shell surrounding the mode as shown in Figure \ref{fig:shell}. In high dimensions, the typical set exhibits the effect of \textit{concentration of measure}~\parencite{ledoux2001concentration} illustrated in Figure \ref{fig:typical_set_scaled}. 

The concept of the typical set is not only important for properly understanding probability distributions, but also for developing new computational methods. The notion of the typical set is originally borrowed from the field of information theory, in which one of the main tasks is to compress and encode a \textit{message} with as few \textit{words} as possible. In probability theory, the typical set defines the most efficient way to compress a probability distribution by focusing on a limited region of parameter space. As we will see in the next sections, the task of developing powerful and effective computational methods comes down to how efficiently we can locate and approximate the typical set of a probability distribution.

\section{Laplace approximation}
\label{sec:laplace_approximation}

\looseness=-1 Before we move on to stochastic estimators of expectation values let us first discuss another simple deterministic method, called \textit{Laplace approximation}, that, unlike \textit{quadrature in a uniform grid}, can extend to higher dimensions~\parencite{tierney1986accurate}. The \textit{Laplace approximation} makes a very strong assumption about the target probability distribution. In particular, it assumes that it can be sufficiently described by a Gaussian probability density, similar to equation \ref{eq:gaussian_pdf}. The mean of the Gaussian density is determined at the point of the mode of the target density
\begin{equation}
    \label{eq:mode_of_target}
    \mu = \underset{\theta}{\mathrm{arg\,max}} ~p(\theta)\,,
\end{equation}
and the precision matrix $\Sigma^{-1}$ (i.e. inverse of the covariance matrix $\Sigma$) is given by the second-order derivatives of the negative logarithm of the target probability density function evaluated at the mode,
\begin{equation}
    \label{eq:precision_of_target}
    \left( \Sigma^{-1}\right)_{ij} = -\frac{\partial^{2}}{\partial\theta_{i}\partial\theta_{j}} \log p(\theta)\bigg\vert_{\theta=\mu}\,.
\end{equation}

\begin{figure}[ht!]
    \centering
	\centerline{\includegraphics[scale=0.65]{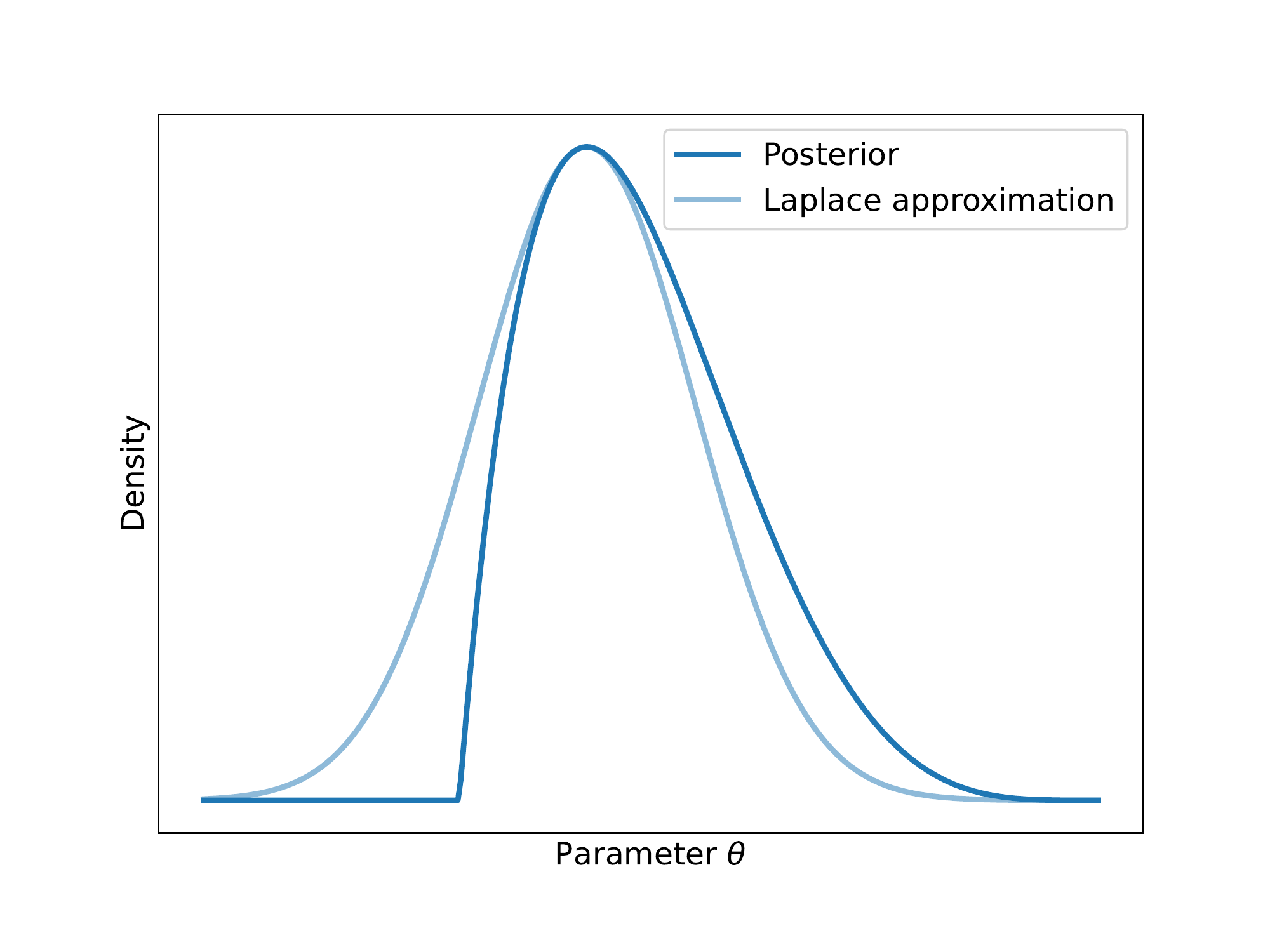}}
    \caption{Illustration of the Laplace approximation to a skewed probability density.}
    \label{fig:laplace}
\end{figure}
The reasoning behind this approach is quite simple, one effectively performs a Taylor expansion of the logarithm of the density, up to second order, around the \textit{maximum a posteriori} point $\mu$,
\begin{equation}
    \log p_{L}(\theta) = \log p(\theta=\mu) -\frac{1}{2}(\theta-\mu)^{T}\Sigma^{-1}(\theta-\mu)+\dots\,,
\end{equation}
where the first order term simply vanishes because we evaluate the expansion around the maximum. For this reason, this very common method is often called the saddle--point approximation. Expectation values can then be determined using the Gaussian density $p_{L}(\theta)=\mathcal{N}(\theta\vert\mu,\Sigma)$ in place of the target density $p(\theta)$ in the formula for the expectation value \ref{eq:expectation_value_integral_formula},
\begin{equation}
    \label{eq:expectation_value_laplace}
    \mathbb{E}_{p_{L}}[f]=\int f(\theta) \mathcal{N}(\theta\vert\mu,\Sigma) d\theta \,.
\end{equation}

The quality of the \textit{Laplace approximation} is determined by the overlap of the typical set of the target distribution with that of the Gaussian approximation. The greater the overlap, the more accurate the approximation will be.

\section{Monte Carlo estimators}

Another type of estimators is \textit{stochastic} estimators, and in particular \textit{Monte Carlo} estimators~\parencite{brooks2011handbook} that rely on a collection of independent points or samples,
\begin{equation}
    \label{eq:samples}
    \lbrace \theta_{1},\dots, \theta_{n} \rbrace \in \Theta\,,
\end{equation}
from the distribution $p(\theta)$, such that the ensemble average of a function $f(\theta)$,
\begin{equation}
    \label{eq:ensemble_average}
    \hat{f}_{n}^{MC} = \frac{1}{n}\sum_{i=1}^{n}f(\theta_{n})\,,
\end{equation}
\textit{asymptotically} converges to the corresponding expectation value
\begin{equation}
    \label{eq:ensemble_average_expectation}
    \lim_{n\to\infty}\hat{f}_{n}^{MC} = \mathbb{E}_{p}[f(\theta)]\,.
\end{equation}

The asymptotic result of equation \ref{eq:ensemble_average_expectation} is not particularly useful as a computational algorithm will never be able to produce infinite samples. Fortunately, the behaviour of Monte Carlo estimators can be quantified even for finite samples.

For any square--integrable (i.e. both $\mathbb{E}_{p}[f]$ and $\mathbb{E}_{p}[f^{2}]$ exist and are finite) real--valued function $f(\theta)$, the Monte Carlo estimator satisfies the \textit{central limit theorem},
\begin{equation}
    \label{eq:monte_carlo_clt}
    \frac{\hat{f}_{n}^{MC} - \mathbb{E}_{p}[f(\theta)]}{\textrm{MC--SE}_{n}[f]}\sim \mathcal{N}(0,1)\,,
\end{equation}
where $\textrm{MC--SE}_{n}[f]$ is the \textit{Monte Carlo Standard Error} defined as,
\begin{equation}
    \label{eq:monte_carlo_standard_error}
    \textrm{MC--SE}_{n}[f] = \sqrt{\frac{\textrm{Var}_{p}[f]}{n}}\,.
\end{equation}
This means that we can estimate the expected number of samples that is required to reach a certain level of precision for our estimates.

\begin{figure}[H]
    \centering
	\centerline{\includegraphics[scale=0.65]{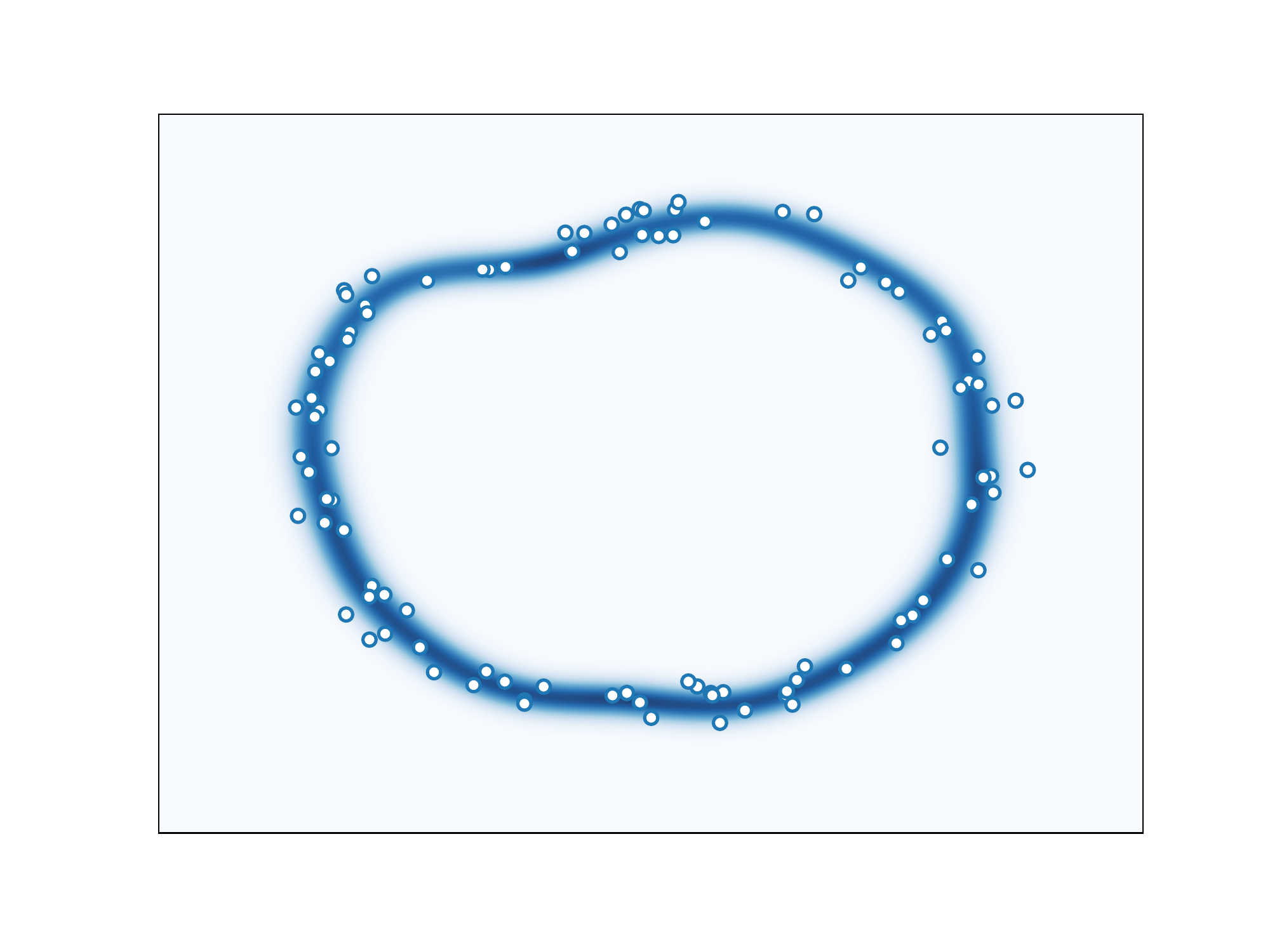}}
    \caption{Illustration of the typical set including samples generated using exact Monte Carlo sampling.}
    \label{fig:shell_samples}
\end{figure}
Another interesting property of Monte Carlo estimators is that their precision, as quantified by the Monte Carlo Standard Error of equation \ref{eq:monte_carlo_standard_error}, does not depend on the dimensionality of the problem but relies only on the number $n$ of samples instead. This means that Monte Carlo estimators can be applied even in high--dimensional problems. This insensitivity to the \textit{curse of dimensionality} is directly related to the fact that the Monte Carlo samples are already distributed in the typical set as shown in Figure \ref{fig:shell_samples}. As we will discover shortly, once we discuss more advanced methods, the difficult part is to get the samples to the typical set in the first place.

We can also think of the Monte Carlo samples as a stochastic grid where the computation is mostly focused in the regions of parameter space that contribute to the computation of the expectation value. Starting from the Monte Carlo estimator,
\begin{equation}
    \label{eq:ensemble_average2}
    \hat{f}_{n}^{MC} = \frac{1}{n}\sum_{i=1}^{n}f(\theta_{n})\,,
\end{equation}
and manipulate it into the quadrature form
\begin{equation}
    \label{eq:quadrature2}
    \hat{f}_{n}^{MC} = \sum_{i=1}^{n}f(\theta_{i})p(\theta_{i})\frac{1}{n p(\theta_{i})}\,,
\end{equation}
where $\Delta\theta_{i}=1/n p(\theta_{i})$ is the effective volume of each sample.

Monte Carlo estimators are very powerful methods assuming that one can generate independent samples from the target distribution. However, in most interesting and realistic cases, this is not feasible. In that case, one has to rely to alternative methods.

\section{Importance sampling}

One alternative method to exact Monte Carlo sampling, that does not rely on exact samples from the target distribution but instead requires an \textit{auxiliary distribution} is \textit{importance sampling}. Importance sampling estimators use samples from the auxiliary distribution and correct for any deviation from the typical set of the target distribution using \textit{weighting factors}. Although~\textcite{kloek1978bayesian} is typically credited with introducing importance sampling to statistics, there are references to it in statistical physics as early as 1949~\parencite{goertzel1949quota, kahn1951estimation}.

In order to derive the \textit{importance weights} necessary for the computation of the expectation values we start with the definition of the expectation value and do some re--arrangements,
\begin{equation}
    \label{eq:expectation_value_auxiliary_rearrangement}
    \begin{split}
        \mathbb{E}_{p}[f(\theta)] &= \int_{\Theta} f(\theta) p(\theta) d\theta \\
        &= \int_{\Theta} f(\theta) \frac{p(\theta)}{q(\theta)} q(\theta) d\theta \\
        &= \mathbb{E}_{q}\left[f(\theta) \times \frac{p(\theta)}{q(\theta)}\right] \,.
    \end{split}
\end{equation}

We can now estimate the expectation value,
\begin{equation}
    \label{eq:expectation_importance_sampling}
    \Hat{f}_{n}^{IS} = \frac{1}{n} \sum_{i=1}^{n} w(\Tilde{\theta}_{i}) f(\Tilde{\theta}_{i})\,,
\end{equation}
using samples from the \textit{auxiliary} distribution,
\begin{equation}
    \label{eq:samples_from_auxiliary}
    \lbrace \Tilde{\theta}_{1}, \dots,\Tilde{\theta}_{n} \rbrace  \sim q (\Tilde{\theta})\,,
\end{equation}
and importance weights given by,
\begin{equation}
    \label{eq:importance_weights}
    w(\Tilde{\theta}) = \frac{p(\Tilde{\theta})}{q(\Tilde{\theta})}\,.
\end{equation}

For any square--integrable real--valued function $f(\theta)$ , the importance sampling estimator satisfies the \textit{central limit theorem},
\begin{equation}
    \label{eq:importance_sampling_clt}
    \frac{\hat{f}_{n}^{IS} - \mathbb{E}_{p}[f(\theta)]}{\textrm{IS--SE}_{n}[f]}\sim \mathcal{N}(0,1)\,,
\end{equation}
where $\textrm{MC--SE}_{n}[f]$ is the \textit{Importance Sampling Standard Error} defined as,
\begin{equation}
    \label{eq:importance_sampling_standard_error}
    \textrm{IS--SE}_{n}[f] = \sqrt{\frac{\textrm{Var}_{q}[w f]}{n}}\,.
\end{equation}
By comparing the expression \ref{eq:importance_sampling_standard_error} for $\textrm{IS--SE}$ to the respective expression \ref{eq:monte_carlo_standard_error} for the \textit{Monte Carlo Standard Error} we can define the \textit{Effective Sample Size (ESS)},
\begin{equation}
    \label{eq:effective_sample_size}
    \textrm{ESS}_{n}[f] = \frac{\textrm{Var}_{q}[w f]}{\textrm{Var}_{p}[f]} n \,,
\end{equation}
as the effective number of exact samples that contain the same amount of information as the $n$ samples and their importance weights.

It is important to mention here that if the target or auxiliary distribution is known only up to a normalisation factor, for instance if the computation of the normalisation constant is very costly, then the importance weights have to be normalised such that,
\begin{equation}
    \label{eq:importance_weight_normalisation}
    \sum_{i=1}^{n} w(\Tilde{\theta}_{i}) = 1\,,
\end{equation}
for the aforementioned estimators to be valid.

\begin{figure}[ht!]
    \centering
	\centerline{\includegraphics[scale=0.65]{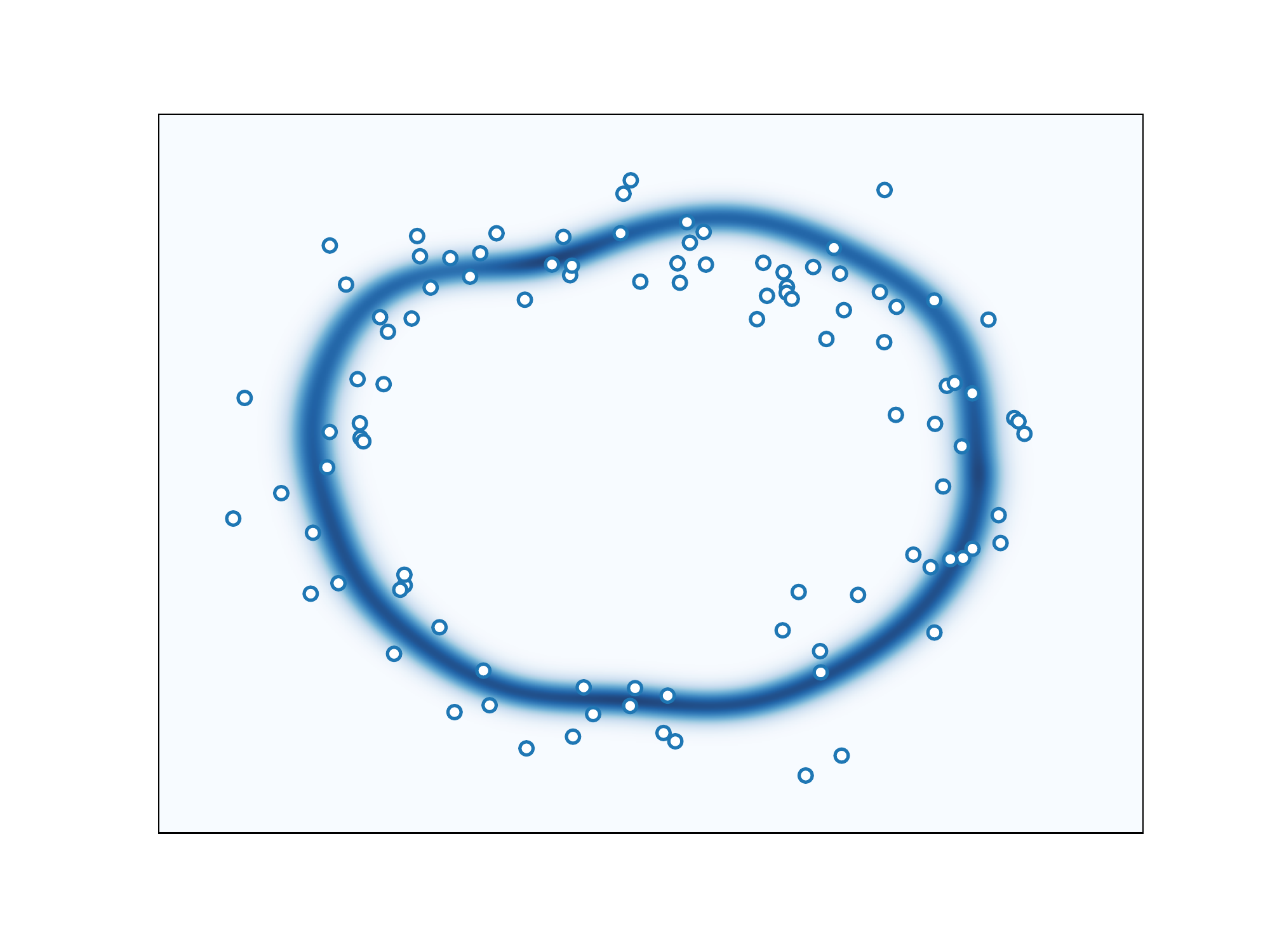}}
    \caption{Illustration of the typical set including samples generated using an unsuitable importance density. In high dimensions the typical set corresponds to a very thin shell and it is difficult to achieve sufficient overlap between the typical set of the auxiliary and target distribution. Here, this is depicted by samples that do not reside in the typical set of the target and will thus have low importance weights.}
    \label{fig:shell_importance_sampling}
\end{figure}

The quality of the importance sampling estimator is determined by the amount of overlap between the auxiliary and target distribution. Samples from the auxiliary distribution residing in regions of high overlap will receive large importance weights and those residing in regions of little or no overlap will receive small importance weights.

As we showed in previous sections, in low dimensions the typical set is broad so we should expect that the construction of importance sampling estimators for low dimensional cases to be a feasible procedure. In higher dimensions however, the typical shell is very thin thus complicating the choice of auxiliary distributions with significant overlap with the target distribution.

It is common in practice to assume that an auxiliary distribution with broader density tails than the target distribution would be sufficient to construct an importance sampling estimator. Although the reasoning of this idea is appealing, it can however be misleading as it does not extend to higher dimensions in which the typical set becomes the central object of interest and not the probability density.

It is useful to define a measure of the quality of an importance sampling estimator. A straightforward choice would be to define the \textit{importance sampling effective sample size},
\begin{equation}
    \label{eq:importance_sampling_effective_sample_size}
    N_{\textrm{eff}} = \frac{\left( \sum_{i=1}^{n}w(\Tilde{\theta}_{i})\right)^{2}}{\sum_{i=1}^{n}w(\Tilde{\theta}_{i})^{2}}\,.
\end{equation}
Equation \ref{eq:importance_sampling_effective_sample_size} is just a heuristic diagnostic and it should not be confused with equation \ref{eq:effective_sample_size}.

\section{Markov chain Monte Carlo}

Importance sampling estimators trade the ability to produce exact samples from the target distribution with weighted samples from an auxiliary distribution. On the other hand, \textit{Markov chain Monte Carlo (MCMC)} estimators replace the exact samples with \textit{correlated} samples generated by a \textit{Markov chain}~\parencite{gilks1995markov, brooks2011handbook}.

Therefore, the key idea in MCMC is to explore the typical set using a sequence of local steps. Starting a point $\theta_{1}$ in parameter space $\Theta$, the next point $\theta_{2}$ is chosen stochastically in the neighbourhood of $\theta_{1}$. Then the process is repeated for the next point $\theta_{3}$ in the neighbourhood of $\theta_{2}$ and so on. At the end, we have generated a chain of $n$ samples that is \textit{Markov}, meaning that each sample conditionally depends only on the previous one,
\begin{equation}
    \label{eq:markov_property}
    P(\theta_{n}\vert \theta_{1},\dots,\theta_{n-1}) = P(\theta_{n}\vert \theta_{n-1})\,.
\end{equation}

More formally, the Markov chain can be generated by repeatedly sampling from a conditional probability distribution on the product space $\Theta \times \Theta$, known as \textit{Markov transition probability} $T(\theta'\vert \theta)$. Given an initial point $\theta_{1}$, sampling from the \textit{Markov transition probability} $T(\theta'\vert \theta_{1})$ returns sample $\theta_{1}$. We can thus construct a sequence of transitions,
\begin{equation}
    \label{eq:sequence_markov_transitions}
    \begin{split}
        \theta_{2} &\sim T(\theta_{2}\vert \theta_{1}) \\
        \theta_{3} &\sim T(\theta_{3}\vert \theta_{2}) \\
        &\dots \\
        \theta_{n} &\sim T(\theta_{n}\vert \theta_{n-1})\,,
    \end{split}
\end{equation}
that constitute the Markov chain $\lbrace \theta_{1}, \theta_{2}, \dots, \theta_{n} \rbrace$. The samples of the Markov chain are not independent, but they are \textit{correlated}. The reason for this is their sequential origin i.e. $\theta_{n}$ depends on $\theta_{n-1}$ which depends on $\theta_{n-2}$ so even samples that are not right next to each other in the Markov chain can be correlated.
\begin{figure}[ht!]
    \centering
	\centerline{\includegraphics[scale=0.65]{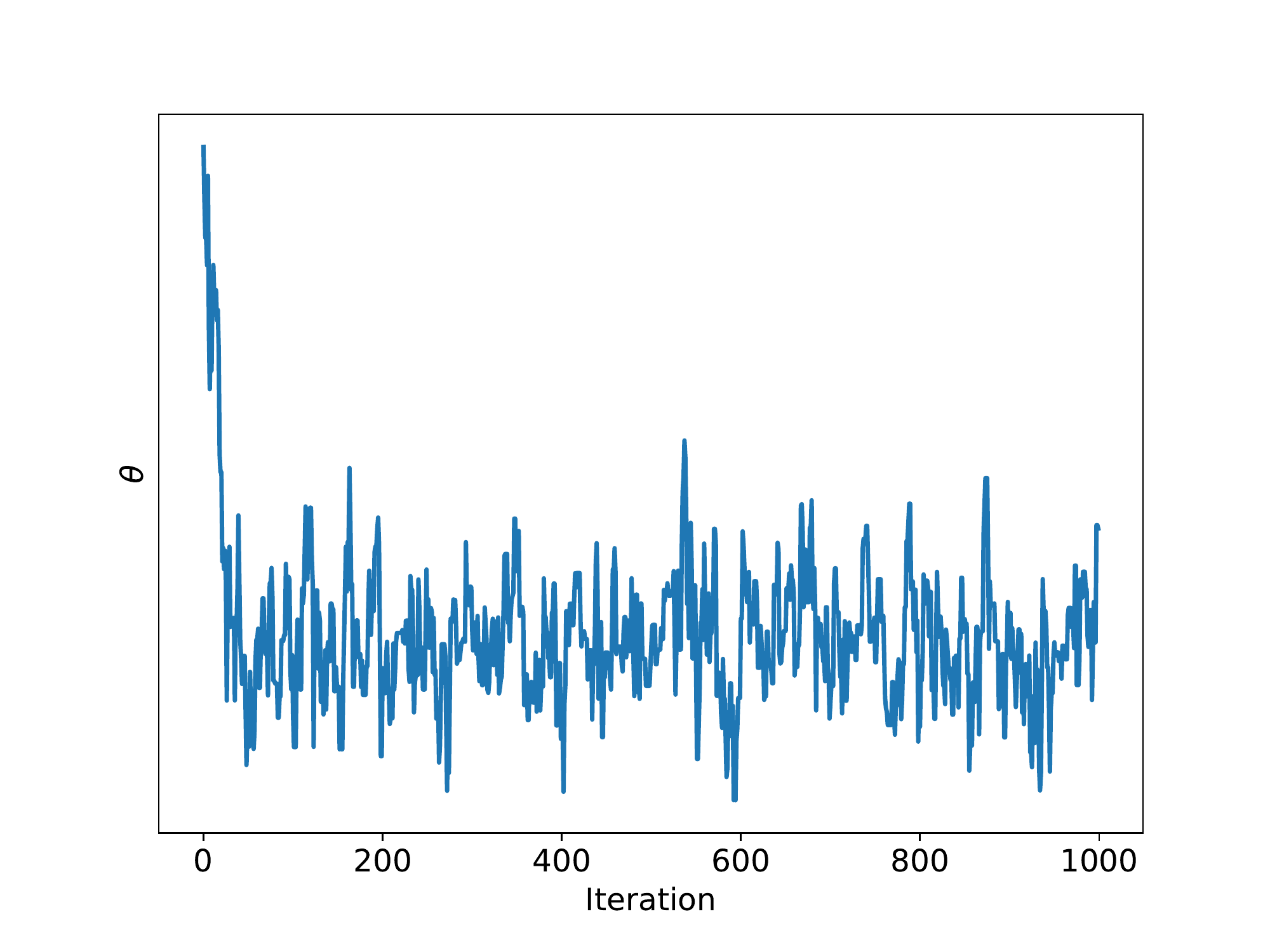}}
    \caption{Example of trace plot of a Markov chain for parameter $\theta$. The chain reaches the stationary state after about $50$ iterations.}
    \label{fig:trace_plot}
\end{figure}

\subsection{Requirements of MCMC}

In general, the possible values $\theta$ of the Markov chain are called the \textit{states} of the Markov chain~\parencite{tierney1994markov, mackay2003information}. For a Markov chain Monte Carlo estimator to generate samples from the target distribution, the Markov chain must satisfy a couple of conditions:
\begin{enumerate}
    \item First of all, the Markov chain has to leave the target distribution $p$ \textit{invariant} or \textit{stationary}, 
    \begin{equation}
        \label{eq:invariance_condition}
        p(\theta ') = \int T(\theta'\vert\theta)p(\theta)d\theta\,.
    \end{equation}
    This means that if we start from a state $\theta$ of $p$, the next state $\theta'$ is also a state of $p$. In practice, a sufficient but not necessary condition is \textit{detailed balance}, which requires that each transition $\theta \rightarrow \theta'$ is reversible. More formally, for any pair of states $\theta$ and $\theta'$ the following relation must hold,
    \begin{equation}
        \label{eq:detailed_balance}
        T(\theta'\vert\theta)p(\theta) = T(\theta\vert\theta')p(\theta')\,,
    \end{equation}
    meaning that the probability of being at state $\theta$ and transitioning to state $\theta'$ is equal to the probability of being at state $\theta'$ and transitioning to state $\theta$.
    
    \item Furthermore, we need to make sure that the \textit{stationary} distribution is unique and that the distribution of states is able to converge to it regardless the starting point $\theta_{1}$. In other words, we need to make sure that the \textit{stationary} distribution is also the \textit{limiting} distribution. This requires two properties, \textit{irreducibility}, that is the ability to visit any state $\theta$ for which $p(\theta)>0$ in a finite number of steps, and \textit{aperiodicity}, meaning that no states are only accessible at certain regularly spaced times. These two properties combined, when met, render the Markov chain \textit{ergodic}.
\end{enumerate}

\subsection{Expected behaviour}

\begin{figure}[!htb]
    \centering
	\centerline{\includegraphics[scale=0.65]{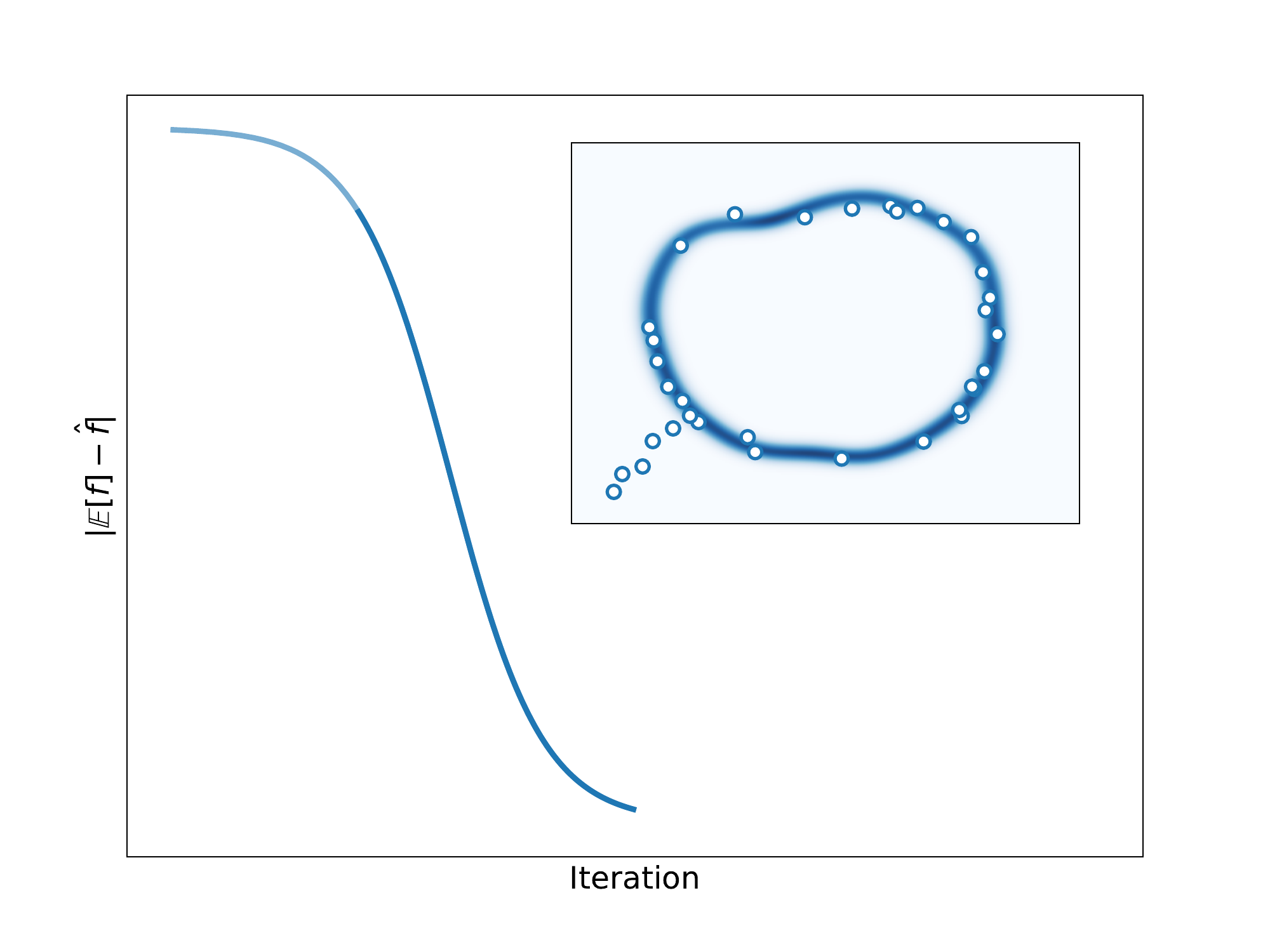}}
    \caption{Initial convergence phase -- the absolute difference between the estimate of $f$ and the expectation value decreases rapidly as the chain approaches explores the typical set for the first time.}
    \label{fig:expectation_initial_convergence_combined}
\end{figure}

\begin{figure}[!htb]
    \centering
	\centerline{\includegraphics[scale=0.65]{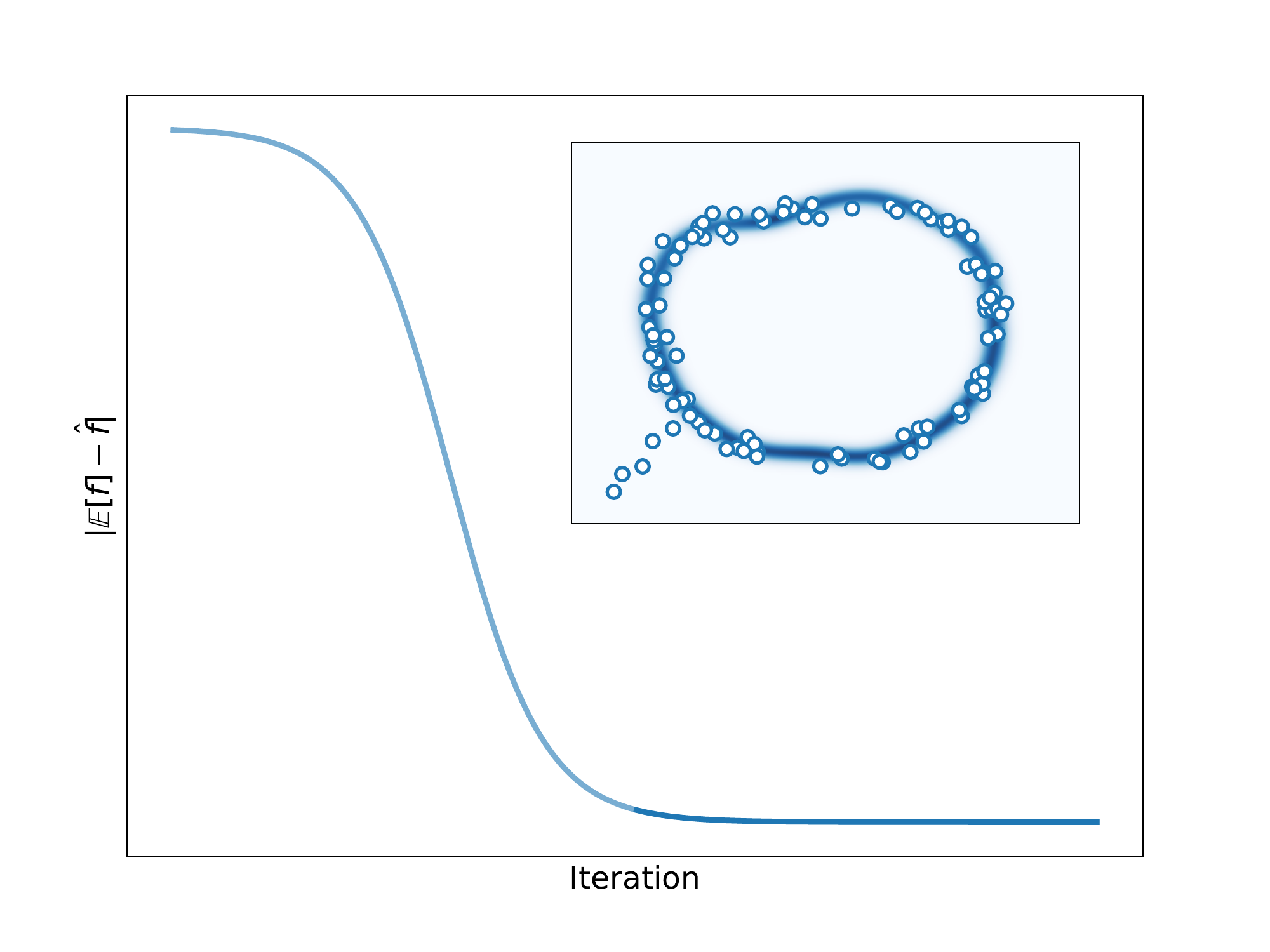}}
    \caption{Stationary or equilibrium phase -- the absolute difference between the estimate of $f$ and the expectation value has reached its minimum value as the chain samples fully populate the typical set.}
    \label{fig:expectation_convergence_combined}
\end{figure}

When all of the aforementioned conditions are obeyed, the Markov chain samples from the target distribution. The behaviour of the Markov chain, in terms of the computed expectation values, passes through four stages that characterise its normal behaviour~\parencite{betancourt2017conceptual}.
\begin{figure}[!htb]
    \centering
	\centerline{\includegraphics[scale=0.65]{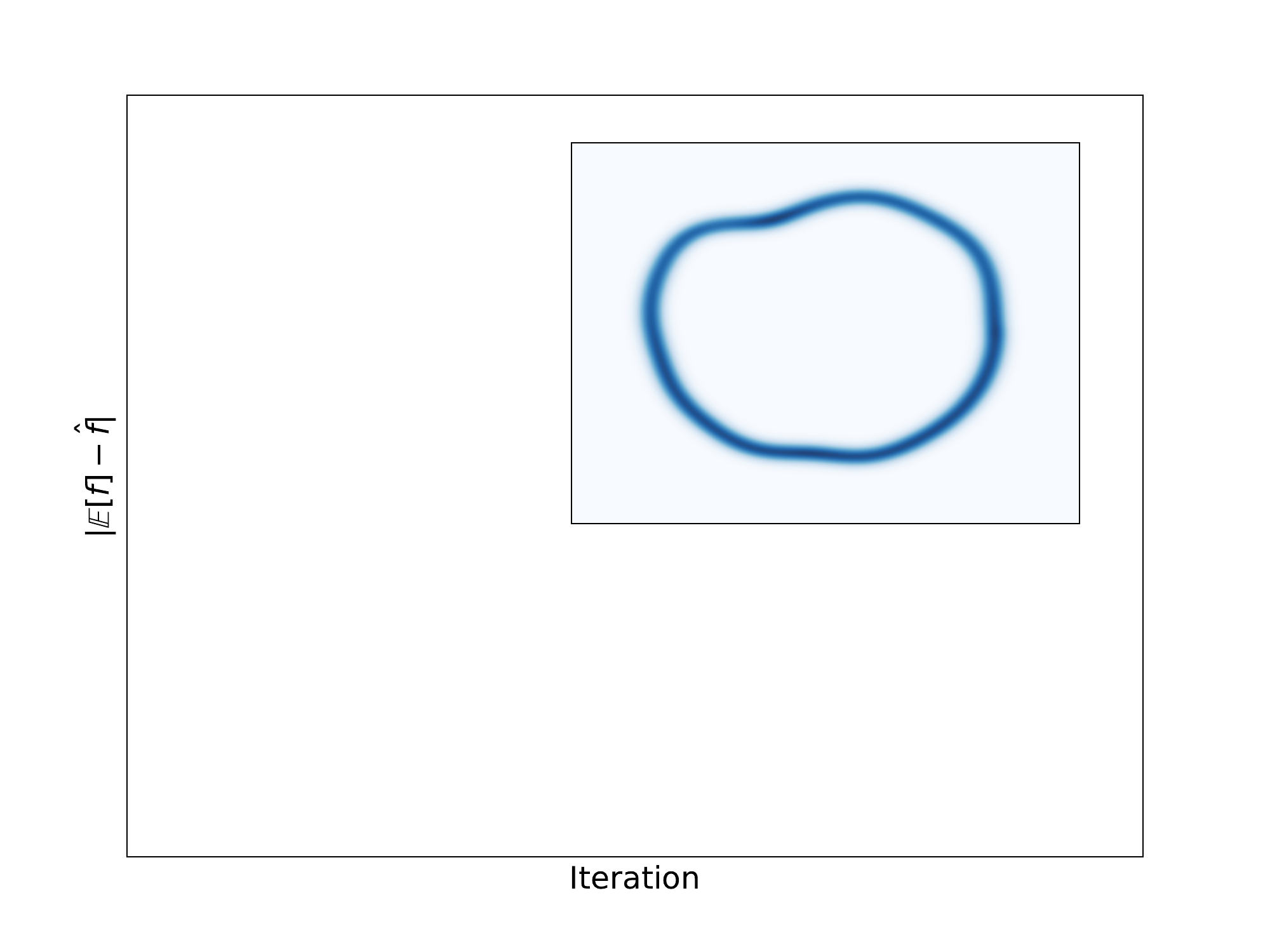}}
    \caption{Initial stage of exploration -- no exploration has taken place.}
    \label{fig:expectation_initial_combined}
\end{figure}
\begin{figure}[!htb]
    \centering
	\centerline{\includegraphics[scale=0.65]{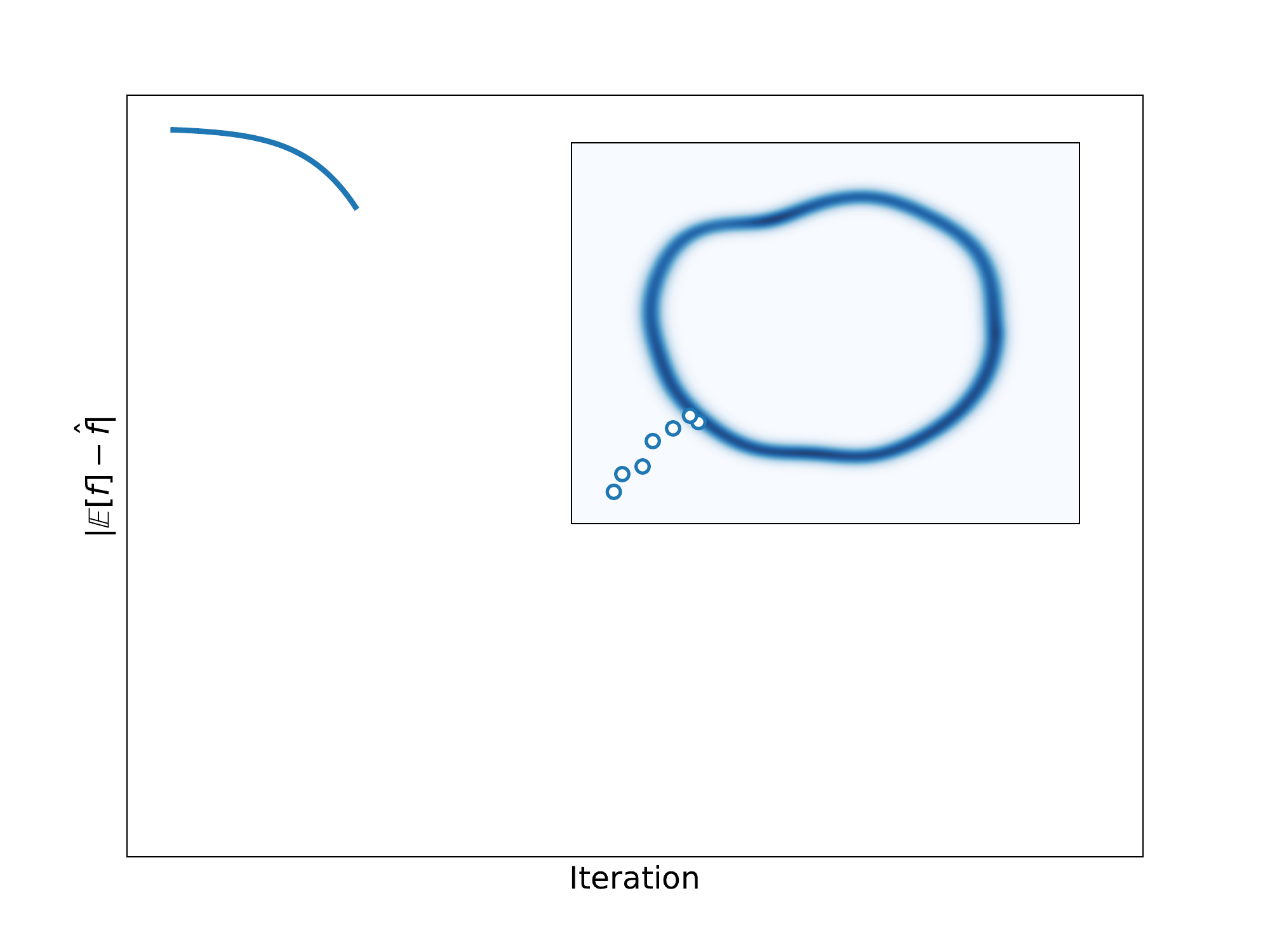}}
    \caption{Burn--in stage of exploration --  the absolute difference between the estimate of $f$ and its expectation value slowly decreases as the chain approaches the typical set.}
    \label{fig:expectation_burnin_combined}
\end{figure}
The first stage, shown in Figure \ref{fig:expectation_initial_combined} consists of the initialisation of the Markov chain. Often we do not know where the typical set resides and thus we set the first state of the Markov chain to some arbitrary point in parameter space. During the second stage, the Markov chain moves towards the typical set as shown in Figure \ref{fig:expectation_burnin_combined}. At the same time the absolute difference of the estimated value $\Hat{f}$ from the expectation value $\mathbb{E}[f]$ slowly decreases. In the third state shown in Figure \ref{fig:expectation_initial_convergence_combined}, the Markov chain starts to explore the typical set. The absolute difference between the estimate of $f$ and its expectation value decreases very rapidly. Finally, in the fourth stage shown in Figure \ref{fig:expectation_convergence_combined} the Markov chain wanders inside the typical set and the standard error of the estimate asymptotically decreases as prescribed by the central limit theorem.

\subsection{Central limit theorem of MCMC}

Markov chain Monte Carlo estimators are particularly useful for many analyses since they obey a \textit{central limit theorem} that allows us to quantify their precision~\parencite{kipnis1986central, geyer1992practical, tierney1994markov}. In particular, given a square--integrable real--valued function $f(\theta)$ and a long enough Markov chain, the following is true,
\begin{equation}
    \label{eq:mcmc_clt}
    \frac{\Hat{f}_{n}^{\textrm{MCMC}}-\mathbb{E}_{p}[f(\theta)]}{\textrm{MCMC--SE}_{n}[f]}\sim \mathcal{N}(0,1)\,,
\end{equation}
where $\textrm{MCMC--SE}_{n}[f]$ is the \textit{Markov chain Monte Carlo Standard Error} given by,
\begin{equation}
    \label{eq:mcmc_se}
    \textrm{MCMC--SE}_{n}[f] = \sqrt{\frac{\textrm{Var}_{p}[f]}{\textrm{ESS}_{n}[f]}}\,.
\end{equation}
Comparing $\textrm{MCMC--SE}_{n}[f]$ with the standard error $\textrm{MC--SE}_{n}[f]$ of the exact Monte Carlo estimator given by equation \ref{eq:monte_carlo_standard_error}, one immediately notices that the number of samples $n$ has been replaced by the term $\textrm{ESS}_{n}[f]$. This term, called the \textit{Effective Sample Size} accounts for the loss of information due to the correlation between samples due to the Markov property of the chain. The effective sample size is given by,
\begin{equation}
    \label{eq:mcmc_ess}
    \textrm{ESS}_{n}[f] = \frac{n}{\tau[f]}\,,
\end{equation}
where $\tau[f]$ is the \textit{relaxation} or \textit{autocorrelation time} of the Markov chain. Less formally, $\tau[f]$ describes the number of steps required for the Markov chain to ``forget'' where it started, meaning that only one out of $\tau[f]$ is actually independent. A method for estimating the autocorrelation time of a Markov chain will be discussed in the next subsection.

\subsection{Autocorrelation}

\begin{figure}[ht!]
    \centering
	\centerline{\includegraphics[scale=0.65]{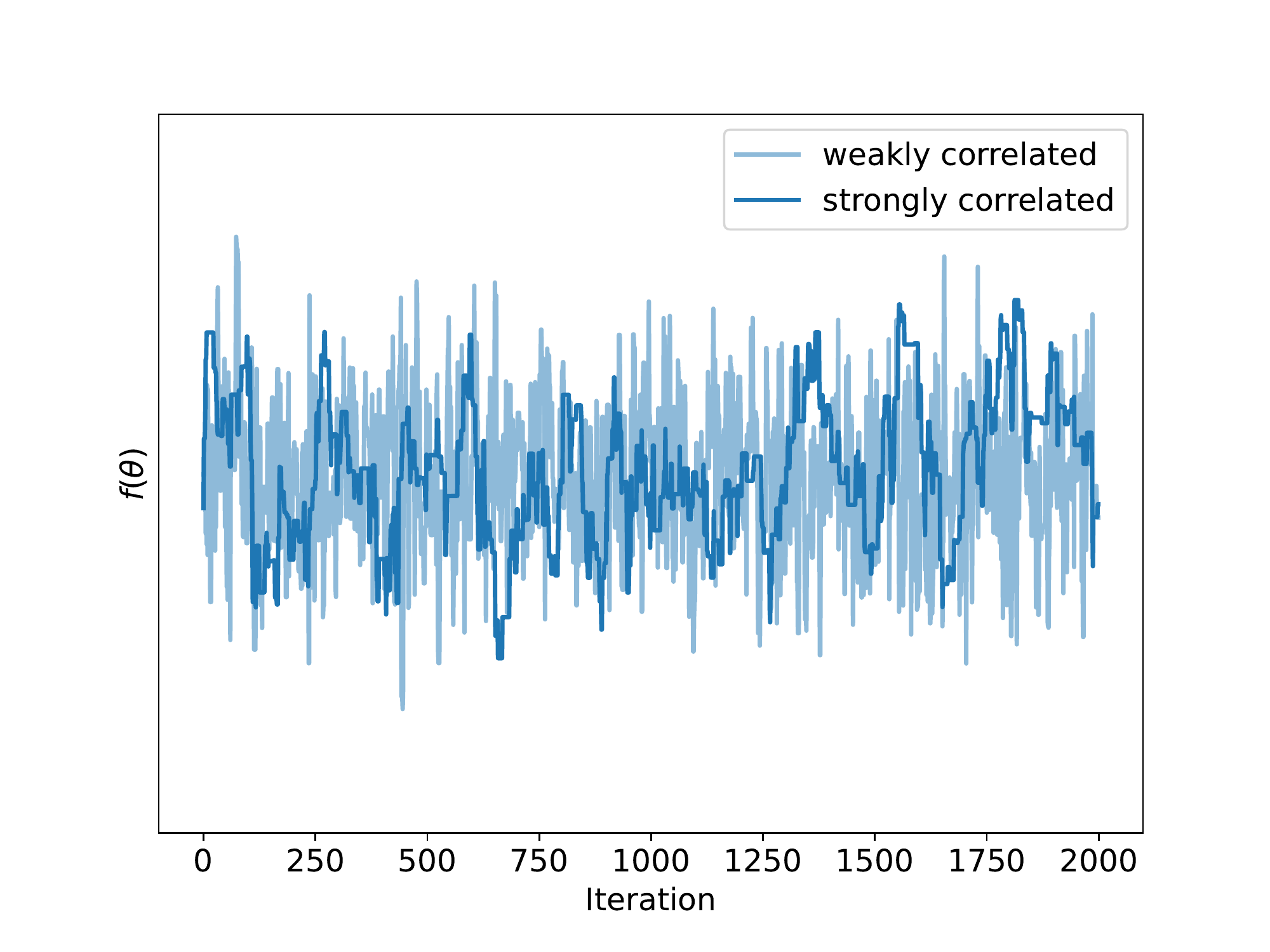}}
    \caption{Markov chains with different degrees of autocorrelation.}
    \label{fig:correlated_chains}
\end{figure}

The autocorrelation of the Markov chains is a necessary evil of MCMC methods and must be properly understood before making any inference~\parencite{geyer1992practical}. Figure \ref{fig:correlated_chains} shows two Markov chains with different levels of autocorrelation. In the weakly correlated chain, large jumps take place from one iteration to the next. On the other hand, the strongly correlated chain is characterised by very short jumps and more rigid trajectories.

In order to quantify and measure the degree of autocorrelation of a Markov chain we need to compare the states of the chain after fixed number of iterations called \textit{lags}. Given an arbitrary function $f(\theta)$ of the states,
\begin{equation}
    \label{eq:autocorrelation_mean}
    \mu_{f} = \mathbb{E}_{p}[f]\,,
\end{equation}
is the mean value of the function expressed as the expectation value over the stationary distribution $p(\theta)$. The value $f(\theta_{i}) - \mu_{f}$ then quantifies the deviation of the $i$--th state of the chain from the mean value $\mu_{f}$. The expectation value of the product of two such deviations defines the \textit{autocovariance} of the chain,
\begin{equation}
    \label{eq:autocovariance}
    c_{ij} = \mathbb{E}_{p}\left[(f_{i} - \mu_{f})(f_{j} - \mu_{f}) \right]\,,
\end{equation}
where $f_{i}=f(\theta_{i})$ and $f_{j}=f(\theta_{j})$. Once the Markov chain has reached the stationary phase, the autocovariance will no longer depend on the particular states, $\theta_{i}$ and $\theta_{j}$, that we are comparing but on the number of iterations, called lag $\ell=j - i$, between them,
\begin{equation}
    \label{eq:autocovariance_invariance}
    c_{ij} = c_{i,i+\ell}=c_{\ell}\,.
\end{equation}
Finally, if we normalise the autocovariance by the variance,
\begin{equation}
    \label{eq:autocorrelation_variance}
    \mathrm{Var}_{p}[f] = \mathbb{E}_{p}\left[ (f-\mu_{f})^{2}\right]\,,
\end{equation}
we get the lag--$\ell$ \textit{autocorrelation function},
\begin{equation}
    \label{eq:autocorrelation_function}
    \rho_{\ell}[f] = \frac{\mathbb{E}_{p}\left[(f_{i+\ell} - \mu_{f})(f_{i} - \mu_{f}) \right]}{\mathrm{Var}_{p}[f]}\,.
\end{equation}
The normalisation ensures that the maximum possible value of $\rho_{\ell}[f]$ is $+1$ for fully correlated states and $-1$ for the completely anti--correlated states. The value of $0$ corresponds to uncorrelated samples. The lag--$\ell$ autocorrelation is always unity as any state is perfectly correlated with itself. Furthermore, the autocorrelation function depends only on the absolute lag and is invariant under changes of the sign, that is, $\rho_{\ell}[f]=\rho_{-\ell}[f]$. For this reason, only the non--negative part of the autocorrelation function is often plotted. 

\begin{figure}[ht!]
    \centering
	\centerline{\includegraphics[scale=0.65]{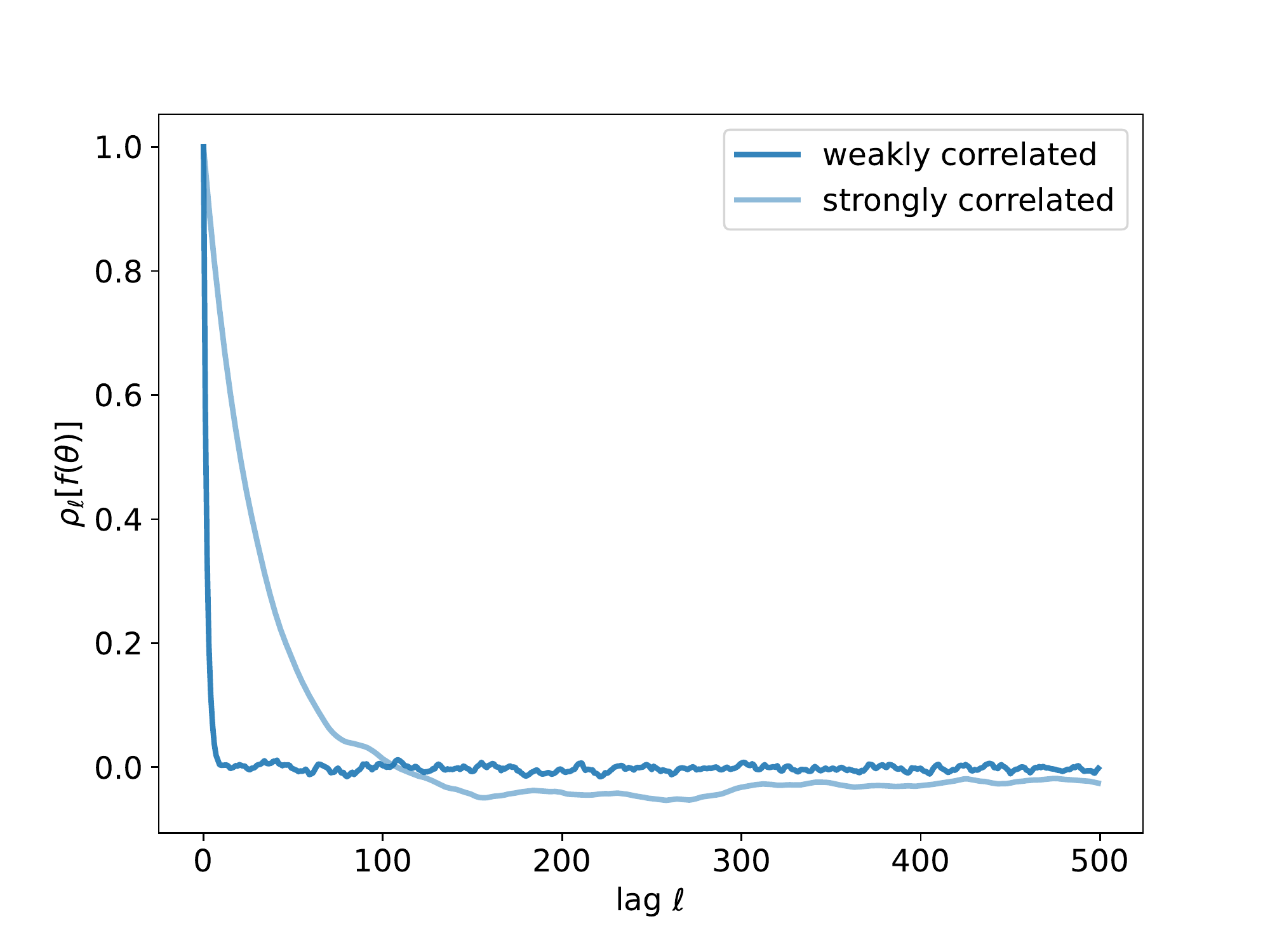}}
    \caption{Autocorrelation as a function of lag $\ell$ for a weakly and a strongly correlated chain.}
    \label{fig:autocorrelation}
\end{figure}

Figure \ref{fig:autocorrelation} shows the autocorrelation function for the weakly and strongly correlated chains of Figure \ref{fig:correlated_chains}. We notice that although both functions begin at the value of $1$ for lag $\ell=0$, they approach the value of $0$ at different rates. In particular, the autocorrelation function of the weakly correlated chain goes to $0$ after only a few lags, whereas the one corresponding to the strongly correlated chain takes much longer. 

The asymptotic variance of an infinitely long chain is defined as,
\begin{equation}
    \label{eq:asymptotic_variance}
    \lim_{n\to\infty} n\times\mathrm{Var}_{p}\left[ \Hat{f}_{n}^{\mathrm{MCMC}}\right] = \mathrm{Var}_{p}[f]\times\tau_{I}[f]\,,
\end{equation}
where,
\begin{equation}
    \label{eq:integrated_autocorrelation_time}
    \tau_{I}[f] = \sum_{\ell=-\infty}^{\ell=+\infty}\rho_{\ell}[f] = 1 + 2\times\sum_{\ell=1}^{\ell=+\infty}\rho_{\ell}[f]\,,
\end{equation}
is the \textit{integrated autocorrelation time} and the last equality hold due to the lag--sign invariance of the autocorrelation function. Equation \ref{eq:asymptotic_variance} implies that the standard error of MCMC is,
\begin{equation}
    \label{eq:standard_error_mcmc}
    \textrm{MCMC--SE}_{n}[f] =\sqrt{ \frac{\mathrm{Var}_{p}[f]}{\mathrm{ESS}_{n}[f]} }\,,
\end{equation}
where we have defined the \textit{effective sample size} as,
\begin{equation}
    \label{eq:effective_sample_size_2}
    \mathrm{ESS}_{n}[f] = \frac{n}{1+2\times\sum_{\ell=1}^{\ell=+\infty}\rho_{\ell}[f]}\,.
\end{equation}

Estimating the integrated autocorrelation time, and thus the effective sample size, in not trivial in practice. The autocorrelation function can be very noisy in large lags, as shown in Figure \ref{fig:autocorrelation}. This means that the sum in equation \ref{eq:integrated_autocorrelation_time} needs to be truncated in practice in order avoid adding noise.

% !TEX TS-program = pdflatex
% !TEX root = ../ArsClassica.tex

%************************************************
\chapter{Simple MCMC methods}
\label{chp:mcmc}
%************************************************

\begin{flushright}
\itshape
Not all those who wander are lost. \\
\medskip
--- J.R.R. Tolkien
\end{flushright}

During the first half of the twentieth century, research efforts were focused on the task of understanding the equilibrium behaviour of thermodynamic systems. Furthermore, it was well understood that this behaviour was described by specific probability distributions (e.g. \textit{canonical} distribution for a system in constant temperature). The physicists of that time showed great interest in methods that produced \textit{exact samples} from such probability distributions. Enrico Fermi, for instance, would exploit such methods to make amazingly--quick predictions of experimental outcomes as early as 1930s~\parencite{metropolis1987beginning}. During the next two decades, \textit{Stan Ulam} and \textit{John von Neumann} developed various such algorithms which, collectively, were anointed with the name ``Monte Carlo'' after the infamous casino.

After the war, \textit{Nicholas Metropolis} lead the group in \textit{Los Alamos} in applying Monte Carlo methods to increasingly complex thermodynamic systems. As \textit{exact sampling} was possible only for a limited number of distributions, Metropolis, along with \textit{Arianna Rosenbluth, Marshall Rosenbluth, Edward Teller}, and \textit{Augusta Teller} introduced the so--called \textit{Metropolis} algorithm that produced correlated samples from a wider range of probability distributions~\parencite{metropolis1953equation}. Arianna implemented the algorithm on the \textit{MANIAC} computer~\parencite{gubernatis2005marshall} and thus she is considered the first person in history to implement a MCMC method.

After decades of empirical success in physics and chemistry, the statistician Hastings~\parencite{hastings1970} generalised the method by realising that by introducing a small modification he could allow for any \textit{proposal} distribution, not just a limited family of symmetric ones. The method is known today as \textit{Metropolis--Hastings}. Despite \textit{Hasting}'s seminal contribution, it was not until 1984 that \textcite{geman1984stochastic} introduced the \textit{Gibbs sampler} for the task of image reconstruction, for the broader statistical community to realise the potential of MCMC methods for parameter inference~\parencite{gelfand1990sampling}.

\section{Metropolis--Hastings}

The key idea of the \textit{Metropolis--Hastings} algorithm is to separate the \textit{Markov transition probability} into two steps, a \textit{proposal} and an \textit{acceptance} step. During the \textit{proposal} step, a new state $\theta'$ is generated conditional on the current state $\theta$,
\begin{equation}
    \label{eq:proposal_distribution}
    \theta' \sim q(\theta'\vert\theta)\,,
\end{equation}
by sampling from a proposal distribution $q(\theta'\vert\theta)$. The aim of this step is to produce a new state that is likely, but not necessary, to reside in the typical set of the target distribution. The form of the conditional proposal distribution can be chosen based on the particular target distribution. As we will see shortly, many of the developments in the field of MCMC focus explicitly on the choice of the proposal distribution.

Once the new state $\theta '$ is generated, its validity (i.e. whether or not it belongs to the typical set) is assessed in the \textit{acceptance} step. In particular, the new state $\theta'$ is accepted with probability,
\begin{equation}
    \label{eq:acceptance_criterion}
    \alpha(\theta',\theta) = \min\left(1, \frac{p(\theta')q(\theta\vert\theta')}{p(\theta)q(\theta'\vert\theta)} \right)\,.
\end{equation}
Equation \ref{eq:acceptance_criterion} is often called the \textit{Metropolis acceptance probability}. If accepted, the new state $\theta'$ is added to the Markov chain and the process is repeated with that as the current state (i.e. $\theta \leftarrow \theta')$. On the other hand, if the state $\theta'$ is rejected, the current state $\theta$ is added (i.e. repeated) on the chain. This acceptance/rejection procedure based on equation \ref{eq:acceptance_criterion} is often referred to as the \textit{Metropolis acceptance criterion}.

It is important to mention here that the \textit{Metropolis acceptance criterion} can be evaluated even if we are only able to compute $p(\theta)$ up to a normalisation constant, as any such factor would cancel out in the ratio $p(\theta')/p(\theta)$ that appears in equation \ref{eq:acceptance_criterion}. This is a very important feature of the algorithm and one of the reasons for its widespread success. In practice, it is very difficult to know the exact value for the \textit{model evidence} $\mathcal{Z}=p(d)$ that acts as the normalisation factor for the posterior distribution $p(\theta\vert d)$ that might be the target distribution.

It is straightforward to show that the \textit{Metropolis--Hastings} algorithm leaves the target distribution $p(\theta)$ \textit{stationary} by first proving that it preserves \textit{detailed balance}. The \textit{Markov transition probability} is simply,
\begin{equation}
    \label{eq:markov_transition_probability_mh}
    T(\theta'\vert\theta) = q(\theta'\vert\theta) \alpha(\theta',\theta)\,,
\end{equation}
that is, the probability of proposing the new state $\theta'$ given the old state, times the probability of accepting it. Therefore, the \textit{Markov transition probability} for the \textit{Metropolis--Hastings} algorithm preserves detailed balance,
\begin{equation}
    \label{eq:mh_preserves_detailed_balance}
    \begin{split}
        T(\theta'\vert\theta)p(\theta) &= q(\theta'\vert\theta) \alpha(\theta',\theta) p(\theta) \\
        &= q(\theta'\vert\theta) \min\left(1, \frac{p(\theta')q(\theta\vert\theta')}{p(\theta)q(\theta'\vert\theta)} \right) p(\theta) \\
        &= \min\left( p(\theta)q(\theta'\vert\theta),  p(\theta')q(\theta\vert\theta')\right) \\
        &= q(\theta\vert\theta') \alpha(\theta,\theta') p(\theta') \\
        &= T(\theta\vert\theta')p(\theta')\,,
    \end{split}
\end{equation}

\begin{algorithm}[ht!]
\caption{Metropolis--Hastings} \algolabel{mh}
\begin{algorithmic}[1]
\REQUIRE{initial state $\theta_{1}$, (unnormalised) target density $f(\theta)\propto p(\theta)$, proposal density $q(\theta'\vert\theta)$, and number of iterations $N$}
\ENSURE{Markov chain $\theta_{1}, \theta_{2}, \dots, \theta_{N}$ that has $p(\theta)$ as its equilibrium distribution}
\FOR{$t=1$ \TO $N$}
\STATE{Draw new state from proposal distribution $\theta'\sim q(\theta'\vert\theta_{t})$}
\STATE{Compute acceptance probability $\alpha = \min\left(1, \frac{f(\theta')q(\theta_{t}\vert \theta')}{f(\theta_{t})q(\theta'\vert \theta_{t})} \right)$}
\STATE{Draw uniform random number $u\sim \mathcal{U}(0,1)$}
\IF{$u < \alpha$}
    \STATE{Accept proposed state and set $\theta_{t+1}\leftarrow \theta'$}
\ELSE
    \STATE{Reject proposed state and set current state as the next $\theta_{t+1}\leftarrow \theta_{t}$}
\ENDIF
\ENDFOR
\end{algorithmic}
\end{algorithm}

\subsection{Random--walk Metropolis}

A very common, and simplifying in practice, choice for the proposal distribution is the conditional normal distribution $q(\theta'\vert\theta)=\mathcal{N}(\theta'\vert\theta,\Sigma)$ centred around the current state $\theta$ with covariance matrix $\Sigma$~\parencite{metropolis1953equation, tierney1994markov}. The probability density has the usual Gaussian functional form,
\begin{equation}
    \label{eq:gaussian_proposal}
    q(\theta'\vert\theta) = \det(2\pi\Sigma)^{-\frac{1}{2}} \exp{\left[-\frac{1}{2}(\theta'-\theta)^{T}\Sigma^{-1}(\theta'-\theta)\right]}\,.
\end{equation}
The symmetry of this proposal distribution,
\begin{equation}
    \label{eq:gaussian_proposal_symmetry}
    q(\theta'\vert\theta) = q(\theta\vert\theta')\,,
\end{equation}
means that the \textit{Metropolis acceptance probability} of equation \ref{eq:acceptance_criterion} is simplified and the $q$ terms drop out,
\begin{equation}
    \label{eq:acceptance_criterion_symmetric}
    \alpha(\theta',\theta) = \min\left(1, \frac{p(\theta')}{p(\theta)} \right)\,.
\end{equation}

\subsection{Independence Metropolis}

Another simple choice of proposal distribution is to make it independent of the current state $\theta$. For instance, one can choose a normal distribution $q(\theta)=\mathcal{N}(\theta\vert\mu,\Sigma)$ with mean $\mu$ and covariance matrix $\Sigma$, both of which must be known \textit{a priori} and can not depend on the current state. In this case, the \textit{Metropolis acceptance probability} reduces to,
\begin{equation}
    \label{eq:acceptance_criterion_independence}
    \alpha(\theta',\theta) = \min\left(1, \frac{p(\theta')q(\theta)}{p(\theta)q(\theta')} \right)\,.
\end{equation}
One of the benefits of \textit{Independence Metropolis}~\parencite{hastings1970, tierney1994markov}, as this approach is called, is that any states produced are independent samples from the target distribution $p(\theta)$. However, it suffers from similar problems to \textit{Importance sampling}. Instead of vanishingly small importance weights, in \textit{Independence Metropolis} we might experience vanishingly small \textit{acceptance probability} when the overlap of the typical set of the proposal distribution $q$ with the target $p$ is small. For this reason, the use of \textit{Independence Metropolis} is wise only when we have good reasons to believe that the proposal distribution is sufficiently close to the target distribution or the dimensionality is low.

\subsection{Metropolis--adjusted Langevin algorithm}

As we have seen, the normal proposal distribution of \textit{Random--walk Metropolis} can utilise only \textit{global} information about the target distribution (i.e. the covariance matrix) in order to achieve efficient sampling. Although sufficient in low to moderate dimensional problems, this strategy can become inefficient as the number of parameters of the target distribution increases. In practice, \textit{Random--walk Metropolis} proposes new states indiscriminately along directions of great covariance without taking into account the local structure of the typical set. This results in either low acceptance probabilities or small proposed steps being accepted as the typical set become thinner in high dimensions.

One way to circumvent this effect and achieve better sampling performance is to capitalise on the knowledge of the gradient of the target distribution in order to ``bias'' the proposed states towards directions that are more likely to lead to higher acceptance probability. \textit{Metropolis--adjusted Langevin algorithm (MALA)}~\parencite{roberts2002langevin} achieves this by using a conditional normal distribution,
\begin{equation}
    \label{eq:mala_proposal}
    q(\theta'\vert\theta) = \mathcal{N}\left(\theta+\tau\Sigma \nabla \log p(\theta), 2\tau\Sigma \right)\,,
\end{equation}
where its mean $\theta+\tau\Sigma \nabla \log p(\theta)$ is shifted from the current state $\theta$ along the direction of the gradient of the logarithm of the target distribution $\nabla \log p(\theta)$. If known, $\Sigma$ can be an approximate covariance matrix that characterises the target distribution, otherwise, a unit--diagonal matrix can be used. $\tau$ is the step size of the method and determines the amount of shift of the proposal distribution. In the limit that $\tau\rightarrow 0$, MALA reduces to RWM. The step size $\tau$ can be modified in order to achieve the theoretically optimal acceptance probability of $0.574$. Despite the fact that the aforementioned acceptance rate has only been proven to be optimal for certain types of target distributions~\parencite{roberts1998optimal}, we expect that values in the range between $0.4$ and $0.8$ would result in a high performance for most applications.

In terms of the typical set, we can think of the gradient of the log probability as a guide that allows for better--informed proposals that are more likely to belong to the typical shell. 

\subsection{Adaptive Metropolis}

Hyperparameters of the proposal distribution, such as the covariance matrix $\Sigma$ of RWM or the step size $\tau$ of MALA, do not have to be chosen \textit{a priori} or based on preliminary MCMC runs but they can instead be adaptively tuned during the run. \textcite{haario2001adaptive} presented a prototype adaptive version of RWM in which the proposal distribution is continuously adapted during the run using all of the collected samples in order to estimate its covariance matrix. The estimation of the covariance matrix is efficient as only incremental updates are required using simple recursive formulas.

In order to achieve this kind of proposal adaptation in practice we need to abandon the Markov property of the chain. In general, this is not a problem as there is nothing special about the Markov property apart from its simplicity. However, continuous tuning of the proposal distribution during the run requires the adaptation to be \textit{diminishing}, with very specific characteristics, in order to preserve the \textit{ergodicity} of the method~\parencite{brooks2011handbook}.

One of the most commonly used algorithms for \textit{diminishing adaptation} is the \textit{stochastic approximation} algorithm of \textcite{robbins1951stochastic}. Suppose that we have a function $f(\lambda_{i})=f_{i}$ , which encodes some aspect of the behaviour of the $i$--th state of chain (e.g. the acceptance probability) as a function of some tunable property $\lambda$ (e.g. the proposal scale), that has expectation value,
\begin{equation}
    \label{eq:robbins_monro_expectation}
    \mathbb{E}[f(\lambda)] = \frac{1}{n}\sum_{i=1}^{n}f(\lambda_{i})\,.
\end{equation}
The solution to the equation $\mathbb{E}[f(\lambda)] = f^{*}$ can be found iteratively, using the recursive formula,
\begin{equation}
    \label{eq:robbins_monro_recursion}
    \lambda_{i+1} = \lambda_{i} - \gamma_{i} \left(f_{i} - f^{*} \right)\,,
\end{equation}
assuming that $f$ is a non--decreasing function of $\lambda$ that is uniformly bounded \parencite{andrieu2008tutorial}. The parameter $\gamma_{i}$ determines the \textit{learning rate} or the rate of convergence of the approximation and has to obey two conditions,
\begin{equation}
    \label{eq:robbins_monro_conditions}
    \sum_{i=1}^{n}\gamma_{i}=\infty\,,\quad \sum_{i=1}^{n}\gamma_{i}^{2}<\infty \,.
\end{equation}
The former condition ensures that any point $\theta$ can eventually be reached, and the latter condition ensures that the fluctuations introduced by new iterations is contained and does not prevent convergence. A commonly used schedule for the learning rate that satisfies the above conditions has the form $\gamma_{i}=i^{-\kappa}$ for $\kappa\in(0.5,1]$.

\begin{algorithm}[ht!]
\caption{Adaptive Metropolis} \algolabel{am}
\begin{algorithmic}[1]
\REQUIRE{initial state $\theta_{1}$, (unnormalised) target density $f(\theta)\propto p(\theta)$, the target acceptance rate $\alpha^{*}$, learning rate schedule (e.g. $g_{t}=1/t$), and number of iterations $N$}
\ENSURE{Markov chain $\theta_{1}, \theta_{2}, \dots, \theta_{N}$ that has $p(\theta)$ as its equilibrium distribution}
\STATE{Initialise $\mu_{1}=0$, $\Sigma_{1} = 1$, $\log\lambda_{1}=0$}
\FOR{$t=1$ \TO $N$}
\STATE{Draw new state from proposal distribution $\theta'\sim \mathcal{N}(\theta'\vert\theta_{t},\lambda_{t}\Sigma_{t})$}
\STATE{Compute acceptance probability $\alpha_{t} = \min\left(1, f(\theta')/f(\theta_{t}) \right)$}
\STATE{Draw uniform random number $u\sim \mathcal{U}(0,1)$}
\IF{$u < \alpha_{t}$}
    \STATE{Accept proposed state and set $\theta_{t+1}\leftarrow \theta'$}
\ELSE
    \STATE{Reject proposed state and set current state as the next $\theta_{t+1}\leftarrow \theta_{t}$}
\ENDIF
\STATE{Update mean estimate $\mu_{t+1}\leftarrow\mu_{t}-\gamma_{t} (\mu_{t} - \theta_{t+1})$}
\STATE{Update covariance estimate 
\[
\Sigma_{t+1}\leftarrow \Sigma_{t} - \gamma_{t}\left[ \Sigma_{t} - (\mu_{t} - \theta_{t+1})(\mu_{t} - \theta_{t+1})^{T}\right]
\]}
\STATE{Update proposal scale estimate $\log\lambda_{t+1}\leftarrow\log\lambda_{t}-\gamma_{t} (\alpha_{t} - \alpha^{*})$}
\ENDFOR
\end{algorithmic}
\end{algorithm}
Let us now go through an example of developing an \textit{adaptive} version of the commonly used RWM, in which we tune the covariance matrix $\Sigma$ of the Gaussian proposal distribution, $\theta'\sim\mathcal{N}(\theta'\vert\theta,\lambda\Sigma)$, using the following \textit{diminishing adaptation} scheme, 
\begin{equation}
    \label{eq:covariance_adaptation}
    \begin{split}
        \mu_{i+1} &= \mu_{i} - \gamma_{i}\left( \mu_{i} - \theta_{i+1}\right)\,, \\
        \Sigma_{i+1} &= \Sigma_{i}- \gamma_{i}\left[ \Sigma_{i} - (\mu_{i} - \theta_{i+1})(\mu_{i} - \theta_{i+1})^{T}\right]\,,
    \end{split}
\end{equation}
where the $\mu$ is the mean value used for the estimation of the covariance $\Sigma$, and $\gamma_{i}=1/i$ is the \textit{learning rate}. At the same time we can also tune the \textit{magnitude} of the proposal scale, $\lambda_{i}$, by attempting to match the acceptance probability $\alpha$ to the theoretically optimal value of $\alpha^{*}=0.234$,
\begin{equation}
    \label{eq:proposal_scale_adaptation}
    \log\lambda_{i+1} = \log\lambda_{i} - \gamma_{i}\left( \alpha_{i} - \alpha^{*}\right)\,.
\end{equation}
Understanding equation \ref{eq:proposal_scale_adaptation} is straightforward, if the observed acceptance rate is greater than the target (i.e. $\alpha_{i} < \alpha^{*}$) then the logarithm of the magnitude of the proposal scale $\log\lambda$ is reduced and \textit{vice versa}. The \textit{Adaptive Metropolis} method presented in this paragraph constitutes a generalisation of the method presented by \textcite{haario2001adaptive}. With the inclusion of the adaptation of the proposal scale using equation \ref{eq:proposal_scale_adaptation} the algorithm resembles that of \textcite{andrieu2008tutorial}.

\section{Gibbs sampling}

Another very popular Markov chain Monte Carlo method, to which we partly owe the widespread use of Bayesian inference today, is \textit{Gibbs sampling}. Initially known as the \textit{heat bath} algorithm in the \textit{statistical physics} literature, the \textit{Gibbs sampler} enjoyed great success in the \textit{statistical community} following the seminal paper by \textcite{geman1984stochastic} that demonstrated its benefits for analysing \textit{Gibbs} distributions on lattices in the context of image processing.

\subsection{Gibbs sampler}

Gibbs sampler attempts to overcome the \textit{curse of dimensionality} using \textit{conditioning}~\parencite{casella1992explaining}. In particular, assuming that exact sampling from the conditional distributions of the target distribution $p(\theta)$ is possible, we can generate samples from the target distribution by sequentially sampling from its full set of conditionals instead. Given an initial state $\theta=(\theta_{1}, \dots, \theta_{D})$, the next state in the Markov chain can be generated as,
\begin{equation}
    \label{eq:gibbs_sampler}
    \begin{split}
        \theta_{1}' &\sim p(\theta_{1}\vert\theta_{2},\dots,\theta_{D}) \\
         & \vdots \\
        \theta_{k}' &\sim p(\theta_{k}'\vert\theta_{1},\dots,\theta_{k-1}',\theta_{k+1},\dots\theta_{D}) \\
         & \vdots \\
        \theta_{D}' &\sim p(\theta_{D}'\vert\theta_{1},\dots,\dots\theta_{D-1}')\,.
    \end{split}
\end{equation}
The current state $\theta$ is then replaced by the new state $\theta'=(\theta_{1}', \dots, \theta_{D}')$ and the process is repeated until enough states are collected in the Markov chain. The order of the state updates of equation \ref{eq:gibbs_sampler} can be either fixed (with a possible reversal after every iteration), as shown above, or randomised  to ensure detailed balance.

\begin{algorithm}[ht!]
\caption{Gibbs sampler} \algolabel{gibbs}
\begin{algorithmic}[1]
\REQUIRE{initial state $\theta^{(1)}=(\theta_{1}^{(0)}, \theta_{2}^{(0)}, \dots, \theta_{D}^{(0)})$, all the conditional distributions $p(\theta_{k}\vert \theta_{1},\dots,\theta_{k-1},\theta_{k+1},\dots,\theta_{D})$ of target $p(\theta)$, and number of iterations $N$}
\ENSURE{Markov chain $\theta^{(1)}, \theta^{(2)}, \dots, \theta^{(N)}$ that has $p(\theta)$ as its equilibrium distribution}
\FOR{$t=1$ \TO $N$}
    \FOR{$k=1$ \TO $D$}
        \STATE{Draw from conditional $\theta_{k}'\sim p(\theta_{k}\vert \theta_{1}',\dots,\theta_{k-1}',\theta_{k+1}^{(t)},\dots,\theta_{D}^{(t)})$}
    \ENDFOR
    \STATE{Set $\theta^{(t+1)}\leftarrow \theta' = (\theta_{1}',\dots,\theta_{D}')$}
\ENDFOR
\end{algorithmic}
\end{algorithm}

\subsection{Metropolis--within--Gibbs sampler}

The \textit{Gibbs sampler} relies on our ability to produce samples from each one of the conditional distributions. This however is not always feasible as some of the components of the full conditional set might not admit an exact sampling solution. Instead of abandoning \textit{Gibbs sampler} altogether, \textcite{muller1991generic,muller1992alternatives} suggested the use of a compromise between the \textit{Gibbs sampler} and the \textit{Metropolis--Hastings} algorithm.

The key idea behind the \textit{Metropolis--within--Gibbs} sampler is to use \textit{Gibbs sampling} for as many of the conditional distributions as possible in order to produce exact samples, and rely on correlated samples generated using \textit{Metropolis--Hastings} for any conditional distributions that \textit{exact sampling} is not possible.

Suppose that we have a \textit{partial} state $(\theta_{1}',\dots,\theta_{k-1}',\theta_{k},\dots,\theta_{D}$) and we have difficulty generating exact samples from the conditional distribution $\theta_{k}'\sim p(\theta_{k}\vert\theta_{1}',\dots,\theta_{k-1}',\theta_{k+1},\dots,\theta_{D})$. We can then treat $p(\theta_{k}\vert\theta_{1}',\dots,\theta_{k-1}',\theta_{k+1},\dots,\theta_{D})$ as the target distribution for a \textit{Metropolis--Hastings} estimator as follows, in order to proceed with the computation,
\begin{enumerate}
    \item First, we have to propose a new sample $\theta_{k}^{*}\sim q(\theta_{k}\vert\theta_{1}',\dots,\theta_{k-1}',\theta_{k+1},\dots,\theta_{D})$ from an arbitrary proposal distribution,
    \item Then, compute the \textit{Metropolis acceptance probability} 
    \begin{equation}
        \label{eq:acceptance_probability_metropolis_within_gibbs}
        \begin{split}
            \alpha(\theta_{k}^{*},\theta_{k}) = \min\bigg(1, &\frac{p(\theta_{k}^{*}\vert\theta_{1}',\dots,\theta_{k-1}',\theta_{k+1},\dots,\theta_{D}) }{p(\theta_{k}\vert\theta_{1}',\dots,\theta_{k-1}',\theta_{k+1},\dots,\theta_{D}) } \\
            \times &\frac{q(\theta_{k}\vert\theta_{1}',\dots,\theta_{k-1}',\theta_{k+1},\dots,\theta_{D})}{q(\theta_{k}^{*}\vert\theta_{1}',\dots,\theta_{k-1}',\theta_{k+1},\dots,\theta_{D})}\bigg)\,,
        \end{split}
    \end{equation}
    \item Finally, accept the new state $\theta_{k}'\leftarrow\theta_{k}^{*}$ with probability $\alpha(\theta_{k}^{*},\theta_{k})$, otherwise reject and keep the previous state $\theta_{k}'\leftarrow\theta_{k}$\,.
\end{enumerate}
Using the algorithm presented above we can replace \textit{exact sampling} from conditional distributions where it is not feasible with \textit{Metropolis--Hastings} estimates.

\begin{algorithm}[ht!]
\caption{Metropolis--within--Gibbs sampler} \algolabel{metropolis_within_gibbs}
\begin{algorithmic}[1]
\REQUIRE{initial state $\theta^{(1)}=(\theta_{1}^{(0)}, \theta_{2}^{(0)}, \dots, \theta_{D}^{(0)})$, proposal distributions $q(\theta_{k}\vert \theta_{1},\dots,\theta_{k-1},\theta_{k+1},\dots,\theta_{D})$, (unnormalised) target distribution $f(\theta)\propto p(\theta)$, the conditional distributions $p(\theta_{\ell}\vert \theta_{1},\dots,\theta_{\ell-1},\theta_{\ell+1},\dots,\theta_{D})$ of target $p(\theta)$ where $\ell \in L$ and $L$ the set of indices for which exact sampling of the conditional is possible, and number of iterations $N$}
\ENSURE{Markov chain $\theta^{(1)}, \theta^{(2)}, \dots, \theta^{(N)}$ that has $p(\theta)$ as its equilibrium distribution}
\FOR{$t=1$ \TO $N$}
    \FOR{$k=1$ \TO $D$}
        \IF{$k\in L$}
            \STATE{Draw from conditional $\theta_{k}'\sim p(\theta_{k}\vert \theta_{1}',\dots,\theta_{k-1}',\theta_{k+1}^{(t)},\dots,\theta_{D}^{(t)})$}
        \ELSE
            \STATE{Draw from proposal $\theta_{k}^{*}\sim q(\theta_{k}\vert     \theta_{1}',\dots,\theta_{k-1}',\theta_{k+1}^{(t)},\dots,\theta_{D}^{(t)})$}
            \STATE{Compute acceptance probability
            \begin{equation*}
                \begin{split}
                    \alpha = \min\bigg( 1, &\frac{p(\theta_{k}^{*}\vert \theta_{1}',\dots,\theta_{k-1}',\theta_{k+1}^{(t)},\dots,\theta_{D}^{(t)})}{p(\theta_{k}\vert \theta_{1}',\dots,\theta_{k-1}',\theta_{k+1}^{(t)},\dots,\theta_{D}^{(t)})} \\
                    \times &\frac{q(\theta_{k}\vert \theta_{1}',\dots,\theta_{k-1}',\theta_{k+1}^{(t)},\dots,\theta_{D}^{(t)})}{q(\theta_{k}^{*}\vert \theta_{1}',\dots,\theta_{k-1}',\theta_{k+1}^{(t)},\dots,\theta_{D}^{(t)})} \bigg)
                \end{split}
            \end{equation*}
            }
            \STATE{Draw uniform number $u\sim\mathcal{U}(0,1)$}
            \IF{$u < \alpha$}
                \STATE{accept new partial state and set $\theta_{k}'\leftarrow \theta_{k}^{*}$}
            \ELSE
                \STATE{reject new partial state and set     $\theta_{k}'\leftarrow \theta_{k}^{(t)}$}
            \ENDIF
        \ENDIF
    \ENDFOR
    \STATE{Set $\theta^{(t+1)}\leftarrow \theta' = (\theta_{1}',\dots,\theta_{D}')$}
\ENDFOR
\end{algorithmic}
\end{algorithm}
% !TEX TS-program = pdflatex
% !TEX root = ../ArsClassica.tex

%************************************************
\chapter{Auxiliary variable MCMC methods}
\label{chp:auxiliary}
%************************************************

\begin{flushright}
\itshape
Natura non facit saltus. \\
\medskip
--- Gottfried Leibniz
\end{flushright}

Auxiliary variable MCMC methods rely on the introduction of one or more additional variables in order to make sampling from the target distribution more efficient.

\section{Simulated annealing}

In metallurgy, \textit{annealing} refers to the thermal process used to harden steel. Initially, the metal is heated to a high temperature and then it is cooled down slowly enough for the atoms to \textit{self--arrange} in an ordered pattern that corresponds to the minimum energy~\parencite{cahn1996physical}. The slow rate of cooling ensures that the energy of the system will reach its global minimum instead of getting trapped in local minima.

Realising that the \textit{Metropolis--Hastings} method can be used to simulate the process of gradually cooling a solid towards a low--temperature equilibrium state, \textcite{kirkpatrick1983optimization} suggested that we should construct a sequence of \textit{Boltzmann} distributions,
\begin{equation}
    \label{eq:boltzmann_distribution}
    p_{i}(\theta) \propto e^{-\frac{H(\theta)}{T_{i}}}\,,
\end{equation}
for a series of temperatures $T_{1}>T_{2}>\dots>T_{m}$ and simulate from each one in succession by performing a number of MCMC steps in each temperature before moving on to the next. The result is a non--homogeneous Markov chain, that is, a Markov chain with time--varying target density. Assuming that $T_{1}$ is high enough and $T_{m}\approx 0$ we can find the global minimum of the energy $E(\theta)$ by simulating the cooling process of a solid.

\begin{figure}[ht!]
    \centering
	\centerline{\includegraphics[scale=0.65]{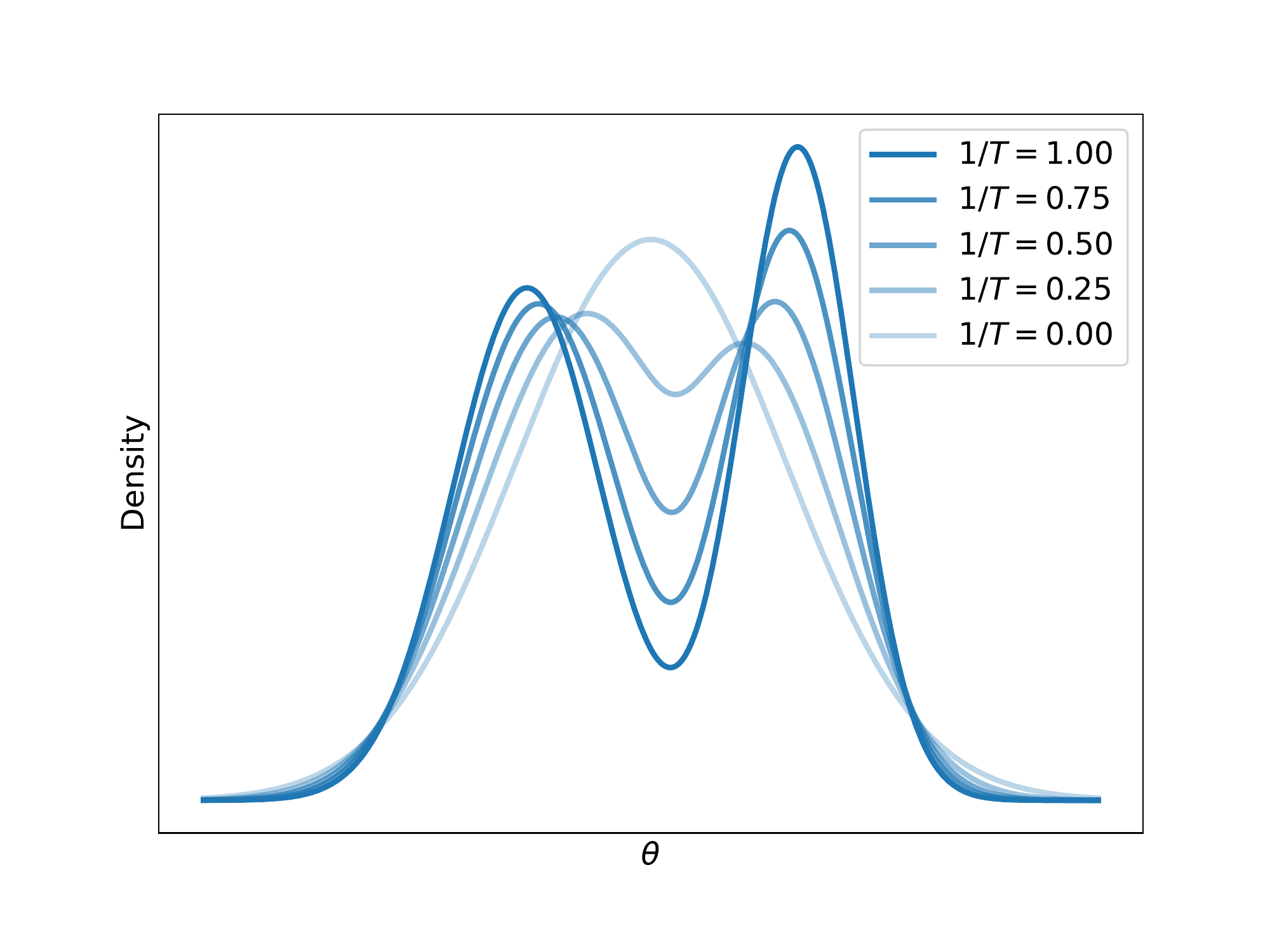}}
    \caption{Illustration of the gradual annealing performed in the posterior distribution. The prior distribution corresponds to $1/T\rightarrow 0$ and the posterior is recovered as $1/T\rightarrow 1$.}
    \label{fig:annealing}
\end{figure}

\textit{Simulated annealing} can be used for sampling too, not just for optimisation. By stopping the cooling process earlier at $T_{m}=1$ we can sample from any target distribution $p(\theta)$, not just \textit{Boltzmann} distributions, simply setting $H(\theta)=-\log p(\theta)$.  Furthermore, for applications in which the target distribution is the posterior distribution we can construct the following sequence of densities,
\begin{equation}
    \label{eq:boltzmann_distribution_bayesian}
    p_{i}(\theta) \propto p(\theta) e^{\frac{\log p(d\vert\theta)}{T_{i}}}\,,
\end{equation}
where $T_{1}>T_{2}>\dots>T_{m} = 1$. In this case, for $T_{1} >> 1$ we effectively sample from the prior distribution,
\begin{equation}
    \label{eq:boltzmann_prior}
    p_{i}(\theta) \propto p(\theta)\,,
\end{equation}
whereas in the limit that $T_{m} = 1$ we acquire samples from the posterior,
\begin{equation}
    \label{eq:boltzmann_prior_2}
    p_{i}(\theta) \propto p(\theta)p(d\vert\theta)\,.
\end{equation}
The number of MCMC steps to perform in each temperature, before moving on to the next one, is arbitrary and different mixing criteria can be utilised (e.g. Gelman--Rubin, autocorrelation thresholds, etc.). The benefit of using simulated annealing for sampling is that by simulating multiple Markov chains, possibly in parallel, through this sequence of densities, the risk of the chains getting trapped in isolated modes of the posterior distribution is minimised. This means that this approach can be used when the probability distribution is strongly multimodal. Furthermore, if the number of temperature levels is large enough and the spacing between them small enough, then the Markov chain is approximately always in equilibrium, meaning that no, or minor, burn--in is required to be discarded.

\begin{algorithm}[ht!]
\caption{Simulated annealing} \algolabel{sa}
\begin{algorithmic}[1]
\REQUIRE{initial state $\theta_{1}^{(1)}$, temperature schedule $T_{1}>T_{2}>\dots>T_{m}=1$, prior density $\pi(\theta)\equiv p(\theta\vert\mathcal{M})$, likelihood function $\mathcal{L}(\theta)\equiv p(d\vert\theta,\mathcal{M})$, and number of MCMC iterations $N$ per temperature}
\ENSURE{Multiple Markov chains $\theta^{(1)}, \theta^{(2)}, \dots, \theta^{(m)}$ with the last one having $p(\theta)$ as its equilibrium distribution}
\FOR{$i=1$ \TO $m$}
    \STATE{Set annealed density $p_{i}(\theta) \propto \pi(\theta)\mathcal{L}(\theta)^{1/T_{i}}$}
    \STATE{Generate Markov chain $\theta_{1}^{(i)},\dots,\theta_{N}^{(i)}$ targeting $p_{i}(\theta)$ (e.g. using Metropolis--Hastings)}
    \STATE{Set last state as the first state for the next annealed density $\theta_{1}^{(i+1)}\leftarrow \theta_{N}^{(i)}$}
\ENDFOR
\end{algorithmic}
\end{algorithm}

%************************************************

\section{Slice sampling}

Slice sampling is another MCMC method that relies on an auxiliary variable in order to make sampling easier~\parencite{besag1993spatial, neal1997markov, neal2003slice}. The method is based on the realisation that sampling from the target distribution with density $p(\theta)$ is equivalent to uniform sampling from the area or volume below the curve or surface of $f(\theta)\propto p(\theta)$. This is equivalent to the introduction of an auxiliary variable $\phi$, called \textit{height}, such that the joint distribution $p(\theta, \phi)$ is uniform over the region,
\begin{equation}
    \label{eq:uniform_region}
    U = \left\lbrace (\theta, \phi)\,:\,0 < \phi < f(\theta) \right\rbrace\,.
\end{equation}
In other words, the joint distribution can be written as,
\begin{equation}
    \label{eq:joint_distribution}
    p(\theta,\phi) = 
    \begin{cases}
    1/\mathcal{Z} & \text{if } 0 < \phi < f(\theta)\,, \\
    0 & \text{otherwise}\,,
    \end{cases}
\end{equation}
where
\begin{equation}
    \label{eq:normalisation_constant}
    \mathcal{Z} = \int f(\theta) d\theta\,.
\end{equation}
To sample from the target distribution $p(\theta)$ we first sample uniformly from $p(\theta, \phi)$ and then marginalise over $\phi$ by dropping the $\phi$--value of each sample and keeping the $\theta$--value. The proof that this results in the marginal density for $\theta$ is straightforward,
\begin{equation}
    \label{eq:marginal_density}
    p(\theta) = \int p(\theta, \phi) d\phi = \int_{0}^{f(\theta)} \frac{1}{\mathcal{Z}} d\phi = \frac{f(\theta)}{\mathcal{Z}}\,.
\end{equation}

Generating independent samples from the uniform joint density $p(\theta, \phi)$ is rarely possible in practice. Instead, one might prefer to construct a Markov chain that leaves the distribution $p(\theta, \phi)$ invariant. One such option is to use \textit{Gibbs sampling}, that is, to sample alternately from the conditional distribution $p(\phi\vert\theta)$, which is uniform over the interval $(0, f(\theta))$, and then from the conditional distribution $p(\theta\vert\phi)$, which is uniform over the region,
\begin{equation}
    \label{eq:slice_definition}
    S = \left\lbrace \theta\,:\,\phi < f(\theta)\right\rbrace\,,
\end{equation}
called the \textit{slice}. Applying this procedure repeatedly will produce a Markov chain that has the joint distribution $p(\theta, \phi)$ as its stationary distribution.

Sampling uniformly from the aforementioned slice is not trivial either. However, the fact that the conditional density $p(\theta\vert\phi)$ is uniform allows us to construct procedures to sample from it which would otherwise would not have worked. \textcite{neal2003slice} proposed the following sequence of steps for univariate probability distributions,
\begin{enumerate}
    \item Uniformly sample a real value $\phi$ in the interval $(0, f(\theta_{0}))$, therefore defining the horizontal slice $S = \left\lbrace \theta\,:\,\phi < f(\theta_{0})\right\rbrace$ that always includes $\theta_{0}$,
    \item Find an interval $I=(L,R)$ around $\theta_{0}$ along the slice that contains all, or much of, the slice,
    \item Sample a new value $\theta_{1}$ from the part of the slice within the interval, that is, from $I\cap S$.
\end{enumerate}
It is important to mention that as we often work with $g(\theta) = \log f(\theta)$, to avoid numerical issues, one can use the variable $\psi = \log(\phi) = g(\theta_{0}) - e$, where $e$ is exponentially distributed with mean one, to define the slice as $S = \lbrace \theta\,:\,\psi<g(\theta_{0}) \rbrace$.

\begin{figure}[ht!]
    \centering
	\centerline{\includegraphics[scale=0.65]{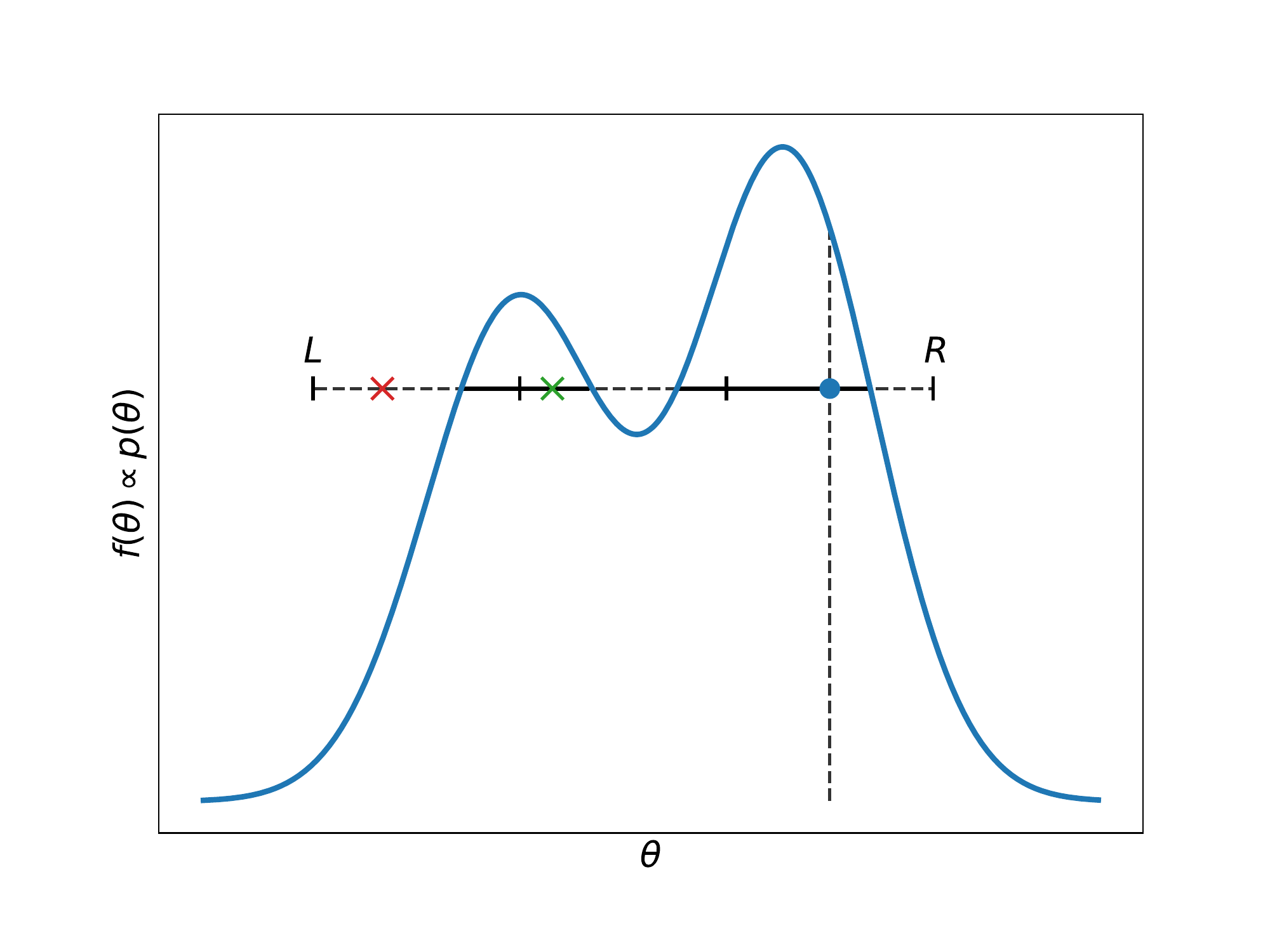}}
    \caption{Illustration of the stepping--out and shrinking procedures used in slice sampling. Given an initial state $\theta$ in the Markov chain, an auxiliary variable $\phi$ is sampled corresponding to the height thus defining the extended state $(\theta, \phi)$ shown here as a \textit{blue} point. An interval of a certain width is placed uniformly around the current point $(\theta, \phi)$ and expanded in steps of size equal to the initial width until both of its ends, $L$ and $R$, are outside the graph. A new state, shown in \textit{red}, is then proposed uniformly along the interval $(L,R)$. Since the proposed state lies above the graph of $f(\theta)$ (i.e. not in the slice shown as a continuous line) it is rejected. A new state, shown in \textit{green}, is the proposed uniformly between the rejected state and $R$. Since the proposed state is below the graph, and thus in the slice, it is accepted and added to the Markov chain. The whole process is then repeated.}
    \label{fig:slice_sampling}
\end{figure}

The first step in the above procedure is trivial, yet steps two and three require more serious consideration. \textcite{neal2003slice} suggested to use the so--called \textit{stepping--out} and \textit{shrinking} procedures for those steps respectively. \textit{Stepping--out} works by uniformly positioning an interval of width $w$ around $\theta_{0}$ such that it includes $\theta_{0}$, and then expanding the interval in steps of size $w$ until both its ends $(L, R)$ are outside the slice $S$. This effectively constructs the interval $I=(L, R)$. It is worth noting that the algorithm is valid even if only a pre--specified number of expansions take place and the interval ends up not covering the entirety of the slice $S$. The \textit{shrinking} procedure that follows functions by uniformly sampling points in the interval $I=(L, R)$ until one of them lies in the slice $S$. Every time a point is rejected, being outside of the slice, the interval $I$ shrinks such that the rejected point now defines one of its two boundaries, determined by whether the rejected point lies left or right of $\theta_{0}$.

The fact that the three--step procedure presented so far describes a slice sampling update from a univariate probability distribution $p(\theta)$ does not prohibit its use in multivariate cases. In particular, there are many ways the aforementioned recipe can be generalised and used in target distribution with more than one parameter. Perhaps the simplest one is to apply this univariate scheme along each coordinate axis in turn, updating one parameter at a time. This corresponds to a \textit{Metropolis--within--Gibbs} scheme. Another option is to apply 1--D updates in random directions. This is more general than the previous one, and there is freedom to choose the distribution of the random directions. The directions can be drawn from a multivariate zero--mean normal distribution with unit--diagonal covariance matrix or a more appropriate non--diagonal covariance matrix that encodes some of the correlations of the parameters of the target distribution. Such a covariance can be configured \textit{a priori}, estimated during a short preliminary run from samples from the target, or adaptively tuned using an appropriate algorithm for \textit{diminishing adaptation}.

One of the great benefits of slice sampling is the fact that it has a single hyper--parameter, the initial width $w$ of the interval $I$. Furthermore, the value of $w$ is adapted continuously by the \textit{stepping--out} and \textit{shrinking} procedures. This sort of \textit{local adaptation} is absent from many MCMC that assume a global proposal scale. Another characteristic of slice sampling is the lack of rejected samples in the Markov chain. Unlike methods that include a Metropolis acceptance criterion, slice sampling always moves to a new state in every iteration.

\begin{algorithm}[ht!]
\caption{Slice sampling} \algolabel{ss}
\begin{algorithmic}[1]
\REQUIRE{initial state $\theta_{1}$, (unnormalised) target density $f(\theta)\propto p(\theta)$, number of maximum expansions $m$, and number of iterations $N$}
\ENSURE{Markov chain $\theta_{1}, \dots, \theta_{N}$ that has $p(\theta)$ as its equilibrium distribution}
\STATE{Draw ``height'' auxiliary variable $\phi \sim \mathcal{U}(0,f(\theta_{1}))$}
\FOR{$t=1$ \TO $N$}
    \STATE{Draw left bound of the $I$ interval $L\sim\mathcal{U}(\theta_{t}-w,\theta_{t})$}
    \STATE{Set right bound $R\leftarrow L + w$}
    \STATE{Draw uniform variable $u\sim\mathcal{U}(0,1)$}
    \STATE{Set maximum number of interval expansions to the left $J\leftarrow \text{Floor}(m\times u)$}
    \STATE{Set maximum number of interval expansions to the right $K\leftarrow (m-1)-J$}
    \WHILE{$J>0$ \AND $\phi<f(L)$}
        \STATE{Expand left boundary $L\leftarrow L - w$ in steps of $w$}
        \STATE{Reduce count $J \leftarrow J - 1$ by one}
    \ENDWHILE
    \WHILE{$K>0$ \AND $\phi<f(R)$}
        \STATE{Expand right boundary $L\leftarrow R + w$ in steps of $w$}
        \STATE{Reduce count $K \leftarrow K - 1$ by one}
    \ENDWHILE
    \REPEAT
        \STATE{Draw state within interval $\theta'\sim \mathcal{U}(L,R)$}
        \IF{$\theta'<\theta_{t}$}
            \STATE{Contract interval $L\leftarrow \theta'$}
        \ELSE
            \STATE{Contract interval $R\leftarrow \theta'$}
        \ENDIF
    \UNTIL{$\phi < f(\theta')$}
    \STATE{Accept new state $\theta_{t+1}\leftarrow \theta'$}
\ENDFOR
\end{algorithmic}
\end{algorithm}

%************************************************

\section{Hamiltonian Monte Carlo}

\textit{Hamiltonian Monte Carlo (HMC)} introduces a \textit{momentum} auxiliary variable and uses the gradient of the target probability density to efficiently explore the typical set. HMC turns the problem of sampling from the target distribution into the approximate simulation of Hamiltonian dynamics with a subsequent \textit{Metropolis} correction step~\parencite{neal2011mcmc}. In the statistical physics literature HMC was suggested as a method of efficiently simulating states from a physical system~\parencite{duane1987hybrid}, which was then employed to statistical inference problems~\parencite{neal1992bayesian, neal1993probabilistic, neal1996, liu2001monte}.

% https://mc-stan.org/docs/2_29/reference-manual/hamiltonian-monte-carlo.html

\begin{figure}[ht!]
    \centering
	\centerline{\includegraphics[scale=0.65]{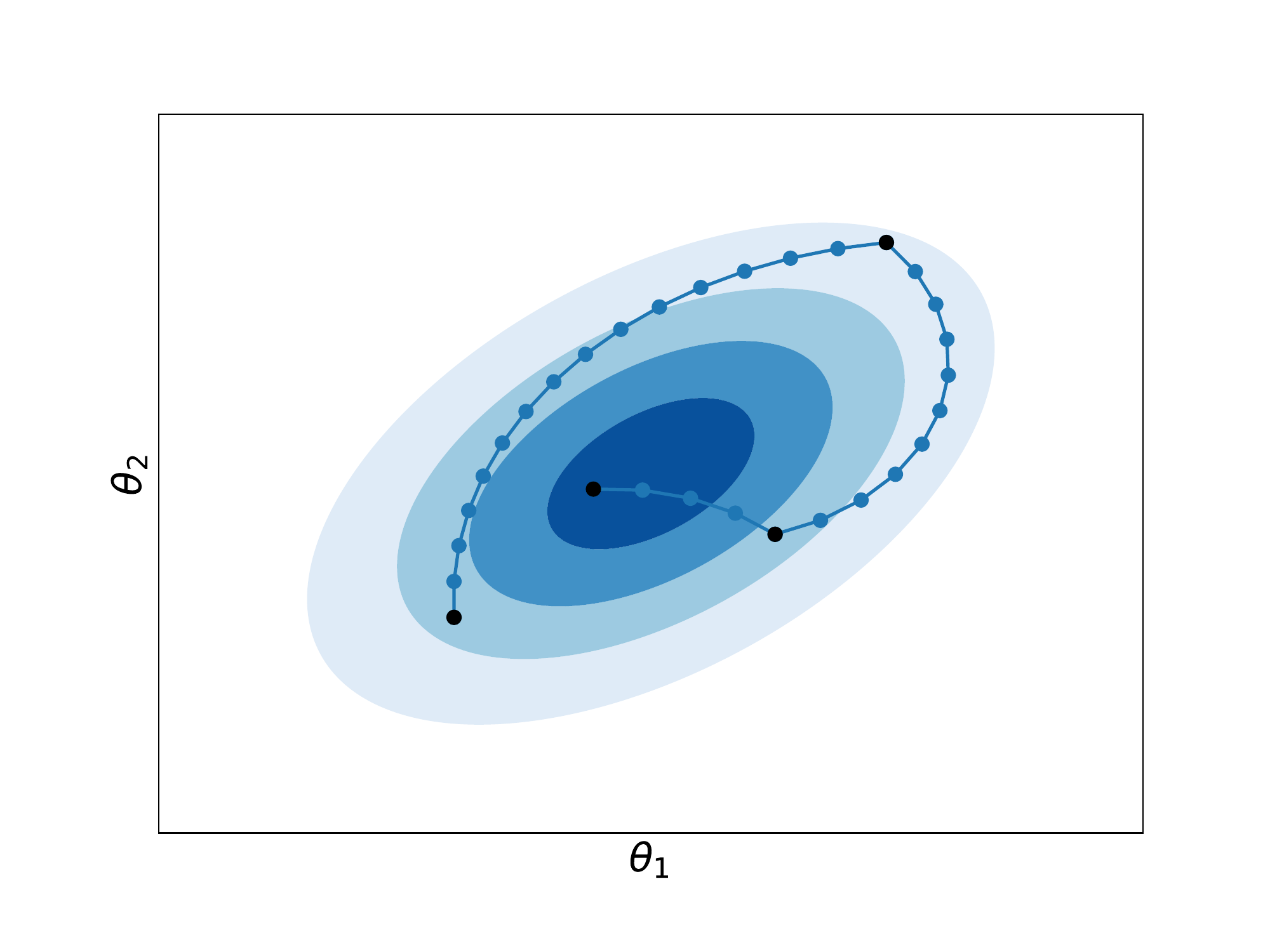}}
    \caption{Illustration of Hamiltonian trajectories in parameter space. The \textit{black} points correspond to the accepted states.}
    \label{fig:hamiltonian}
\end{figure}

\subsection{Auxiliary momentum variable}
HMC introduces an auxiliary variable $\rho$ and samples from the joint probability density,
\begin{equation}
    \label{eq:hmc_joint_distribution}
    p(\rho, \theta) = p(\rho\vert\theta)p(\theta)\,.
\end{equation}
In most applications of HMC, the momentum variable $\rho$ chosen to be Gaussian--distributed,
\begin{equation}
    \label{eq:hmc_momentum}
    \rho \sim \mathcal{N}(0,M)\,,
\end{equation}
and its probability density function to be independent of the state variable $\theta$ (i.e. $p(\rho\vert\theta)=p(\theta)$). $M$ is the symmetric, \textit{positive definite} \textit{mass matrix} that has the role of the \textit{Euclidean metric}, that is to define the relative \textit{length scales} between parameters. In practice, $M$ can be chosen to be $\Sigma^{-1}$, meaning the inverse of the sample covariance matrix that characterises the target distribution assuming that it is known or easy to estimate.

\subsection{The Hamiltonian}
HMC treats sampling from the joint distribution $p(\rho,\theta)$ as a problem of solving the \textit{Hamiltonian dynamics} given the \textit{Hamiltonian},
\begin{equation}
    \label{eq:hamiltonian}
    \begin{split}
        \mathcal{H}(\rho, \theta) &= - \log p(\rho, \theta) \\
         &= -\log p(\rho\vert\theta) - \log p(\theta) \\
         &= T(\rho\vert\theta) + V(\theta) \,,
    \end{split}
\end{equation}
where
\begin{equation}
    \label{eq:kinetic_energy}
    T(\rho\vert\theta) = -\log p(\rho\vert\theta)\,,
\end{equation}
is the \textit{kinetic energy}, and,
\begin{equation}
    \label{eq:potential_energy}
    V(\theta) = - \log p(\theta)\,,
\end{equation}
is the \textit{potential energy}.

\subsection{Hamilton's equations}
The dynamics of a system (i.e. its evolution in time) that is characterised by the \textit{Hamiltonian} of  equation \ref{eq:hamiltonian} are given by solving \textit{Hamilton's equations},
\begin{equation}
    \label{eq:hamiltons_equations}
    \begin{split}
        \frac{d\theta}{dt} &= \frac{\partial\mathcal{H}}{\partial\rho} = \frac{\partial T}{\partial\rho}\,,\\
        \frac{d\rho}{dt} &= -\frac{\partial\mathcal{H}}{\partial\theta} = -\frac{\partial T}{\partial\theta}-\frac{\partial V}{\partial\theta}\,,
    \end{split}
\end{equation}
or, in the case that the momentum variable $\rho$ is independent of the state variable $\theta$, that is $p(\rho\vert\theta)=p(\rho)$,
\begin{equation}
    \label{eq:hamiltons_equations_2}
    \begin{split}
        \frac{d\theta}{dt} &= \frac{\partial\mathcal{H}}{\partial\rho} = \frac{\partial T}{\partial\rho}\,,\\
        \frac{d\rho}{dt} &= -\frac{\partial\mathcal{H}}{\partial\theta} = -\frac{\partial V}{\partial\theta}\,.
    \end{split}
\end{equation}
Therefore, given an initial state $(\rho, \theta)$, the system's evolution in time is completely determined by equations \ref{eq:hamiltons_equations_2}.

\subsection{Leapfrog integration}
Solving \textit{Hamilton's equations} analytically is only feasible for very simple systems that correspond to simple target probability distributions. In practice, however, we aim to solve equations \ref{eq:hamiltons_equations_2} for systems of arbitrary complexity. To this end, we turn to numerical methods for integrating this system of differential equations.

The most commonly used numerical method is the \textit{leapfrog integration algorithm}~\parencite{leimkuhler2004simulating} that begins by sampling a value for the momentum variable $\rho$ according to equation \ref{eq:hmc_momentum} and then proceeds by applying $L$ times the following steps,
\begin{equation}
    \label{eq:leapfrog_integration}
    \begin{split}
        \rho &\leftarrow \rho - \frac{\epsilon}{2} \frac{\partial V}{\partial\theta}\,, \\
        \theta &\leftarrow \theta + \epsilon M^{-1} \rho\,, \\
        \rho &\leftarrow \rho - \frac{\epsilon}{2} \frac{\partial V}{\partial\theta}\,,
    \end{split}
\end{equation}
where $\epsilon$ is the \textit{integration step size} that determines the smallest time interval. The length of the trajectory will then be $\epsilon L$ and the new state of the system is denoted as $(\rho',\theta')$. The numerical error introduced into the calculation by the \textit{leapfrog} algorithm is of the order of $\epsilon^{3}$ per step and $\epsilon^{2}$ globally~\parencite{leimkuhler2004simulating}.

\subsection{Metropolis acceptance criterion}
If the \textit{leapfrog algorithm} were perfect and did not introduce any numerical error, we would not have to do anything more than re--sample the \textit{momentum} variable every $L$ integration steps. However, the \textit{leapfrog integrator} is far from this which means that we need to account for the numerical error that it introduces before it accumulates. To this end, we only accept and add the new state $(\rho',\theta')$ into the Markov chain with probability
\begin{equation}
    \label{eq:hmc_acceptance}
    \alpha(\theta',\theta) = \min\left(1, \frac{p(\rho',\theta')}{p(\rho,\theta)} \right)\,,
\end{equation}
and reject it otherwise by adding $(\rho,\theta)$ into the chain. Equation \ref{eq:hmc_acceptance} is simply the \textit{Metropolis acceptance probability} for HMC. Therefore, we see that HMC is essentially a case of \textit{Metropolis--Hastings} with symmetric proposal distribution in the augmented state space of $(\rho,\theta)$.

\begin{algorithm}[ht!]
\caption{Hamiltonian Monte Carlo} \algolabel{hmc}
\begin{algorithmic}[1]
\REQUIRE{initial state $\theta_{1}$, potential energy $V(\theta)=-\log p(\theta)$ up to an additive constant, kinetic energy definition $T(\rho)=\rho^{T}M^{-1}\rho / 2$, number of leapfrog steps $L$, integration step size $\epsilon$, and number of iterations $N$}
\ENSURE{Markov chain $\theta_{1}, \dots, \theta_{N}$ that has $p(\theta)$ as its equilibrium distribution}
\STATE{Draw momentum variable $\rho_{t}\sim \mathcal{N}(0,M)$}
\STATE{Set proposed state $\theta'\leftarrow \theta_{t}$}
\STATE{Set proposed momentum $\rho'\leftarrow \rho$}
\FOR{$=1$ \TO $L$}
    \STATE{Update momentum $\rho' \leftarrow \rho' - \frac{\epsilon}{2}\frac{\partial V}{\partial\theta}$}
    \STATE{Update position $\theta'\leftarrow \theta' + \epsilon M^{-1} \rho'$}
    \STATE{Update momentum $\rho' \leftarrow \rho' - \frac{\epsilon}{2}\frac{\partial V}{\partial\theta}$}
\ENDFOR
\STATE{Reverse momentum $\rho' \leftarrow -\rho'$}
\STATE{Compute acceptance probability 
\[
\alpha = \min\left( 1, \exp\left[ V(\theta) - V(\theta') + T(\theta) - T(\theta') \right]\right)
\]}
\STATE{Draw uniform number $u\sim\mathcal{U}(0, 1)$}
\IF{$u < \alpha$}
    \STATE{Accept proposed state and set $\theta_{t+1}\leftarrow\theta'$}
\ELSE
    \STATE{Reject proposed state and set $\theta_{t+1}\leftarrow\theta$}
\ENDIF
\end{algorithmic}
\end{algorithm}

\subsection{Performance and tuning}

The sampling performance of HMC is very sensitive to its tuning~\parencite{neal2011mcmc, hoffman2014no} and many efforts have been made to develop heuristics and automated tuning procedures for the two hyperparameters, $\epsilon$ and $L$, that the method relies upon. The step size $\epsilon$ can be adaptively tuned by trying to match the observed acceptance rate to the theoretically optimal value of $0.65$. Tuning the number of steps $L$ is more cumbersome in practice. In principle, $L$ can be tuned by minimising the autocorrelation time of the Markov chain. In practice this requires running multiple preliminary runs with different values of $L$ in order to determine the most efficient one. For this reason, other approaches, such as \textit{Empirical HMC}~\parencite{wu2018faster} and the \textit{No U-Turn Sampler (NUTS)}~\parencite{hoffman2014no}, have been proposed that automate the use of HMC for many applications.

%************************************************

% !TEX TS-program = pdflatex
% !TEX root = ../ArsClassica.tex

%************************************************
\chapter{Ensemble MCMC methods}
\label{chp:ensemble}
%************************************************

\begin{flushright}
\itshape
As for me, I am tormented with an everlasting itch for things remote.\\ I love to sail forbidden seas, and land on barbarous coasts. \\
\medskip
--- Herman Melville, Moby--Dick or, the Whale
\end{flushright}

In order to avoid issues caused by multimodality or the need for tuning the proposal distribution, ensemble MCMC methods rely on an ensemble of parallel samplers, often called \textit{walkers}, that sample from an extended probability distribution. A common way to construct such an extended probability distribution is using the product density,
\begin{equation}
    \label{eq:extended_density}
    \pi(\lbrace\theta_{k}\rbrace_{k=1}^{K}) = \prod_{k=1}^{K} p_{k}(\theta_{k})\,,
\end{equation}
where $p_{k}(\theta_{k})$ are the individual densities, one of which can correspond to the target distribution of interest (e.g. the posterior), and $K$ is the number of walkers. It is important to note here that $\theta_{k}$ is not the $k$--th component of a vector, but a $D$--dimensional vector itself.

The simplest product density that we can construct based on equation \ref{eq:extended_density} is to assume that $p_{k}(\theta_{k})=p(\theta_{k})$ for all $k$, meaning that the product density is just the product of $K$ identical copies of the target distribution $p(\theta_{k})$. A natural question to ask is then why would anyone want to do this? Why sample $K$ copies of the same distribution instead of just one? The answer is that the walkers sampling each copy do not have to be independent of each other and instead are allowed to exchange information about their current state. For instance, the proposal distribution for a single walker can depend on the current positions of the rest of the walkers in the ensemble. This allows for effective proposals that take into account the relative length--scales and positions of the modes of the target distribution.

Of course, other product densities, that do not rely on the simplifying assumption that $p_{k}(\theta_{k})=p(\theta_{k})$ for all $k$, can also be defined as we will see in the case of the \textit{parallel tempering} algorithm in Section \ref{sec:parallel_tempering}. In those cases, the goal is not usually to construct effective proposal distribution but rather to deal with the challenge of multimodality.

\section{Gaussian ensemble}

Perhaps the simplest way to construct an ensemble MCMC method that limits the requirement for tuning, to some extent, its proposal distribution is the \textit{Gaussian ensemble (GE)} algorithm. GE uses an ensemble of $K$ walkers that target a product density of the form of equation \ref{eq:extended_density}, where all copies $p_{k}(\theta_{k})$ are identical and correspond to the target distribution of interest (e.g. posterior), and the proposal distribution of each walker is simply a normal distribution informed by the positions of the rest of the walkers in the ensemble~\parencite{speagle2019conceptual}.

\begin{figure}[ht!]
    \centering
	\centerline{\includegraphics[scale=0.65]{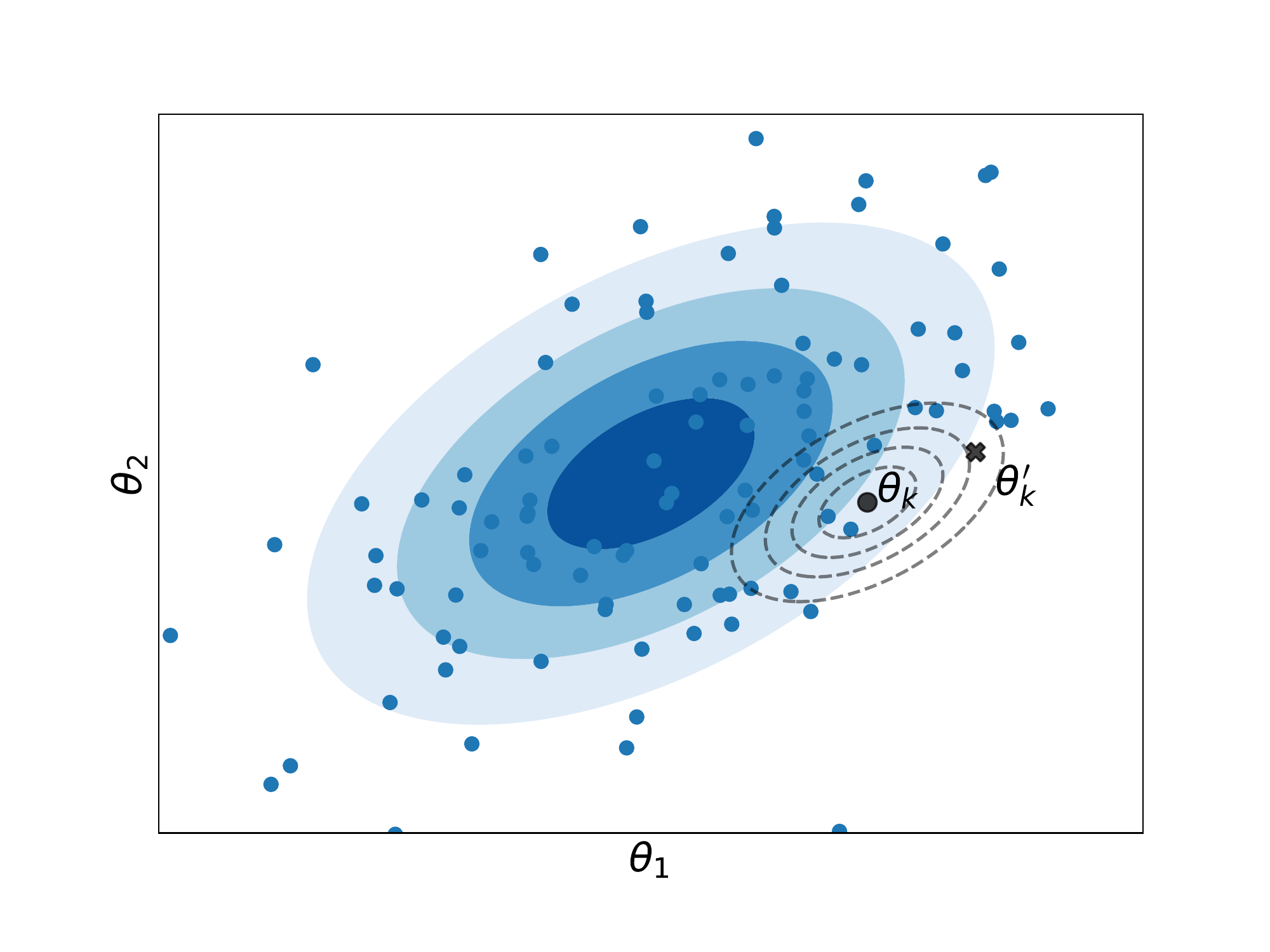}}
    \caption{Illustration of the Gaussian ensemble MCMC method. A new state $\theta_{k}'$ is proposed in the vicinity of the position $\theta_{k}$ of the walker that is updated using an rescaled version of the sample covariance matrix of the rest of the walkers (i.e. excluding $\theta_{k}$) for the normal proposal distribution.}
    \label{fig:gaussian_ensemble}
\end{figure}
In particular, in a given iteration $t$ of the method, the algorithm performs a loop over the $K$ walkers updating each walker in turn. A new position $\Tilde{\theta}_{k}$ is proposed from a normal distribution,
\begin{equation}
    \label{eq:ensemble_gaussian_proposal}
    \theta_{k}' \sim \mathcal{N}\left(\theta\vert \theta_{k}^{(t-1)}, \gamma\Sigma_{-k}\right)\,,
\end{equation}
centred on the current state $\theta_{k}^{(t-1)}$ of the $k$--th walker and $\gamma$ is a multiplying factor used to scale the covariance matrix in order to achieve the optimal acceptance rate (e.g. $\gamma = 2.38^{2}/D$). The covariance matrix $\Sigma_{-k}$ of the proposal distribution is simply the sample covariance estimated using the positions of the ensemble $\lbrace\theta_{1}^{(t)},\allowbreak \dots,\allowbreak \theta_{k-1}^{(t)},\allowbreak \theta_{k+1}^{(t-1)},\allowbreak\dots ,\allowbreak \theta_{K}^{(t-1)}\rbrace$ which excludes the $k$--th walker. It is important to notice also that all the walkers up to and excluding the $k$--th have already been updated and it is their updated positions that are used to compute the proposal covariance. This is essentially a \textit{Metropolis--within--Gibbs} scheme in disguise. The new point $\Tilde{\theta}_{k}$ is then accepted or rejected based on the usual Metropolis criterion and the process continuous with the next walker until all of them have been updated.

\begin{algorithm}[ht!]
\caption{Gaussian ensemble} \algolabel{ge}
\begin{algorithmic}[1]
\REQUIRE{initial state for the ensemble $\theta^{(1)}=(\theta_{1}^{(1)},\dots,\theta_{K}^{(1)})$, (unnormalised) target density $f(\theta)\propto p(\theta)$, covariance scaling factor (e.g. $\gamma = 2.38^{2}/D$), and number of iterations $N$}
\ENSURE{Markov chain $\theta_{1}, \dots, \theta_{N}$ that has $p(\theta)$ as its equilibrium distribution}
\FOR{$t=1$ \TO $N$}
    \FOR{$k=1$ \TO $K$}
        \STATE{Compute ensemble mean $\mu_{-k}^{(t)}=\frac{1}{K-1}\sum_{i\neq k}\theta_{i}^{(t)}$ excluding the $k$th state}
        \STATE{Compute ensemble covariance matrix $\Sigma_{-k}^{(t)}=\frac{1}{K-1}\sum_{i\neq k}(\theta_{i}^{(t)}-\mu_{-k}^{(t)})(\theta_{i}^{(t)}-\mu_{-k}^{(t)})^{T}$ excluding the $k$th state}
        \STATE{Draw proposal $\theta_{k}'\sim \mathcal{N}(\theta_{k}^{(t)}, \gamma\Sigma_{-k}^{(t)})$}
        \STATE{Compute acceptance probability $\alpha_{k} = \min\left(1, f(\theta_{k}')/f(\theta_{k}^{(t)})\right)$}
        \STATE{Draw uniform number $u\sim\mathcal{U}(0,1)$}
        \IF{$u < \alpha_{k}$}
            \STATE{Accept proposed state and set $\theta_{k}^{(t+1)}\leftarrow \theta_{k}'$}
        \ELSE
            \STATE{Reject proposed state and set $\theta_{k}^{(t+1)}\leftarrow \theta_{k}^{(t)}$}
        \ENDIF
    \ENDFOR
\ENDFOR
\end{algorithmic}
\end{algorithm}
\looseness=-1 GE solves the problem of tuning the proposal, up to the scaling factor $\gamma$ of the covariance matrix, but still assumes a Gaussian proposal. This means that we do not expect that GE will perform better than $K$ parallel well--tuned \textit{Random--walk Metropolis} samplers. As we will discuss in the next couple of sections, there are ways to relax this limitation and allow for more flexible proposals. Last but not least, the estimation of the proposal covariance matrix requires that the absolute minimum size of the ensemble to be $D+1$ for the covariance to be non--singular.

\section{Affine--invariant stretch move}

The \textit{affine--invariant ensemble sampler} and in particular the \textit{stretch move} introduced by \textcite{goodman2010ensemble} is perhaps the most popular ensemble MCMC method in the astronomical literature, made available in the \textit{Python} implementation \textit{emcee} \parencite{foreman2013emcee}. The \textit{stretch move} algorithm relaxes the limitation of the Gaussian proposal and instead updates each walker in turn along the direction of a different uniformly selected walker sampled from the rest of the ensemble. As we will discuss this change introduces both benefits and challenges.

\begin{figure}[!ht]
    \centering
	\centerline{\includegraphics[scale=0.65]{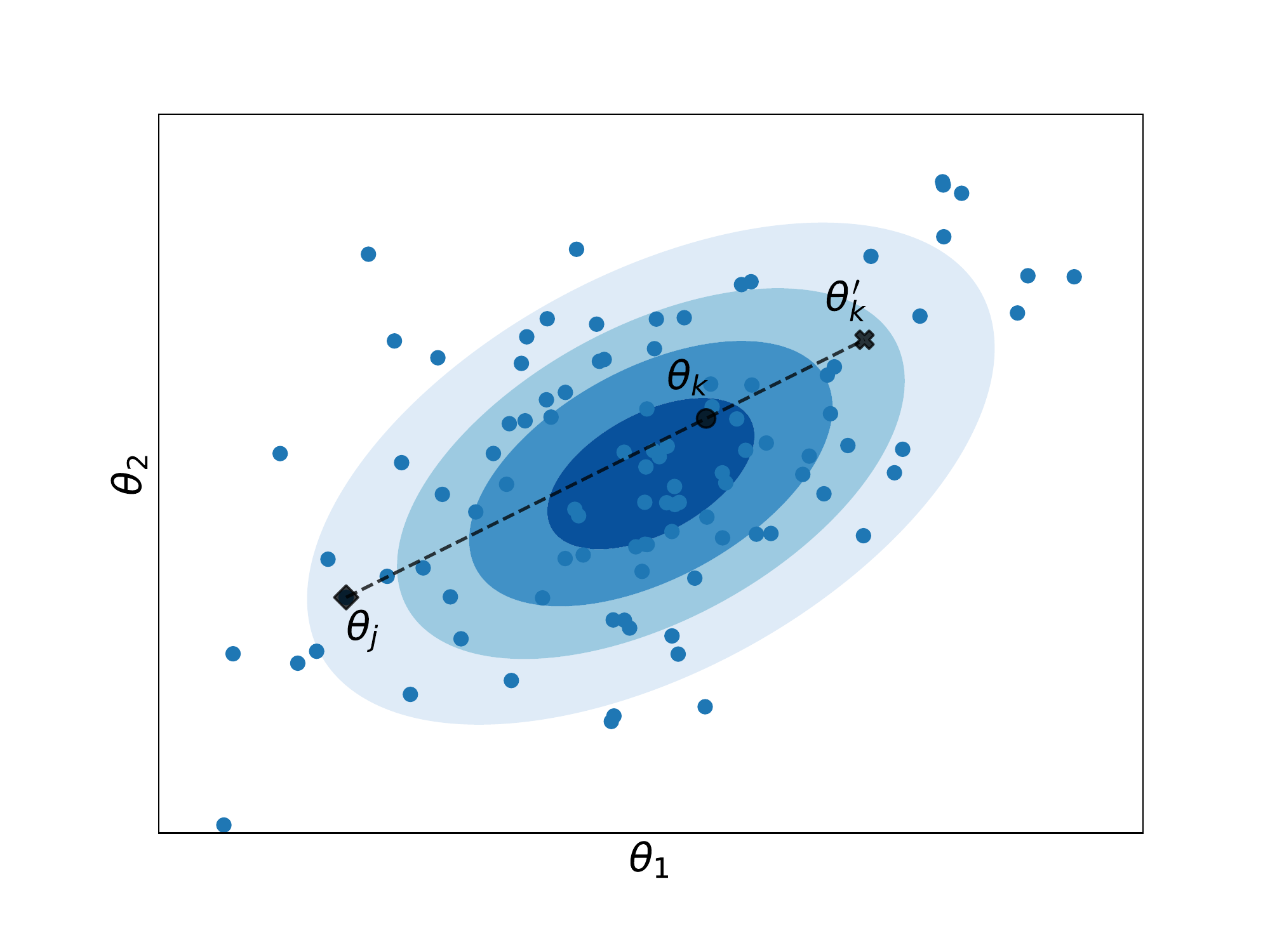}}
    \caption{Illustration of the affine--invariant stretch move. The selected walker $\theta_{k}$ is moved to its new position $\theta_{k}'$ along the line defined by $\theta_{j}$ and $\theta_{k}$. $\theta_{j}$ is a walker that is uniformly selected from the rest of the ensemble (i.e. excluding $\theta_{k}$).}
    \label{fig:stretch}
\end{figure}
In particular, in a given iteration $t$ of the method, the algorithm performs a loop over the $K$ walkers updating each one in turn. In the so--called \textit{stretch move}, we move a walker $\theta_{k}$ using a uniformly selected walker $\theta_{j}$ from the complementary ensemble $S_{-k} = \lbrace\theta_{1}^{(t)},\allowbreak \dots,\allowbreak \theta_{k-1}^{(t)},\allowbreak \theta_{k+1}^{(t-1)},\allowbreak \dots, \allowbreak \theta_{K}^{(t-1)}\rbrace$ that excludes $\theta_{k}$. The $\theta_{j}$ walker acts as an \textit{anchor point} for the move that consists of a proposal of the form
\begin{equation}
    \label{eq:stretch_proposal}
    \theta_{k}' = \zeta \theta_{k} + (1-\zeta) \theta_{j}\,,
\end{equation}
where $\zeta$ is a scaling variable with a probability density $g$ that satisfies the symmetry condition,
\begin{equation}
    \label{eq:symmetry_condition_zeta}
    g\left( \frac{1}{\zeta}\right) = \zeta g(\zeta)\,,
\end{equation}
such that the move expressed by equation \ref{eq:stretch_proposal} is symmetric in the \textit{Metropolis} sense. A particular density that obeys this condition is
\begin{equation}
    \label{eq:zeta_density}
    g(\zeta) \propto 
    \begin{cases}
    \frac{1}{\sqrt{\zeta}} & \mathrm{if}\;\zeta\in\left[\frac{1}{\alpha}, \alpha \right]\,, \\
    0 & \mathrm{otherwise}\,,
    \end{cases}
\end{equation}
where $\alpha > 1$ is a parameter that can be tuned to enhance the performance. The default value is usually set to $\alpha=2$. The new state $\theta_{k}'$ is then accepted with \textit{Metropolis} probability,
\begin{equation}
    \label{eq:stretch_metropolis_probability}
    \alpha(\theta_{k}',\theta_{k}) = \min\left( 1, \zeta^{D-1}\frac{p(\theta_{k}')}{p(\theta_{k})}\right)\,,
\end{equation}
where the $\zeta^{D-1}$ comes from the fact that the update takes place along a straight line. The process is then repeated for the next walker, until all the walkers are updated for the current iteration $t$ before the algorithm moves to its next iteration.

\begin{algorithm}[ht!]
\caption{Affine--invariant stretch move} \algolabel{aism}
\begin{algorithmic}[1]
\REQUIRE{initial state for the ensemble $\theta^{(1)}=(\theta_{1}^{(1)},\dots,\theta_{K}^{(1)})$, (unnormalised) target density $f(\theta)\propto p(\theta)$, and number of iterations $N$}
\ENSURE{Markov chain $\theta_{1}, \dots, \theta_{N}$ that has $p(\theta)$ as its equilibrium distribution}
\FOR{$t=1$ \TO $N$}
    \FOR{$k=1$ \TO $K$}
        \STATE{Draw a walker $\theta_{j}$ from the complementary ensemble $S_{-k} = \left\lbrace \theta_{1}^{(t+1)}, \dots, \theta_{k-1}^{(t+1)}, \theta_{k+1}^{(t)}, \dots, \theta_{K}^{(t)} \right\rbrace$}
        \STATE{Draw random number $\zeta\sim g(\zeta)$}
        \STATE{Compute proposed state $\theta_{k}' \leftarrow \zeta \theta_{k} + (1-\zeta)\theta_{j}$}
        \STATE{Compute acceptance probability $\alpha_{k} = \min\left(1, \zeta^{D-1}f(\theta_{k}')/f(\theta_{k}^{(t)})\right)$}
        \STATE{Draw uniform number $u\sim\mathcal{U}(0,1)$}
        \IF{$u < \alpha_{k}$}
            \STATE{Accept proposed state and set $\theta_{k}^{(t+1)}\leftarrow \theta_{k}'$}
        \ELSE
            \STATE{Reject proposed state and set $\theta_{k}^{(t+1)}\leftarrow \theta_{k}^{(t)}$}
        \ENDIF
    \ENDFOR
\ENDFOR
\end{algorithmic}
\end{algorithm}
One of the strict requirements of this method is the minimum number of walkers to be $D+1$ for it to be \textit{ergodic} and avoid the risk of walkers getting \textit{trapped} in some hyper--plane of lower than $D$ dimensions. Practically, the actual number of walkers required is much larger as it determines the plethora of possible new directions along which updates take place in each iteration. In this sense, the initial positions of the walkers and the number of them are the only free hyperparameters of this method. A great benefit of this method is that it is \textit{affine--invariant}, that is, its performance is insensitive to any linear correlations between the parameters of the target distribution. As the astronomical community has witnessed during the past few years, this offers a great advantage over other methods.

\section{Differential evolution}

Another ensemble method in the spirit of the \textit{stretch move} is the \textit{differential evolution} MCMC~\parencite{ter2006markov, ter2008differential}. Unlike the \textit{stretch move} that requires another single walker to act as an \textit{anchor point} for a proposal, \textit{differential evolution} involves two. We will discuss shortly how this difference can affect the performance and alter the characteristics of the method.

\begin{figure}[ht!]
    \centering
	\centerline{\includegraphics[scale=0.65]{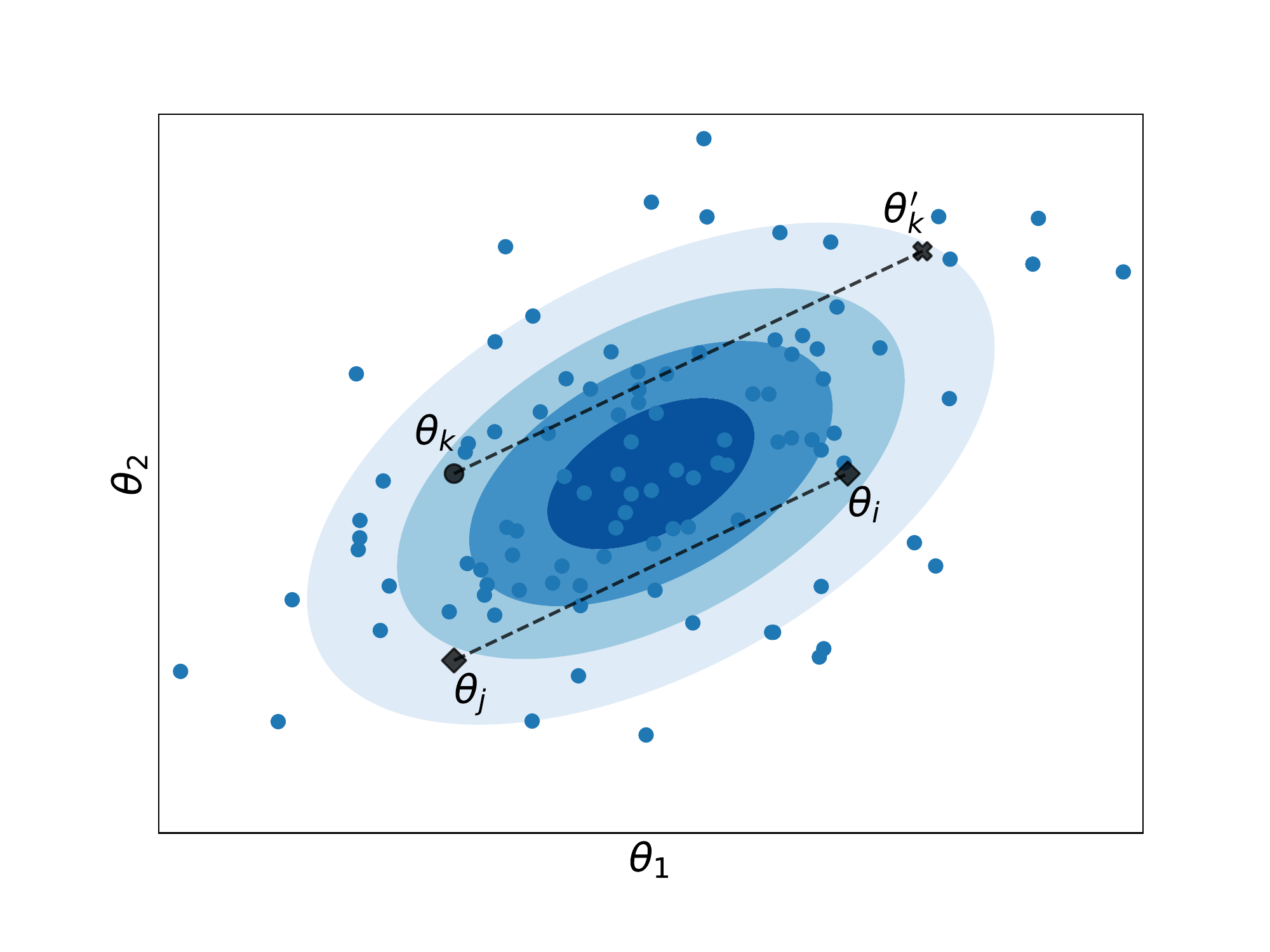}}
    \caption{Illustration of the differential evolution Monte Carlo. The selected walker $\theta_{k}$ is moved to its new position $\theta_{k}'$ parallel to the line defined by $\theta_{i}$ and $\theta_{j}$. The latter are two walkers that are uniformly selected from the rest of the ensemble (i.e. excluding $\theta_{k}$).}
    \label{fig:differential_evolution}
\end{figure}
An update of the ensemble works as follows: the algorithm performs a loop over the $K$ walkers updating each one in turn. Assuming that the current walker to be updated is $\theta_{k}$, the algorithm uniformly selects two walkers (without replacement), $\theta_{i}$ and $\theta_{j}$,  from the complementary ensemble $S_{-k} = \lbrace\theta_{1}^{(t)},\allowbreak \dots,\allowbreak \theta_{k-1}^{(t)},\allowbreak \theta_{k+1}^{(t-1)},\allowbreak \dots, \allowbreak \theta_{K}^{(t-1)}\rbrace$ that excludes $\theta_{k}$. The vector $\theta_{i}-\theta_{j}$ connecting the two auxiliary walkers defines the direction along which a move is proposed. The move consists of a proposal of the form
\begin{equation}
    \label{eq:differential_evolution_proposal}
    \theta_{k}' = \theta_{k} + \gamma \times \left( \theta_{i}-\theta_{j}\right) + \epsilon\,,
\end{equation}
where $\gamma$ is a non--zero scaling factor and $\epsilon\sim\mathcal{N}(0, \sigma^{2})$ is some optional Gaussian noise. The value of $\gamma$ determines the scale of the proposal. Its default value is often set to $\gamma = 2.38 / \sqrt{2D}$ which results in the optimal acceptance rate (i.e. $23.4\%$) for normal target distributions. In practice, one can adapt $\gamma$ using some diminishing adaptation scheme during the run. The proposed update of equation \ref{eq:differential_evolution_proposal} is then accepted with Metropolis acceptance probability
\begin{equation}
    \label{eq:differential_evolution_acceptance}
    \alpha(\theta',\theta) = \min\left(1, \frac{p(\theta')}{p(\theta)}\right)\,.
\end{equation}

\begin{algorithm}[ht!]
\caption{Differential evolution} \algolabel{de}
\begin{algorithmic}[1]
\REQUIRE{initial state for the ensemble $\theta^{(1)}=(\theta_{1}^{(1)},\dots,\theta_{K}^{(1)})$, (unnormalised) target density $f(\theta)\propto p(\theta)$, proposal scale parameter (e.g. $\gamma = 2.38/\sqrt{2D}$), optional Gaussian noise standard deviation (e.g. $\sigma = 10^{-3})$, and number of iterations $N$}
\ENSURE{Markov chain $\theta_{1}, \dots, \theta_{N}$ that has $p(\theta)$ as its equilibrium distribution}
\FOR{$t=1$ \TO $N$}
    \FOR{$k=1$ \TO $K$}
        \STATE{Draw walkers $\theta_{i}$ and $\theta_{j}$ without replacement from the complementary ensemble $S_{-k} = \left\lbrace \theta_{1}^{(t+1)}, \dots, \theta_{k-1}^{(t+1)}, \theta_{k+1}^{(t)}, \dots, \theta_{K}^{(t)} \right\rbrace$}
        \STATE{Draw random noise $\epsilon\sim\mathcal{N}(0,\sigma^{2})$}
        \STATE{Compute proposed state $\theta_{k}' \leftarrow \theta_{k} + \gamma(\theta_{i}-\theta_{j}) + \epsilon$}
        \STATE{Compute acceptance probability $\alpha_{k} = \min\left(1, f(\theta_{k}')/f(\theta_{k}^{(t)})\right)$}
        \STATE{Draw uniform number $u\sim\mathcal{U}(0,1)$}
        \IF{$u < \alpha_{k}$}
            \STATE{Accept proposed state and set $\theta_{k}^{(t+1)}\leftarrow \theta_{k}'$}
        \ELSE
            \STATE{Reject proposed state and set $\theta_{k}^{(t+1)}\leftarrow \theta_{k}^{(t)}$}
        \ENDIF
    \ENDFOR
\ENDFOR
\end{algorithmic}
\end{algorithm}
The advantage of \textit{differential evolution} over the \textit{stretch moves} comes down to the flexibility of their proposals. The direction along which a walker moves in the context of the \textit{stretch moves} is determined by a single walker. This means that at any given iteration, the number of equally possible directions is $K-1$. On the other hand, \textit{differential evolution} moves each walker along a direction defined by two walkers. This implies that the total number of possible directions is given by the binomial combination $\binom{K-1}{2}$. The latter increases much faster with the number size of the ensemble $K$ than the former, offering a larger variety of possible trajectories for the walkers. In other words, \textit{differential evolution} is expected to perform better even with a lower number of walkers.

\section{Parallel tempering}
\label{sec:parallel_tempering}

So far we have only discussed ensemble methods that target a trivial product density given by the product of $K$ copies of the target distribution as shown in equation \ref{eq:extended_density}. The main rationale for attempting to do this was to reduce the tuning requirements of MCMC. If we focus on addressing the challenge of multimodality, that is, the existence of multiple peaks in the target distribution, then we have to introduce a different product density as the extended target distribution.

\looseness=-1 One such choice is,
\begin{equation}
    \label{eq:product_density_pt}
    p^{*}\left(\lbrace\theta_{k}\rbrace_{k=1}^{K}\right) = \prod_{k=1}^{K} p_{k}(\theta_{k})\,,
\end{equation}
where
\begin{equation}
    \label{eq:tempered_posterior}
    p_{k}(\theta_{k}) \propto p^{\beta_{k}}(d\vert\theta,\mathcal{M})p(\theta\vert\mathcal{M})\,,
\end{equation}
is the \textit{annealed or tempered posterior} that offers a simple interpolation between the prior $p(\theta\vert \mathcal{M})$ and the unnormalised posterior density $p(d\vert\theta,\mathcal{M})\allowbreak p(\theta\vert \mathcal{M})$ for different monotonically--increasing values of $\beta_{k}\in[0,1]$. In the limit that $\beta_{k}=1$ for all values of $k$, equation \ref{eq:product_density_pt} reduces to the usual product density of equation \ref{eq:extended_density}.

\begin{figure}[!htb]
    \centering
	\centerline{\includegraphics[scale=0.65]{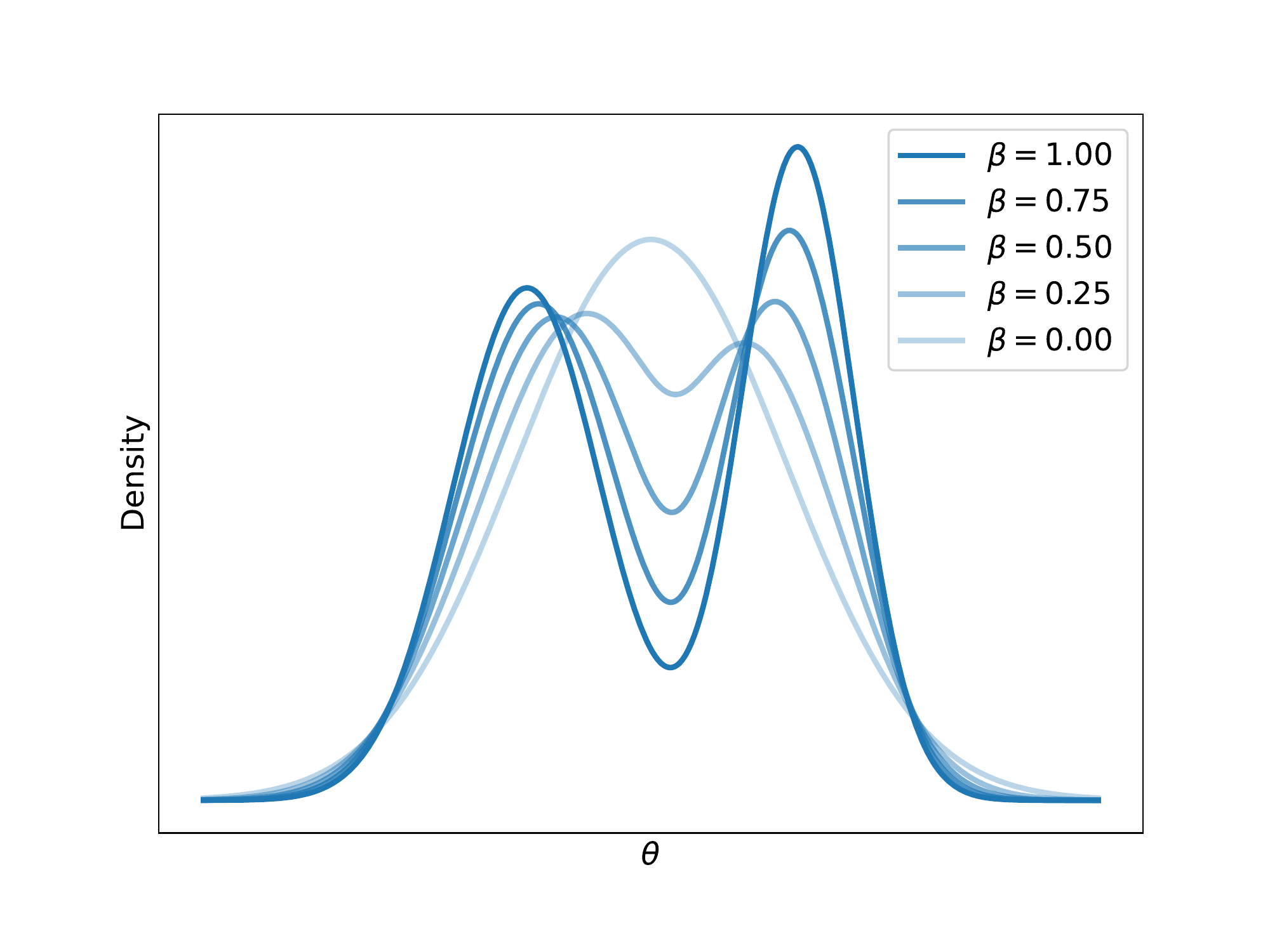}}
    \caption{Illustration of the gradual tempering performed in the posterior distribution. The prior distribution corresponds to $\beta\rightarrow 0$ and the posterior is recovered as $\beta\rightarrow 1$.}
    \label{fig:tempering}
\end{figure}

\looseness=-1 The method of \textit{parallel tempering (PT)}~\parencite{earl2005parallel}, also known as \textit{replica exchange Monte Carlo (REMC)}~\parencite{swendsen1986replica, hukushima1996exchange} or \textit{Metropolis--coupled Markov chain Monte Carlo (MC$^{3}$)}~\parencite{geyer1991markov}, relies on $K$ parallel Markov chains, each one targeting a different tempered density. The $\beta_{k}$ values are usually chosen \textit{a priori} using a heuristic rule (e.g. $\beta_{k}=(k-1)^{3}/(K-1)^3$), or are set adaptively during the run using some diminishing adaptation scheme. The choice of MCMC method used for each different $\beta_{k}$ is completely arbitrary and it can be anything from simple \textit{Random--walk Metropolis} to \textit{Hamiltonian Monte Carlo} or even an ensemble MCMC method.

So far, PT might look very similar to a parallel version of the sequential \textit{simulated annealing} method in which $\beta_{k}=1/T_{k}$ has the role of the inverse temperature. The crucial difference that makes PT so powerful is the fact that one can also perform \textit{between--chain exchange moves}. Either periodically (e.g. once every $10$ steps) or randomly (e.g. with probability $10\%$) a swap can take place between two states $\theta_{i}$ and $\theta_{j}$ that belong to different tempered posteriors (i.e. $\beta_{i}\neq \beta_{j}$). The reason that exchange/swap moves are desirable is that they enable the transfer of information from states of low $\beta$ to those of higher $\beta$. 

\begin{figure}[!htb]
    \centering
	\centerline{\includegraphics[scale=0.65]{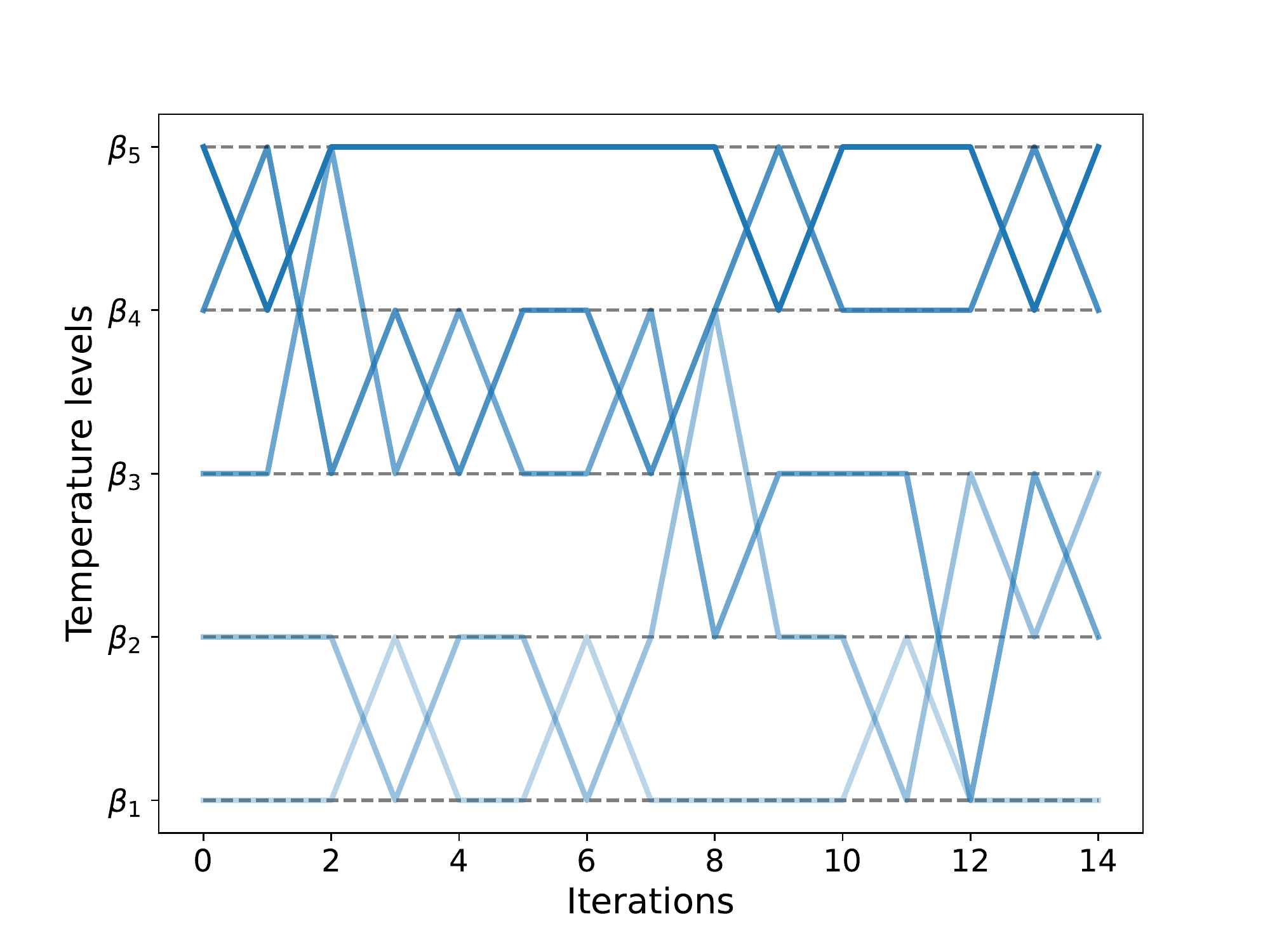}}
    \caption{Illustration of the parallel tempering swaps performed between adjacent temperature levels.}
    \label{fig:beta_swaps}
\end{figure}
To understand how to perform a swap in practice let us consider the extended state,
\begin{equation}
    \label{eq:extended_state_before_swap}
    \lbrace \theta_{k} \rbrace_{k=1}^{K} = \lbrace\theta_{1}, \dots, \theta_{i}, \dots, \theta_{j}, \dots, \theta_{K}\rbrace\,,
\end{equation}
prior to the swap, where $\theta_{i}$ and $\theta_{j}$ are the two states that we want to exchange. This means that the proposed new state will be,
\begin{equation}
    \label{eq:extended_state_after_swap}
    \lbrace \Tilde{\theta}_{k} \rbrace_{k=1}^{K} = \lbrace\theta_{1}, \dots, \theta_{j}, \dots, \theta_{i}, \dots, \theta_{K}\rbrace\,.
\end{equation}
Notice that the rest of the states, with the exception of $\theta_{i}$ and $\theta_{j}$, are left unaffected by this exchange proposal. The \textit{Metropolis acceptance probability} for this proposal is,
\begin{equation}
    \label{eq:exchange_metropolis_acceptance}
    \begin{split}
        \alpha_{ij} &= \min\left( 1, \frac{p^{*}\left( \lbrace \Tilde{\theta}_{k} \rbrace_{k=1}^{K}\right)}{p^{*}\left( \lbrace \theta_{k} \rbrace_{k=1}^{K}\right)} \right) \\
        &= \min\left( 1, \frac{p_{i}(\theta_{j})p_{j}(\theta_{i})}{p_{i}(\theta_{i})p_{j}(\theta_{j})}\right) \\
        &= \min\left( 1, \frac{p^{\beta_{i}}(d\vert\theta_{j},\mathcal{M})p(\theta_{j}\vert\mathcal{M})p^{\beta_{j}}(d\vert\theta_{i},\mathcal{M})p(\theta_{i}\vert\mathcal{M})}{p^{\beta_{i}}(d\vert\theta_{i},\mathcal{M})p(\theta_{i}\vert\mathcal{M})p^{\beta_{j}}(d\vert\theta_{j},\mathcal{M})p(\theta_{j}\vert\mathcal{M})} \right) \\
        &= \min\left( 1, \frac{p^{\beta_{i}}(d\vert\theta_{j},\mathcal{M})p^{\beta_{j}}(d\vert\theta_{i},\mathcal{M})}{p^{\beta_{i}}(d\vert\theta_{i},\mathcal{M})p^{\beta_{j}}(d\vert\theta_{j},\mathcal{M})} \right) \\
        &= \min\left[ 1, \left( \frac{p(d\vert\theta_{i},\mathcal{M})}{p(d\vert\theta_{j},\mathcal{M})}\right)^{(\beta_{j}-\beta_{i})} \right]\,,
    \end{split}
\end{equation}

The chains are usually chosen to be in adjacent $\beta_{k}$ levels (i.e. $i=j- 1$) such that the overlap between the typical sets of $p_{i}(\theta_{i})$ and $p_{j}(\theta_{j})$ is large, leading to high acceptance probabilities. \textcite{atchade2011towards} estimated that the optimal acceptance rate is $23.4\%$. The exchange updates are typically performed after the local MCMC updates are completed in all $\beta$ levels for a given iteration. Furthermore, there are different strategies for proposing swaps between adjacent temperature levels~\parencite{lingenheil2009efficiency}. One option is to randomly select a pair of adjacent temperature levels per iteration. Another strategy involves proposing to swap all adjacent pairs starting from the lowest or highest $\beta$ level and progressively moving towards the other end of the ladder. Finally, strategies that involve two steps, for instance, proposing to swap all even pairs in one iteration and all odd pairs in the next, have also been suggested in the literature~\parencite{lingenheil2009efficiency}.

\begin{algorithm}[ht!]
\caption{Parallel tempering} \algolabel{pt}
\begin{algorithmic}[1]
\REQUIRE{initial state for the ensemble $\theta^{(1)}=(\theta_{1}^{(1)},\dots,\theta_{K}^{(1)})$, prior probability density $\pi(\theta) \equiv p(\theta\vert\mathcal{M})$, likelihood function $\mathcal{L}(\theta)\equiv p(d\vert\theta,\mathcal{M})$, temperature ladder (e.g. $\beta_{k} = (k-1)^{3}/(K-1)^{3}$), local MCMC kernel $\theta'\leftarrow\mathcal{K}(\theta; f(\theta))$ (e.g. a single random--walk Metropolis update), and number of iterations $N$}
\ENSURE{$K$ Markov chains that each has $p_{t}(\theta)\propto \pi(\theta)\mathcal{L}(\theta)^{\beta_{k}}$ as its equilibrium distribution}
\FOR{$t=1$ \TO $N$}
    \FOR{$k=1$ \TO $K$}
        \STATE{Update state using local MCMC update $\theta_{k}'\leftarrow \mathcal{K} (\theta_{k}^{(t)}; \pi(\theta)\mathcal{L}(\theta)^{\beta_{k}})$}
    \ENDFOR
    \STATE{Draw random value of $k$ uniformly $k\sim\mathcal{U}(1, K-1)$}
    \STATE{Compute acceptance probability $\alpha_{k} = \min\left(1, \left[\frac{\mathcal{L}\left(\theta_{k}^{(t+1)}\right)}{\mathcal{L}\left(\theta_{k+1}^{(t+1)}\right)}\right]^{\beta_{k+1}-\beta_{k}}\right)$}
    \STATE{Draw uniform number $u\sim\mathcal{U}(0,1)$}
    \IF{$u < \alpha_{k}$}
        \STATE{Accept proposed swap and set $\theta_{k}^{(t+1)}\leftarrow \theta_{k+1}'$ and $\theta_{k+1}^{(t+1)}\leftarrow \theta_{k}'$}
    \ELSE
        \STATE{Reject proposed swap and set $\theta_{k}^{(t+1)}\leftarrow \theta_{k}'$ and $\theta_{k+1}^{(t+1)}\leftarrow \theta_{k+1}'$}
    \ENDIF
\ENDFOR
\end{algorithmic}
\end{algorithm}
% !TEX TS-program = pdflatex
% !TEX root = ../ArsClassica.tex

%************************************************
\chapter{Evidence and Bayes factor computation}
\label{chp:evidence}
%************************************************

\begin{flushright}
\itshape
There is nothing more deceptive than an obvious fact. \\
\medskip
--- Arthur Conan Doyle
\end{flushright}

\section{Naive Monte Carlo estimator}

The simplest estimator for the evidence we can construct is just the expectation value of the likelihood function with respect to the prior distribution~\parencite{hammersley1964percolation, raftery1991three}. The, so--called, \textit{Naive Monte Carlo (NMC)} estimator can be computed as the sum
\begin{equation}
    \label{eq:naive_monte_carlo}
    \Hat{p}_{\mathrm{NMC}}(d\vert\mathcal{M})=\frac{1}{n}\sum_{i=1}^{n}p(d\vert \theta_{i},\mathcal{M})\,,\quad \mathrm{with}\quad \theta_{i}\sim p(\theta\vert \mathcal{M})\,.
\end{equation}
Although simple and unbiased, this approach can become extremely inefficient and result in a high variance in higher dimensions as the probability mass concentrates in the typical set that occupies a negligible fraction of the prior volume~\parencite{newton1994approximate}. For this reason, this technique is only recommended for low--dimensional problems (i.e. $D\leq 3$).

\section{Importance sampling estimator}

A more general strategy for the unbiased estimation of the evidence is importance sampling using samples from an auxiliary distribution $q(\theta)$. A simple estimator can then be constructed as,
\begin{equation}
    \label{eq:importance_sampling_evidence}
    \Hat{p}_{\mathrm{IS}}(d\vert\mathcal{M})=\frac{1}{n}\sum_{i=1}^{n}\frac{p(\theta_{i}\vert d,\mathcal{M})}{q(\theta_{i})}\,,\quad \mathrm{with}\quad \theta_{i}\sim q(\theta)\,.
\end{equation}
These estimators share the same difficulty as most methods based upon importance sampling, that is, a large overlap between the typical set of the proposal and posterior distribution must be achieved for the method to be effective. Constructing effective proposal distributions becomes increasingly unmanageable as the number of dimensions increases and thus the application of this method on its own is limited to low dimensions. Finally, the importance sampling estimator reduces to the NMC one when the proposal distribution is chosen to be the prior.

\section{Harmonic mean estimator}

The \textit{harmonic mean (HM)} estimator is another variation of the importance sampling estimator in which the posterior is used as the proposal and the prior as the target distribution~\parencite{newton1994approximate}. This suggests the following estimator,
\begin{equation}
    \label{eq:harmonic_mean}
    \Hat{p}_{\mathrm{HM}}(d\vert\mathcal{M}) = \frac{1}{\frac{1}{n}\sum_{i=1}^{n}\frac{1}{p(d\vert\theta_{i}\mathcal{M})}}\,,\quad \mathrm{with}\quad \theta_{i}\sim p(\theta\vert d, \mathcal{M})\,.
\end{equation}
The possible occurrence of samples with small likelihood value renders the variance of this estimator infinite~\parencite{neal2008harmonic}. This pathology can be addressed by using a mixture $q(\theta)=\delta p(\theta\vert\mathcal{M})+(1-\delta)p(\theta\vert d, \mathcal{M})$  between the prior and the posterior as the proposal distribution, where $\delta$ is very small (e.g. $\delta=0.05$). The resulting method is then called the \textit{stabilised harmonic mean (SHM)} estimator~\parencite{newton1994approximate}.

\section{Laplace estimator}

As discussed in detail in Section \ref{sec:laplace_approximation}, for a sufficiently Gaussian target distribution $p(\theta)$ we can use the \textit{Laplace approximation}, that is, a second order expansion around the mode, to estimate expectation values~\parencite{tierney1986accurate}. Assuming that the target distribution is the unnormalised posterior $p(d\vert\theta \mathcal{M})p(\theta\vert \mathcal{M})$, the Gaussian approximation's mean is given by,
\begin{equation}
    \label{eq:laplace_mean_posterior}
    \mu = \underset{\theta}{\mathrm{arg\,max}} ~\left[p(d\vert\theta, \mathcal{M})p(\theta\vert \mathcal{M})\right]\,,
\end{equation}
following equation \ref{eq:mode_of_target}, and the inverse covariance is given by,
\begin{equation}
    \label{eq:laplace_covariance_posterior}
    \left( \Sigma^{-1}\right)_{ij} = - \frac{\partial^{2}}{\partial\theta_{i}\partial\theta_{j}} \left[\log p(d\vert\theta, \mathcal{M})+\log p(\theta\vert \mathcal{M}) \right]\,,
\end{equation}
following equation \ref{eq:precision_of_target}. Then, the model evidence is approximated by the normalising constant of the Gaussian, or in other words,
\begin{equation}
    \label{eq:laplace_evidence}
    \begin{split}
        \Hat{p}_{L}(d\vert \mathcal{M}) &= \int e^{-\frac{1}{2}(\theta-\mu)^T\Sigma^{-1}(\theta-\mu)}d\theta \\
        &= (2\pi)^{D/2}\det (\Sigma)^{1/2} p(d\vert\theta=\mu,\mathcal{M}) p(\theta=\mu\vert\mathcal{M})\,.
    \end{split}
\end{equation}
As with any method, this result is only as good as the assumptions entering its calculation. The closer the posterior resembles a normal distribution, the better the outcome of the \textit{Laplace} estimator will be.

\section{Bridge sampling}
Originally, \textcite{meng1996simulating} introduced \textit{bridge sampling (BS)} as a way to directly estimate the \textit{Bayes factor} of two models, $\mathcal{M}_{1}$ and $\mathcal{M}_{2}$. However, in this section we present a version of BS that targets the model evidence of a single model $\mathcal{M}$. BS follows from the basic identity,
\begin{equation}
    \label{eq:bridge_sampling_identity}
    1 = \frac{\int p(d\vert\theta,\mathcal{M})p(\theta\vert\mathcal{M})\alpha(\theta)q(\theta)}{\int p(d\vert\theta,\mathcal{M})p(\theta\vert\mathcal{M})\alpha(\theta)q(\theta)}\,,
\end{equation}
where $q(\theta)$ is the \textit{proposal distribution} and $\alpha(\theta)$ is the so--called \textit{bridge function} the \textit{support} of which encompasses that of both the target posterior and of the proposal distribution.

Multiplying both sides of equation \ref{eq:bridge_sampling_identity} with the model evidence $p(d\vert \mathcal{M})$ results in
\begin{equation}
    \label{eq:bridge_sampling_proof}
    \begin{split}
        p(d\vert \mathcal{M}) &= \frac{\int p(d\vert\theta,\mathcal{M})p(\theta\vert\mathcal{M})\alpha(\theta)q(\theta)}{\int \frac{p(d\vert\theta,\mathcal{M})p(\theta\vert\mathcal{M})}{p(d\vert \mathcal{M})}\alpha(\theta)q(\theta)} \\
        &= \frac{\int p(d\vert\theta,\mathcal{M})p(\theta\vert\mathcal{M})\alpha(\theta)q(\theta)}{\int p(\theta\vert d, \mathcal{M})\alpha(\theta)q(\theta)} \,,
    \end{split}
\end{equation}
which can be written as,
\begin{equation}
    \label{eq:bridge_sampling_expectations}
    p(d\vert \mathcal{M}) = \frac{\mathbb{E}_{q(\theta)}\left[p(d\vert\theta,\mathcal{M})p(\theta\vert\mathcal{M})\alpha(\theta) \right]}{\mathbb{E}_{p(\theta\vert d, \mathcal{M})}\left[ \alpha(\theta)q(\theta) \right]} \,,
\end{equation}
in terms of expectation values. The model evidence can then be approximated as,
\begin{equation}
    \label{eq:bridge_sampling_estimator}
    \Hat{p}(d\vert \mathcal{M}) = \frac{\frac{1}{n_{2}}\sum_{i=1}^{n_{2}}p(d\vert\Tilde{\theta}_{i},\mathcal{M})p(\Tilde{\theta}_{i}\vert\mathcal{M})\alpha(\Tilde{\theta}_{i})}{\frac{1}{n_{1}}\sum_{j=1}^{n_{1}}\alpha(\theta_{j}^{*})q(\theta_{j}^{*})}\,,
\end{equation}
where $\Tilde{\theta}_{i}$ are samples from the proposal distribution,
\begin{equation}
    \label{eq:bridge_sampling_proposal_samples}
    \Tilde{\theta}_{i} \sim q(\theta)\,,
\end{equation}
and $\theta_{j}^{*}$ are samples from the posterior distribution,
\begin{equation}
    \label{eq:bridge_sampling_posterior_samples}
    \theta_{j}^{*} \sim p(\theta\vert d, \mathcal{M})\,.
\end{equation}

It is clear from the above discussion that BS relies on samples from both the proposal distribution $q(\theta)$, which plays the role of an \textit{importance density}, and the posterior distribution $p(\theta\vert d, \mathcal{M})$. Often, the proposal distribution is some distribution that is easy to sample from and its \textit{typical set} has a large overlap with the one of the posterior distribution. A common proposal used in practice is a normal distribution with its first two moments matching those of the posterior distribution.

Although highly arbitrary, the choice of the \textit{bridge function} $\alpha (\theta)$ can have a significant impact on the precision of the method for a given proposal distribution. For instance, setting $\alpha(\theta)=[q(\theta)]^{-1}$ the BS estimator reduces to the naive Monte Carlo estimator, whereas setting $\alpha(\theta)=[p(d\vert \theta,\mathcal{M})p(\theta\vert \mathcal{M})q(\theta)]^{-1}$ leads to the \textit{harmonic mean estimator}. \textcite{meng1996simulating} showed that the \textit{optimal bridge function}, that is, the one that minimises the mean--square--error, is,
\begin{equation}
    \label{eq:optimal_bridge_function}
    \alpha(\theta) = \frac{C}{s_{1}p(d\vert \theta, \mathcal{M})p(\theta\vert \mathcal{M}) + s_{2}p(d\vert \mathcal{M})q(\theta)}\,,
\end{equation}
where $s_{1}=n_{1}/(n_{1}+n_{2})$ and $s_{2}=n_{2}/(n_{1}+n_{2})$ and $C$ is a constant that cancels out and its value does not affect the outcome in any way. The \textit{bridge function} of equation \ref{eq:optimal_bridge_function} depends on the model evidence $p(d\vert\mathcal{M})$, the same quantity that we are trying to approximate. We can resolve this issue by employing an iterative scheme,
\begin{equation}
    \label{eq:bridge_sampling_update}
    \Hat{p}_{\mathrm{BS}}^{(t+1)}(d\vert \mathcal{M}) = \frac{\frac{1}{n_{2}}\sum_{i=1}^{n_{2}}\frac{p(d\vert\Tilde{\theta}_{i},\mathcal{M})p(\Tilde{\theta}_{i}\vert\mathcal{M})}{s_{1}p(d\vert\Tilde{\theta}_{i},\mathcal{M})p(\Tilde{\theta}_{i}\vert\mathcal{M}) + s_{2}\Hat{p}_{\mathrm{BS}}^{(t)}(d\vert \mathcal{M})q(\Tilde{\theta}_{i})}}{\frac{1}{n_{1}}\sum_{j=1}^{n_{1}}\frac{q(\theta_{j}^{*})}{s_{1}p(d\vert \theta_{j}^{*}, \mathcal{M})p(\theta_{j}^{*}\vert \mathcal{M}) + s_{2}\Hat{p}_{\mathrm{BS}}^{(t)}(d\vert \mathcal{M})q(\theta_{j}^{*})}}\,,
\end{equation}
starting from some initial guess of the value of the model evidence $\Hat{p}_{\mathrm{BS}}^{(0)}(d\vert \mathcal{M})$ and keep updating it until the estimate has converged for some arbitrary tolerance level. Rearranging the terms on the right hand side, the aforementioned estimator can be written in the simpler form
\begin{equation}
    \label{eq:bridge_sampling_update_simple}
    \Hat{p}_{\mathrm{BS}}^{(t+1)}(d\vert \mathcal{M}) = \frac{\frac{1}{n_{2}}\sum_{i=1}^{n_{2}}\frac{\ell_{2,i}}{s_{1}\ell_{2,i}+s_{2}\Hat{p}_{\mathrm{BS}}^{(t)}(d\vert \mathcal{M})}}{\frac{1}{n_{1}}\sum_{j=1}^{n_{1}}\frac{1}{s_{1}\ell_{1,j}+s_{2}\Hat{p}_{\mathrm{BS}}^{(t)}(d\vert \mathcal{M})}}\,,
\end{equation}
where we have defined
\begin{equation}
    \label{eq:bridge_sampling_l1}
    \ell_{1,j} = \frac{p(d\vert \theta_{j}^{*}, \mathcal{M})p(\theta_{j}^{*}\vert \mathcal{M})}{q(\theta_{j}^{*})}\,,
\end{equation}
and
\begin{equation}
    \label{eq:bridge_sampling_l2}
    \ell_{2,i} = \frac{p(d\vert\Tilde{\theta}_{i},\mathcal{M})p(\Tilde{\theta}_{i}\vert\mathcal{M})}{q(\Tilde{\theta}_{i})}\,.
\end{equation}
Furthermore, the numerical stability of equation \ref{eq:bridge_sampling_update_simple} can be improved and overflow issues avoided if we define
\begin{equation}
    \label{eq:bridge_sampling_r}
    \Hat{p}_{\mathrm{BS}}^{(t)}(d\vert \mathcal{M}) = \Hat{r}^{(t)}\exp(\ell^{*})\,,
\end{equation}
and use the iterative formula
\begin{equation}
    \label{eq:bridge_sampling_update_stable}
    \Hat{r}^{(t+1)} = \frac{\frac{1}{n_{2}}\sum_{i=1}^{n_{2}}\frac{\exp[\log(\ell_{2,i})-\ell^{*}]}{s_{1}\exp[\log(\ell_{2,i})-\ell^{*}]+s_{2}\Hat{r}^{(t)}}}{\frac{1}{n_{1}}\sum_{j=1}^{n_{1}}\frac{1}{s_{1}\exp[\log(\ell_{1,j})-\ell^{*}]+s_{2}\Hat{r}^{(t)}}},
\end{equation}
where $\ell^{*}$ is a constant that we can choose in order to make the sums numerically tractable, for instance $\ell^{*} = \mathrm{median}[\log(\ell_{1,j})]$.

Compared to other methods such as importance sampling or the harmonic mean estimator, BS estimates are more robust in cases in which the overlap between the \textit{typical sets} of the proposal and posterior distribution is far from perfect.

\section{Thermodynamic integration}

A large body of work in \textit{statistical physics} is concerned with methods for the estimation of normalising constants and \textit{partition functions} in particular. The method of \textit{thermodynamic integration (TI)} was developed for exactly this purpose~\parencite{gelman1998simulating}. \textcite{friel2008marginal} studied the particular case in which the normalising constant that is estimated using TI is the model evidence. To this end, they introduced the notion of the \textit{power posterior},
\begin{equation}
    \label{eq:power_posterior}
    p(\theta\vert d,\beta, \mathcal{M})\propto p^{\beta}(d\vert\theta,\mathcal{M})p(\theta\vert \mathcal{M})\,,
\end{equation}
in which $\beta$ is an auxiliary variable in the interval $[0,1]$. By construction, the normalising constant of the \textit{power posterior} is simply,
\begin{equation}
    \label{eq:power_posterior_normalisation}
    p(d\vert \beta, \mathcal{M}) = \int p^{\beta}(d\vert\theta,\mathcal{M})p(\theta\vert \mathcal{M}) d\theta \,,
\end{equation}
where $p(d\vert \beta=1,\mathcal{M})$ is the model evidence and $p(d\vert \beta=0,\mathcal{M})$ is the integral over the prior which is simply equal to $1$. Furthermore, the logarithm of the model evidence is,
\begin{equation}
    \label{eq:thermodynamic_integration_logz}
    \begin{split}
        \log p(d\vert \mathcal{M}) &= \log\left[ \frac{p(d\vert \beta=1,\mathcal{M})}{p(d\vert \beta=0,\mathcal{M})}\right] \\
        &= \int_{0}^{1}\mathbb{E}_{p(\theta\vert d, \beta, \mathcal{M})}[\log p(d\vert\theta,\beta,\mathcal{M})]d\beta \,,
    \end{split}
\end{equation}
that is, the integral over $\beta$ of the expectation value of the likelihood with respect to the posterior for each value of $\beta$. To prove the above identity we first need to notice that,
\begin{equation}
    \label{eq:thermodynamic_integration_proof}
    \begin{split}
        \frac{d\log p(d\vert\beta,\mathcal{M})}{d\beta} &= \frac{1}{p(d\vert\beta,\mathcal{M})} \times\frac{d p(d\vert\beta,\mathcal{M})}{d\beta} \\
        &= \frac{1}{p(d\vert\beta,\mathcal{M})} \int \frac{d}{d\beta} p^{\beta}(d\vert\theta,\mathcal{M})p(\theta\vert \mathcal{M}) d\theta \\
        &= \int \log p(d\vert\theta,\mathcal{M}) \frac{p^{\beta}(d\vert\theta,\mathcal{M})p(\theta\vert \mathcal{M})}{p(d\vert\beta,\mathcal{M})}d\theta \\
        &= \int \log p(d\vert\theta,\mathcal{M}) p(\theta\vert d,\beta,\mathcal{M})d\theta \\
        &= \mathbb{E}_{p(\theta\vert d,\beta,\mathcal{M})}\left[ \log p(d\vert\theta,\mathcal{M})\right]\,,
    \end{split}
\end{equation}
Integrating both sides with respect to $\beta$ leads to equation \ref{eq:thermodynamic_integration_logz} and completes the proof.

Using equation \ref{eq:thermodynamic_integration_logz} to estimate the model evidence often requires the discretisation of the integral. A sequence $0=\beta_{1}<\beta_{2}<\dots<\beta_{m}=1$ must be chosen \textit{a priori} or based on \textit{diminishing adaptation} scheme. The model evidence can then be approximated using the \textit{trapezoidal rule},
\begin{equation}
    \label{eq:thermodynamic_integration_trapezoidal}
    \Hat{p}_{\mathrm{TI}}(d\vert \mathcal{M}) = \sum_{i=1}^{m}(\beta_{i+1}-\beta_{i})\frac{E_{i}+E_{i+1}}{2}\,,
\end{equation}
where,
\begin{equation}
    \label{eq:thermodynamic_integration_average_likelihood}
    E_{i}=\mathbb{E}_{p(\theta\vert d,\beta_{i},\mathcal{M})}\left[ \log p(d\vert\theta,\mathcal{M})\right]\,,
\end{equation}
is the expected likelihood at $\beta_{i}$.

There are two sources of error in the above approximation. The first one is the Monte Carlo error that originates from the estimation of equation \ref{eq:thermodynamic_integration_average_likelihood} using a finite number of samples. The second type has to do with the choice of the discretisation of $\beta$. \textcite{calderhead2009estimating} showed that the discretisation error depends on the \textit{Kullback--Leibler (KL) divergence} between subsequent densities $p(\theta\vert d, \beta_{i},\mathcal{M})$ and $p(\theta\vert d, \beta_{i+1},\mathcal{M})$. This means that the optimal discretisation sequence of $\beta$ values is the one minimising the KL   divergence between subsequent power posteriors. Of course, knowing the optimal scheme is \textit{a priori} hardly ever possible and thus we must rely on \textit{ad hoc} choices (e.g. $\beta_{i}=(i-1)^{3}/(m-1)^{3}$) or \textit{diminishing adaptation} strategies.

Thermodynamic integration can be combined with many different MCMC methods in order to estimate the model evidence. Perhaps the simplest one is to run $m$ independent chains, either in parallel or serially, and then estimate the evidence using equation \ref{eq:thermodynamic_integration_trapezoidal} where the expected likelihood of each discrete $\beta$ value is computed with equation \ref{eq:thermodynamic_integration_average_likelihood} for each chain. Furthermore, the chains do not even have to be independent for this method to work. Lastly, a parallel tempering approach can be followed as it is often done in practice.

\section{Annealed importance sampling}

\textit{Annealed importance sampling (AIS)} is another method that relies on a sequence of annealed or tempered distributions in order to construct an importance sampling estimator for the model evidence~\parencite{neal2001annealed}, similarly to \textit{simulated annealing} and \textit{parallel tempering}.

The basic idea is to use MCMC transitions in order to push a collection of $n$ particles through a series of $m$ \textit{intermediate distributions}
\begin{equation}
    \label{eq:annealed_posterior}
    p(\theta\vert d,\beta, \mathcal{M})\propto p^{\beta}(d\vert\theta,\mathcal{M})p(\theta\vert \mathcal{M})\,,
\end{equation}
where $0=\beta_{1}<\beta_{2}<\dots<\beta_{m}=1$, connecting the prior for $\beta_{1}=0$ to the posterior for $\beta_{m}=1$. The particles are initialised by drawing samples from the prior
\begin{equation}
    \label{eq:sample_prior}
    \theta_{i}^{(1)}\sim p(\theta\vert\mathcal{M})\,,
\end{equation}
and assigned (unnormalised) importance weights
\begin{equation}
    \label{eq:initial_importance_weights}
    w_{i}^{(1)} = 1\,,
\end{equation}
for $i\in\lbrace 1, \dots, n\rbrace$ the particle index.

A number of $N$ MCMC steps is then performed for each particle before the value of $\beta$ is updated to the next value in the predefined sequence. The number $N$ of MCMC steps is chosen such that the Markov chains defined by the particle trajectories have enough time to reach the stationary distribution.
The critical difference between AIS and simulated annealing is that the associated importance weights $w_{i}$ are updated during the run every time we move from one intermediate distribution to the next,
\begin{equation}
    \label{eq:importance_weight_update}
    w_{i}^{(t+1)} = w_{i}^{(t)} \times \frac{p(\theta_{i}\vert d, \beta_{t+1}, \mathcal{M})}{p(\theta_{i}\vert d, \beta_{t}, \mathcal{M})} = w_{i}^{(t)} \times \frac{p^{\beta_{t+1}}(d\vert\theta_{i},\mathcal{M})}{p^{\beta_{t}}(d\vert\theta_{i},\mathcal{M})}\,.
\end{equation}
In practice, the logarithm of the weights is used in order to avoid numerical issues. Once the final distribution (i.e. the posterior) is reached and the particle weights are updated accordingly, the model evidence can be estimated as
\begin{equation}
    \label{eq:ais_estimator}
    \Hat{p}_{\mathrm{AIS}}(d\vert\mathcal{M}) = \frac{1}{n} \sum_{i=1}^{n}w_{i}^{(m)}\,.
\end{equation}
Furthermore, the samples $\theta_{i}^{(m)}$ combined with their respective weights $w_{i}^{(m)}$ can be used to compute arbitrary expectation values
\begin{equation}
    \label{eq:ais_expectation_values}
    \mathbb{E}_{p(\theta\vert d,\mathcal{M})}\left[f(\theta)\right] = \frac{\sum_{i=1}^{n}w_{i}^{(m)}f\left(\theta_{i}^{(m)}\right)}{\sum_{i=1}^{n}w_{i}^{(m)}}\,.
\end{equation}

Assuming that the annealing process is slow enough (i.e. large number $m$ of $\beta$ levels and number $N$ of MCMC steps) and a large enough collection of particles is used, then AIS yields unbiased estimates of the model evidence and weighted posterior samples, even in high dimensions. In the limit that the number of MCMC steps $N$ goes to zero, the AIS estimator reduces to the usual importance sampling estimator.

\section{Savage--Dickey density ratio}

Suppose now that we have two models or hypotheses and their respective parameters, $\mathcal{M}_{0}:\theta$ and $\mathcal{M}_{1}:\theta,\phi$, such that $\mathcal{M}_{0}$ is nested inside $\mathcal{M}_{1}$. This means that the more complex model, $\mathcal{M}_{1}$, is reduced to the simpler one, $\mathcal{M}_{0}$, for some specific choice of one or more of its parameters, $\phi=\phi_{0}$. This specific parameter choice is often called a \textit{point--null hypothesis} as it is associated with zero probability mass in the context of the $\mathcal{M}_{1}$ model.

Evaluating the plausibility of this hypothesis can be done by computing the \textit{Bayes factor} between the two models. The \textit{Savage--Dickey density ratio (SDDR)} is a method that aims to do exactly this that was introduced by \textcite{dickey1970weighted}, \textcite{dickey1971weighted}, \textcite{gunel1974bayes}, and \textcite{dickey1976approximate} who in turn attributed the origin of the method to \textit{Leonard Jimmie Savage}.

\begin{figure}[H]
    \centering
	\centerline{\includegraphics[scale=0.65]{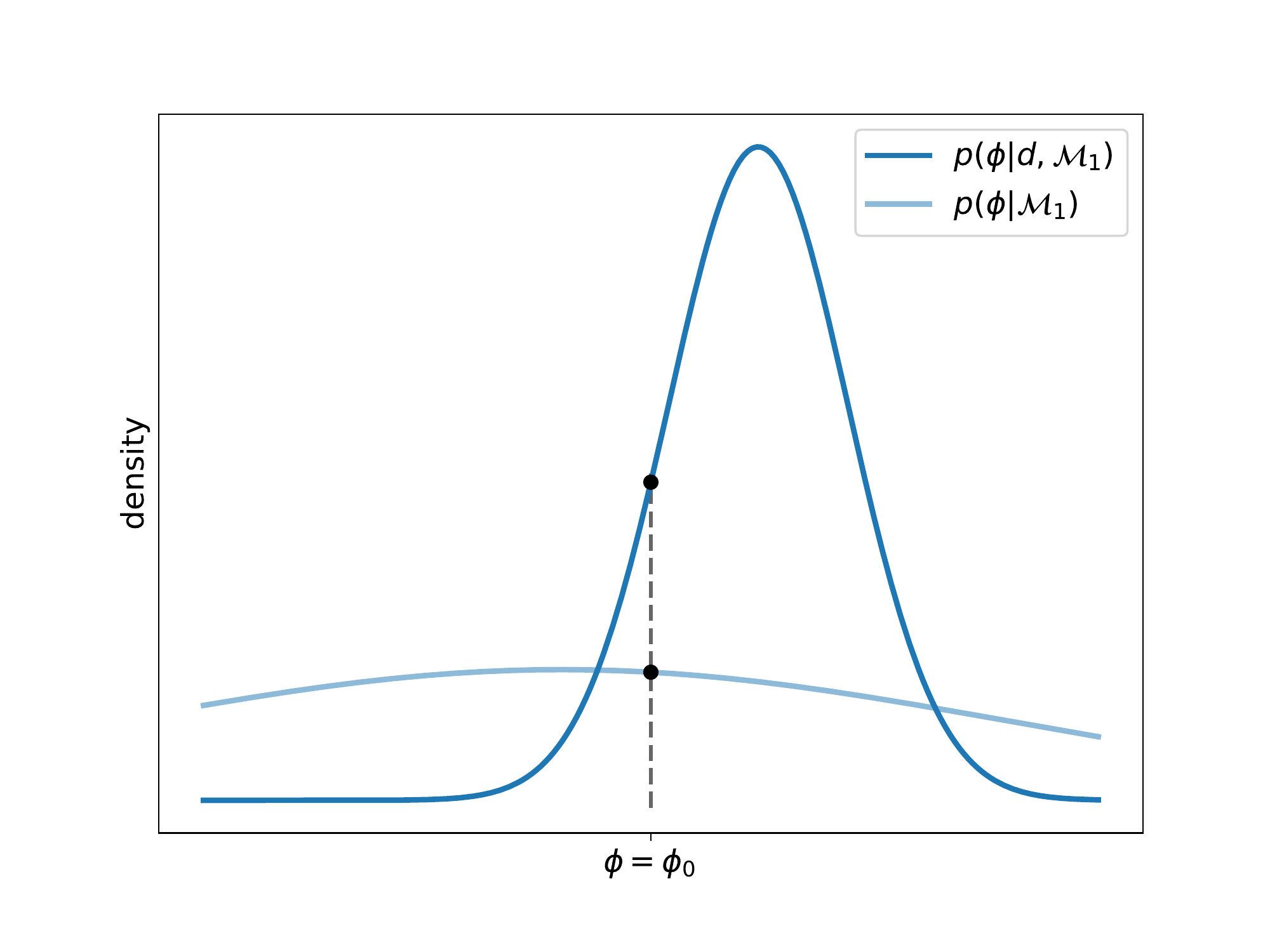}}
    \caption{The \textit{Savage--Dickey density ratio} expresses the Bayes factor $\mathrm{BF}_{01}$ as the ratio of marginal posterior to the prior density at the point $\phi_{0}$ in which model $M_{1}$ reduces to $M_{0}$.}
    \label{fig:savage_dickey}
\end{figure}

Although the SDDR can be only applied to nested models, it has the advantage that it is simple to compute, given some posterior samples, without making any assumptions about the Gaussianity of the posterior distribution. In particular, the \textit{Bayes factor} of $\mathcal{M}_{0}$ over $\mathcal{M}_{1}$ is simply,
\begin{equation}
    \label{eq:savage_dickey_bayes_factor}
    \mathrm{BF}_{01} = \frac{p(d\vert\mathcal{M}_{0})}{p(d\vert\mathcal{M}_{1})} = \frac{p(\phi=\phi_{0}\vert d,\mathcal{M}_{1})}{p(\phi=\phi_{0}\vert\mathcal{M}_{1})}\,,
\end{equation}
where the numerator of the right--hand--side ratio is just the \textit{marginal} posterior of $\phi$ for $\mathcal{M}_{1}$ evaluated at $\phi=\phi_{0}$, and the denominator is the prior of $\phi$ for $\mathcal{M}_{1}$ evaluated at $\phi=\phi_{0}$. In other words, the \textit{Bayes factor} is simply the marginal posterior to prior ratio for $\mathcal{M}_{1}$ evaluated at $\phi=\phi_{0}$. This means that only the parameters $\phi$ determine the value of the \textit{Bayes factor}, and the nuisance parameters $\theta$, that are common among the two models, are irrelevant. A schematic representation of SDDR is depicted in Figure \ref{fig:savage_dickey}.

The validity of this method relies on two conditions. First, that the likelihood function of $\mathcal{M}_{1}$ has to reduce to that of $\mathcal{M}_{0}$
\begin{equation}
    \label{eq:savage_dickey_likelihood_condition}
    p(d\vert \theta, \phi=\phi_{0}, \mathcal{M}_{1}) = p(d\vert \theta, \mathcal{M}_{0})\,,
\end{equation}
and the same must be true for the prior
\begin{equation}
    \label{eq:savage_dickey_prior_condition}
    p(\theta\vert \phi=\phi_{0},\mathcal{M}_{1}) = p(\theta\vert \mathcal{M}_{0})\,.
\end{equation}
The condition of equation \ref{eq:savage_dickey_prior_condition} is satisfied by separable priors,
\begin{equation}
    \label{eq:savage_dickey_separable_priors}
    p(\theta, \phi\vert \mathcal{M}_{1}) = p(\theta\vert \mathcal{M}_{1}) p(\phi\vert \mathcal{M}_{1})\,.
\end{equation}
The proof of equation \ref{eq:savage_dickey_bayes_factor} is straightforward, starting with,
\begin{equation}
    \label{eq:evidence_0}
    \begin{split}
        p(d\vert \mathcal{M}_{0}) &= \int p(d\vert \theta, \mathcal{M}_{0}) p(\theta \vert \mathcal{M}_{0}) d\theta \\
        &= \int p(d\vert \theta, \phi=\phi_{0}, \mathcal{M}_{1}) p(\theta \vert\phi=\phi_{0} \mathcal{M}_{1}) d\theta \\
        &= p(d\vert \phi=\phi_{0} \mathcal{M}_{1})\,,
    \end{split}
\end{equation}
in which we used the fact that $\mathcal{M}_{0}$ is nested in $\mathcal{M}_{1}$ for $\phi=\phi_{0}$. The next step is simply to employ Bayes' theorem
\begin{equation}
    \label{eq:bayes_theorem_savage_dickey}
    p(\phi=\phi_{0}\vert d,\mathcal{M}_{1}) = \frac{p(d\vert \phi=\phi_{0} \mathcal{M}_{1})p(\phi=\phi_{0}\vert \mathcal{M}_{1})}{p(d\vert \mathcal{M}_{1})}\,,
\end{equation}
and solve for the ratio of model evidences by first substituting equation \ref{eq:evidence_0} into it to compete the proof.

Practical use of equation \ref{eq:savage_dickey_bayes_factor} requires the evaluation the marginal posterior of $\mathcal{M}_{1}$ at $\phi=\phi_{0}$. As the closed--form expression for the marginal posterior is rarely available, one can use samples from posterior (e.g. generated using MCMC) to create a density histogram for $\phi$. Even better, \textit{Kernel Density Estimation (KDE)}~\parencite{silverman2018density} can be used to approximate the marginal posterior from samples as,
\begin{equation}
    \label{eq:savage_dickey_kde}
    \Hat{p}_{\mathrm{KDE}}(\phi\vert d, \mathcal{M}_{1}) = \frac{1}{n} \sum_{i=1}^{n}K_{h}\left(\phi-\phi_{i}\right)\,,
\end{equation}
where $K_{h}$ is the \textit{kernel} and $h$ is the \textit{bandwidth}, a parameter that controls the smoothing. The \textit{kernel} is generally a non--negative function, and most commonly it is chosen to be a simple Gaussian,
\begin{equation}
    \label{eq:gaussian_kernel}
    K_{h}(\phi)=\frac{1}{(2\pi)^{D/2}h^{D}} e^{-\frac{\phi^{2}}{2 h^{2}}}\,,
\end{equation}
where $D$ is the dimensionality of $\phi$ (i.e. the number of elements of the $\phi$ vector). Finally, the value of $h$ can either be determined on the basis of trial--and--error, or heuristics such as,
\begin{equation}
    \label{eq:silverman_kde}
    h = 1.06 \Hat{\sigma} n^{-1/5}\,,
\end{equation}
for the $1$--D case where $\Hat{\sigma}$ is the standard deviation of the samples~\parencite{silverman2018density}.
% !TEX TS-program = pdflatex
% !TEX root = ../ArsClassica.tex

%************************************************
\chapter{Advanced methods}
\label{chp:advanced}
%************************************************

\begin{flushright}
\itshape
Look on my works, ye Mighty, and despair! \\
\medskip
--- Percy Shelley, Ozymandias
\end{flushright}

This chapter introduces two advanced Monte Carlo methods, \textit{Sequential Monte Carlo} and \textit{Nested sampling}, which combine different previously introduced methods, such as MCMC and importance sampling, in order to provide samples from posterior distributions and estimate the model evidence. What distinguishes those two methods from all the previous ones introduced in this thesis, is their level of complexity and their reliance on multiple individual algorithms as their constituent parts.

\section{Sequential Monte Carlo}

\textit{Sequential Monte Carlo (SMC)} is, from a physics point of view, conceptually related to the notion of \textit{thermodynamic reversibility}. For a physical process starting from a state $A$ and ending in a state $B$, to be thermodynamically reversible, the transition has to be slow enough such that each intermediate state of the system is approximately in equilibrium.

\subsection{Background}

The basic idea of SMC is to slowly guide a population of $n$ particles $\lbrace\theta_{i}^{(t)} \rbrace_{i=1}^{n}$, drawn from a known probability distribution $\rho(\theta)$, through a series of intermediate distributions which create a path from $\rho(\theta)$ to the target distribution of interest $p(\theta)$~\parencite{liu1998sequential}. In the context of SMC, the rate of this transition is governed by the number of intermediate distributions bridging $\rho(\theta)$ to $p(\theta)$. Just like in \textit{annealed importance sampling (AIS)}, SMC relies on a number of MCMC steps performed in each intermediate step by every particle. This aims to equilibrate the particles by letting them reach the equilibrium distribution of each step. Furthermore, when transitioning from an intermediate distribution to the next, the particle distribution is adjusted using importance sampling. This guarantees that the particle distribution at any stage is the correct equilibrium distribution. 

The main difference between SMC and AIS is the use of resampling in the case of SMC. During the run, the particle distribution might experience \textit{weight degeneracy}, that is, only a few of the particles have non--negligible importance weights with the rest of them being vanishingly small. This high weight--variance can substantially affect any expectation values. In order to address this issue, SMC performs regular resampling steps, in which the particle distribution is resampled according to their weights, and the importance weights are re-initialised to be equal.

SMC methods are particularly suited for challenging target distributions which exhibit multiple modes. Furthermore, modifications of the main algorithm that we will present here can also be used for tasks of \textit{online learning} in which the data arrive sequentially. These algorithms are most often called by the name of \textit{particle filters}~\parencite{naesseth2019elements}.

\subsection{Bridging the prior and the posterior}

A common way to construct such a sequence of intermediate distributions that bridge a known density $\rho(\theta )$ to the target density $p(\theta )$ is to interpolate between the two densities
\begin{equation}
    \label{eq:geometric_interpolation}
    p_{t}(\theta) \propto \rho^{1-\beta_{t}}(\theta) p^{\beta_{t}}(\theta)\,,\quad t=1,\dots,m\,,
\end{equation}
where $\beta_{t}$ is a temperature annealing ladder, such that
\begin{equation}
    \label{eq:temperature_ladder}
    0 = \beta_{1} < \beta_{2} < \dots < \beta_{m} = 1\,.
\end{equation}
In the Bayesian context, a natural choice is to set the prior as the auxiliary density $\rho(\theta) = p(\theta\vert\mathcal{M})$ and the posterior as the target density $p(\theta) = p(\theta\vert d,\mathcal{M})$. Equation \ref{eq:geometric_interpolation} then reduces to the usual annealed or tempered interpolation
\begin{equation}
    \label{eq:annealed_path}
    p_{t}(\theta) \propto p^{\beta_{t}}(d\vert\theta\mathcal{M}) p(\theta\vert\mathcal{M})\,.
\end{equation}
Although we will focus on this case, the algorithm is valid for any pair of distributions as long as the support of the auxiliary density encompasses that of the target.

\subsection{Correction -- Selection -- Mutation}

\begin{figure}[!ht]
    \centering
	\centerline{\includegraphics[scale=1.14]{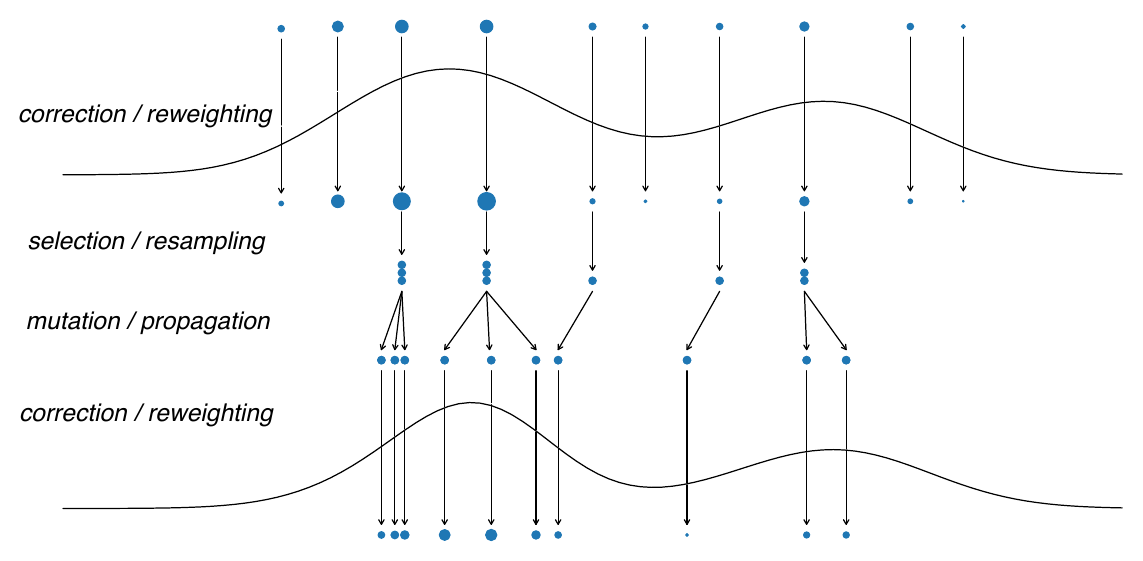}}
    \caption{Illustration of the \textit{Sequential Monte Carlo} algorithm with its three fundamental steps. During the correction step the particles are reweighted to represent the next probability distribution. Selection removes the particles with the smaller important weights and multiplies those with larger weights. Finally, mutation diversifies the particles by moving them.}
    \label{fig:smc}
\end{figure}
Given the initial positions of the particles $\lbrace\theta_{i}^{(1)} \rbrace_{i=1}^{n}$ drawn from the prior distribution, as well as the initial weights $\lbrace W_{i}^{(1)}\rbrace _{i=1}^{n}=1/n$, SMC proceeds by the sequential application of the following three steps, selection, mutation, and correction, until the posterior density is reached. The procedure that takes place in a single iteration $t$ is illustrated in Figure \ref{fig:smc} and involves the steps:
\begin{enumerate}
    \item \textbf{Correction / reweighting} -- During this stage, the weights of the particles are updated according to
    \begin{equation}
        \label{eq:update_particle_weights}
        w_{i}^{(t)}=  W_{i}^{(t-1)}\times \frac{p_{t}(\theta_{i}^{(t-1)})}{p_{t-1}(\theta_{i}^{(t-1)})} = W_{i}^{(t-1)}\times \left[p(d\vert\theta_{i}^{(t-1)},\mathcal{M})\right]^{\beta_{t}-\beta_{t-1}}\,,
    \end{equation}
    where with $w_{i}$ we denote the unnormalised weights and with $W_{i}=w_{i}/\sum_{i=1}^{n}w_{i}$ the normalised ones.
    
    The reweighing step accounts and corrects for any deviations of the particle distribution from the typical set of the target $p_{t}$. The ratio of the normalisation constants is estimated as
    \begin{equation}
        \label{eq:ratio_of_norm_constants}
        \frac{\mathcal{Z}_{t}}{\mathcal{Z}_{t-1}} = \sum_{i=1}^{n}w_{i}^{(t)}\,.
    \end{equation}
    Assuming the density for $t=1$ corresponds to the prior for which $\mathcal{Z}_{1}=1$, equation \ref{eq:ratio_of_norm_constants} will eventually lead to the estimation of the model evidence $\mathcal{Z}_{m} = p(d\vert\mathcal{M})$. 
    
    \item \textbf{Selection / resampling} -- The particle positions $\lbrace\theta_{i}^{(t-1)} \rbrace_{i=1}^{n}$ are resampled according to their weights $\lbrace W_{i}^{(t-1)} \rbrace_{i=1}^{n}$. The weights are then set again to be equal, $W_{i}^{(t-1)}=1/n$. Their new, resampled, positions are denoted as  $\lbrace\Tilde{\theta}_{i}^{(t-1)} \rbrace_{i=1}^{n}$. Particles with small weight values are removed and those with large importance weights are multiplied.
    
    Resampling can be done using simple \textit{multinomial resampling}, in which we draw $n$ new particles, with replacement, with probabilities given by their weights, or using more advanced schemes characterised by lower variance~\parencite{li2015resampling}. This process can be performed in each iteration, or only when some criterion is triggered (e.g. when the effective sample size of the weights drops below a threshold). 
    
    Finally, caution must be taken when applying resampling too frequently. This could lead to the phenomenon of \textit{weight impoverishment} in which there is no diversity between the particle positions. Fortunately, weight impoverishment is also reduced by the next step.
    
    \item \textbf{Mutation / propagation} -- Finally, the population of particles $\lbrace\Tilde{\theta}_{i}^{(t-1)} \rbrace_{i=1}^{n}$ is updated and the particles move to their new positions $\lbrace\theta_{i}^{(t)} \rbrace_{i=1}^{n}$ by performing a number of MCMC steps targeting the density $p_{t}(\theta)$. 
    
    The purpose of this step is to diversify the particles and allow their distribution to approach the stationary distribution. An advantage of SMC is that the particle distribution from the previous iteration can be used to construct efficient proposal distributions for MCMC for the current density. Furthermore, as $n$ particles are updated at once, this step can be done in parallel. Any MCMC method can be used in this step and there is no requirement for the final/new positions to be uncorrelated from the initial ones, although in practice this helps reduce the variance of the estimates.
    
    A common approach is to use the particle covariance $\Sigma^{(t-1)}$ to construct a normal proposal distribution $q(\theta'\vert\theta)=\mathcal{N}(\theta'\vert\theta,\Sigma^{(t-1)})$.
\end{enumerate}

Once all three steps are completed, the value of $\beta$ is updated and the process is repeated again until $\beta$ reaches the value of one. The names of those three steps are inspired by natural selection and evolutionary programming. The reason is the apparent analogy with genetic algorithms~\parencite{koza1994genetic}. More specifically, reweighting, resampling and propagation have the roles of correction, selection and mutation in genetic algorithms, in which the particle positions are the \textit{genes} and the importance weights play the role of the so--called \textit{fitness}. A critical difference with most genetic algorithms is the fact that SMC solves a sampling task, not an optimisation one, and thus the solution is represented by the distribution of the particles and not by any particle individually.

\begin{algorithm}[ht!]
\caption{Sequential Monte Carlo} \algolabel{smc}
\begin{algorithmic}[1]
\REQUIRE{initial state for the ensemble $\theta^{(1)}=(\theta_{1}^{(1)},\dots,\theta_{K}^{(1)})$, prior probability density $\pi(\theta) \equiv p(\theta\vert\mathcal{M})$, likelihood function $\mathcal{L}(\theta)\equiv p(d\vert\theta,\mathcal{M})$, and a local MCMC kernel $\theta'\leftarrow\mathcal{K}(\theta; f(\theta))$ (e.g. $N$ steps of random--walk Metropolis update)}
\ENSURE{Posterior samples and estimate of the model evidence $\mathcal{Z}$}
\STATE{Initialise temperature parameter $\beta_{1} = 1$}
\STATE{Initialise estimate of evidence $\mathcal{Z} = 1$}
\FOR{$i = 1$ \TO $n$}
    \STATE{Draw particle positions from the prior $\theta_{i}^{(1)}\sim \pi(\theta)$}
    \STATE{Initialise particle weights $W_{i}^{(1)} = 1 / n$}
\ENDFOR
\WHILE{$\beta_{t}\neq 1$}
    \STATE{Update iteration index $t \leftarrow t + 1$}
    \STATE{Set temperature $\beta_{t}$ solving $\left( \sum_{i=1}^{n}w_{i}^{(t)}(\beta_{t})\right)^{2}/\sum_{i=1}^{n}\left(w_{i}^{(t)}(\beta_{t})\right)^{2} = \alpha \times n$ where the importance weights are computed as $w_{i}^{(t)}\leftarrow W_{i}^{(t-1)}\mathcal{L}(\theta_{i}^{(t-1)})^{\beta_{t}-\beta_{t-1}}$}
    \STATE{Update evidence estimate $\mathcal{Z}\leftarrow \mathcal{Z}\times n^{-1}\sum_{i=1}^{n}w_{i}^{(t)}$}
    \STATE{$\left\lbrace \Tilde{\theta}_{i}^{(t-1)}\right\rbrace_{i=1}^{n}\leftarrow$ resample $\left\lbrace \theta_{i}^{(t-1)}\right\rbrace_{i=1}^{n}$ according to $\left\lbrace w_{i}^{(t)}\right\rbrace_{i=1}^{n}$}
    \FOR{$i=1$ \TO $n$}
        \STATE{Reset weights $W_{i}^{(t)}\leftarrow 1/n$}
    \ENDFOR
    \STATE{Update particles using MCMC 
    \[
    \left\lbrace \theta_{i}^{(t)}\right\rbrace_{i=1}^{n}\leftarrow\mathcal{K}\left(\left\lbrace \Tilde{\theta}_{i}^{(t-1)}\right\rbrace_{i=1}^{n}; \pi(\theta)\mathcal{L}(\theta)^{\beta_{t}}\right)
    \]
    }
\ENDWHILE
\end{algorithmic}
\end{algorithm}

\subsection{Effective sample size}

A common measure of the quality of the importance weights of the particles, at any iteration of the SMC run, is the \textit{effective sample size (ESS)}
\begin{equation}
    \label{eq:effective_sample_size_smc}
    \text{ESS}_{t} = \frac{\mathbb{E}_{p_{t}}\left[w^{(t)}\right]^{2}}{\mathbb{E}_{p_{t}}\left[\left(w^{(t)}\right)^{2}\right]}\,.
\end{equation}
which can be estimated as:
\begin{equation}
    \label{eq:effective_sample_size_smc_estimator}
    \Hat{\text{ESS}}_{t} = \frac{\left( \sum_{i=1}^{n}w_{i}^{(t)}\right)^{2}}{\sum_{i=1}^{n}(w_{i}^{(t)})^{2}} = \frac{1}{\sum_{i=1}^{n}(W_{i}^{(t)})^{2}}\,.
\end{equation}

\subsection{Setting the temperature ladder}

A $\beta$ ladder can be specified \textit{a priori} or determined adaptively during the run~\parencite{gelman1998simulating}. In the first case, the resampling step is usually triggered whenever the ESS drops below a prespecified threshold value (e.g. $50\% - 95\%$). In the latter case, the next value of $\beta$ is chosen adaptively such that the ESS has an approximately constant fraction $\alpha$ (e.g. $50\% - 95\%$) of the number of particles $n$ throughout the duration of the SMC run. Numerically, this can be done by solving
\begin{equation}
    \label{eq:bisection_method}
    \frac{\left[ \sum_{i=1}^{n}w_{i}^{(t)}(\beta_{t})\right]^{2}}{ \sum_{i=1}^{n}\left[w_{i}^{(t)}(\beta_{t})\right]^{2}}=\alpha \times n\,,
\end{equation}
for the next $\beta_{t}$ such that $\beta_{t-1}<\beta_{t}\leq 1$ using, for instance, the \emph{bisection method}~\parencite{burden2015numerical}.

\section{Nested sampling}

\textit{Nested sampling (NS)}, originally developed by \textcite{skilling2004nested, skilling2006nested}, is a method for estimating the model evidence $\mathcal{Z} = p(d\vert\mathcal{M})$. The basic idea is to approximate the evidence by integrating the prior in \textit{nested} shells of constant likelihood. Despite its original purpose to estimate the model evidence, NS can also provide weighted samples from the posterior distribution as an optional byproduct. Therefore, the method is suitable for both tasks of parameter estimation and model comparison~\parencite{ashton2022nested}. Over the years, many variants of NS have emerged, with each one aiming to improve a different aspect of the original version~\parencite{brewer2011diffusive, feroz2013importance, higson2019dynamic}.

\subsection{Multi--dimensional integration}

\begin{figure}[H]
    \centering
	\centerline{\includegraphics[scale=0.53]{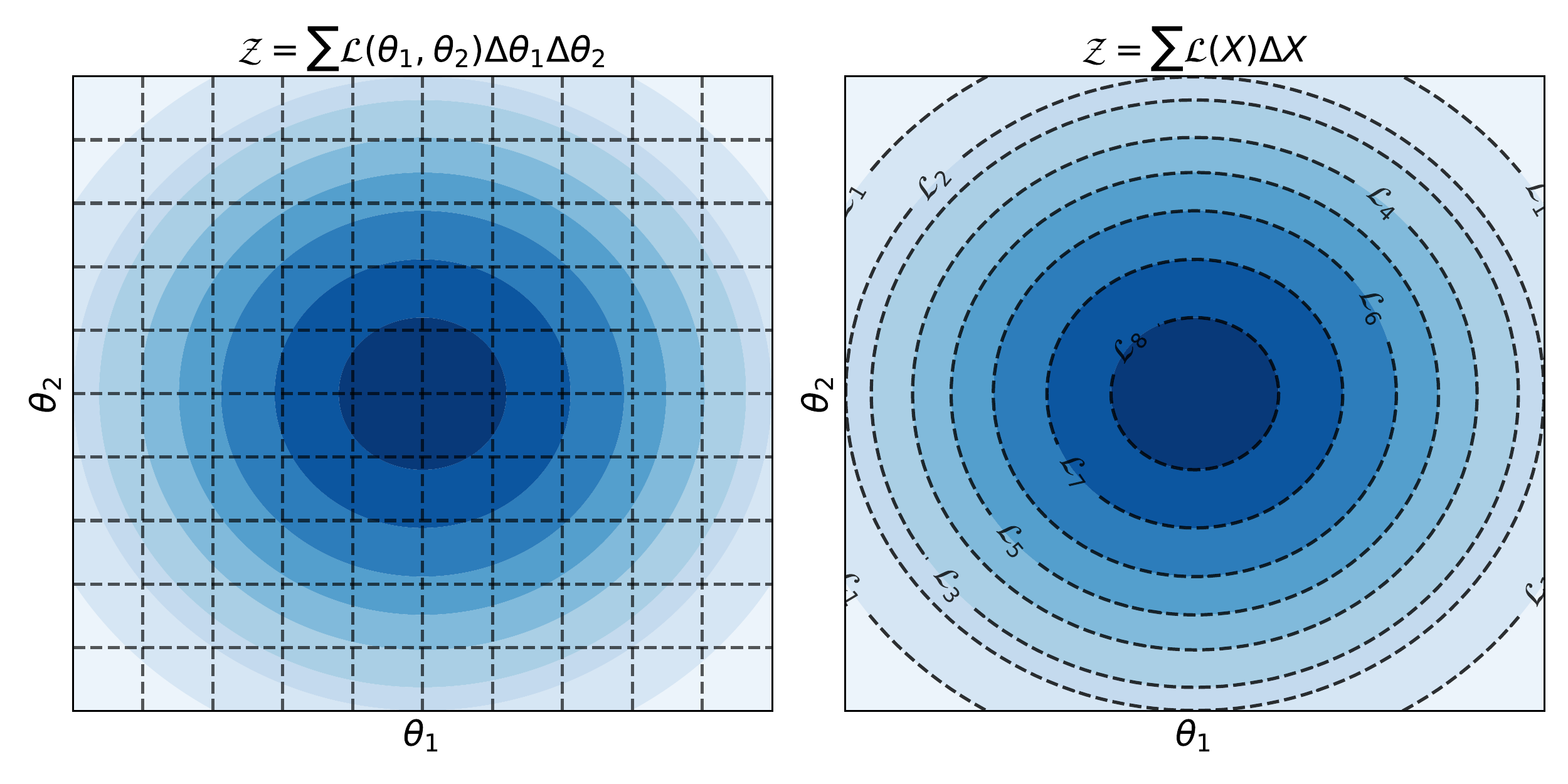}}
    \caption{Illustration comparing two ways which one can use to approximate the model evidence integral. The left panel shows the direct multi--dimensional integration over the parameters. The right panel shows the one--dimensional integration over the prior volume $X$ enclosed in the iso--likelihood contours.}
    \label{fig:nested_sampling_integral}
\end{figure}
\begin{figure}[H]
    \centering
	\centerline{\includegraphics[scale=0.53]{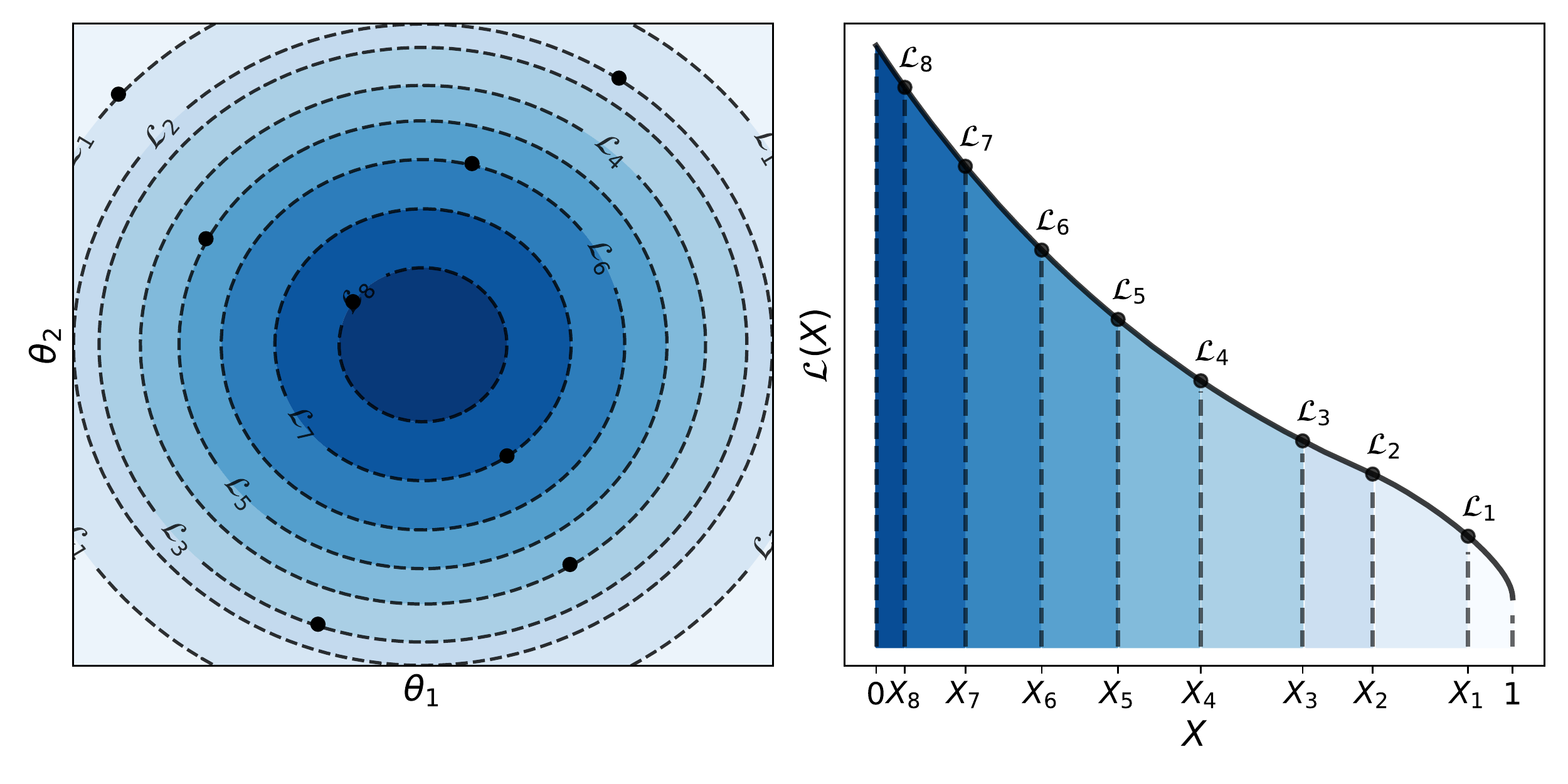}}
    \caption{Illustration of $8$ samples drawn uniformly from the prior with their respective iso--likelihood contours (\textit{left}), along with their corresponding contributions to the evidence integral (\textit{right}).}
    \label{fig:nested_sampling_samples}
\end{figure}

NS attempts to compute the evidence integral,
\begin{equation}
    \label{eq:evidence_integral}
    \mathcal{Z} = \int \mathcal{L}(\theta) \pi(\theta) d\theta \,,
\end{equation}
where $\mathcal{L}(\theta) = p(d\vert\theta,\mathcal{M})$ is the likelihood function and $\pi(\theta)=p(\theta\vert \mathcal{M})$ is the prior, by transforming it into a one--dimensional integral over the prior volume
\begin{equation}
    \label{eq:prior_volume}
    X(\lambda) = \int _{\mathcal{L}(\theta)>\lambda}\pi(\theta)d\theta,
\end{equation}
enclosed in the iso--likelihood contour defined by $\mathcal{L}(\theta) = \lambda$. Equation \ref{eq:evidence_integral} can then be written as
\begin{equation}
    \label{eq:evidence_integral_transformed}
    \mathcal{Z} = \int _{0}^{+\infty}X(\lambda)d\lambda = \int_{0}^{1} \mathcal{L}(X)dX\,,
\end{equation}
assuming that $\mathcal{L}(X(\lambda)) = \lambda$ exists. Figure \ref{fig:nested_sampling_integral} illustrates the equivalency between the integrals of equations \ref{eq:evidence_integral} and \ref{eq:evidence_integral_transformed}. Unlike equation \ref{eq:evidence_integral}, the above integral is now 1--dimensional and can be approximated using standard numerical integration techniques (e.g. quadrature),
\begin{equation}
    \label{eq:evidence_quadrature}
    \mathcal{Z} = \sum_{i=1}^{m}\mathcal{L}_{i}w_{i}\,,
\end{equation}
where,
\begin{equation}
    \label{eq:ns_weights}
    w_{i} = \frac{X_{i-1}-X_{i+1}}{2}\,.
\end{equation}
This of course assumes that we are able to evaluate the iso--likelihood contours $\mathcal{L}_{i}=\mathcal{L}(X_{i})$ associated with an ordered collection of samples with prior volume $1 > X_{1} > X_{2} > \dots > X_{m} > 0$. This is illustrated in Figure \ref{fig:nested_sampling_samples} for a collection of $8$ samples that are uniformly distributed in the prior volume.

Using the simpler weights $w_{i}=X_{i}-X_{i+1}$ in equation \ref{eq:evidence_quadrature} a lower bound on the evidence can be estimated as
\begin{equation}
    \label{eq:evidence_integral_lower_bound}
    \mathcal{Z} \geq \sum_{i=1}^{m}\mathcal{L}_{i} \left(X_{i}-X_{i+1}\right)\,.
\end{equation}
Similarly, an upper bound also exists, using $w_{i}=X_{i-1}-X_{i}$, which can be written as
\begin{equation}
    \label{eq:evidence_integral_upper_bound}
    \mathcal{Z} \leq \sum_{i=1}^{m}\mathcal{L}_{i}\, \left(X_{i-1}-X_{i}\right)+X_{m}\mathcal{L}_{\mathrm{max}}\,,
\end{equation}
where $\mathcal{L}_{\mathrm{max}}$ is the maximum likelihood value to be found as $X\to 0$.

Soon after its original inception, it was realised that a NS run can also be used for the task of parameter estimation without any additional computation. In particular, the collected samples combined with their normalised weights
\begin{equation}
    \label{eq:posterior_weights}
    p_{i} = \frac{\mathcal{L}_{i}w_{i}}{\mathcal{Z}}\,,
\end{equation}
correspond to weighted samples from the posterior distribution and thus can be used to compute expectation values
\begin{equation}
    \label{eq:ns_expectation_values}
    \mathbb{E}_{p}\left[ f(\theta)\right] = \sum_{i}^{n}p_{i}f(\theta_{i})\,.
\end{equation}

\subsection{Sampling procedure}

\begin{figure}[H]
    \centering
	\centerline{\includegraphics[scale=0.53]{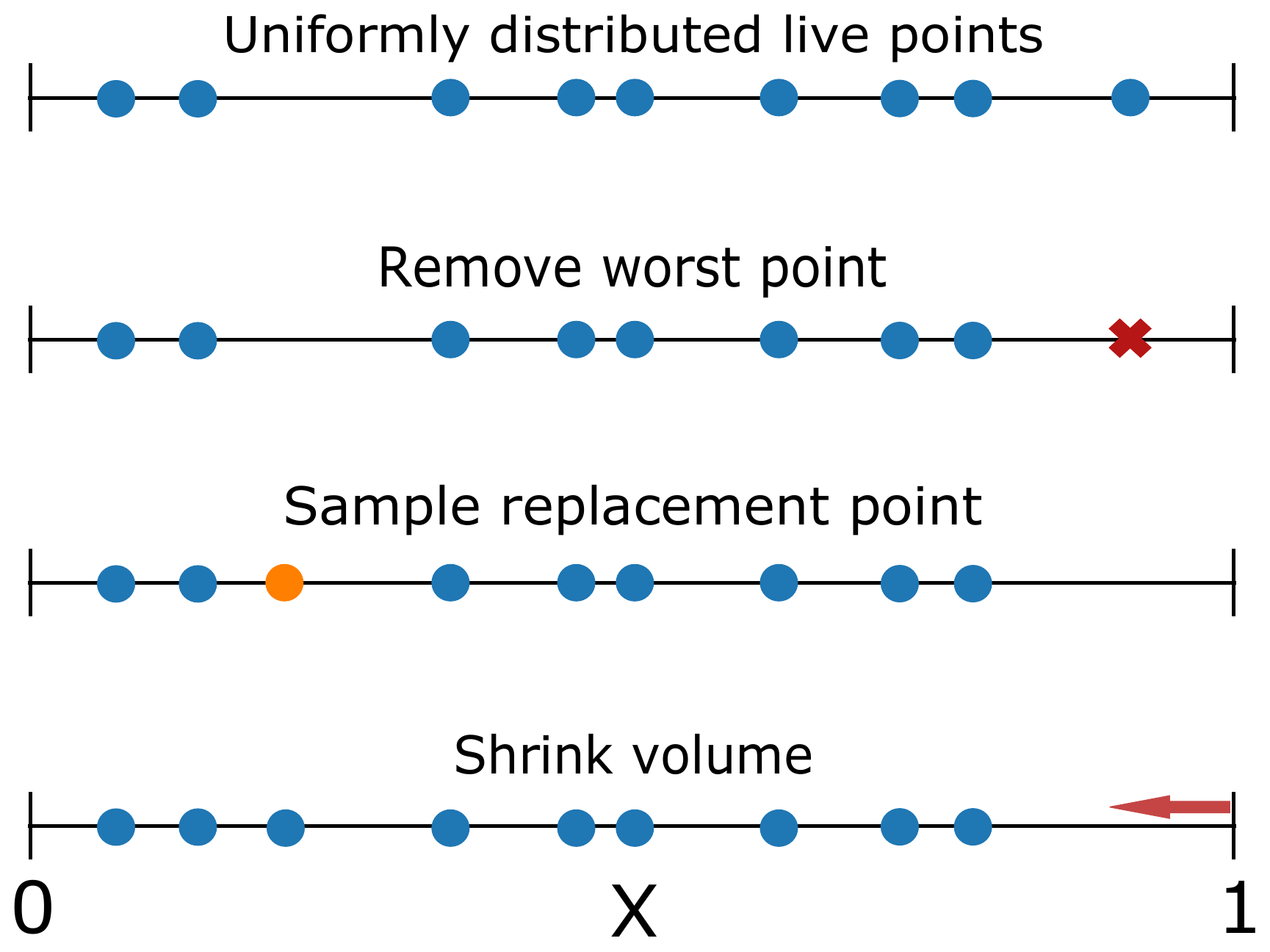}}
    \caption{Illustration of the nested sampling procedure. Given some uniformly distributed points from the prior, we identify and remove the worst point, that is, the point with the minimum likelihood value. $\mathcal{L}_{min}$, and replace it a new point sampled from the prior subject to the likelihood constrain $\mathcal{L} > \mathcal{L}_{min}$. Finally, the volume is contracted to account for the removal of the worst point.}
    \label{fig:nested_sampling_step}
\end{figure}
The NS algorithm begins by drawing a collection of points $\lbrace \theta_{i}\rbrace_{i=1}^{n}$ uniformly from the prior, often called \textit{live points}. We can associate each live point $\theta_{i}$ with a prior volume $X$--value, namely the volume that would be enclosed by the iso--likelihood contour $\mathcal{L}_{i}=\mathcal{L}(\theta_{i})$. On average, we expect roughly half of the live points to fall inside the iso--likelihood contour corresponding to half prior volume $X=1/2$, one quarter to $X=1/4$, one eighth to $X=1/8$ and so on. In other words, since the live points are uniformly distributed under the prior, the corresponding $X$--values are uniformly distributed between $0$ and $1$. This is illustrated in Figure \ref{fig:nested_sampling_samples} and the top panel of Figure \ref{fig:nested_sampling_step}.

What we described so far is only the first step of the algorithm, and one still needs a way to propagate the live points into regions of smaller prior volume (i.e. lower $X$) in order to probe iso--likelihood contours corresponding to higher likelihood values. NS achieves this by first identifying the live point with the lowest likelihood value $\mathcal{L}^{*}=\mathcal{L}_{1}$, corresponding to volume $X_{1}$ and removing it. The remaining live points are now distributed over a compressed volume $X_{1}$. On average, the volume compression factor is 
\begin{equation}
    \label{eq:average_volume_compression}
    t = e^{-1/n}\,,
\end{equation}
such that the compressed volume is $X_{1}=t\times X_{0}$, where $X_{0}=1$ is the initial total volume. Finally, we sample a new live point to replace the one that we removed. The new point is sampled uniformly from the prior subject to the constrain $\mathcal{L} > \mathcal{L}^{*}$, that is, from the likelihood--constrained prior
\begin{equation}
    \label{eq:likelihood_constrained_prior}
    \pi^{*}(\theta) \propto
    \begin{cases}
    \pi(\theta) & \mathrm{if }\mathcal{L}(\theta) > \mathcal{L}^{*}\,, \\
    0 & \text{otherwise}\,.
    \end{cases}
\end{equation}
This whole process, that is shown in Figure \ref{fig:nested_sampling_step}, is repeated multiple times until a criterion for termination is met. In each iteration, the volume shrinks on average by the compression factor of equation \ref{eq:average_volume_compression}.

\subsection{Termination criterion}

During an NS run, the remaining prior volume $X$ asymptotically approaches $0$. The fact that we can only perform a finite number of steps means that it is unavoidable to introduce a truncation error into the evidence estimate of equation \ref{eq:evidence_integral_transformed}. A common way of determining when to stop is to approximately estimate the amount of remaining evidence and terminate the run when this can be considered negligible for the purpose of the analysis. 

Perhaps the simplest way to roughly estimate the remaining evidence is by utilising the upper bound of equation \ref{eq:evidence_integral_upper_bound}. In this case, the remaining evidence is approximated as $\Delta\mathcal{Z}=\mathcal{L}_{\mathrm{max}}X_{m}$, where $\mathcal{L}_{\mathrm{max}}$ is the maximum likelihood value, estimated from the current population of live points, and $X_{m}$ is simply the estimate of the remaining prior volume. An alternative would be to use the mean likelihood of the live points and get $\Delta\mathcal{Z}=\Bar{\mathcal{L}}X_{m}$. The run then terminates when $\Delta \mathcal{Z}/\mathcal{Z}$ drops below a prespecified threshold.

Of course, neither of these approaches guarantees that the run will terminate early and that beyond lies a ``spike'' of huge likelihood. Upon deciding to stop, however, the current estimate of the model evidence is approximately corrected by either adding $\Delta\mathcal{Z}$ or removing the live points one--by--one in accordance with the NS procedure and adding their respective evidence contributions $\mathcal{L}_{i}X_{i}$, but without replacing them with new ones.

\subsection{Uncertainty}

So far we have assumed that the compression factor is given by equation \ref{eq:average_volume_compression}, however, this is simply the mean compression factor associated with the removal of the outermost or lowest--likelihood live point. In truth, the prior volume bounded by the iso--likelihood contour of that point can be slightly different from what the mean compression factor predicts. The compression in volume $t$ associated with the removal of the outermost of $n$ live points follows a $\mathrm{Beta}(n,1)$ probability distribution with density,
\begin{equation}
    \label{eq:beta_n1}
    p(t) = n t^{n-1}\,,
\end{equation}
where the first $n$ factor comes from the fact that any live point could be the outermost, and the second factor from the fact that the remaining $n-1$ live points lie uniformly distributed above the outermost at $t$.

The compression factors can therefore be sampled from the probability distribution of equation \ref{eq:beta_n1} instead of just assuming their expected value of equation \ref{eq:average_volume_compression}. Furthermore, we can use equation \ref{eq:beta_n1} to compute the expectation values
\begin{equation}
    \label{eq:logt_expectations}
    \mathbb{E}[\log t] = -\frac{1}{n}\,,\quad\,\mathrm{Var}[\log t] = \frac{1}{n^{2}}\,.
\end{equation}
Since the individual $\log t$ are independent, we expect that after $i$ steps, the prior volume to have shrunk to
\begin{equation}
    \label{eq:contracted_prior_volume}
    \log X_{i} = -\frac{i\pm\sqrt{i}}{n}\,.
\end{equation}
What the above expression means is that there is uncertainty in the estimates of the compression factor which enter into the prior volume estimates too. In other words, there is uncertainty in the number of steps $i$ required for the prior volume to shrink to a certain value $X_{i}$.

The uncertainty originating from the noisy estimates of the compression factor $t$ also propagates into the estimate of the model evidence. To quantify this we need to consider the information gained when transitioning from the prior $\pi(\theta)$ to the posterior $p(\theta)$, given by the \textit{Kullback--Leibler (KL) divergence},
\begin{equation}
    \label{eq:prior_to_posterior_information_gain}
    H = \int p(\theta)\log\left( \frac{p(\theta)}{\pi(\theta)}\right)d\theta\,.
\end{equation}
We can rewrite the above equation in terms of the prior volume $X$, as
\begin{equation}
    \label{eq:prior_to_posterior_information_gain_transformed}
    \begin{split}
        H &= \int p(X)\log p(X) dX \\
         &= -\int p(X)\log X dX + \int p(\log X) \log p(\log X) d \log X\,,
    \end{split}
\end{equation}
where $p(X)=\mathcal{L}(X)/\mathcal{Z}$ is the volume posterior density. Ignoring the second term on the right hand size, which is subdominant, we thus get that the KL divergence provides a measure of the compression we require to reach the bulk of the posterior mass,
\begin{equation}
    \label{eq:posterior_bulk_compression}
    H = - \log X\,.
\end{equation}
Comparing equation \ref{eq:contracted_prior_volume} and \ref{eq:posterior_bulk_compression} we roughly expect $n H \pm \sqrt{n H}$ steps to reach the bulk of the posterior mass. Equivalently, the uncertainty introduced into the estimate of the model evidence is
\begin{equation}
    \label{eq:evidence_uncertainty}
    \Delta \log\mathcal{Z} = \sqrt{\frac{H}{n}}\,.
\end{equation}
Of course, the above expression does not include any sources of numerical error such as truncation error.

\subsection{Likelihood--constrained prior sampling}

The efficient application of the NS algorithm requires sampling from the prior distribution $\pi(\theta)$ subject to the likelihood constrain $\mathcal{L} > \mathcal{L}^{*}$. Unfortunately, drawing points from the prior until the likelihood criterion is met is not feasible in practice, as the volume contained in the constrained prior shrinks exponentially with each iteration. For this reason, two different approaches, \textit{region} and \textit{step} samplers are often employed in order to produce samples from the likelihood--constrained prior.

For the sake of simplicity, both samplers usually operate in the \textit{latent} parameter space that the prior is uniform over the unit hypercube. In this case, the practitioner specifies their prior preference, not by providing a (log--) probability density function, but by defining the \textit{inverse--cumulative density function} $\Phi^{-1}$ that transforms points $\phi$ in the hypercube to points $\theta$ in the original parameter space. For instance, let us assume that we require a normal prior on a parameter $\theta\sim\mathcal{N}(\mu,\sigma^{2})$. We can transform a unit hypercube parameter $\phi\sim\mathcal{U}(0,1)$, using the standard normal distribution's \textit{inverse--cumulative density function} $\Phi^{-1}$, such that,
\begin{equation}
    \label{eq:unit_to_theta}
    \theta = \mu + \Phi^{-1}(\phi) \sigma\,.
\end{equation}

\subsubsection{Region samplers}

The basic idea behind \textit{region} samplers is to construct a hypersurface that bounds a given iso--likelihood contour. In practice, this is done using simple geometric shapes (e.g. spheres, ellipses, etc.). The hypersurface must encompass the current distribution of live points and at least contain the currently estimated volume. One can then sample uniformly from within the volume enclosed by the hypersurface until the likelihood--constraint is satisfied. In order to reduce the risk of missing parts of the currently estimated volume, the bounding region is usually expanded by a prespecified factor or using cross--validation of the live points.

Most region samplers attempt to construct such a bounding region by wrapping the current generation of live points with one or multiple ellipsoids. Using multiple ellipsoids offers some flexibility in the case of multimodal posterior distributions. The most popular such sampler is the \textit{MultiNest} algorithm that determines the shape and location of the ellipsoids based on the mean and covariance of the live points, by first estimating the number of distinct modes, and thus required ellipsoids, using a clustering algorithm~\parencite{feroz2009multinest}.

Region samplers have to face serious challenges when the complexity of the posterior or the dimensionality of the parameter space increases. In the first case, the ability to accurately bound the current volume depends on the number of live points, with a higher number often resulting in better bounding regions. The second limitation arises from the curse of dimensionality. As the number of dimensions increases, most of the volume of the bounding shape concentrates near its edges, and given that the bounding region is often chosen to be significantly broader than the encompassing likelihood--constrained volume to guarantee that no parts are encroached, the total number of samples until one is found to lie within the iso--likelihood contour increases exponentially. As a consequence, region samplers are more efficient and appropriate for low--dimensional problems, that is, $D\lesssim 10 - 20$. 

\subsubsection{Step samplers}

On the other hand, \textit{step} samplers do not rely on a bounding region and thus bypass some of the pathologies of \textit{region} samplers. Instead, they evolve a randomly chosen live point through a sequence of local steps to an approximately independent position. This is usually achieved using some MCMC method targeting the likelihood--constrained prior of equation \ref{eq:likelihood_constrained_prior} as the target distribution. The advantage of using MCMC methods in the context of NS is that, in each iteration, one can use the distribution of the live points to construct effective proposal distributions for the MCMC sampler.

Although \textit{step} samplers enjoy a better scaling with the number of dimensions than region samplers, there are still challenges in their use. First of all, determining the minimum number of steps to perform for the new point to be independent of its starting position (i.e. the randomly chosen live point) is not trivial. Although small correlations can be effectively ignored, larger violations can have catastrophic results and lead to substantial bias in the final estimates of NS. Furthermore, the step sampler must be tuned appropriately to achieve good sampling performance. Adaptation during a given iteration has to be diminishing in order to avoid spurious effects and biases.

Although any MCMC method  (e.g. random walk Metropolis, slice sampling, etc.) can in principle be used as a step sampler, there are also methods that are naturally suited and have been developed for use in the context of sampling from the likelihood--constrained prior. One such example is \textit{Galilean Monte Carlo (GMC)}~\parencite{feroz2013exploring, skilling2012bayesian, skilling2019galilean} that samples by moving consistently along a direction until a proposed point is rejected, by being outside the iso--likelihood contour. In this case, the sampler reflects off the current iso--likelihood boundary.

\subsection{Parallelisation}

Parallelising NS is not as straightforward as with other Monte Carlo algorithms (e.g. Sequential Monte Carlo), as the method relies on updating a single (worst) point at a time. In general, we would like to generate as many candidate points per step as the number of available CPUs (i.e. $n_{\mathrm{CPU}}$) and evaluate their likelihoods in parallel. In this case, there are three strategies that one can follow:
\begin{enumerate}
    \item Replace a single live point and discard as many as $n_{\mathrm{CPU}}-1$ acceptable live points. This scheme is quite wasteful, particularly in cases in which it is likely that more than one candidate point satisfies the likelihood constraint.
    \item Replace the worst (i.e. lowest likelihood) $n_{\mathrm{CPU}}$ live points in a single step. This results in linear speed--up with respect to the number of CPUs but increases the variance of the evidence estimate by a factor of $\sqrt{n_{\mathrm{CPU}}}$.
    \item Replace a single live point and consider the other $n_{\mathrm{CPU}}-1$ candidates for subsequent iterations. This results in a speed--up of $n\log(1+n_{\mathrm{CPU}}/n)$ which is approximately linear for $n_{\mathrm{CPU}}<<n$. The ``diminishing returns'' represented by the logarithmic factor in this expression originate from the fact that the likelihood threshold increases as the run progresses, and thus the points might not be valid for a subsequent iteration. This strategy is the most widely employed in practice.
\end{enumerate}

Finally, it is important to note that apart from parallelising a single NS run, it is also possible to combine different, possibly parallel, independent NS runs into a joint one, thus achieving linear scaling. In order to combine two or more runs together, we collect the points from all runs as live points and begin by removing the worst point, which with no loss of generality we assume that it belongs to run A. Then, as a replacement that satisfies the likelihood constraint, we simply take the replacement that was originally used in run A. We then proceed with the next worst point until all points are accounted for.

\begin{algorithm}[ht!]
\caption{Nested sampling} \algolabel{ns}
\begin{algorithmic}[1]
\REQUIRE{termination criterion (e.g. $\Delta \log \mathcal{Z} \leq \epsilon)$, number of live points $n$, an estimate of the compression factor e.g. $t=\exp(-1/n)$, prior distribution $\pi(\theta)$, likelihood function $p_{i}=\mathcal{L}(\theta)$)}
\ENSURE{Estimate of model evidence $\mathcal{Z}$, posterior samples $\Tilde{\theta}_{i}$ with weights $\mathcal{L}_{i}w_{i}/\mathcal{Z}$}
\STATE{Initialise volume $X = 1$ and evidence $\mathcal{Z}=0$}
\STATE{Draw $n$ live points from the prior $\theta_{1}, \theta_{2}, \dots, \theta_{n}\sim\pi(\theta)$}
\REPEAT
\STATE{Find the minimum likelihood value of the live points $\mathcal{L}^{*}\leftarrow \min\left( \mathcal{L}(\theta_{1}), \dots, \mathcal{L}(\theta_{n}) \right)$}
\STATE{Replace live point $\theta^{*}$ corresponding to $\mathcal{L}^{*}$ with a new point from the prior satisfying $\mathcal{L}>\mathcal{L}^{*}$}
\STATE{Set $w^{*}\leftarrow \Delta X$ where $\Delta X = (1-t)X$}
\STATE{Update estimate of the evidence $\mathcal{Z}\leftarrow \mathcal{Z} + \mathcal{L}^{*}w^{*}$}
\STATE{Store values of $w_{i}\leftarrow w^{*}$, $\mathcal{L}_{i}\leftarrow \mathcal{L}^{*}$,  and $\Tilde{\theta}_{i}\leftarrow \theta^{*}$}
\STATE{Contract volume $X \leftarrow t X$}
\UNTIL{termination criteria satisfied}
\STATE{Update evidence $\mathcal{Z}\leftarrow \mathcal{Z}+\frac{1}{n}\sum_{j=1}^{n}\mathcal{L}(\theta_{j})\,X$}
\end{algorithmic}
\end{algorithm}
% Third Part : My papers
\cleardoublepage
% !TEX TS-program = pdflatex
% !TEX root = ../ArsClassica.tex

%************************************************
\part{Novel Developments}
\label{prt:recent}
%************************************************
% !TEX TS-program = pdflatex
% !TEX root = ../ArsClassica.tex

%************************************************
\chapter{Ensemble Slice Sampling}
\label{chp:ess}
%************************************************
 
\lstset{numbers=left,
    numberstyle=\scriptsize,
    stepnumber=1,
    numbersep=8pt
}    

This chapter presents \textit{Ensemble Slice Sampling} which is the main contribution introduced in the paper titled \textit{Ensemble Slice Sampling: Parallel, black--box, and gradient--free inference} that was published in the journal \textit{Statistics and Computing} in 2021~\parencite{karamanis2021ensemble}. The content of the chapter is almost identical to that included in the aforementioned publication with the exception of minor text and figure formatting differences.

\begin{center}
\rule{0.5\textwidth}{.4pt}
\end{center}
\vspace{8pt}

Slice Sampling has emerged as a powerful Markov Chain Monte Carlo algorithm that adapts to the characteristics of the target distribution with minimal hand-tuning. However, Slice Sampling's performance is highly sensitive to the user-specified initial length scale hyperparameter and the method generally struggles with poorly scaled or strongly correlated distributions. This paper introduces Ensemble Slice Sampling (ESS), a new class of algorithms that bypasses such difficulties by adaptively tuning the initial length scale and utilising an ensemble of parallel walkers in order to efficiently handle strong correlations between parameters. These affine--invariant algorithms are trivial to construct, require no hand-tuning, and can easily be implemented in parallel computing environments. Empirical tests show that Ensemble Slice Sampling can improve efficiency by more than an order of magnitude compared to conventional MCMC methods on a broad range of highly correlated target distributions. In cases of strongly multimodal target distributions, Ensemble Slice Sampling can sample efficiently even in high dimensions. We argue that the parallel, black-box and gradient-free nature of the method renders it ideal for use in scientific fields such as physics, astrophysics and cosmology which are dominated by a wide variety of computationally expensive and non-differentiable models.

\section{Introduction}
\emph{Bayesian inference and data analysis} has become an integral part of modern science. This is partly due to the ability of Markov Chain Monte Carlo (MCMC) algorithms to generate samples from intractable probability distributions. MCMC methods produce a sequence of samples, called a \emph{Markov chain}, that has the target distribution as its equilibrium distribution. The more samples are included, the more closely the distribution of the samples approaches the target distribution. The Markov chain can then be used to numerically approximate expectation values (e.g. parameter uncertainties, marginalised distributions).

Common MCMC methods entail a significant amount of time spent hand-tuning the hyperparameters of the algorithm to optimize its efficiency with respect to a target distribution. The emerging and routine use of such mathematical tools in science calls for the development of black-box MCMC algorithms that require no hand-tuning at all. This need led to the development of adaptive MCMC methods like the Adaptive Metropolis algorithm \parencite{haario2001adaptive} which tunes its proposal scale based on the sample covariance matrix. Unfortunately, most of those algorithms still include a significant number of hyperparameters (e.g. components of the covariance matrix)  rendering the adaptation noisy. Furthermore, the tuning is usually performed on the basis of prior knowledge, such as one or more long preliminary runs which further slow down the sampling. Last but not least, there is no reason to believe that a single Metropolis proposal scale is optimal for the whole distribution (i.e. the appropriate scale could vary from one part of the distribution to another). Another approach to deal with those issues would be to develop methods that by construction require no or minimal hand-tuning. An archetypal such method is the Slice Sampler \parencite{neal2003slice}, which has only one hyperparameter, the initial length scale.

It should be noted that powerful adaptive methods that require no hand-tuning (although they do require preliminary runs) already exist. Most notable of them is the No U-Turn Sampler (NUTS) \parencite{hoffman2014no}, an adaptive extension of Hamiltonian Monte Carlo (HMC) \parencite{neal2011mcmc}. However, such methods rely on the gradient of the log probability density function. This requirement is the reason why these methods are limited in their application in quantitative fields such as physics, astrophysics and cosmology, which are dominated by computationally costly non--differentiable models. Thus, our objective in this paper is to introduce a parallel, black-box and gradient--free method that can be used in the aforementioned scientific fields.

This paper presents Ensemble Slice Sampling (ESS), an extension of the Standard Slice Sampling method. ESS naturally inherits most of the benefits of Standard Slice Sampling, such as the acceptance rate of $1$, and most importantly the ability to adapt to the characteristics of a target distribution without any hand-tuning at all. Furthermore, we will show that ESS's performance is insensitive to linear correlations between the parameters, thus enabling efficient sampling even in highly demanding scenarios. We will also demonstrate ESS's performance in strongly multimodal target distributions and show that the method samples efficiently even in high dimensions. Finally, the method can easily be implemented in parallel taking advantage of multiple CPUs thus facilitating Bayesian inference in cases of computationally expensive models.

Our implementation of ESS is inspired by \textcite{tran2015reunderstanding}. However, our method improves upon that by extending the direction choices (e.g. Gaussian and global move), adaptively tuning the initial proposal scale, and parallelising the algorithm. \textcite{nishihara2014parallel} developed a general algorithm based on the elliptical slice sampling method~\parencite{murray2010elliptical} and a Gaussian mixture approximation to the target distribution. ESS utilises an ensemble of parallel and interacting chains, called walkers. Other methods that are based on the ensemble paradigm include the Affine Invariant Ensemble Sampler \parencite{goodman2010ensemble} and the Differential Evolution MCMC \parencite{ter2006markov} along with its various extensions \parencite{ter2008differential, vrugt2009accelerating}, as well as more recent approaches that are based on langevin diffusion dynamics \parencite{garbuno2020interacting, garbuno2020affine} and the time discretization of stochastic differential equations \parencite{leimkuhler2018ensemble} in order to achieve substantial speedups.

In Section \ref{sec_ess:slice}, we will briefly discuss the Standard Slice Sampling algorithm. In Section \ref{sec_ess:ensemble}, we will introduce the Ensemble Slice Sampling method. In Section \ref{sec_ess:empirical} we will investigate the empirical evaluation of the algorithm. We reserve Sections \ref{sec_ess:discussion} and \ref{sec_ess:conclusion} for discussion and conclusion, respectively.

\section{Standard Slice Sampling}
\label{sec_ess:slice}
\textit{Slice Sampling} is based on the idea that sampling from a distribution $p(x)$ whose density is proportional to $f(x)$ is equivalent to uniformly sampling from the region underneath the graph of $f(x)$. More formally, in the univariate case, we introduce an auxiliary variable, the height $y$, thus defining the joint distribution $p(x,y)$, which is uniform over the region $U = \{ (x,y) : 0 < y < f(x) \}$. To sample from the marginal density for $x$, $p(x)$, we sample from $p(x,y)$ and then we ignore the $y$ values.

Generating samples from $p(x,y)$ is not trivial, so we might consider defining a Markov chain that will converge to that distribution. The simplest, in principle, way to construct such a Markov chain is via Gibbs sampling. Given the current $x$, we sample $y$ from the conditional distribution of $y$ given $x$, which is uniform over the range $(0, f(x) )$. Then we sample the new $x$ from the \textit{slice} $S=\{ x : y < f(x)\}$. 

Generating a sample from the slice $S$ may still be difficult, since we generally do not know the exact form of $S$. In that case, we can update $x$ based on a procedure that leaves the uniform distribution of $S$ invariant. \textcite{neal2003slice} proposed the following method:
\begin{description}
\item Given the current state $x_{0}$, the next one is generated as:
\begin{enumerate}
    \item Draw $y_{0}$ uniformly from $(0,f(x_{0}))$, thus defining the horizontal slice $S = \{ x : y_{0} < f(x)\}$,
    \item Find an interval $I = (L,R)$ that contains all, or much, of $S$ (e.g. using the stepping-out procedure defined below),
    \item Draw the new point $x_{1}$ uniformly from $I \cap S$.
\end{enumerate}
\end{description}

\begin{figure}
    \centering
    \includegraphics[scale=0.65]{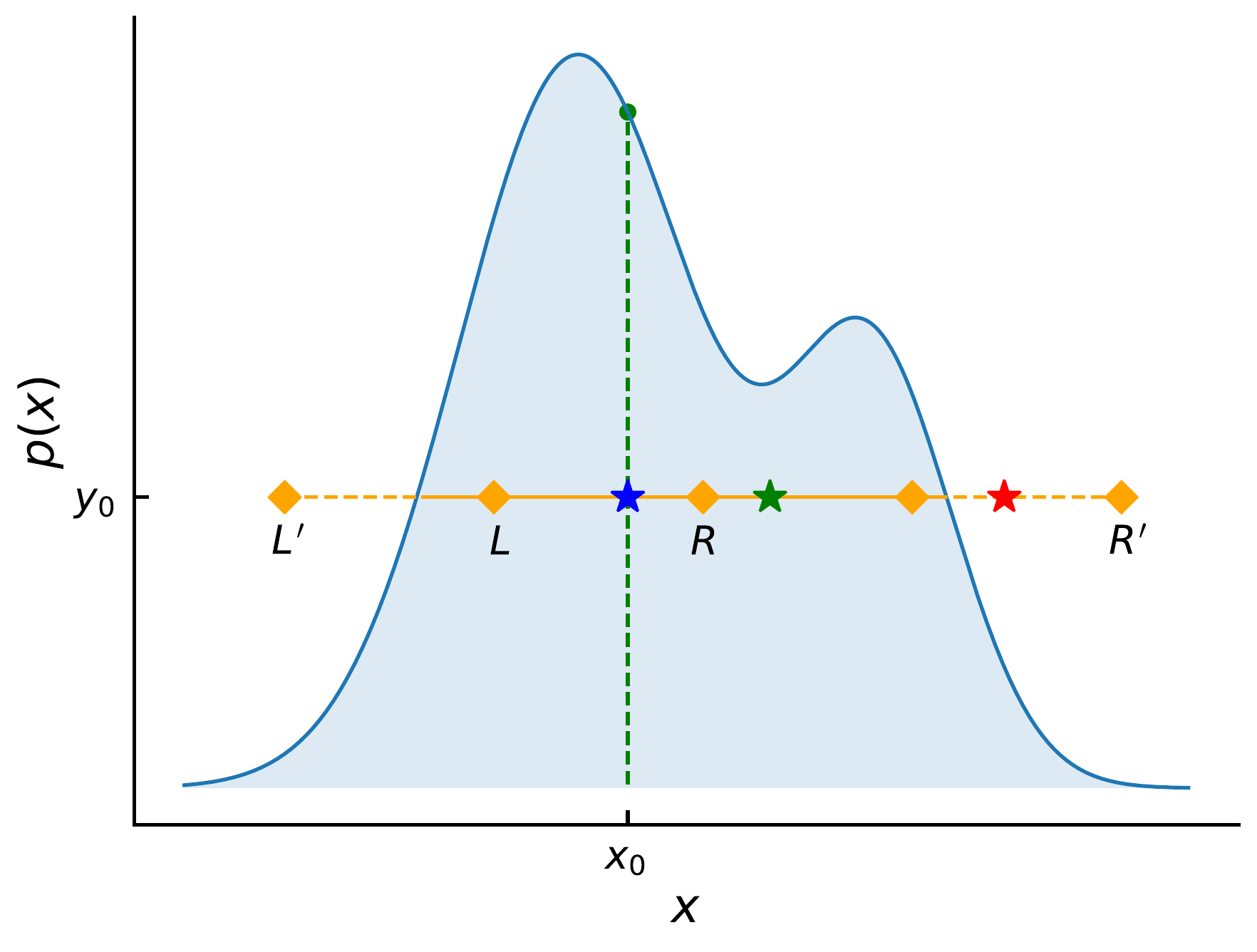}
    \caption{The plot shows the univariate slice sampling method. Given an initial value $x_{0}$, a value $y_{0}$ is uniformly sampled along the vertical slice $(0,f(x_{0}))$ (green dashed line) thus defining the initial point (blue star). An interval $(L,R)$ is randomly positioned horizontally around the initial point, and then it is expanded in steps of size $\mu=R-L$ until both of its ends $L', R'$ are outside the slice. The new point (green star) is generated by repeatedly sampling uniformly from the expanded interval $(L',R')$ until a point is found inside the slice. Points outside the slice (e.g. the red star) are used to shrink the interval $(L',R')$ by moving $L'$ or in this case $R'$ to that point and accelerate the sampling procedure.}
\label{fig_ess:slice}
\end{figure}

In order to find the interval $I$, \textcite{neal2003slice} proposed to use the \emph{stepping-out} procedure that works by randomly positioning an interval of length $\mu$ around the point $x_{0}$ and then expanding it in steps of size $\mu$ until both ends are outside of the slice. The new point $x_{1}$ is found using the \emph{shrinking} procedure, in which points are uniformly sampled from $I$ until a point inside $S$ is found. Points outside $S$ are used to shrink the interval $I$. The stepping-out and shrinking  procedures are illustrated in Figure \ref{fig_ess:slice}. By construction, the stepping-out and shrinking procedures can adaptively tune a poor estimate of the length scale $\mu$ of the initial interval. The length scale $\mu$ is the only free hyperparameter of the algorithm. For a detailed review of the method we direct the reader to \cite{neal2003slice} and \cite{mackay2003information} (also Exercise 30.12 in that text).

It is important to mention here that for multimodal distributions there is no guarantee that the slice would cross any of the other modes, especially if the length scale is underestimated initially. Ideally, in order to provide a large enough initial value of the scale factor $\mu$, prior knowledge of the distance between the modes is required. As we will show in the next section, Ensemble Slice Sampling does not suffer from this complication and can handle strongly multimodal distributions efficiently.

\section{Ensemble Slice Sampling}
\label{sec_ess:ensemble}
The univariate slice sampling scheme can be used to sample from multivariate distributions by sampling repeatedly along each coordinate axis in turn (one parameter at a time) or by sampling along randomly selected directions \parencite{mackay2003information}. Using either of those choices, the Standard Slice Sampler performs acceptably in cases with no strong correlations in parameter space. The overall performance of the algorithm generally depends on the number of expansions and contractions during the stepping-out and shrinking procedures, respectively. Ideally we would like to minimize that number. A reasonable initial estimate of the length scale is still required in order to reduce the amount of time spent expanding or contracting the initial interval. 

However, when strong correlations are present two issues arise. First, there is no single value of the initial length scale that minimizes the computational cost of the stepping-out and shrinking procedures along all directions in parameter space. The second problem concerns the choice of direction. In particular, neither the component-wise choice (one parameter at a time) nor the random choice is suitable in strongly correlated cases. Using such choices results in highly autocorrelated samples.

Our approach would be to target each of those two issues individually. The resulting algorithm, Ensemble Slice Sampling (ESS), is invariant under affine transformations of the parameter space, meaning that its performance is not sensitive to linear correlations. Furthermore, ESS minimizes the computational cost of finding the slice by adaptively tuning the initial length scale. Last but not least, unlike most MCMC methods, ESS is trivially parallelizable, thus enabling the data analyst to take advantage of modern high performance computing facilities with multiple CPUs.

\subsection{Adaptively tuning the length scale}
\label{sec_ess:approx}

Let us first consider the effect of the initial length scale on the performance of the univariate slice sampling method. For instance, if the initial length scale is $\lambda$ times smaller than the actual size of the slice, then the stepping-out procedure would require $\mathcal{O}(\lambda)$ steps in order to fix this. However, in this case, since the final interval is an accurate approximation of the slice there would probably be no contractions during the shrinking phase. On the other hand, when the initial length scale is larger than the actual slice then the number of expansions would be either one or zero. In this case though, there would be a number of contractions.\\

\subsubsection{Stochastic approximation}
As the task is to minimize the total number of expansions and contractions we employ and adapt the \emph{Robbins--Monro} stochastic approximation algorithm \parencite{robbins1951stochastic} of \textcite{tibbits2014automated}. Ideally, based on the reasoning of the previous paragraph, only one expansion and one contraction will take place. Therefore, the target ratio of number of expansions to total number of expansions and contractions is $1/2$. To achieve this, we update the length scale $\mu$ based on the following recursive formula:
\begin{equation}
    \label{eq_ess:approx}
    \mu^{(t+1)} = 2 \mu^{(t)}\frac{N_{e}^{(t)}}{N_{e}^{(t)}+N_{c}^{(t)}}\, ,
\end{equation}
where $N_{e}^{(t)}$ and $N_{c}^{(t)}$ are the number of expansions and contractions during iteration $t$. It is easy to see that when the fraction $N_{e}^{(t)}/(N_{e}^{(t)}+N_{c}^{(t)})$ is larger than $1/2$ the length scale $\mu$ will be increased. In the case where the fraction is smaller than $1/2$ the length scale $\mu$ will be decreased accordingly. The optimization can stop either when the fraction is close to $1/2$ within a threshold or when a maximum number of tuning steps has been completed. The pseudocode for the first case is shown in \Algo{approximate}. In order to preserve detailed balance it is important to be sure that the adaptation stops after a finite number of iterations. In practice this happens after $\mathcal{O}(10)$ iterations. An alternative would be to use diminishing adaptation \parencite{roberts2007coupling} but we found that our method is sufficient in practice (see Section 4.3 for more details).

\begin{algorithm}
\caption{Function to tune the length scale $\mu$.}
    \algolabel{approximate}
\begin{algorithmic}[1]
\STATE{\textbf{function} TuneLengthScale($t$, $\mu^{(t)}$, $N_{e}^{(t)}$,  $N_{c}^{(t)}$, $M^{\text{adapt}}$)}
\IF{$t\leq M^{\text{adapt}}$}
    \STATE{Compute $\mu^{(t+1)}$ using Equation \ref{eq_ess:approx},}
    \STATE{\bf{return} $\mu^{(t+1)}$}
\ELSE
    \STATE{\bf{return} $\mu^{(t)}$}
\ENDIF
\end{algorithmic}
\end{algorithm}

\subsection{The choice of direction and parallelisation}
\label{sec_ess:direction}

In cases where the parameters are correlated we can accelerate mixing by moving more frequently along certain directions in parameter space. One way of achieving this is to exploit some prior knowledge about the covariance of the target distribution. However, such an approach would either require significant hand-tuning or noisy estimations of the sample covariance matrix during an initial run of the sampler. For that reason we employ a different approach to exploit the covariance structure of the target distribution and preserve the hand-tuning-free nature of the algorithm.

\subsubsection{Ensemble of walkers}
Following the example of \textcite{goodman2010ensemble} we define an ensemble of parallel chains, called walkers. In our case though, each walker is an individual slice sampler. The sampling proceeds by moving one walker at a time by slice sampling along a direction defined by a subset of the rest of the walkers of the ensemble. As long as the aforementioned direction does not depend on the position of the current walker, the resulting algorithm preserves the detailed balance of the chain. Moreover, assuming that the distribution of the walkers resembles the correlated target distribution, the chosen direction will prefer directions of correlated parameters. 

We define an ensemble of $N$ parallel walkers as the collection $S = \lbrace \mathbf{X_{1}}, \dots,  \mathbf{X_{N}}\rbrace$. The position of each individual walker $\mathbf{X_{k}}$ is a vector $\mathbf{X_{k}}\in \mathbb{R}^{D}$ and therefore we can think of the ensemble $S$ as being in $\mathbb{R}^{N D}$. Assuming that each walker is drawn independently from the target distribution with density $p$, then the target distribution for the ensemble would be the product
\begin{equation}
    \label{eq_ess:product}
    P(\mathbf{X_{1}}, \dots,  \mathbf{X_{N}}) = \prod_{k=1}^{N}p(\mathbf{X_{k}}) \,.
\end{equation}
The Markov chain of the ensemble would preserve the product density of equation \ref{eq_ess:product} without the individual walker trajectories being Markov. Indeed, the position of $\mathbf{X_{k}}$ at iteration $t+1$ can depend on $\mathbf{X_{j}}$ at iteration $t$ with $j\neq k$.

Given the walker $\mathbf{X_{k}}$ that is to be updated there are arbitrary many ways to define a direction vector from the complementary ensemble $S_{[k]}=\lbrace \mathbf{X_{j}},\: \forall j\neq k\rbrace$. Here we will discuss a few of them. Following the convention in the ensemble MCMC literature we call those recipes of defining direction vectors, \emph{moves}. Although the use of the ensemble might seem equivalent to that of a sample covariance matrix in the Adaptive Metropolis algorithm \parencite{haario2001adaptive} the first has a higher information content as it encodes both linear and non-linear correlations. Indeed, having an ensemble of walkers allows for arbitrary many policies for choosing the appropriate directions along which the walkers move in parameter space. As we will shortly see, one of the choices (i.e. the Gaussian move, introduced later in this Section) is indeed the slice sampling analogue of a covariance matrix. However, other choices (i.e. Differential move or Global move) can take advantage of the non-Gaussian nature of the ensemble distribution and thus propose more informative moves. As will be discussed later in this section, those advanced moves make no assumption of Gaussianity for the target distribution. Furthermore, as we will show in the last part of this section, the ensemble can also be easily parallelised. \\

\begin{algorithm}
\caption{Function to return a differential move direction vector.}
    \algolabel{differential}
\begin{algorithmic}[1]
\STATE{\textbf{function} DifferentialMove($k$, $\mu$, $S$)}
\STATE{Draw two walkers $\mathbf{X_{l}}$, and $\mathbf{X_{m}}$ uniformly and without replacement from the complementary ensemble $S$},
\STATE{Compute direction vector $\bm{\eta}_{k}$ using Equation \ref{eq_ess:diff},}
\STATE{\bf{return} $\bm{\eta}_{k}$}
\end{algorithmic}
\end{algorithm}

\subsubsection{Affine transformations and invariance}
Affine invariance is a property of certain MCMC samplers first introduced in the MCMC literature by \textcite{goodman2010ensemble}. An MCMC algorithm is said to be affine invariant if its performance is invariant under the bijective mapping $g:\mathbb{R}^{D}\rightarrow \mathbb{R}^{D}$ of the form $\mathbf{Y}=A \mathbf{X} + b$ where $A\in \mathbb{R}^{D\times D}$ is a matrix and $b\in\mathbb{R}^{D}$ is a vector. Linear transformations of this form are called affine transformations and describe rotations, rescaling along specific axes as well as translations in parameter space. Assuming that $\mathbf{X}$ has the probability density $p(\mathbf{X})$, then $\mathbf{Y}=A\mathbf{X}+b$ has the probability density
\begin{equation}
    \label{eq_ess:affinedensity}
    p_{A,b}(\mathbf{Y})=p(A\mathbf{X}+b)\propto p(\mathbf{X})\,.
\end{equation}

Given a density $p$ as well as an MCMC transition operator $\mathcal{T}$ such that $\mathbf{X}(t+1) = \mathcal{T} \big(\mathbf{X}(t);p\big)$ for any iteration $t$ we call the operator $\mathcal{T}$ affine invariant if
\begin{equation}
    \label{eq_ess:invariance}
    \mathcal{T}\big(A\mathbf{X}+b;p_{A,b}\big) = A\:\mathcal{T}\big(\mathbf{X};p\big)+b
\end{equation}
for $\forall A\in \mathbb{R}^{D\times D}$ and $\forall b \in \mathbb{R}^{D}$. In case of an ensemble of walkers we define an affine transformation from $\mathbb{R}^{N D}$ to $\mathbb{R}^{N D}$ as
\begin{equation}
    \label{eq_ess:transformensemble}
    S = \lbrace \mathbf{X_{1}}, \dots, \mathbf{X_{N}}\rbrace \xrightarrow{A, b} \lbrace A \mathbf{X_{1}}+b, \dots, A\mathbf{X_{N}}+b\rbrace \,.
\end{equation}

The property of affine invariance is of paramount importance for the development of efficient MCMC methods. As we have discussed already, proposing samples more frequently along certain directions can accelerate sampling by moving further away in parameter space. Given that most realistic applications are highly skewed or anisotropic and are characterised by some degree of correlation between their parameters, affine invariant methods are an obvious choice of a tool that can be used in order to achieve high levels of efficiency.

\subsubsection{Differential move}
\begin{figure}[t!]
    \centering
    \includegraphics[scale=0.55]{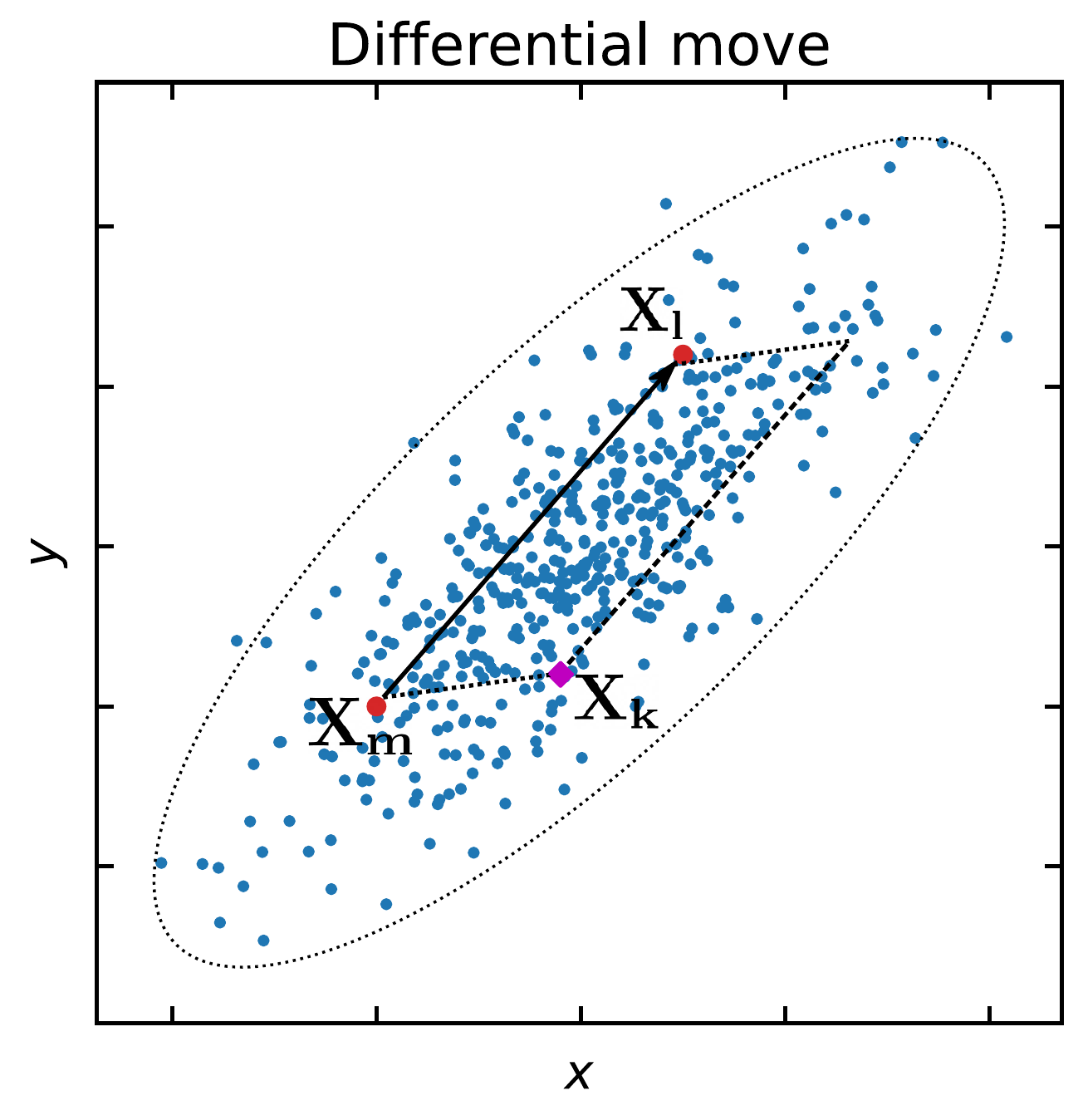}
    \caption{The plot shows the differential direction move. Two walkers (red) are uniformly sampled from the complementary ensemble (blue). Their positions define the direction vector (solid black). The selected walker (magenta) then moves by Slice Sampling along the parallel direction (dashed black).}
\label{fig_ess:diff}
\end{figure}

The differential direction choice works by moving the walker $\mathbf{X}_{k}$ based on two randomly chosen walkers $\mathbf{X}_{l}$ and $\mathbf{X}_{m}$ of the complementary ensemble $S_{[k]}=\lbrace \mathbf{X_{j}},\: \forall j\neq k\rbrace$ \parencite{gilks1994adaptive}, see Figure \ref{fig_ess:diff} for a graphical explanation. In particular, we move the walker $\mathbf{X}_{k}$ by slice sampling along the vector $\bm{\eta}_{k}$ defined by the difference between the walkers $\mathbf{X}_{l}$ and $\mathbf{X}_{m}$. It is important to notice here that the vector $\bm{\eta}_{k}$ is not a unit vector and thus carries information about both the length scale and the optimal direction of movement. It will also prove to be more intuitive to include the initial length scale $\mu$ in the definition of the direction vector in the following way:
\begin{equation}
    \label{eq_ess:diff}
    \bm{\eta}_{k}= \mu \big( \mathbf{X}_{l}-\mathbf{X}_{m}\big)\, .
\end{equation}

The pseudocode for a function that, given the value of $\mu$ and the complementary ensemble $S$, returns a differential direction vector $\bm{\eta}_{k}$ is shown in \Algo{differential}. Furthermore, the Differential move is clearly affine invariant. Assuming that the distribution of the ensemble of walkers follows the target distribution and the latter is highly elongated or stretched along a certain direction then the proposed direction given by equation \ref{eq_ess:diff} will share the same directional asymmetry.   \\

\subsubsection{Gaussian move}
The direction vector $\bm{\eta}_{k}$ can also be drawn from a normal distribution with the zero mean and the covariance matrix equal to the sample covariance of the complementary ensemble $S_{[k]}$,
\begin{equation}
    \label{eq_ess:cov}
    \mathbf{C}_{S}=\frac{1}{|S|}\sum_{j\in S}\big(\mathbf{X}_{j}-\bar{\mathbf{X}}_{S}\big)\big(\mathbf{X}_{j}-\bar{\mathbf{X}}_{S}\big)^{t}\, .
\end{equation}
We chose to include the initial length scale $\mu$ in this definition as well:
\begin{equation}
    \label{eq_ess:gaussian}
    \frac{\bm{\eta}_{k}}{2\mu} \sim \mathcal{N}\big(\mathbf{0},\mathbf{C}_{S} \big)\, .
\end{equation}
The factor of $2$ is used so that the magnitude of the direction vectors are consistent with those sampled using the differential direction choice in the case of Gaussian-distributed walkers.

\begin{figure}[t!]
    \centering
    \includegraphics[scale=0.55]{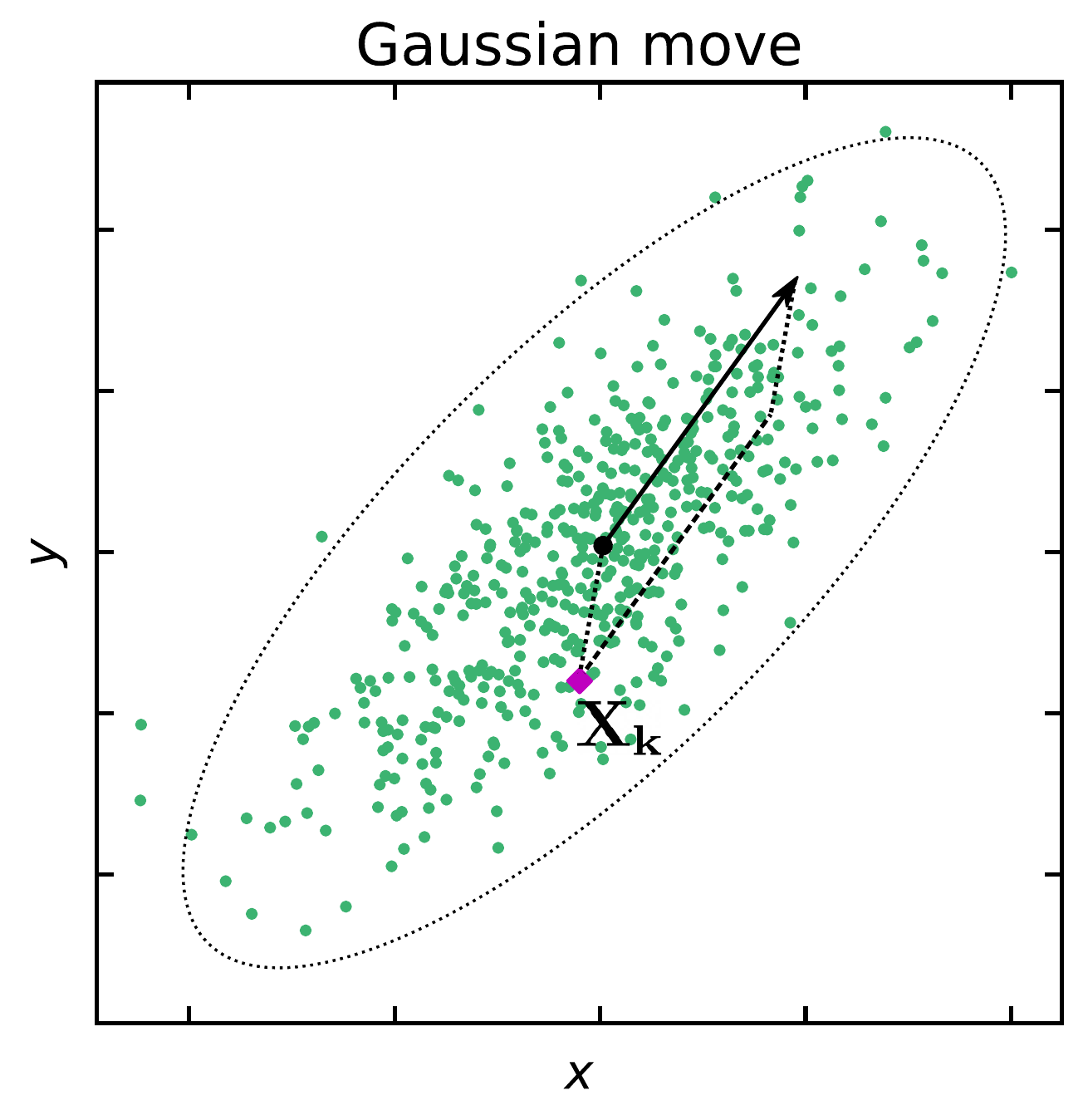}
    \caption{The plot shows the Gaussian direction move. A direction vector (solid black) is sampled from the Gaussian-approximated distribution of the walkers of the complementary ensemble (green). The selected walker (magenta) then moves by Slice Sampling along the parallel direction (dashed black).}
\label{fig_ess:gauss}
\end{figure}

The pseudocode for a function that, given the value of $\mu$ and the complementary ensemble $S$, returns a Gaussian direction vector $\bm{\eta}_{k}$ is shown in \Algo{gaussian}. See Figure \ref{fig_ess:gauss} for a graphical explanation of the method. Moreover, just like the Differential move, the Gaussian move is also affine invariant. In the limit in which the number of walkers is very large and the target distribution is normal, the first reduces to the second. Alternatively, assuming that the distribution of walkers follows the target distribution then the covariance matrix of the ensemble would be the same as that of independently drawn samples from the target density. Therefore any anisotropy characterising the target density would also be present in the distribution of proposed directions given by equation \ref{eq_ess:gaussian}.
\begin{algorithm}
\caption{Function to return a Gaussian Move direction vector.}
    \algolabel{gaussian}
\begin{algorithmic}[1]
\STATE{\textbf{function} GaussianMove($k$, $\mu$, $S$)}
\STATE{Estimate sample covariance $\mathbf{C}_{S}$ of the walkers in the complementary ensemble $S$ using Equation \ref{eq_ess:cov},}
\STATE{Sample $\bm{\eta}_{k}/(2\mu)\sim \mathcal{N}\big(\mathbf{0},\mathbf{C}_{S} \big)$,}
\STATE{\bf{return} $\bm{\eta}_{k}$}
\end{algorithmic}
\end{algorithm}

\subsubsection{Global move}
ESS and its variations described so far (i.e. differential move, Gaussian move) have as much difficulty traversing the low probability regions between modes/peaks in multimodal distributions as most local MCMC methods (e.g. Metropolis, Hamiltonian Monte Carlo, Slice Sampling, etc.). Indeed, multimodal distributions are often the most challenging cases to sample from. Fortunately, Ensemble Slice Sampling's flexibility allows to construct advanced moves which are specifically designed to handle multimodal cases even in moderate to high dimensional parameter spaces. The \emph{global move} is such an example.

We first fit a \emph{Gaussian Mixture} to the distribution of the walkers of the complementary ensemble $S_{[k]}$ using \emph{Variational Inference}. To avoid defining the number of components of the Gaussian Mixture we use a \emph{Dirichlet process} as the prior distribution for the Gaussian Mixture weights\footnote{To this end we use the Scikit-Learn implementation of the Dirichlet process Gaussian mixture.} \parencite{gorur2010dirichlet}. The exact details of the construction of the Dirchlet process Gaussian mixture (DPGM) are beyond the scope of this work and we direct the reader to \textcite{gorur2010dirichlet} and \textcite{bishop2006pattern} for more details. One of the major benefits of fitting the DPGM using variational inference compared to the expectation--maximisation (EM) algorithm~\parencite{dempster1977maximum} that is often used is the improved stability. In particular, the use of priors in the variational Bayesian treatment guarantees that Gaussian components do not collapse into specific data points. This regularisation due to the priors leads to component covariance matrices that do not diverge even when the number of data points (i.e. walkers in our case) in a component is lower than the number of dimensions. In our case, this  means that even if the number of walkers located in a mode of the target distribution is small DPGM would still identify that mode correctly. In such cases, the covariance of the  component that corresponds to that mode would be over--estimated. This however does not affect the performance of the Global move as the latter does not rely on exact estimates of the component covariance matrices.\footnote{Indeed the covariance matrix of a component only enters through equation \ref{eq_ess:globalB} but then it is re--scaled by the factor $\gamma$.}

In practice, we recommend using more than the minimum number of walkers in cases of multimodal distributions (e.g. at least two times as many in bimodal cases). We found that the computational overhead introduced by the variational fitting of the DPGM is negligible compared to the computational cost of the evaluation of the model and posterior distribution in common problems in physics, astrophysics and cosmology. Indeed the cost is comparable, and only a few times higher than the Differential or Gaussian move. The reason for that is the relatively small number of walkers (i.e. $\mathcal{O}(10-10^3)$) that simplifies the fitting procedure.

Once fitting is done, we have a list of the means and covariance matrices of the components of the Gaussian Mixture. As the ensemble of walkers traces the structure of the target distribution, we can use the knowledge of the means and covariance matrices of the Gaussian Mixture to construct efficient direction vectors. Ideally, we prefer direction vectors that connect different modes. This way, the walkers will be encouraged to move along those directions that would otherwise be very unlikely to be chosen.

We uniformly select two walkers of the complementary ensemble and identify the Gaussian components to which they belong, say $i$ and $j$. There are two distinct cases and we will treat them as such. In case A, $i = j$, meaning that the selected walkers originate from the same component. In case B, $i \neq j$, meaning that the two walkers belong to different components and thus probably different peaks of the target distribution. 

As we will show next, only in case B, we can define a direction vector that favors mode-jumping behaviour. In case A, we can sample a direction vector from the Gaussian component that the two select walkers belong to\footnote{In practice we use uniformly sample two walkers from the list of walkers that DPGM identified in that mode. This step removes any dependency on covariance matrix estimates.}:
\begin{equation}
    \label{eq_ess:globalA}
    \frac{\bm{\eta}_{k}}{2\mu} \sim \mathcal{N}\big( \bm{0}, \bm{C}_{i=j} \big)\, ,
\end{equation}
where $\bm{C}_{i=j}$ is the covariance matrix of the i$_{\rm th}$ (or equivalently j$_{\rm th}$) component. Just as in the Gaussian move, the mean of the proposal distribution is zero so that we can interpret $\bm{\eta}$ as a direction vector.

In case B, where the two selected walkers belong to different components, $i \neq j$, we will follow a different procedure to facilitate long jumps in parameter space. We will sample two vectors, one from each component:
\begin{equation}
    \label{eq_ess:globalB}
    \bm{\eta}_{k, n} \sim \mathcal{N}\big( \bm{\mu}_{n}, \gamma \bm{C}_{n} \big)\, ,
\end{equation}
for $n=i$ or $n=j$. Here, $\bm{\mu}_{n}$ is the mean of the nth component and $\bm{C}_{n}$ is its covariance matrix. In practice, we also re-scale the covariance by a factor of $\gamma = 0.001$, which results in direction vectors with lower variance in their orientation. $\gamma < 1$ ensures that the chosen direction vector is close to the vector connecting the two peaks of the distribution. Finally, the direction vector will be defined as:
\begin{equation}
    \label{eq_ess:global}
    \bm{\eta}_{k} = 2 \big( \bm{\eta}_{k,i} - \bm{\eta}_{k,j} \big)\, .
\end{equation}
The factor of $2$ here is chosen to better facilitate mode-jumping. There is also no factor of $\mu$ in the aforementioned expression since in this case there is no need for the scale factor to be tuned.

\begin{figure}[t!]
    \centering
    \includegraphics[scale=0.55]{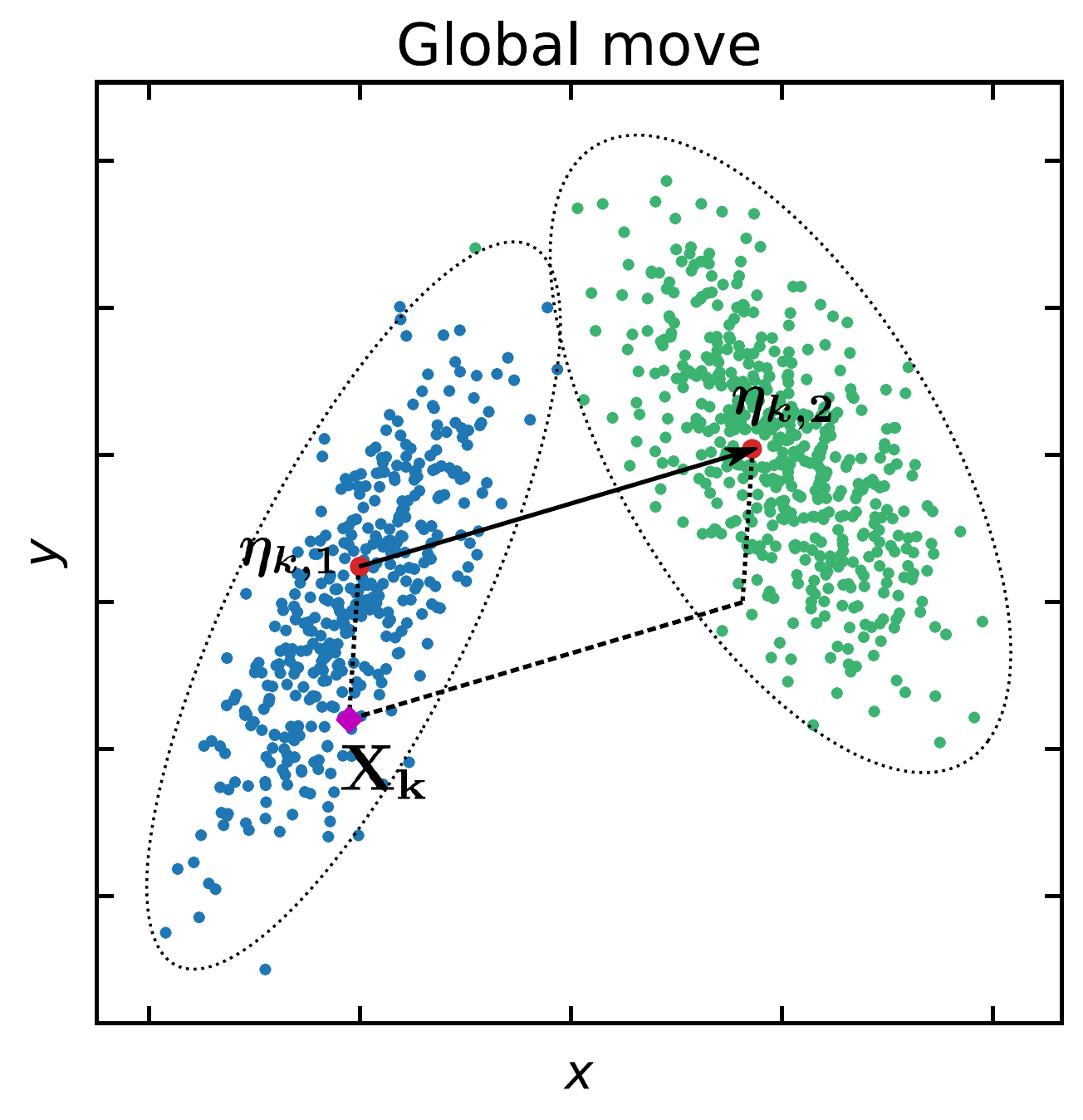}
    \caption{The plot shows the global direction move assuming that the uniformly selected pair of walkers of the complementary ensemble belongs to different components (blue and green). A position (red) is sampled from each component (using the re-scaled by $\gamma$ covariance matrix). Those two points (red) define the direction vector (black) connecting the two modes (blue and green). The selected walker (magenta) then moves by slice sampling along the parallel direction (dashed).}
\label{fig_ess:global}
\end{figure}

The pseudocode for a function that, given the complementary ensemble $S$, returns a Global direction vector $\bm{\eta}_{k}$ is shown in \Algo{global}. See Figure \ref{fig_ess:global} for a graphical explanation of the method. It should be noted that for the global move to work at least one walker needs to be present on each well separated mode.
\begin{algorithm}
\caption{Function to return a global move direction vector.}
    \algolabel{global}
\begin{algorithmic}[1]
\STATE{\textbf{function} GlobalMove($k$, $\mu$, $S$)}
\STATE{Fit Dirichlet process Gaussian mixture (DPGM) to the complementary ensemble $S_{[k]}$,}
\STATE{If $N$ is the number of components of the DPGM then select two components $i, j$ uniformly such that $i \neq j$,}
\IF{$i = j$}
    \STATE{Sample $\bm{\eta}_{k}/(2\mu) \sim \mathcal{N}\big( \bm{0}, \bm{C}_{i=j} \big)$,}
\ELSE
    \STATE{Sample $\bm{\eta}_{k, n} \sim \mathcal{N}\big( \bm{\mu}_{n}, \gamma \bm{C}_{n} \big)$ for $n=i, j$,}
    \STATE{Compute direction vector $\bm{\eta}_{k}$ using Equation \ref{eq_ess:global},}
\ENDIF
\STATE{\bf{return} $\bm{\eta}_{k}$}
\end{algorithmic}
\end{algorithm}

Here we introduced three general and distinct moves that can be used in a broad range of cases. In general, the global move requires a higher number of walkers than the differential or Gaussian move in order to perform well. We found that the differential and Gaussian moves are good choices for most target distributions whereas the global move is only necessary in highly dimensional and multimodal cases. One can use the information in the complementary ensemble to construct more moves tailor-made for specific problems. Such additional moves might include Kernel Density Estimation or Clustering methods and as long as the information used comes from the complementary ensemble (and not from the walker that would be updated) the detailed balance is preserved.

\subsubsection{Parallelizing the ensemble}
Instead of evolving the ensemble by moving each walker in turn we can do this in parallel. A naive implementation of this would result in a subtle violation of detailed balance. We can avoid this by splitting the ensemble into two sets of walkers \parencite{foreman2013emcee} of $n_{\text{Walkers}} / 2$ each. We can now update the positions of all the walkers in the one set in parallel along directions defined by the walkers of the other set (the complementary ensemble). Then we can perform the same procedure for the other set. In accordance with equation \ref{eq_ess:product}, the stationary distribution of the split ensemble would be
\begin{equation}
    \label{eq_ess:split}
    P(\mathbf{X_{1}},\dots,\mathbf{X_{N}}) = \prod_{k=1}^{N/2}p(\mathbf{X_{k}}) \prod_{k=1+N/2}^{N}p(\mathbf{X_{k}})\,.
\end{equation}
The method generates samples from the target distribution by simulating a Markov chain which leaves this product distribution invariant. The transition operator $\mathcal{T}_{1}$ that updates the walkers of the first set (i.e. $k=1,\dots,N/2$) uses the walkers of the complementary ensemble (i.e. $k=1+N/2,\dots,N$) and vice versa for the transition operator $\mathcal{T}_{2}$ that acts on the second set. In the context of ESS the aforementioned transition operators correspond to a single iteration of \Algo{final} coupled with one of the moves (e.g. Differential move).

It follows from the ensemble splitting technique that the maximum number of CPUs used without any of them being idle is equal to the total number of walkers updated concurrently, that is $n_{\text{Walkers}} / 2$.  We will also verify this empirically in Section \ref{sec_ess:empirical}. Of course, this does not mean that if there are more CPUs available they cannot be used as we can always increase the size of the ensemble to match the available CPUs.

Combining this technique with the stochastic approximation solution of Subsection \ref{sec_ess:approx} and the choices (moves) of direction and ensemble-splitting technique of this subsection leads to the Ensemble Slice Sampling method of \Algo{final}\footnote{Perhaps a small detail, but we have included the length scale in the definition of the direction vector $\mathbf{\eta}$ and therefore it does not appear in the definition of the $(L, R)$ interval.}. Of course, another move (e.g. Gaussian, global) can be used instead of the differential move in \Algo{final}. Finally, the minimum number of walkers used should be twice the number of parameters. Using fewer walkers than that could lead to erroneous sampling from a lower dimensional parameter space \parencite{ter2006markov}.

In general, parallelizing a slice sampler is not trivial (e.g. as it is for Metropolis) because each update requires an unknown number of probability density evaluations. However, because of the affine invariance (i.e. performance unaffected by linear correlations) induced by the existence of the ensemble, all iterations require on average the same number of probability density evaluations (i.e. usually $5$ if the stochastic approximation for the length scale $\mu$ is used). Therefore, the parallelization of Ensemble Slice Sampling is very effective in practice. Furthermore, the benefit of having parallel walkers instead of parallel independent chains (e.g. such as in Metropolis sampling) is clear, the walkers share information about the covariance structure of the distribution thus accelerating mixing.

\begin{algorithm}[ht!]
\caption{Single Iteration $t$ of Ensemble Slice Sampling.}
    \algolabel{final}
\begin{algorithmic}[1]
\STATE{Given $t$, $f$, $\mu^{(t)}$, $S_{[0]}$, $S_{[1]}$, $M^{\rm adapt}$:}
\STATE{Initialise $N_{e}^{(t)} = 0$ and $N_{c}^{(t)} = 0$},
\FOR{$i=0, 1$}
    \FOR{$k=1, ..., N/2$}
        \STATE{$k \leftarrow k + i N/2$}
        \STATE{Compute direction vector $\bm{\eta}_{k}\leftarrow$ DifferentialMove($k$, $\mu^{(t)}$, $S_{[i]}$)}
        \STATE{Sample $Y \sim \text{Uniform}(0,f(\mathbf{X_{k}}^{(t)}))$}
        \STATE{Sample $U \sim \text{Uniform}(0,1)$}
        \STATE{Set $L \leftarrow - U$, and $R \leftarrow L + 1$}
        \WHILE{$Y < f(L)$}
            \STATE{$L \leftarrow L - 1$}
            \STATE{$N_{e}^{(t)} \leftarrow N_{e}^{(t)} + 1$}
        \ENDWHILE
        \WHILE{$Y < f(R)$}
            \STATE{$R \leftarrow R + 1$}
            \STATE{$N_{e}^{(t)} \leftarrow N_{e}^{(t)} + 1$}
        \ENDWHILE
        \WHILE{True}
            \STATE{Sample $X' \sim \text{Uniform}(L,R)$}
            \STATE{Set $Y' \leftarrow f(X'\bm{\eta}_{k} + \mathbf{X_{k}}^{(t)})$}
            \IF{$Y<Y'$}
                \STATE{\bf{break}}
            \ENDIF
            \IF{$X'<0$}
                \STATE{$L \leftarrow X'$}
                \STATE{$N_{c}^{(t)} \leftarrow N_{c}^{(t)} + 1$}
            \ELSE
                \STATE{$R \leftarrow X'$}
                \STATE{$N_{c}^{(t)} \leftarrow N_{c}^{(t)} + 1$}
            \ENDIF
        \ENDWHILE
        \STATE{Set $\mathbf{X_{k}}^{(t+1)} \leftarrow X' \bm{\eta}_{k} + \mathbf{X_{k}}^{(t)}$}
    \ENDFOR
\ENDFOR

\STATE{$\mu^{(t+1)} \leftarrow$ TuneLengthScale($t$, $\mu^{(t)}$, $N_{e}^{(t)}$,  $N_{c}^{(t)}$, $M^{adapt}$),}
\end{algorithmic}
\end{algorithm}

\section{Empirical evaluation}
\label{sec_ess:empirical}

To empirically evaluate the sampling performance of the Ensemble Slice Sampling algorithm we perform a series of tests. In particular, we compare its ability to sample from two demanding target distributions, namely the \emph{autoregressive process of order 1} and the \emph{correlated funnel}, against the Metropolis and Standard Slice Sampling algorithms. The Metropolis' proposal scale was tuned to achieve the optimal acceptance rate, whereas the initial length scale of Standard Slice Sampling was tuned using the stochastic scheme of \Algo{approximate}. Ensemble Slice Sampling significantly outperforms both of them. These tests help establish the characteristics and advantages of Ensemble Slice Sampling. Since our objective was to develop a gradient-free black-box method we then proceed to compare Ensemble Slice Sampling with a list of gradient-free ensemble methods such as \emph{Affine Invariant Ensemble Sampling} (AIES), \emph{Differential Evolution Markov Chain} (DEMC) and \emph{Kernel Density Estimate Metropolis} (KM) on a variety of challenging target distributions. Moreover, we are also interested in assessing the convergence rate of the length scale $\mu$ during the first iterations as well as the parallel scaling of the method in the presence of multiple CPUs. Unless otherwise specified we use the differential move for the tests. Unlike ESS that has an acceptance rate of $1$, AIES's and DEMC's acceptance rate is related to the number of walkers. For that reason, and for the sake of a fair comparison, we made sure the selected number of walkers in all examples would yield the optimal acceptance rate for AIES and DEMC. As we will discuss further in Section \ref{sec_ess:discussion} it makes sense to increase the number of walkers in cases of multimodal distributions or strong non-linear correlations. In general though, we recommend using the minimum number of walkers (i.e. twice the number of dimensions) as the default choice and increase it only if it is required by a specific application. For more rules and heuristics about the initialisation and number of walkers we direct the interested reader to Section \ref{sec_ess:discussion}.

\subsection{Performance tests}

\subsubsection{Autoregressive process of order 1}
In order to investigate the performance of ESS. in high dimensional and correlated scenarios we chose a highly correlated Gaussian as the target distribution. More specifically, the target density is a discrete-time \emph{autoregressive process of order 1}, also known as AR(1). This particular target density is ideally suited for benchmarking MCMC algorithms since the posterior density in many scientific studies often approximates a correlated Gaussian. Apart from that, the AR(1) is commonly used as a prior for time-series analysis.

The AR(1) distribution of a random vector $\bm{X}=(X_{1},...,X_{N})$ is defined recursively as follows:
\begin{equation}
    \label{eq_ess:ar1}
    \begin{split}
        X_{1} \sim &\;\mathcal{N}(0,1)\, , \\ 
        X_{2}|X_{1} \sim &\;\mathcal{N}(\alpha X_{1},\beta^{2})\, , \\
        &\vdots \\
        X_{N}|X_{N-1} \sim &\;\mathcal{N}(\alpha X_{N-1},\beta^{2})\, ,
    \end{split}
\end{equation}
where the parameter $\alpha$ controls the degree of correlation between parameters and we chose it to be $\alpha = 0.95$. We set $\beta = \sqrt{1-\alpha^{2}}$ so that the marginal distribution of all parameters is $\mathcal{N}(0,1)$. We also set the number of dimensions to $N=50$. 

\begin{figure*}[t!]
    \centering
    \includegraphics[width=.3\textwidth]{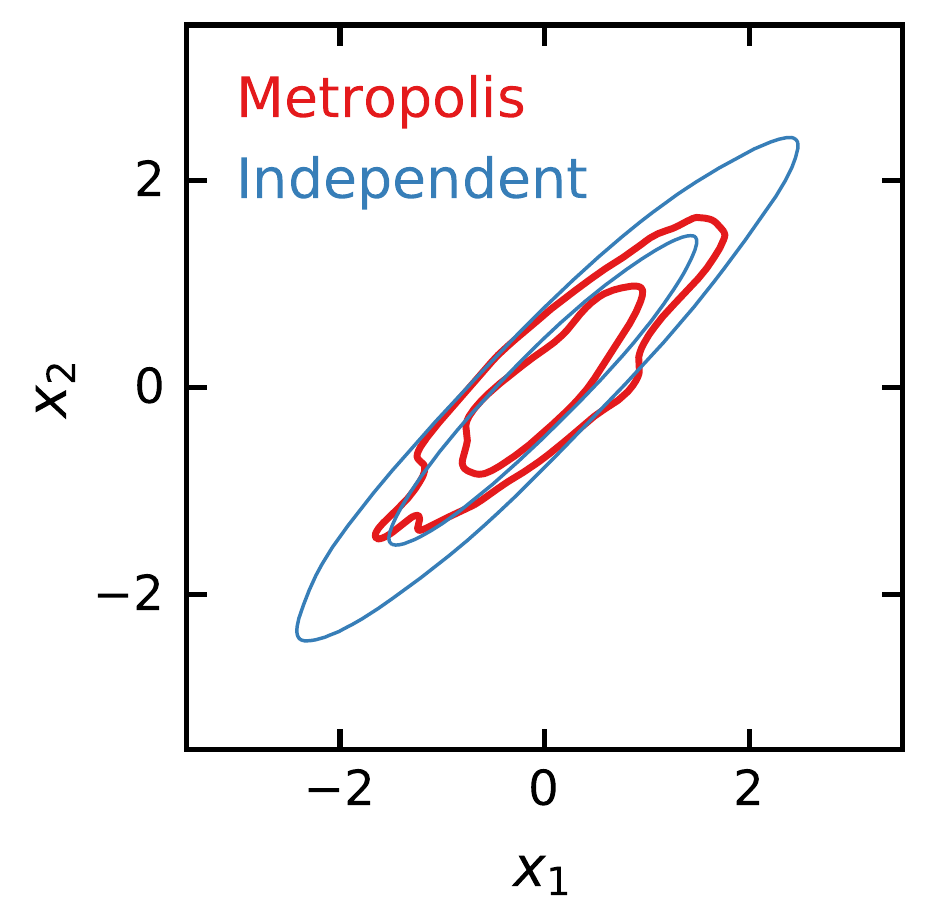}
    \includegraphics[width=.3\textwidth]{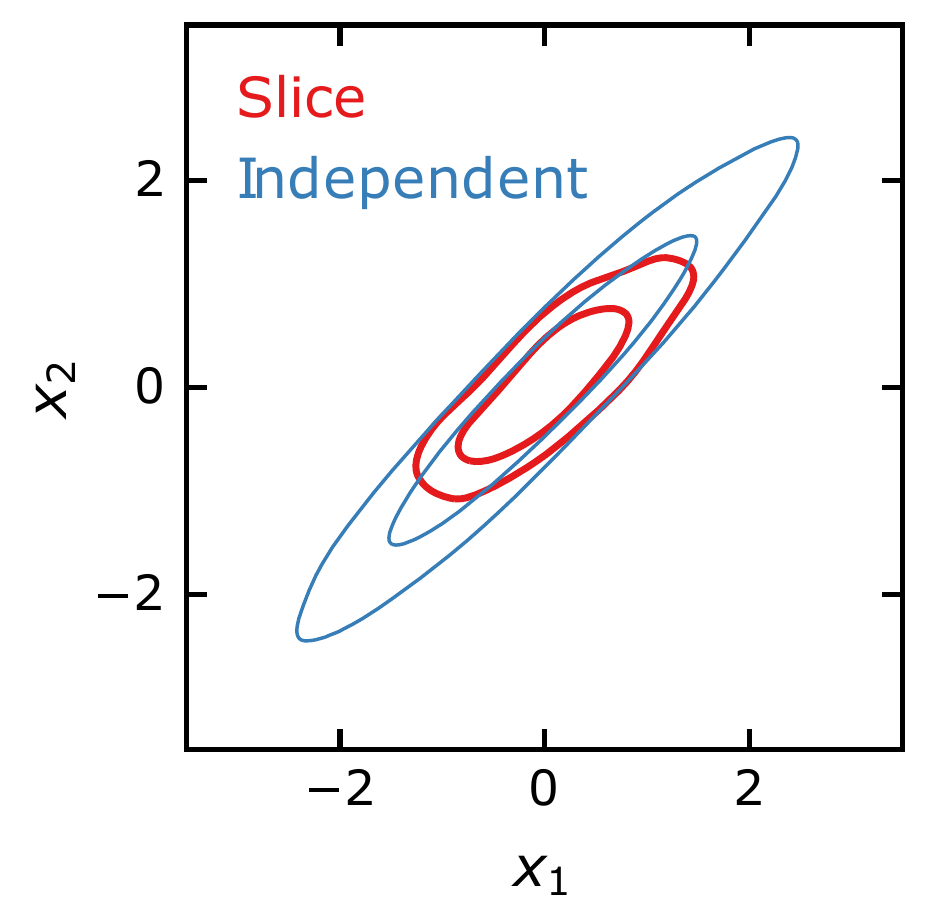}
    \includegraphics[width=.3\textwidth]{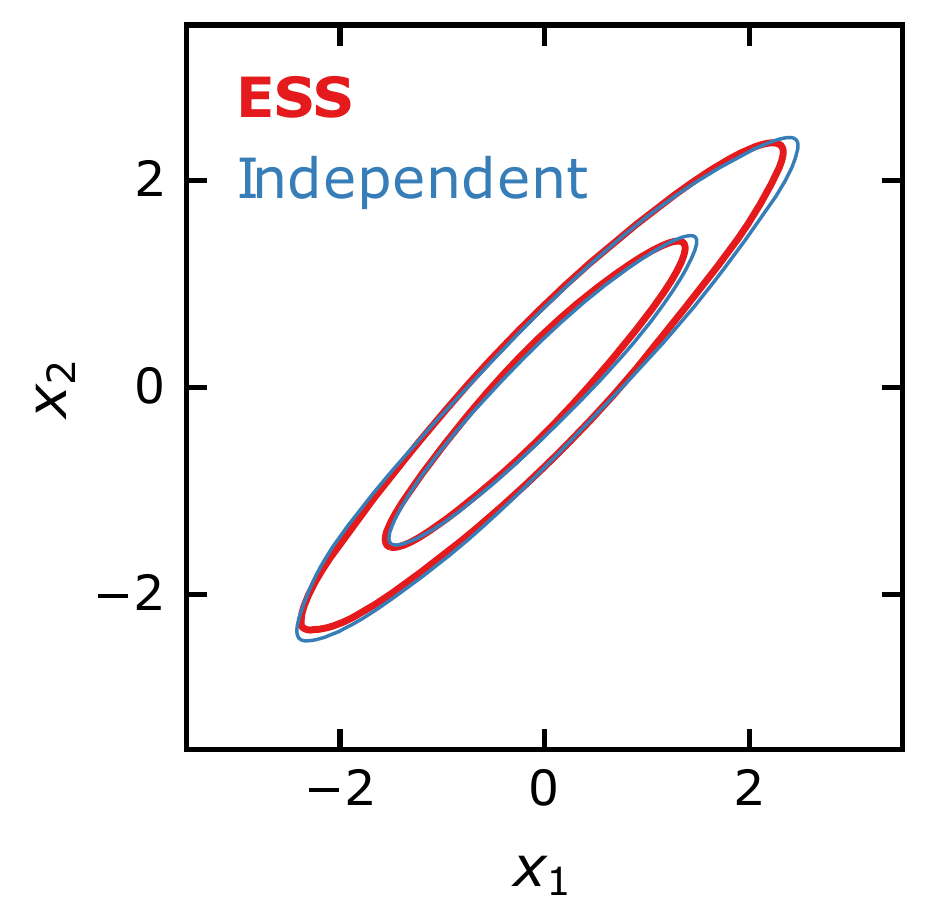}
\caption{The plots compare the 1-sigma and 2-sigma contours generated by the optimised random-walk Metropolis (left), Standard Slice (centre) and Ensemble Slice Sampling (right) methods to those obtained by Independent Sampling (blue) for the AR(1) distribution. All samplers used the same number of probability density evaluations, $3\times 10^{5}$. Only the first two dimensions are shown here.}
\label{fig_ess:ar1}
\end{figure*}

\begin{table}[ht!]
    \centering
    \caption{The table shows a comparison of the optimally tuned Metropolis, Standard Slice, and Ensemble Slice Sampling with the differential move (ESS-D) and the Gaussian move (ESS-G) respectively in terms of the integrated autocorrelation time (IAT) and the number of effective samples per evaluation of the probability density (efficiency) multiplied by $10^4$. These metrics are formally defined in Appendix \ref{app:ess}. The target distributions are the 50--dimensional autoregressive process of order 1 and the 25--dimensional correlated funnel distribution. The total number of iterations was set to $10^{7}$.}
    \def\arraystretch{1.1}
    \begin{tabular}{lccccc}
        \toprule[0.75pt]
         & Metropolis   & Slice   & \textbf{ESS-D} & \textbf{ESS-G} \\
        \midrule[0.5pt]
        \multicolumn{4}{l}{Autoregressive process of order 1} \\
        \midrule[0.5pt]
        IAT          &    4341    &    2075    &   $\mathbf{111}$ &   $\mathbf{107}$   \\
        efficiency   &    2.3    &    1.0    & $\mathbf{17.5}$ & $\mathbf{17.8}$  \\
        \midrule[0.5pt]
        \multicolumn{4}{l}{Correlated funnel distribution} \\
        \midrule[0.5pt]
        IAT          &    -    &    3905    &   $\mathbf{129}$ &   $\mathbf{141}$   \\
        efficiency   &    -    &    0.5    &  $\mathbf{15.3}$ &  $\mathbf{14.0}$ \\
        \bottomrule[0.75pt]
        \end{tabular}
    \label{tab_ess:table1}
\end{table}

For each method, we measured the mean \emph{integrated autocorrelation time} (IAT), and the number of effective samples per evaluation of the probability density function, also termed \emph{efficiency} (see Appendix \ref{app:ess} for details). For this test we ran the samplers for $10^{7}$ iterations. In this example we used the minimum number of walkers (i.e. 100 walkers) for ESS and the equivalent number of probability evaluations for Metropolis and Slice Sampling with each walker initialised at a position sampled from the distribution $\mathcal{N}(0,1)$. The results are presented in Table \ref{tab_ess:table1}. The chain produced by Ensemble Slice Sampling has a significantly shorter IAT ($20-40$ times) compared to either of the other two methods. Furthermore, Ensemble Slice Sampling, with either Differential or Gaussian move, generates an order of magnitude greater number of independent samples per evaluation of the probability density. In this example the Differential and Gaussian moves have achieved almost identical IAT values and efficiencies.

To assess the mixing rate of Ensemble Slice Sampling, we set the maximum number of probability density evaluations to $3\times 10^{5}$ and show the results in Figure \ref{fig_ess:ar1}. We compare the results of Ensemble Slice Sampling with those obtained via the optimally tuned Metropolis and Standard Slice Sampling methods. Ensemble Slice Sampling significantly outperforms both of them, being the only one with a chain resembling the target distribution in the chosen number of probability evaluations.\\

\subsubsection{Correlated funnel}
The second test involves a more challenging distribution, namely the correlated funnel distribution adapted from \textcite{neal2003slice}. The funnel, tornado like, structure is common in Bayesian hierarchical models and possesses characteristics that render it a particularly difficult case. The main difficulty originates from the fact that there is a region of the parameter space where the volume of the region is low but the probability density is high, and another region where the opposite holds. 

Suppose we want to sample an N--dimensional vector $\bm{X}=(X_{1},...,X_{N})$ from the correlated funnel distribution. The marginal distribution of $X_{1}$ is Gaussian with mean zero and unit variance. Conditional on a value of $X_{1}$, the vector $\bm{X}_{2-N}=(X_{2},...,X_{N})$ is drawn from a Gaussian with mean zero and a covariance matrix in which the diagonal elements are $\exp(X_{1})$, and the non-diagonal equal to $\gamma\exp(X_{1})$. If $\gamma=0$, the parameters $X_{2}$ to $X_{N}$ conditional on $X_{1}$ are independent and the funnel distribution resembles the one proposed by \textcite{neal2003slice}. The value of $\gamma$ controls the degree of correlation between those parameters. When $\gamma = 0$ the parameters are uncorrelated. For the following test we chose this to be $\gamma = 0.95$.  We set the number of parameters $N$ to $25$.

Using $10^{7}$ iterations, we estimated the IAT and the efficiency of the algorithms for this distribution as shown in Table \ref{tab_ess:table1}. Just like in the AR(1) case we used the minimum number (i.e. 50) of walkers for ESS with each walker initialised at a position sampled from the distribution $\mathcal{N}(0,1)$. Since the optimally-tuned Metropolis fails to sample from this particular distribution, we do not quote any results. The Metropolis sampler is unable to successfully explore the region of parameter space with negative $X_{1}$ values. The presence of strong correlations renders the Ensemble Slice Sampler $30$ times more efficient than the Standard Slice Sampling algorithm on this particular example. In this example, the Differential move outperforms the Gaussian move in terms of efficiency, albeit by a small margin. In general, we expect the former to be more flexible than the latter since it makes no assumption about the Gaussianity of the target-distribution and recommend it as the default configuration of the algorithm.

\begin{figure*}[t!]
    \centering
    \includegraphics[width=.3\textwidth]{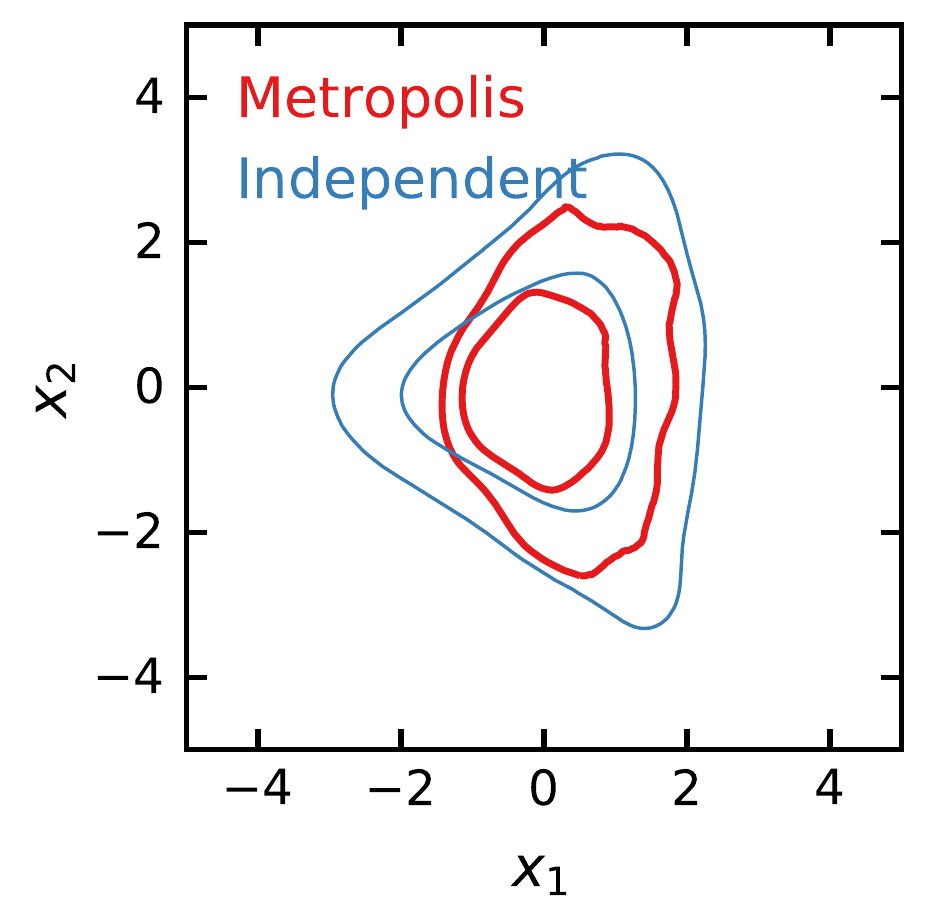}
    \includegraphics[width=.3\textwidth]{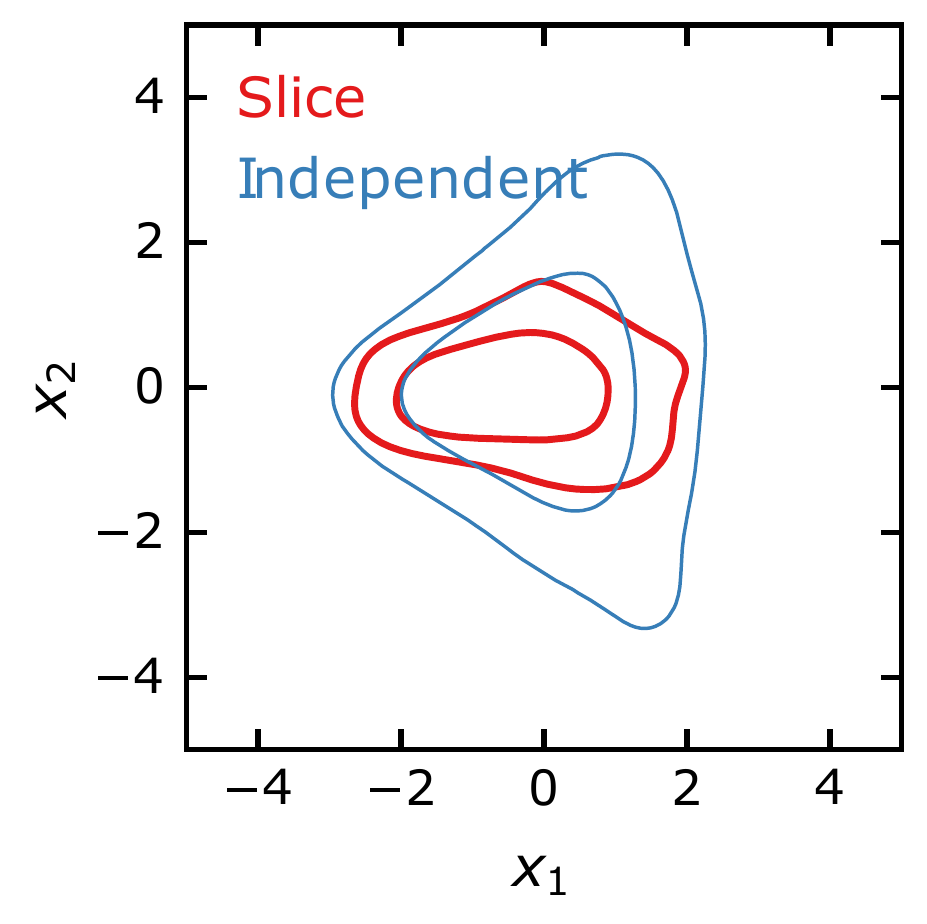}
    \includegraphics[width=.3\textwidth]{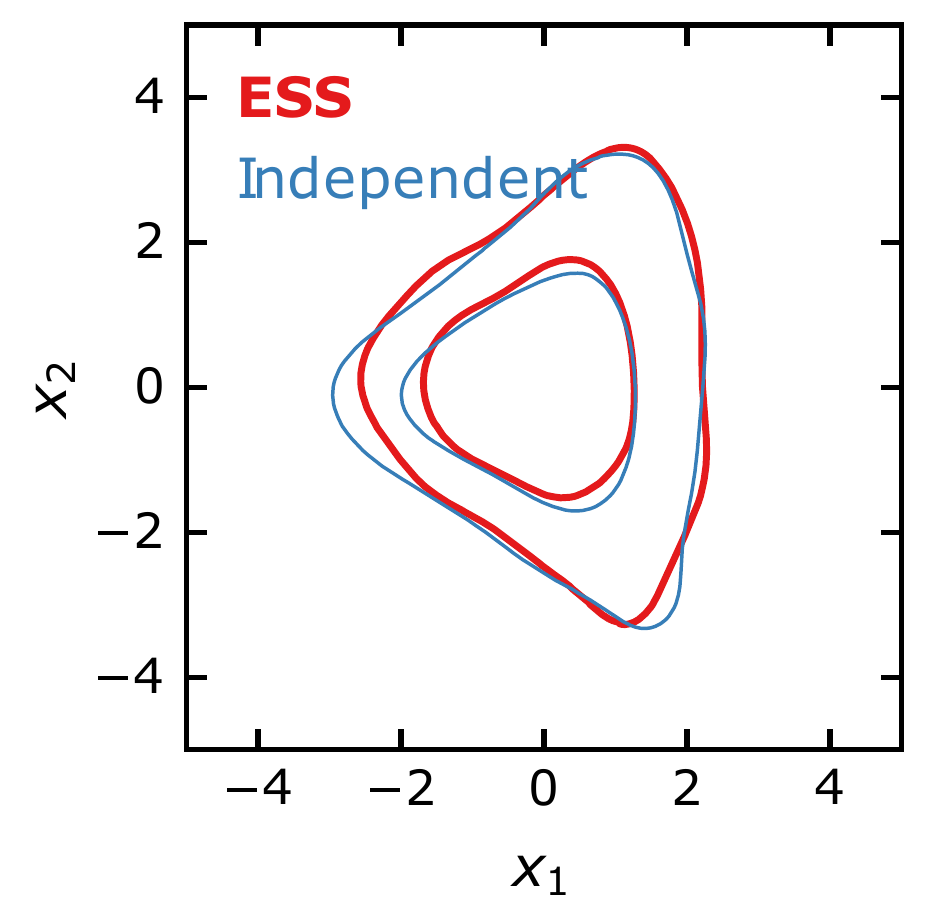}
\caption{The plots compare the 1-sigma and 2-sigma contours generated by the optimised random-walk Metropolis (left), Standard Slice (centre) and Ensemble Slice Sampling (right) methods to those obtained by Independent Sampling (blue) for the correlated funnel distribution. All samplers used the same number of probability density evaluations, $3\times 10^{5}$. Only the first two dimensions are shown here.}
\label{fig_ess:funnel}
\end{figure*}

To assess the mixing rate of the algorithm on this demanding case, we set the maximum number of evaluations of the probability density function to $3\times 10^{5}$. As shown in Figure \ref{fig_ess:funnel}, the Ensemble Slice Sampling is the only algorithm out of the three whose outcome closely resembles the target distribution. The results of Metropolis were incorrect for both, the limited run with $3\times 10^{5}$ iterations and the long run with $10^{7}$ iterations. In particular, the chain produced using the Metropolis method resemble a converged chain but in fact it is biased in favour of positive values of $x_{1}$. The problem arises because of the vanishing low probability of accepting a point with highly negative value of $x_{1}$. This indicates the inability of Metropolis to handle this challenging case. For a more detailed discussion of this problem we direct the reader to Section 8 of \cite{neal2003slice}. In general, the correlated funnel is a clear example of a distribution in which a single Metropolis proposal scale is not sufficient for all the sampled regions of parameter space. The locally adaptive nature of ESS solves this issue.

\subsection{Comparison to other ensemble methods}

So far we have demonstrated Ensemble Slice Sampling's performance in simple, yet challenging, target distributions. The tests performed so far demonstrate ESS's capacity to sample efficiently from highly correlated distributions compared with standard methods such as Metropolis and Slice Sampling. Although the use of Metropolis and Slice Sampling is common, these methods are not considered to be state-of-the-art. For this reason, we will now compare ESS with state-of-the-art gradient-free ensemble MCMC methods.

By far, the two most popular choices\footnote{For instance, in the fields of Astrophysics and Cosmology where most models are not differentiable and gradient methods (e.g. Hamiltonian Monte Carlo or NUTS) are not applicable the default choice is the Affine-Invariant Ensemble Sampler (AIES) \parencite{goodman2010ensemble} as implemented in emcee.} of gradient-free ensemble methods are the Affine-Invariant Ensemble Sampling (AIES) \parencite{goodman2010ensemble} method and the Differential Evolution Monte Carlo (DEMC) \parencite{ter2006markov} algorithm supplemented with a Snooker update \parencite{ter2008differential}.

In cases of strongly multimodal target distributions we will also test our method against Sequential Monte Carlo\footnote{As there are many different flavours of SMC, we decided to use the one implemented in \texttt{PyMC3} which utilises importance sampling, simulated annealing and Metropolis sampling.} (SMC) \parencite{liu1998sequential, del2006sequential} and Kernel Density Estimate Metropolis (KM) \parencite{kombine} which are particle methods specifically designed to handle strongly multimodal densities. \\

\subsubsection{Ring distribution}
Although, all three of the compared methods (i.e. ESS, AIES, DEMC) are affine invariant and thus unaffected by linear correlations, they do however differ significantly in the way they handle non-linear correlations. In particular, only Ensemble Slice Sampling (ESS) is locally adaptive because of its stepping-out procedure and therefore able to handle non-linear correlations efficiently.

To illustrate ESS's performance in a case of strong non-linear correlations we will use the 16--dimensional ring distribution defined by:
\begin{equation}
\begin{split}
    \ln \mathcal{L} = & - \Bigg[ \frac{(x_{n}^{2} + x_{1}^{2} - a)^{2}}{b}\Bigg]^{2} \\ & -\sum_{i=1}^{n-1} \Bigg[ \frac{(x_{i}^{2} + x_{i+1}^{2} - a)^{2}}{b}\Bigg]^{2}\, ,
\end{split}
\label{eq_ess:ring}
\end{equation}
where $a=2$, $b=1$ and $n=16$ is the total number of parameters. We also set the number of walkers to be $64$ and run the samplers for $10^{7}$ steps discarding the first half of the chains. Here we followed the heuristics discussed at the beginning of this section and increased the number of walkers from the minimum of $2\times 16$ to $4\times 16$ due to the presence of strong non-linear correlations in order to achieve the optimal acceptance rate for AIES and DEMC. The number of iterations is large enough for all samplers to converge and provide accurate estimates of the autocorrelation time.

The results are shown in Table \ref{tab_ess:table2} and verify that ESS' performance is an order of magnitude better than that of the other methods. 

\begin{table}[ht!]
    \centering
    \caption{The table shows a comparison of the Affine Invariant Ensemble Sampling (AIES), Differential Evolution Markov Chain (DEMC), and Ensemble Slice Sampling methods in terms of the integrated autocorrelation time (IAT) and the number of effective samples per evaluation of the probability density (efficiency) multiplied by $10^5$. These metrics are formally defined in Appendix \ref{app:ess}. The target distributions are the 16--dimensional ring distribution, the 10--dimensional Gaussian shells distribution and the 13--dimensional hierarchical Gaussian process regression distribution. In all cases the total number of iterations was set to $10^{7}$. It should be noted that in the case of the Gaussian shells the global move was used instead of the differential move.}
    \def\arraystretch{1.1}
    \begin{tabular}{lccc}
        \toprule[0.75pt]
         & AIES   & DEMC   & \textbf{ESS}  \\
        \midrule[0.5pt]
        \multicolumn{4}{l}{Ring distribution} \\
        \midrule[0.5pt]
        IAT          &    49470    &    91128    &   $\mathbf{1675}$   \\
        efficiency   &    2.0    &    1.1    & $\mathbf{12.2}$  \\
        \midrule[0.5pt]
        \multicolumn{4}{l}{Gaussian shells distribution} \\
        \midrule[0.5pt]
        IAT          &    33046    &    2760    &   $\mathbf{89}$   \\
        efficiency   &    3.0    &    36.0    &  $\mathbf{731.0}$ \\
        \midrule[0.5pt]
        \multicolumn{4}{l}{Hierarchical Gaussian process regression} \\
        \midrule[0.5pt]
        IAT          &    55236    &    30990    &   $\mathbf{547}$   \\
        efficiency   &    1.8    &    3.2    &  $\mathbf{38.0}$ \\
        \bottomrule[0.75pt]
        \end{tabular}
    \label{tab_ess:table2}
\end{table}

\subsubsection{Gaussian shells distribution}
Another example that demonstrates ESS's performance in cases of non-linear correlations is the Gaussian Shells distribution defined as:
\begin{equation}
    \mathcal{L}(\mathbf{\Theta}) = \text{circ}(\mathbf{\Theta}|\mathbf{c}_{1}, r_{1}, w_{1})+\text{circ}(\mathbf{\Theta}|\mathbf{c}_{2}, r_{2}, w_{2}),
    \label{eq_ess:shells}
\end{equation}
where
\begin{equation}
    \text{circ}(\mathbf{\Theta}|\mathbf{c}, r, w) = \frac{1}{\sqrt{2\pi}w} \exp \Bigg[-\frac{1}{2} \frac{(|\Theta - \mathbf{c}| - r)^{2}}{w^{2}}\Bigg].
    \label{eq_ess:shell}
\end{equation}
We choose the centres, $\mathbf{c}_{1}$ and $\mathbf{c}_{2}$ to be $-3.5$ and $3.5$ in the first dimension respectively and zero in all others. We take the radius to be $r=2.0$ and the width $w=0.1$. In two dimensions, the aforementioned distribution corresponds to two equal-sized Gaussian Shells. In higher dimensions the geometry of the distribution becomes more complicated and the density becomes multimodal.

For our test, we set the number of dimensions to $10$ and the number of walkers to $40$ due to the existence of two modes. Since this target distribution exhibits some mild multimodal behaviour we opt for the global move instead of the default differential move although the latter also performs acceptably in this case. The total number of iterations was set to $10^{7}$ and the first half of the chains was discarded. The results are presented in Table \ref{tab_ess:table2}. ESS's autocorrelation time is $2-3$ orders of magnitude lower than that of the other methods and the efficiency is higher by $1-2$ orders of magnitude respectively. \\

\subsubsection{Hierarchical Gaussian process regression}
To illustrate ESS's performance in a real-world example we will use a modelling problem concerning the concentration of $CO_{2} $ in the atmosphere adapted from Chapter 5 of \cite{rasmussen2003gaussian}. The data consist of monthly measurements of the mean $CO_{2}$ concentration in the atmosphere measured at the \emph{Mauna Loa Observatory} \parencite{keeling2004atmospheric} in \emph{Hawaii} since 1958. Our goal is to model the concentration of $CO_{2}$ as a function of time. To this end, we will employ a \emph{hierarchical Gaussian process} model with a composite covariance function designed to take care of the properties of the data. In particular, the covariance function (kernel) is the sum of following four distinct terms:
\begin{equation}
    k_1(r) = \theta_1^2  \exp \left(-\frac{r^2}{2\theta_2} \right)\, ,
    \label{eq_ess:kernel1}
\end{equation}
where $r=x-x'$ that describes the smooth trend of the data,
\begin{equation}
    k_2(r) = \theta_3^2 \exp \left[-\frac{r^2}{2 \theta_4}
                                            -\theta_5\sin^2\left(
                                            \frac{\pi r}{\theta_6}\right)
                                           \right]\, ,
    \label{eq_ess:kernel2}
\end{equation}
that describes the seasonal component,
\begin{equation}
    k_3(r) = \theta_7^2  \left [ 1 + \frac{r^2}{2\theta_8 \theta_9}
                                \right ]^{-\theta_8}\, ,
    \label{eq_ess:kernel3}
\end{equation}
which encodes medium-term irregularities, and finally:
\begin{equation}
    k_4(r) = \theta_{10}^2  \exp \left(-\frac{r^2}{2 \theta_{11}} \right) + \theta_{12}^2\delta_{ij}\, ,
    \label{eq_ess:kernel4}
\end{equation}
that describes the noise. We also fit the mean of the data, having in total 13 parameters to sample.

We sample this target distribution using $36$ walkers for $10^{7}$ iterations and we discard the first half of the chains. The number of walkers that was used corresponds to $1.5$ times the minimum number. We found that this value results in the optimal acceptance rate for AIES and DEMC. For this example we use the differential move of ESS. The results are presented in Table \ref{tab_ess:table2}. The integrated autocorrelation time of ESS is $2$ orders of magnitude lower than that of the other methods and its efficiency is more than an order of magnitude higher. The performance is weakly sensitive to the choice of the number of walkers.

\subsubsection{Bayesian object detection}
Another real world example with many applications in the field of \emph{astronomy} is \emph{Bayesian object detection}. The following model adapted from \cite{feroz2008multimodal} can be used with a few adjustments to detect astronomical objects in telescope images often hidden in background noise.

We assume that the 2D circular objects present in the image are described by the Gaussian profile:
\begin{equation}
    \mathbf{G}(x,y; \bm{\theta})= A \exp\bigg[-\frac{(x-X)^{2}+(y-Y)^{2}}{2 R^{2}} \bigg]\, ,
    \label{eq_ess:template}
\end{equation}
where $\mathbf{\theta}=(X, Y, A, R)$ are parameters that define the coordinate position, the amplitude and the size of the object, respectively.
Then the data can be described as:
\begin{equation}
    \mathbf{D}=\mathbf{N} + \sum_{i=1}^{n_{\text{Obj}}} \mathbf{G}(\bm{\theta_{i}})\, ,
    \label{eq_ess:data}
\end{equation}
where $n_{\text{Obj}}$ is the number of objects in the image and $\mathbf{N}$ is an additive Gaussian noise term.

\begin{figure}[thb!]
    \centering
    \includegraphics[scale=0.65]{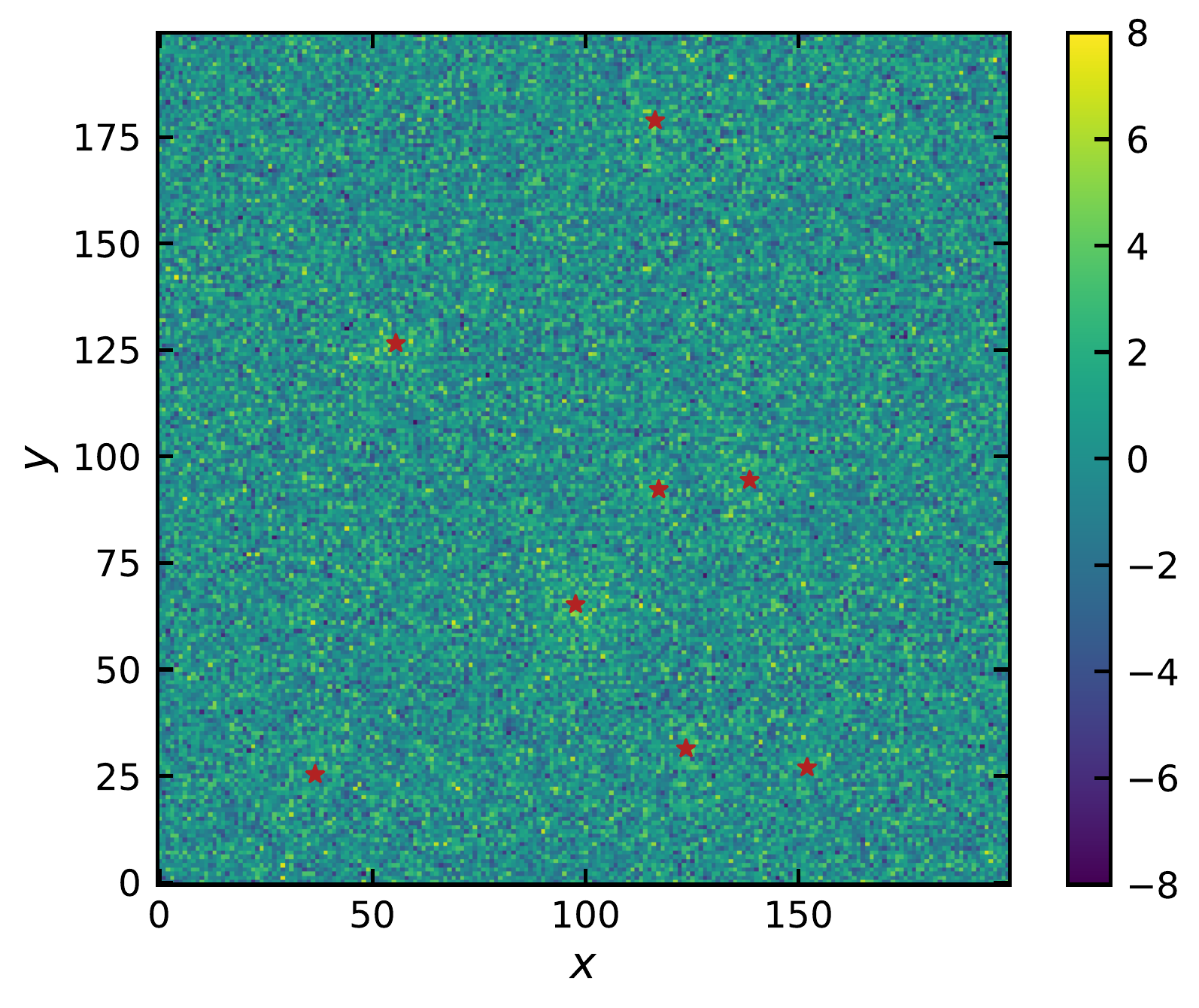}
\caption{The plot shows a simulated image used in the Bayesian object detection exercise. There are $8$ circular objects included here. As the objects are hardly visible due to the background noise their centres are marked with red stars.}
\label{fig_ess:objects}
\end{figure}

Assuming a $200 \times 200$ pixel-wide image, we can create a simulated dataset by sampling the coordinate positions $(X, Y)$ of the objects from $\mathcal{U}(0, 200)$ and their amplitude $A$ and size $R$ from $\mathcal{U}(1, 2)$ and $\mathcal{U}(3, 7)$, respectively. We sample $n_{\text{Obj}} = 8$ objects in total. Finally, we sample the noise $\mathbf{N}$ from $\mathcal{N}(0,4)$. In practice we create a dataset of $100$ such images and one such example is shown in Figure \ref{fig_ess:objects}. Notice that the objects are hardly visible as they are obscured by the background noise, this makes the task of identifying those objects very challenging.

Following the construction of the simulated dataset, the posterior probability density function is defined as:
\begin{equation}
    P(\bm{\theta} | \mathbf{D}) \propto \exp \bigg\{\frac{[\mathbf{G}(\bm{\theta})-\mathbf{D}]^{2}}{2 \sigma^{2}}\bigg\} P(\bm{\theta})\, ,
    \label{eq_ess:obj_post}
\end{equation}
where $\sigma = 2$ is the standard deviation of the $\mathbf{N}$ noise term. The prior $P(\bm{\theta})$ can be decomposed as the product of prior distributions of $X$, $Y$, $A$, and $R$. We used uniform priors for all of these parameters with limits $(0,200)$ for $X$ and $Y$, $(1,2)$ for $A$, and $(2,9)$ for $R$. It is important to mention here that the posterior does not include any prior information about the exact or maximum number of objects in the data. In that sense, the sampler is agnostic about the exact number, positions and characteristics (i.e. amplitude and size) of the objects that it seeks to detect.

We sampled the posterior distribution using $200$ walkers (initialised from the prior distribution) for each image in our dataset (i.e. 100 images in total) using Ensemble Slice Sampling (ESS), Affine Invariant Ensemble Sampling (AIES), and Differential Evolution Markov Chain (DEMC). Although the posterior distribution is multimodal (i.e. $8$ modes) we used the differential move since the number of dimensions is low and there is no reason to use more sophisticated moves like the global move. We used a large enough ensemble of walkers due to the potential presence of multiple modes so that all three samplers are able to resolve them.

We ran each sampler for $10^{4}$ iterations in total and we discarded the first half of the chains. We found that, on average for the 100 images, ESS identifies correctly $7$ out of $8$ objects in the image, whereas AIES and DEMC identify $4$ and $5$, respectively. 

In cases where the objects are well-separated ESS often identifies correctly $8$ out of $8$. Its accuracy falls to $7/8$ in cases where two of the objects are very close to each other or overlap. In those cases ESS identifies the merged object as a single object. In this context, by identification of an object, we mean that at least one walker has sampled the posterior mode which corresponds to that object.

\subsubsection{Gaussian Mixture}
One strengths of ESS is its ability to sample from strongly multimodal distributions in high dimensions. To demonstrate this, we will utilise a Gaussian Mixture of two components centred at $\mathbf{-0.5}$ and $\mathbf{+0.5}$ with standard deviation of $\mathbf{0.1}$. We also put $1/3$ of the probability mass in one mode and $2/3$ in the other.

We first set this distribution at $10$ dimensions and we sample this using $80$ walkers for $10^{5}$ steps. The distance between the two modes in this case is approximately $32$ standard deviations. We then increase the number of dimensions to $50$ and we sample it using $400$ walkers for $10^{5}$ iterations. In this case, the actual distance between the two modes is approximately $71$ standard deviations. The total number of iterations was set to $10^{7}$ for all methods but the SMC.

This problem consists of two, well separated, modes and thus requires using at least twice the minimum number of walkers (i.e. at least 40 for the 10--dimensional case and 200 for the 50--dimensional one). Although the aforementioned configuration was sufficient for ESS to provide accurate estimates, we opted instead for twice that number (i.e. 80 walkers for the 10--dimensional cases and 400 for the 50--dimensional one) in order to satisfy the requirements of the other samplers, mainly the Kernel Density Estimate Metropolis (KM), but also AIES and DEMC. For the Sequential Monte Carlo (SMC) sampler we used $2000$ and $20000$ independent chains for the low and high dimensional case respectively. The temperature ladder that interpolates between the prior and posterior distribution was chosen adaptively guaranteeing an effective sample size of $90\%$ the physical size of the ensemble. Our implementation of SMC was based on that of \texttt{PyMC3} using an independent Metropolis mutation kernel.

\begin{landscape}
\begin{figure*}[thb!]
    \centering
    \includegraphics[scale=0.65]{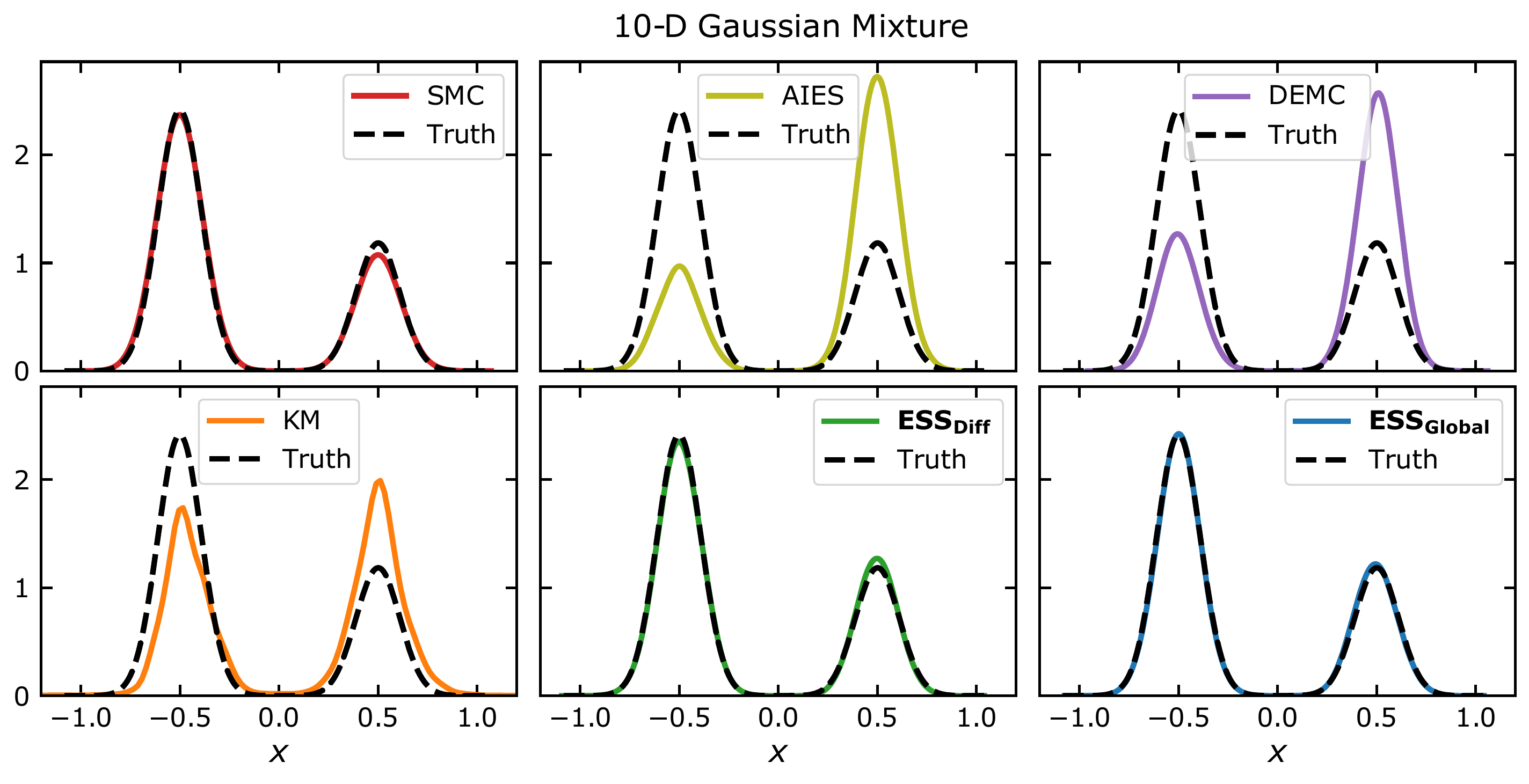}
\caption{The plot compares the results of 6 samplers, namely Sequential Monte Carlo (SMC, red), Affine-Invariant Ensemble Sampling (AIES, yellow), Differential Evolution Markov Chain (DEMC, purple), Kernel Density Estimate Metropolis (KM, orange), Ensemble Slice Sampling using the differential move (ESS, green), and Ensemble Slice Sampling using the global move (ESS, blue). The target distribution is a 10--dimensional Gaussian Mixture. The figure shows the 1D marginal distribution for the first parameter of the 10.}
\label{fig_ess:10dmixture}
\end{figure*}
\end{landscape}

\begin{landscape}
\begin{figure*}[t!]
    \centering
    \includegraphics[scale=0.65]{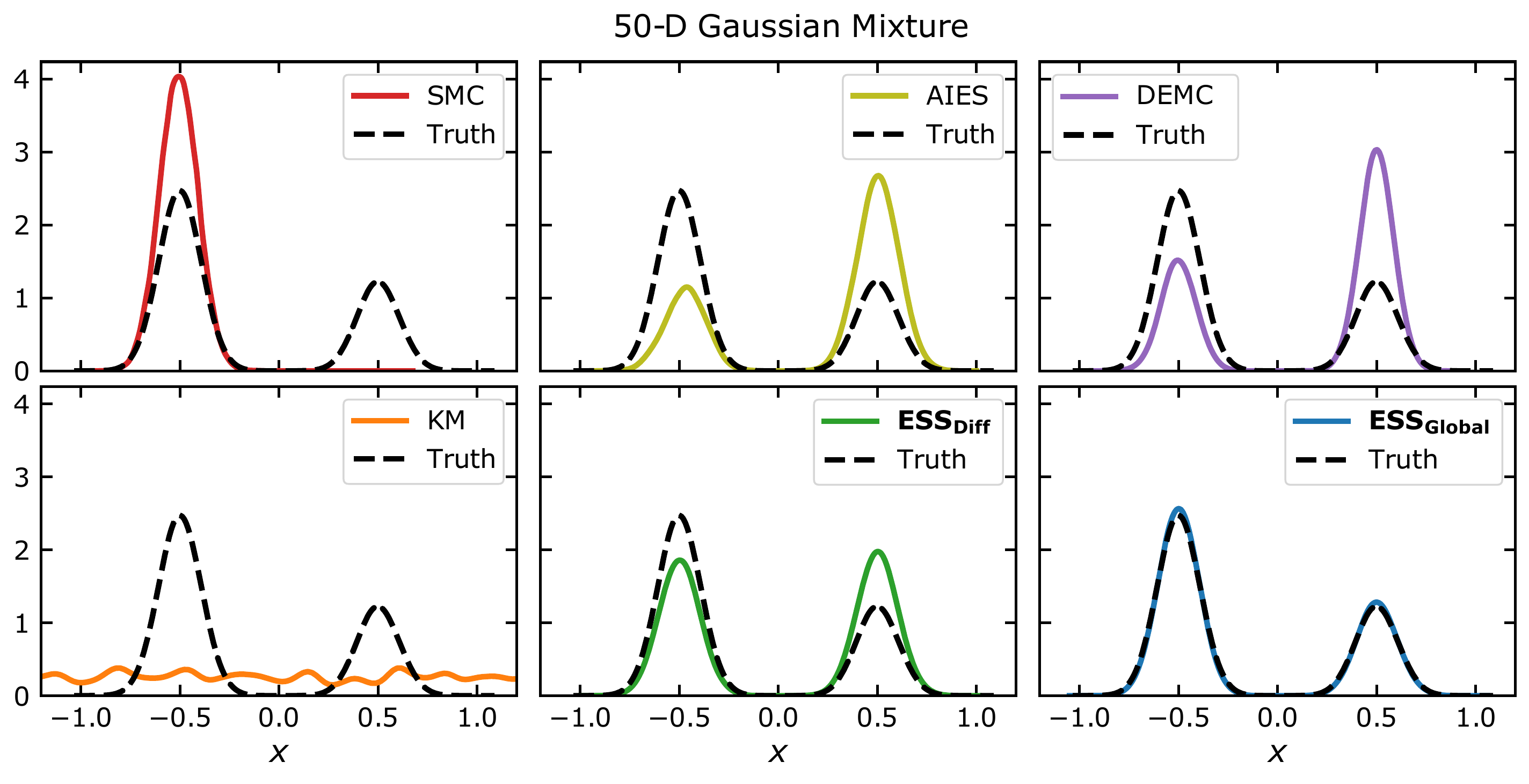}
\caption{The plot compares the results of 6 samplers, namely Sequential Monte Carlo (SMC, red), Affine-Invariant Ensemble Sampling (AIES, yellow), Differential Evolution Markov Chain (DEMC, purple), Kernel Density Estimate Metropolis (KM, orange), Ensemble Slice Sampling using the differential move (ESS, green), and Ensemble Slice Sampling using the global move (ESS, blue). The target distribution is a 50--dimensional Gaussian Mixture.The figure shows the 1D marginal distribution for the first parameter of the 50.}
\label{fig_ess:50dmixture}
\end{figure*}
\end{landscape}

The results for the 10--dimensional and 50--dimensional cases are plotted in Figures \ref{fig_ess:10dmixture} and \ref{fig_ess:50dmixture}, respectively. In the 10--dimensional case, both ESS (differential and global move) and SMC managed to sample from the target whereas AIES, DEMC and KM failed to do so. In the 50--dimensional case, only the Ensemble Slice Sampling with the global move manages to sample correctly from this challenging target distribution. In practice $\text{ESS}_{G}$ is able to handle similar cases in even higher number of dimensions and with more than $2$ modes.

\subsection{Convergence of the Length Scale \texorpdfstring{$\mu$}{u}}

\begin{figure}[t!]
    \centering
    \includegraphics[scale=0.65]{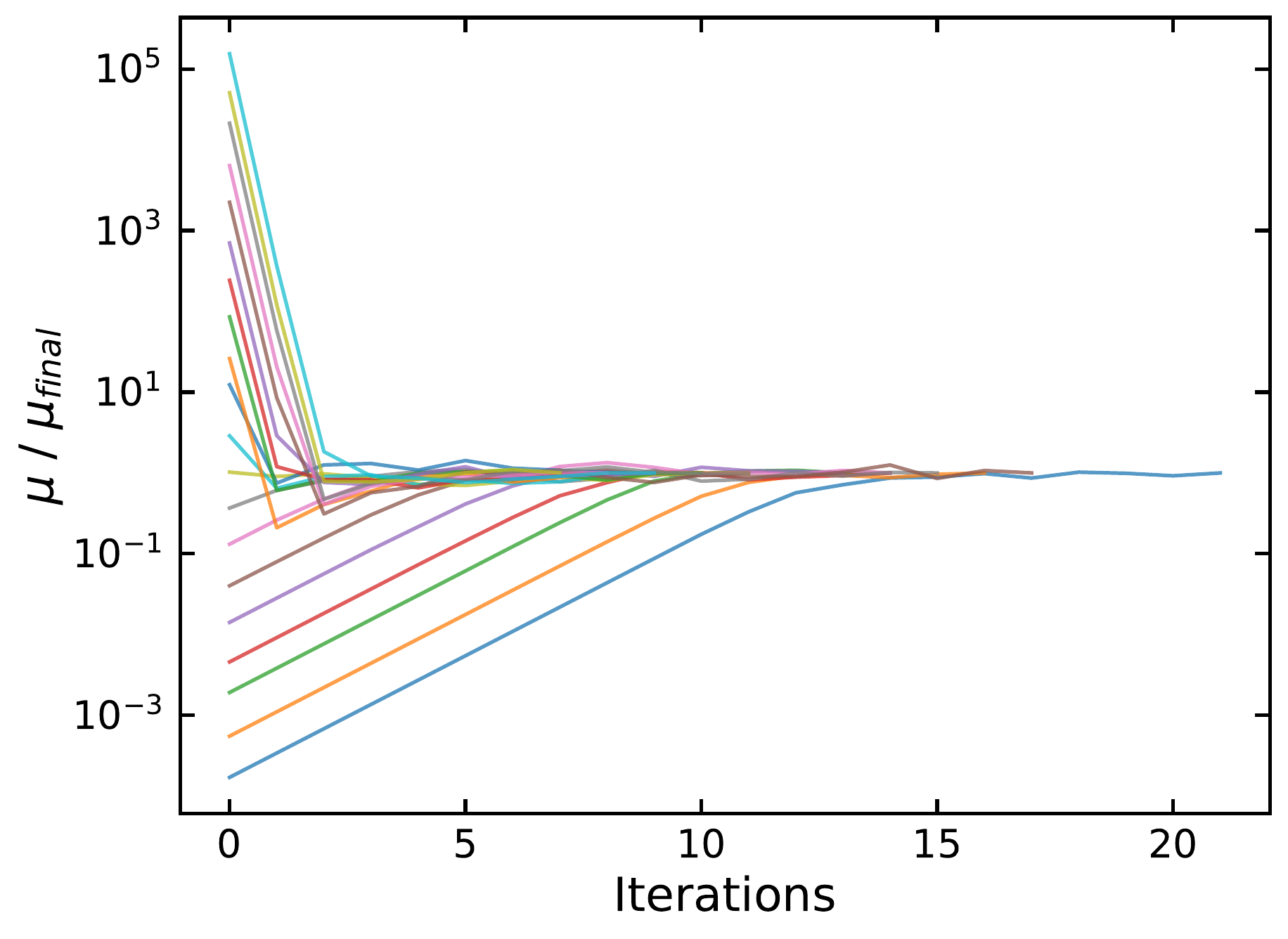}
    \caption{The plot shows the adaptation of the length scale $\mu$ as a function of the number of iterations and starting from a wide range of initial values. Each trace is an independent run and the y-axis shows the value of $\mu$ divided by the final value of $\mu$. The target distribution in this example is a 20--dimensional correlated normal distribution. Starting from larger $\mu$ values leads to significantly faster adaptation.}
\label{fig_ess:scale}
\end{figure}

Figure \ref{fig_ess:scale} plots the convergence of the length scale during the first 20 iterations. The target distribution in this example is a 20--dimensional correlated normal distribution. The length scale $\mu$ was initialised from a wide range of possible values. Adaptation is significantly faster when the initial length scale is larger than the optimal one rather than smaller. Another benefit of using a larger initial estimate would be the reduced number of probability evaluations during the first iterations. This is due to the fact that the shrinking procedure is generally faster than the stepping-out procedure.

\subsection{Parallel Scaling}

\begin{figure}[t!]
    \centering
    \includegraphics[scale=0.65]{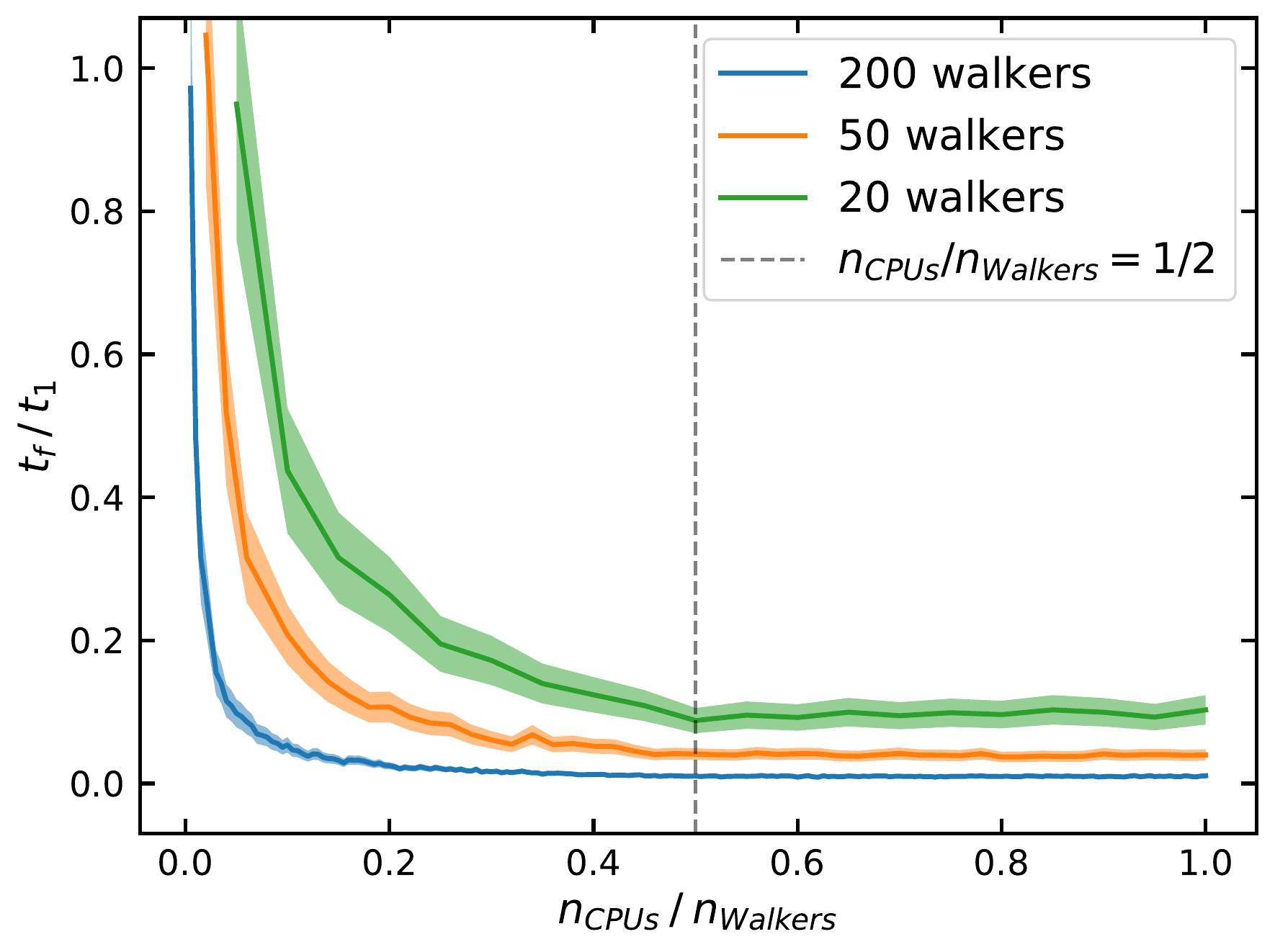}
    \caption{The plot shows the time $t_{f}$ required for ESS to complete a pre-specified number of iterations as a function of the ratio of the number of available CPUs $n_{\rm CPUs}$ to the total number of walkers $n_{\rm Walkers}$. The results are normalised with respect to the single CPU case $t_{1}$. The method scales as $\mathcal{O}(1/n_{\text{CPUs}})$ as long as $n_{\text{CPUs}}\leq n_{\text{Walkers}}/2$ (dashed line). The shaded areas show the $2-\sigma$ intervals.}
\label{fig_ess:parallel}
\end{figure}

By construction, Ensemble Slice Sampling can be used in parallel computing environments by parallelising the ensemble of walkers as discussed in Section \ref{sec_ess:direction}. The maximum number of CPUs used without any of them being idle is equal to the size of complementary ensemble,  $n_{\text{Walkers}}/2$. In order to verify this empirically and investigate the scaling of the method for any number of CPUs, we sampled a 10--dimensional Normal distribution for $10^{5}$ iterations with varying number of walkers. The results are plotted in Figure \ref{fig_ess:parallel}. We sampled the aforementioned distribution multiple times in order to get estimates of the confidence integrals shown in Figure \ref{fig_ess:parallel}. The required time to do the pre-specified number of iterations scales as $\mathcal{O}(1/n_{\text{CPUs}})$ as long as $n_{\text{CPUs}}\leq n_{\text{Walkers}}/2$. This result does not depend on the specific distribution. We can always use all the available CPUs by matching the size of the complementary ensemble (i.e. half the number of walkers) to the number of CPUs.

\section{Discussion}
\label{sec_ess:discussion}

In Section \ref{sec_ess:empirical} we provided a quantitative comparison of the efficiency of Ensemble Slice Sampling compared to other methods. In this Section we will provide some qualitative arguments to informally demonstrate the advantages of Ensemble Slice Sampling over other methods. Furthermore, we will briefly discuss some general aspects of the algorithm and place our work in the context of other related algorithms.

After the brief adaptation period is over and the length scale $\mu$ is fixed, the Ensemble Slice Sampling algorithm performs on average $5$ evaluations of the probability density per walker per iteration, assuming that either the differential or Gaussian move is used. This is in stark contrast with Metropolis-based MCMC methods that perform $1$ evaluation of the probability density per iteration. However, the non-rejection nature of Ensemble Slice Sampling more than compensates for the higher number of evaluations as shown in Section \ref{sec_ess:empirical}, thus yielding a very efficient scheme.

One could think of the number of walkers as the only free hyperparameter of Ensemble Slice Sampling. However, choosing the number of walkers is usually trivial. As we mentioned briefly at the end of Section \ref{sec_ess:ensemble}, there is a minimum limit to that number. In particular, in order for the method to be ergodic, the ensemble should be made of at least $2\times D$ walkers\footnote{The reason that the minimum limit is $2\times D$ instead of $D+1$ has to do with the ensemble splitting procedure that we introduced in order to make the method parallel. Splitting the ensemble into two equal parts means that each walker is updated based on the relative displacements of half the ensemble.}, where $D$ is the number of dimensions of the problem. Assuming that the initial relative displacements of the walkers span the parameter space (i.e. they do not belong to a lower-than-$D$-dimensional space) the resulting algorithm would be ergodic. As shown in Section \ref{sec_ess:empirical}, using a value close to the minimum number of walkers, meaning twice the number of parameters, is generally a good choice. Furthermore, we suggest to increase the number of walkers by a multiplicative factor equal to the number of well separated modes (e.g. four times the number of dimensions in a bimodal density). Other cases in which increasing the number of walkers can improve the sampling efficiency include target distributions with strong non-linear correlations between their parameters.

Regarding the initial positions of the walkers, we found that we can reduce the length of the burn-in phase by initialising the walkers from a tight sphere (i.e. Normal distribution with a very small variance) close to the \emph{Maximum a Posteriori} (MAP) estimate. In high dimensional problems, the MAP estimate will not reside in the typical set and the burn-in phase might be longer. We found that the tight sphere initialisation is still an efficient strategy compared to a more dispersed initialisation \parencite{foreman2013emcee}. Other approaches include initialising the walkers by sampling from the prior distribution or the \emph{Laplace approximation} of the posterior distribution. In multimodal cases, a prior initialisation is usually a better choice. A brief simulated annealing phase can also be very efficient, particularly in cases with many well separated modes.

Recent work on the No U-Turn Sampler \parencite{hoffman2014no} has attempted to reduce the hand-tuning requirements of Hamiltonian Monte Carlo \parencite{betancourt2017conceptual} using the dual averaging scheme of \textcite{nesterov2009primal}. In order to achieve a similar result, we employed the much simpler stochastic approximation method of \textcite{robbins1951stochastic} to tune the initial length scale $\mu$. The Affine Invariant Ensemble Sampler \parencite{goodman2010ensemble} and the Differential Evolution MCMC \parencite{ter2006markov} use an ensemble of walkers to perform Metropolis updates. Our method differs by using the information from the ensemble to perform Slice Sampling updates. So why does ESS perform better, as demonstrated, compared to those other methods? The answer lies in the locally adaptive and non-rejection nature of the algorithm (i.e. stepping out and shrinking) that enables both efficient exploration of non-linear correlations and large steps in parameter space (e.g. using the global move)\footnote{Indeed, large steps like the ones in the 50--dimensional Gaussian Mixture example would not have been possible without the non-rejection aspect of the method as most attempts to jump to the other mode would have missed it using Metropolis updates.}.

For all numerical benchmarks in this paper we used the publicly available, open source \texttt{Python} implementation of Ensemble Slice Sampling called  \texttt{zeus}\footnote{The code is available at \url{https://github.com/minaskar/zeus}.}~\parencite{karamanis2021zeus}.

\section{Conclusion}
\label{sec_ess:conclusion}

We have presented Ensemble Slice Sampling (ESS), an extension of Standard Slice Sampling that eliminates the latter's dependence on the initial value of the length scale hyperparameter and augments its capacity to sample efficiently and in parallel from highly correlated and strongly multimodal distributions. 

In this paper we have compared Ensemble Slice Sampling with the optimally-tuned Metropolis and Standard Slice Sampling algorithms. We found that, due to its affine invariance, Ensemble Slice Sampling generally converges faster to the target distribution and generates chains of significantly lower autocorrelation. In particular, we found that in the case of AR(1), Ensemble Slice Sampling generates an order of magnitude more independent samples per evaluation of the probability density than Metropolis and Standard Slice Sampling. Similarly, in the case of the correlated funnel distribution, Ensemble Slice Sampling outperforms Standard Slice Sampling by an order of magnitude in terms of efficiency. Furthermore, in this case, Metropolis-based proposals fail to converge at all, demonstrating that a single Metropolis proposal scale is often not sufficient.

When compared to state-of-the-art ensemble methods (i.e. AIES, DEMC) Ensemble Slice Sampling outperforms them by $1-2$ orders of magnitude in terms of efficiency for target distributions with non-linear correlations (e.g. the Ring and Gaussian shells distributions). In the real world example of hierarchical Gaussian process regression, ESS's efficiency is again superior by $1-2$ orders of magnitude. Furthermore, in the Bayesian object detection example ESS achieved higher accuracy compared to AIES and DEMC. Finally, in the strongly multimodal case of the Gaussian Mixture, ESS outperformed all other methods (i.e. SMC, AIES, DEMC, KM) and was the only sampler able to produce reliable results in $50$ dimensions.

The consistent high efficiency of the algorithm across a broad range of different problems along with its parallel, black-box and gradient-free nature, renders Ensemble Slice Sampling ideal for use in scientific fields such as physics, astrophysics and cosmology, which are dominated by a wide range of computationally expensive and almost always non-differentiable models. The method is flexible and can be extended further using for example tempered transitions~\parencite{iba2001extended} or subspace sampling~\parencite{vrugt2009accelerating}.

\section{Appendix: Estimating the Effective Sample Size}
\label{app:ess}

% Note: in this sample, the section number is hard-coded in. Following
% proper LaTeX conventions, it should properly be coded as a reference:

%In this appendix we prove the following theorem from
%Section~\ref{sec_ess:textree-generalization}:

Assuming that the computational bottleneck of a MCMC analysis is the evaluation of the probability density function, which is usually a valid assumption in scientific applications, the \emph{efficiency} can be formally defined as the ratio of the \textit{Effective Sample Size} $N_{\rm Eff}$ to the total number of probability evaluations for a given chain. 

The $N_{\rm Eff}$ quantifies the number of effectively independent samples of a chain, and it is defined as
\begin{equation}
    \label{eq_ess:ESS}
    N_{\rm Eff} = \frac{n}{\text{IAT}}\, ,
\end{equation}
where $n$ is the actual number of samples in the chain, and IAT is the \textit{integrated autocorrelation time}. The latter describes the number of steps that the sampler needs to do in order to forget where it started and it is defined as
\begin{equation}
    \label{eq_ess:act}
    \text{IAT} = 1 + 2\sum_{k=1}^{\infty}\rho(k)\, ,
\end{equation}
where $\rho(k)$ is the \textit{normalised autocorrelation function} at lag $k$. In practice, we truncate the above summation in order to remove noise from the estimate \parencite{sokal1997monte}.

Given a chain $X(k)$ with $k=1,2,...,n$ the normalised autocorrelation function $\Hat{\rho}(k)$ at lag $k$ is estimated as 
\begin{equation}
    \Hat{\rho}(k) = \frac{\Hat{c}(k)}{\Hat{c}(0)}\, ,
\end{equation}
where
\begin{equation}
    \Hat{c}(k) = \frac{1}{n-k}\sum_{m=1}^{n-k}\big[ X(k+m)- \Bar{X} \big]\big[  X(m)-\Bar{X} \big]\, ,
\end{equation}
and $\Bar{X}$ is the mean of the samples.

In the case of ensemble methods, the IAT of an ensemble of chains is computed by first concatenating the chain from each walker into a single long chain. We found this estimator has lower variance than the \textcite{goodman2010ensemble} estimator and the \textcite{noauthor_autocorrelation_nodate} estimator. 
% !TEX TS-program = pdflatex
% !TEX root = ../ArsClassica.tex

%************************************************
\chapter{Zeus}
\label{chp:zeus}
%************************************************
 
\lstset{numbers=left,
    numberstyle=\scriptsize,
    stepnumber=1,
    numbersep=8pt
}    

This chapter presents \textit{zeus} which is the main contribution introduced in the paper titled \textit{zeus: A Python implementation of Ensemble Slice Sampling for efficient Bayesian parameter inference} that was published in the journal \textit{Monthly Notices of the Royal Astronomical Society} in 2021~\parencite{karamanis2021zeus}. The content of the chapter is almost identical to that included in the aforementioned publication with the exception of minor text and figure formatting differences.

\begin{center}
\rule{0.5\textwidth}{.4pt}
\end{center}
\vspace{8pt}

We introduce \texttt{zeus}, a well-tested \texttt{Python} implementation of the Ensemble Slice Sampling (ESS) method for Bayesian parameter inference. ESS is a novel Markov chain Monte Carlo (MCMC) algorithm specifically designed to tackle the computational challenges posed by modern astronomical and cosmological analyses. In particular, the method requires only minimal hand--tuning of $1-2$ hyper-parameters that are often trivial to set; its performance is insensitive to linear correlations and it can scale up to 1000s of CPUs without any extra effort. Furthermore, its locally adaptive nature allows to sample efficiently even when strong non-linear correlations are present. Lastly, the method achieves a high performance even in strongly multimodal distributions in high dimensions. Compared to \texttt{emcee}, a popular MCMC sampler, \texttt{zeus} performs $9$ and $29$ times better in a cosmological and an exoplanet application respectively.

\section{Introduction}
\label{sec_zeus:intro}

Over the past few decades the volume of astronomical and cosmological data has increased substantially. In response to that, a variety of astrophysical models have been developed to explain the plethora of observations. Markov chain Monte Carlo (MCMC) has been established as the standard procedure of inferring the model parameters subject to the available data in a Bayesian framework. Within the Bayesian context, the object that quantifies the probability distribution of the model parameters $\theta$ given the data $D$ and model $\mathcal{M}$ is the posterior distribution $\mathcal{P}(\theta)\equiv P(\theta | D, \mathcal{M})$ which is defined using Bayes's theorem:
\begin{equation}
    \label{eq_zeus:bayes}
    \mathcal{P}(\theta) = \frac{\mathcal{L}(\theta) \pi (\theta)}{\mathcal{Z}},
\end{equation}
where $\mathcal{L}(\theta) \equiv P(D|\theta, \mathcal{M})$ is the likelihood function, $\pi (\theta) \equiv P (\theta | \mathcal{M})$ is the prior distribution of the model parameters $\theta$, and $\mathcal{Z}\equiv P(D|\mathcal{M})$ is the, so called, Bayesian model evidence or marginal likelihood and in this context can be treated as a simple normalisation constant.

MCMC does not in general require knowing the value of the model evidence and it only depends on the ability to evaluate the unnormalised posterior distribution for arbitrary values of $\theta$. MCMC methods can then be used to generate (Markov) chains of samples from the posterior distribution. Those samples can be used to calculate integrals (e.g. parameter uncertainties, marginal distributions etc.) that are paramount for modern astronomical and cosmological analyses.

The most commonly used MCMC methods are variants of the Metropolis-Hastings (MH) algorithm \parencite{metropolis1953equation, hastings1970}. MH consists of two steps. First, given the last sample in the chain, a new sample is proposed and then the Metropolis criterion determines whether or not that new sample should be accepted and thus added to the chain. The resulting chain is Markovian in the sense that each sample is proposed based only on the previous sample. The purpose of the Metropolis acceptance criterion is to bias the chain so that the time spent in a region of the parameter space would be proportional to the posterior probability in that region. In other words, the stationary distribution of the Markov chain is the target distribution i.e. the posterior distribution. For a detailed introduction to MCMC methods we direct the reader to \textcite{mackay2003information} and for an intuitive introduction to Bayesian inference to \textcite{jaynes2003probability}.

Arguably, the most difficult part of the MH algorithm is the proposal step. There are many ways of choosing a new sample and the efficiency of the method depends on this choice. By far the simplest one is the use of a normal (Gaussian) distribution, centred around the previous sample to generate the new proposed sample. The resulting method is often called Random Walk Metropolis algorithm and its performance is highly sensitive to the $n(n+1)/2$ elements that form its covariance matrix. Those elements generally need to be chosen \textit{a priori} or be adaptively tuned. More efficient methods utilise the gradient of the target distribution \parencite{2017arXiv170102434B} or an ensemble of parallel and communicating chains \parencite{gilks1994adaptive,ter2006markov, ter2008differential, goodman2010ensemble}.

Out of the methods mentioned in the previous paragraph we will focus our attention on the last one, the ensemble or population MCMC variety. The reason is simple: the Random Walk Metropolis algorithm requires a great amount of tuning (or \textit{a priori} knowledge) for it to perform efficiently and even then there is no guarantee that the proposal covariance matrix is optimal for the whole parameter space. On the other hand, gradient based methods, although very powerful, are in general unsuitable for astronomical applications in which the models that are used are almost always not differentiable.

One benefit of ensemble MCMC over its alternatives is that the ensemble of parallel chains (also known as walkers) collectively sample the posterior, thus information about their distribution can be shared and used to make better educated proposals. Other advantages include the lack of hand-tuning of hyper-parameters and their capacity for parallel implementation. For the aforementioned reasons, ensemble MCMC methods have dominated astronomical analyses. The most common ones are affine--invariant ensemble sampling (AIES) \parencite{goodman2010ensemble} and differential evolution MCMC (DEMC) \parencite{ter2006markov, ter2008differential}, both implemented in the popular \texttt{Python} package \texttt{emcee} \parencite{foreman2013emcee,foreman2019emcee}.

In this paper we introduce \texttt{zeus}, a stable and well-tested \texttt{Python} implementation of Ensemble Slice Sampling (ESS) \parencite{karamanis2021ensemble}. ESS is a method based on the ensemble MCMC paradigm, with the crucial difference being that its proposals are performed via Slice Sampling updates \parencite{neal2003slice} instead of Metropolis-Hastings ones. As we will thoroughly demonstrate in Section \ref{sec_zeus:tests}, this subtle difference leads to substantial improvements in terms of sampling efficiency and robustness. \texttt{zeus} is a user-friendly tool that does not require any hand-tuning or preliminary runs and can scale up to 1000s of CPUs without any extra effort from the user.

\texttt{zeus} has been used in various astronomical and cosmological analyses, including cosmological tests of gravity \parencite{Tamosiunas2020}, relativistic effects and primordial non-Gaussianity \parencite{Wang2020}, 21cm intensity mapping \parencite{Umeh2021}, and has been implemented as part of the \texttt{CosmoSIS} package \parencite{Zuntz2015}.

\texttt{zeus} is open source software that is publicly available at \url{https://github.com/minaskar/zeus} under the \texttt{GPL-3 Licence}. Detailed documentation and examples on how to get started are available at \url{https://zeus-mcmc.readthedocs.io}.

\section{Ensemble Slice Sampling}
\label{sec_zeus:ess}

\texttt{zeus} is a \texttt{Python} implementation of the Ensemble Slice Sampling (ESS) method presented in \textcite{karamanis2021ensemble}. 
Here we will provide a high-level description of the method and will refer to the accompanying paper for more details about the underlying algorithmic structure and mathematics.

ESS combines the ensemble MCMC paradigm with slice sampling. Since the use of slice sampling in astronomical parameter inference is rare we will start by explaining its function and how it differs from Metropolis updates. Then we will move on to discuss how it can be efficiently combined with ensemble MCMC.

\subsection{Slice sampling}

Slice sampling is based on the idea that sampling from a distribution with density $P(x)$ is equivalent to uniform sampling from the area under the plot of $f(x)\propto P(x)$. To this end, we introduce an auxiliary variable $y$, called height, such that the joint distribution $P(x,y)$ is uniform over the region $U=\lbrace (x,y): 0<y<f(x) \rbrace$. To sample from the marginal distribution $P(x)$, we first sample from $P(x,y)$ and then we marginalise by dropping the $y$ value of each sample.

In order to generate samples from $P(x,y)$ we utilise the following scheme \parencite{neal2003slice}:
\begin{enumerate}
    \item Given the current state $x_{0}$, draw $y_{0}$ uniformly from $(0,f(x_{0}))$.
    \item Find an interval $I=(L,R)$ that contains all, or at least part, of the slice $s=\lbrace x: y_{0}<f(x)\rbrace$.
    \item Draw the new sample $x_{1}$ uniformly from $I\cap S$.
\end{enumerate}

\begin{figure}[H]
    \centering
	\centerline{\includegraphics[scale=0.9]{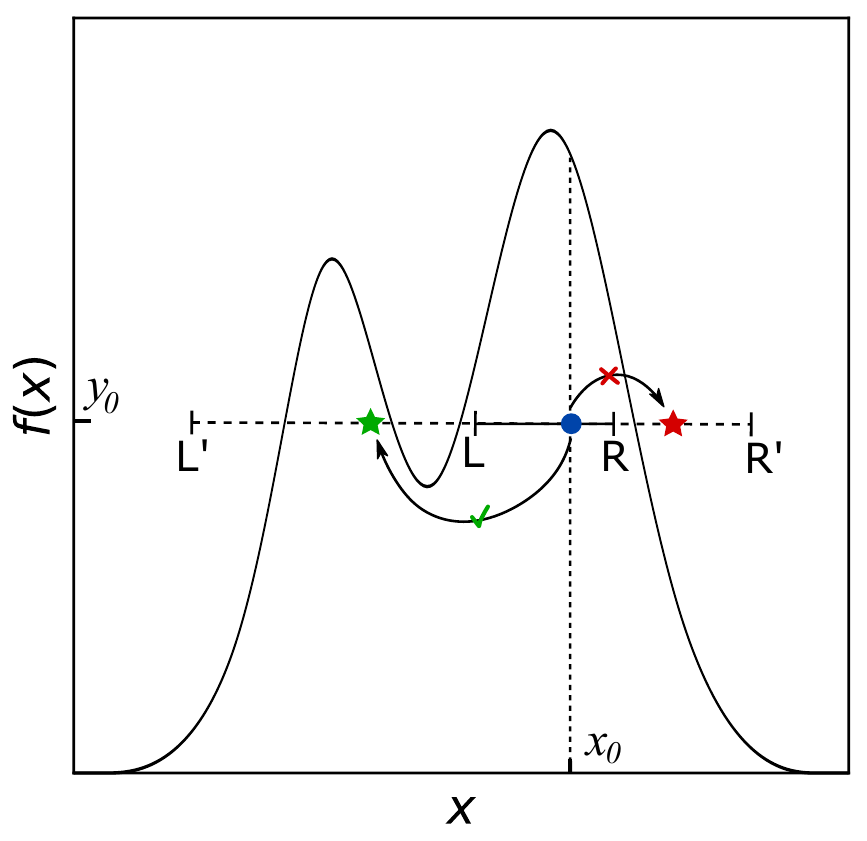}}
    \caption{Illustration of the univariate slice sampling update. Given the current sample $x_{0}$, a value $y_{0}$ is uniformly sampled along the vertical slice $(0,f(x_{0}))$ (dashed line) thus defining the initial point (blue). An interval $(L,R)$ is uniformly positioned horizontally around $(x_{0},y_{0})$ and it is expanded in steps of size $R-L$ until both its ends are outside the slice. The new sample is generated by  repeatedly sampling (uniformly) from the interval $(L',R')$ until a sample (green star) is found inside the slice. Samples outside of the slice (red star) are rejected and they are instead used to shrink $(L',R')$.}
    \label{fig_zeus:slice}
\end{figure}

To construct the interval $I$ (step ii), \textcite{neal2003slice} introduced the stepping-out procedure that works by randomly positioning an interval of length $\mu$ around the sample $x_{0}$ (i.e. blue dot in Figure \ref{fig_zeus:slice}) and then expanding it in steps of size $\mu$ until both its ends (i.e. $L'$ and $R'$) are outside the slice. To obtain $x_{1}$ (i.e. green star in Figure \ref{fig_zeus:slice}) we then use the shrinking procedure in which candidates are sampled uniformly from $I$ until a point inside the slice $S$ is found. Samples outside of the slice are used to shrink the interval $I$. The two procedures are shown in Figure \ref{fig_zeus:slice}.

The length scale $\mu$ is the only free hyperparameter of slice sampling and although its choice can reduce or increase the computational cost of the method it generally does not affect its mixing properties (e.g. convergence rate, autocorrelation time, etc.). \texttt{zeus} utilises a stochastic optimization algorithm similar to \textcite{tibbits2014automated} and based on the \textcite{robbins1951stochastic} optimisation scheme in order to tune $\mu$ to its optimal value (see Section 3.1 of \textcite{karamanis2021ensemble} for more details).

It is important to note here that for multimodal target distributions there is no guarantee that the approximate slice would cross any of the other modes. In particular, if the initial estimate of the length scale $\mu$ is low then the probability of missing the other peaks, assuming that they are located far away, is also low. As we will show in Section \ref{sec_zeus:tests}, unlike simple slice sampling, ESS and thus \texttt{zeus} does not suffer from this effect.

\subsection{Walkers, moves and parallelism}

\begin{figure}
    \centering
	\centerline{\includegraphics[scale=0.9]{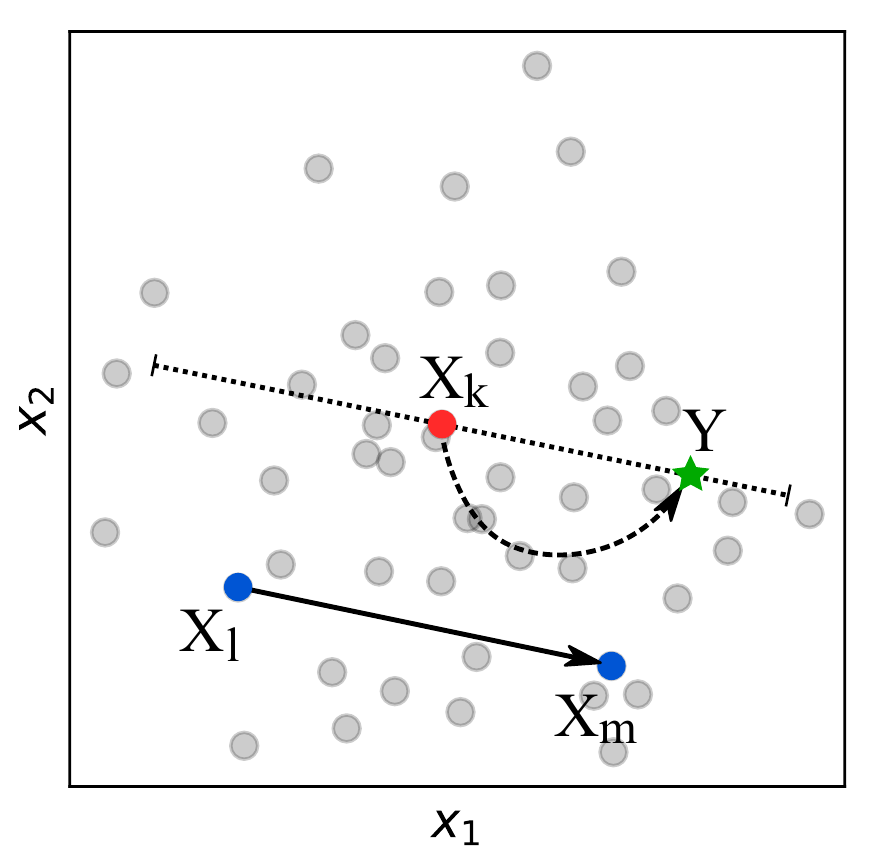}}
    \caption{The figure illustrates the differential move in the context of Ensemble Slice Sampling. The walker $X_{k}$ to be updated is shown in red. Two walkers, $X_{l}$ and $X_{m}$, (blue) are uniformly selected from the complementary ensemble (grey). The approximate slice (dotted line) is constructed parallel to the two walkers $X_{l}$ and $X_{m}$ using the stepping-out procedure. The new position $Y$ (green) of $X_{k}$ is sampled using the shrinking procedure along the approximate slice.}
    \label{fig_zeus:move}
\end{figure}

The slice sampling update described in the previous paragraphs is a univariate update scheme. For it to be used to sample from multivariate target distributions it needs to be generalised accordingly. Perhaps the simplest such generalisation in a multivariate setting is the use of slice sampling to sample along each coordinate axis in turn (i.e. component-wise slice sampling) or to sample along randomly selected directions in parameter space \parencite{mackay2003information}. Although valid, both of these approaches are unsuitable in cases of correlated parameters in which the proper choice of direction can substantially accelerate mixing.

To address this issue, \textcite{tibbits2014automated} proposed to orthogonalise the parameter space using the sample covariance, thus getting rid of linear correlations between parameters. We will instead follow a different, perhaps more flexible, approach to construct an efficient slice sampler. Our aim is to utilise an ensemble of parallel chains/walkers that can exchange information about the covariance structure of the target distribution and thus by-pass the difficulties posed by correlations.

As hinted in the introduction, the ensemble of walkers collectively sample the target distribution and thus their positions encode information about the correlations between the parameters. One way to take advantage of this information is to use it to construct direction vectors along which slice sampling can take place. Many moves that generate direction vectors from the complementary ensemble are possible. \texttt{zeus} offers a collection of them, including some that utilise clustering algorithms and density estimation methods. As we will show in Section \ref{sec_zeus:tests}, such moves can help accelerate sampling in difficult cases such as strongly multimodal distributions. Any distribution of the complementary ensemble can be used as a valid proposal to generate such direction vectors and \texttt{zeus} offers a highly flexible interface for the user to define such a move or choose one (or a mixture) from the ones that are already implemented and tested. Here is a list of the currently implemented moves in \texttt{zeus}:

\begin{itemize}
    \item \textbf{Differential move:} This is the default move used by \texttt{zeus} and shown in Figure \ref{fig_zeus:move}. Using the differential move, Ensemble Slice Sampling updates the position of each walker in the ensemble by slice sampling along a direction defined by the difference between two uniformly selected walkers from the rest of the ensemble (i.e. the complementary ensemble).
    \item \textbf{Gaussian move:} The Gaussian move samples the direction vectors along which slice sampling is performed from a normal distribution that shares the same covariance structure as the complementary ensemble. This approach is very efficient in cases in which the target distribution is close to normal.
    \item \textbf{Global move:} The Global move utilises a Dirichlet Process Gaussian Mixture to fit the complementary ensemble and proposes directions along different peaks of the target distribution in cases of strong multi-modality.
    \item \textbf{KDE move:} The KDE move samples the direction vectors from a Gaussian Kernel Density Estimate of the complementary ensemble. This can be useful in cases of highly non-Gaussian target distributions.
    \item \textbf{Random move:} The Random move performs slice sampling along isotropic directions. This is equivalent of standard multivariate slice sampling and it is mostly offered for testing purposes as it cannot handle correlations efficiently.
\end{itemize}
For more information on how those moves work as well as a comparison of the Differential, Gaussian and Global moves we direct the interested reader to \textcite{karamanis2021ensemble}. Unless stated otherwise the Differential move will be used for the following examples.

To parallelise this process and capitalise on the availability of multiple CPUs we randomly split the ensemble into two sets of walkers (i.e. active and passive sets)~\parencite{foreman2013emcee} and choose to update the positions of the active walkers along direction vectors defined by passive walkers. Then the passive becomes active and \textit{vice versa} and the process is repeated. The ensemble splitting technique is required in order to parallelise the algorithm without violating detailed balance. Parallelisation is achieved in practice using either \texttt{multiprocessing} or \texttt{MPI} using the implemented \texttt{ChainManager} utility that can distribute both multiple ensembles and multiple chains in parallel computing environments at the same time. Heuristics to determine the number of required walkers per application are discussed in Section \ref{sec_zeus:discussion}.

\section{Empirical Evaluation}
\label{sec_zeus:tests}

For the empirical evaluation of \texttt{zeus} we use five toy examples that manifest significant aspects of real astronomical applications\footnote{For additional demonstrations on similarly common structures (e.g. the funnel) we direct the reader to the accompanying paper \parencite{karamanis2021ensemble}.} (i.e. linear and non-linear correlations, multimodality, heavy tails, hard boundaries) and two real-world astronomical examples characteristic of modern astronomical analyses.

\subsection{Toy examples}
In order to understand the behaviour of \texttt{zeus} in various sampling scenarios, it is important to study its performance in different toy examples that demonstrate different characteristics of common target distributions that arise in astronomical applications. For that reason, we chose five such toy examples. The first one is a normal (Gaussian) distribution which by definition is characterised only by the linear correlation between its parameters. The second toy problem is the ring distribution, a characteristic example of strong non-linear correlations. The third example is a Gaussian mixture with two components. While the purpose of the first two examples is to study the behaviour of the algorithm in the presence of linear and non-linear correlations respectively, the goal of the third example is to demonstrate the ability of \texttt{zeus} to sample efficiently from multimodal target distributions. The fourth toy example investigates the effect that heavy tails have on the sampling efficiency and the fifth shows the effects that hard boundaries have on sampling.

We compare \texttt{zeus} with two popular alternatives offered by \texttt{emcee}, namely affine--invariant ensemble sampling with the \textit{stretch} move (\texttt{emcee}/AIES) and the differential evolution move (\texttt{emcee}/DEMC). The main goal of this analysis is to justify our choice of slice sampling as the basis of \texttt{zeus} instead of Metropolis updates through the use of simple yet instructive toy examples.

For all three toy examples discussed below we adopt the same analysis procedure, where we initialise the walkers by sampling from a normal distribution $\mathcal{N}(\mathbf{0}, \mathbf{I})$ where $\mathbf{I}$ is the identity covariance matrix and we discarded $10^4$ iterations as burn--in. 

The main metric that we use to investigate the behaviour of the samplers in those toy examples and to compare their performance is the distribution of steps performed by the walkers. As a step, we define the distance spanned in parameter space by a single walker in a single iteration. This is a fundamental measure of the efficiency of an MCMC method and it is directly related to the expected squared jump distance (ESJD)~\parencite{pasarica2010adaptively} given by:
\begin{equation}
    \label{eq_zeus:esjd}
    \text{ESJD} = \mathbf{E}\left[|\theta_{t+1}-\theta_{t}|^{2}\right]= 2 \,( 1 - \rho_{1}) \cdot\text{Var}_{(\pi)}(\theta_{t})\,,
\end{equation}
where $\theta_t$ are the chain samples, $\rho_{1}$ is the first-order autocorrelation, and $\text{Var}_{(\pi)}(\theta_{t})$ is a function of the stationary distribution only. Assuming that the higher-order autocorrelations $\rho_{2}, \rho_{3}, \dots$ are monotonically decreasing with respect to $\rho_{1}$, then maximising the ESJD leads to minimisation of the autocorrelation between chain elements and thus maximisation of the sampling efficiency. In other words, the further away (i.e. the greater the ESJD) the walkers jump per iteration,  the higher the sampling efficiency of the method. A benefit of using ESJD instead of the autocorrelation time as a metric is that the former, as an expectation value, is more accurate when computed using short chains.

In order to account for the different computational costs (i.e. different number of model evaluations per iteration) between \texttt{zeus} and \texttt{emcee} we thinned the chains of the latter method according to the average number of model evaluations of \texttt{zeus}. This allowed us to compare the distribution of steps of the three samplers as shown in Figures \ref{fig_zeus:gaussian_sep}, \ref{fig_zeus:ring_sep}, \ref{fig_zeus:bimodal_sep}, \ref{fig_zeus:student_sep}, and \ref{fig_zeus:truncated_sep} for the five toy examples respectively.

\subsubsection{The correlated normal distribution}

%\begin{landscape}
\begin{figure*}[!htb]
    \centering
	\centerline{\includegraphics[scale=0.45]{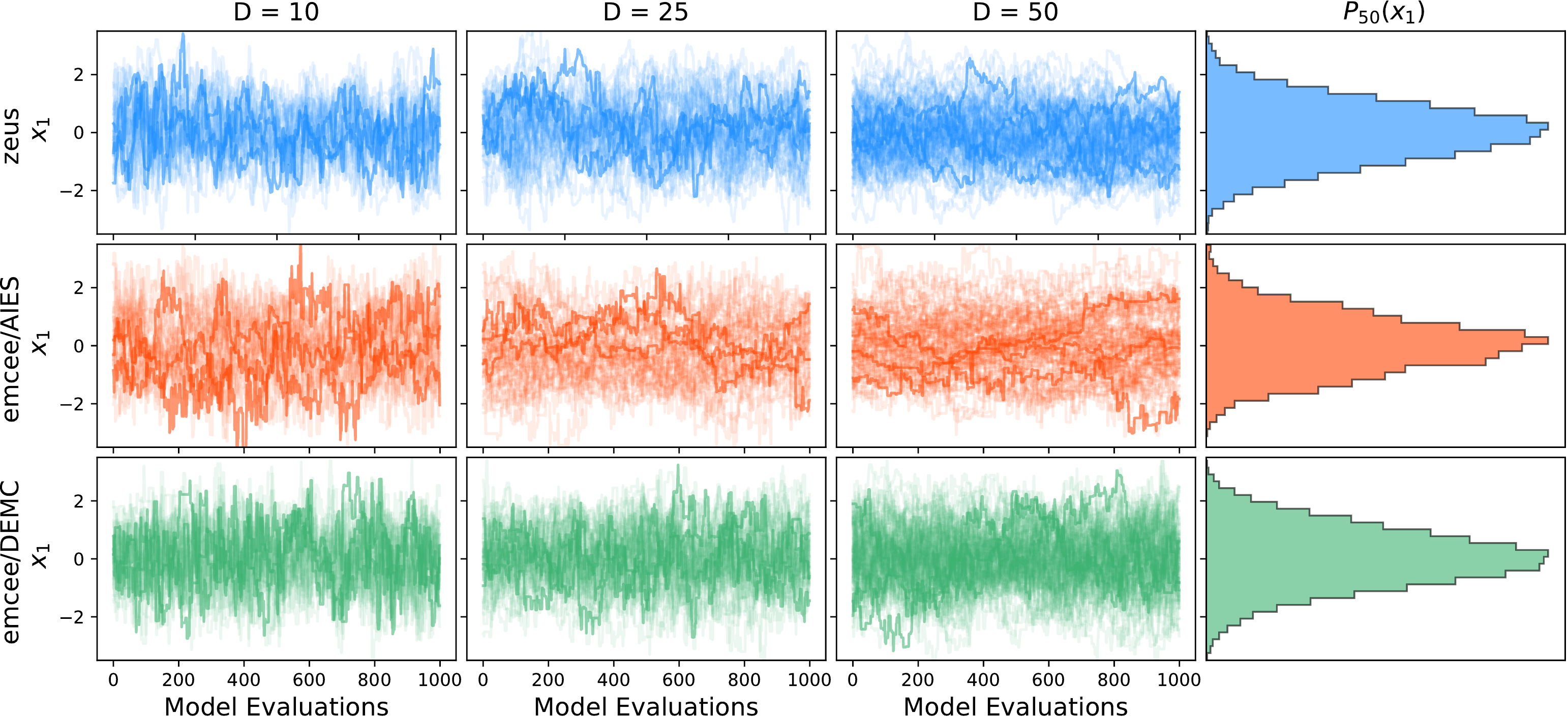}}
    \caption{The figure shows numerical results (i.e. walker trajectories/chains for the first parameter) demonstrating the performance of the three ensemble MCMC methods in the case of a normal (Gaussian) target distribution in $10, 25$ and $50$ dimensions respectively. The last column illustrates the 1-D marginal posterior corresponding to the first parameter $x_{1}$ estimated directly from the samples for the 50-dimensional case.}
    \label{fig_zeus:gaussian}
\end{figure*}
%\end{landscape}
Starting with the normal target distribution it is important to note here that all three of the methods used in the comparison are affine--invariant\footnote{Differential evolution Metropolis is only approximately affine--invariant due to the jitter that it is often added to its proposal. This however has a negligible effect.}, meaning that their performance is immune to any linear correlations between the parameters. Since the normal distribution incorporates, by construction, only linear correlations (i.e. the 2D marginal distribution contours look like ellipses), it is the perfect testing ground to assess the effect that high dimensionality has on the three methods independently of other complications. For our example, we used a zero-mean normal distribution with a covariance matrix in which the diagonal elements are set to $1$ and the off-diagonal ones are equal to $0.95$. We then proceed by sampling the aforementioned distribution in $10$, $25$ and $50$ dimensions. Based on Figure \ref{fig_zeus:gaussian} one can see that the walkers of \texttt{emcee}/AIES dissolve into an inefficient random walk characterised by low step size and high autocorrelation time as the number of parameters increases. \texttt{zeus} and \texttt{emcee}/DEMC are not so severely affected by the high number of parameters exhibiting a substantially lower autocorrelation.

\begin{figure}[ht!]
    \centering
	\centerline{\includegraphics[scale=0.65]{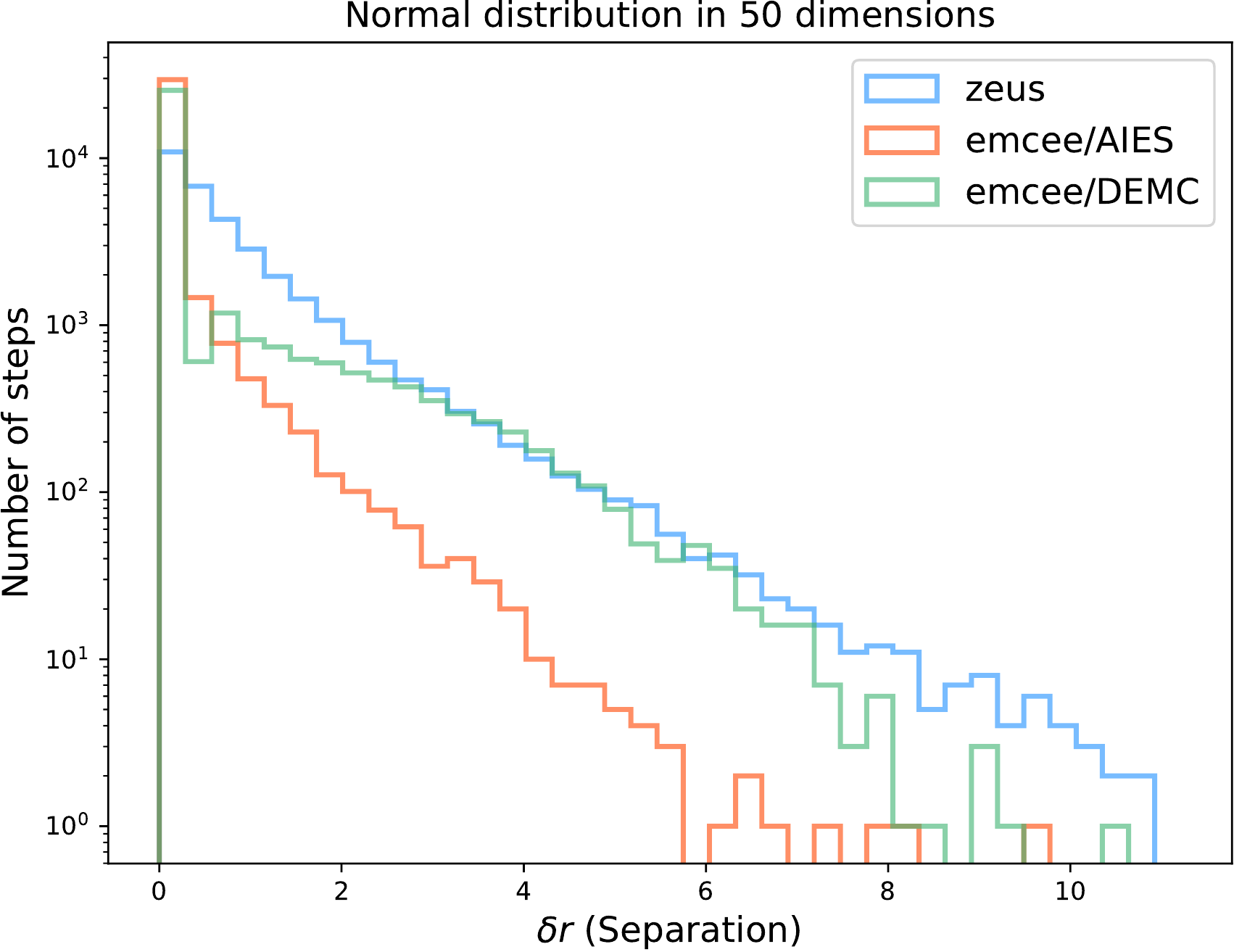}}
    \caption{This figure shows the distribution of step sizes of walkers for the three different samplers in the case of a normal (Gaussian) target distribution in $D=50$. It is important to note here that both \texttt{emcee} algorithms exhibit a peak at zero separation; \texttt{zeus} on the other hand does not due to its non-rejection nature.}
    \label{fig_zeus:gaussian_sep}
\end{figure}
Let us now try to explain this difference in behaviour by looking into the distribution of the steps of the walkers in Figure \ref{fig_zeus:gaussian_sep}. One thing to notice here is that the distribution of the steps of \texttt{zeus}'s walkers extends significantly further away than those of \texttt{emcee}/AIES and \texttt{emcee}/DEMC. This should come as no surprise since the construction of the approximate slice allows for larger steps than Metropolis updates as shown in Table \ref{tab_zeus:table1}. This is because when a proposal is rejected in slice sampling the approximate slice shrinks and another sample is proposed instead. In this way, \texttt{zeus}'s walkers always move and the chance of staying fixed is zero -- unlike MH-based updates in which frequent rejection of samples is a necessity. This aforementioned procedure leads to greater steps in parameter space. The difference between \texttt{emcee}/AIES and \texttt{emcee}/DEMC is attributed to the fact that DEMC uses a proposal scale\footnote{The proposal scale $\gamma$ is similar to $\mu$ used in ESS in the sense that its value determines the length scale of the proposed jumps in parameter space. A high value would lead to large steps that are often rejected and a low value would lead to small steps that are often accepted but do not carry the walkers far. For such methods, a balance must be found.} $\gamma = 2.38 / \sqrt{D}$ that guarantees a constant acceptance rate accounting for the number of dimensions $D$. This proposal scale is however optimal only in the case of a normal target distribution such as the one that we are studying here and there is no guarantee that it would return acceptable results in non-Gaussian distributions. For the case of \texttt{emcee}/AIES, the relevant proposal scale $\gamma$ is allowed to vary in the range between $1/\alpha$ and $\alpha$ where $\alpha = 2$ is often taken as the typical value. It is clear that in the latter case $\gamma$ does not possess the desired scaling $\gamma\propto 1/\sqrt{D}$ and thus, although the method generates proposals in the right overall direction, most of the samples do not reside in the typical set \parencite{speagle2019conceptual}. In other words, the lack of proper scaling of the proposal scale with the number of dimensions leads to \texttt{emcee}/AIES ``overshooting'' the typical set where most of the posterior mass is located.

%\begin{landscape}
\begin{figure}[!ht]
    \centering
	\centerline{\includegraphics[scale=0.45]{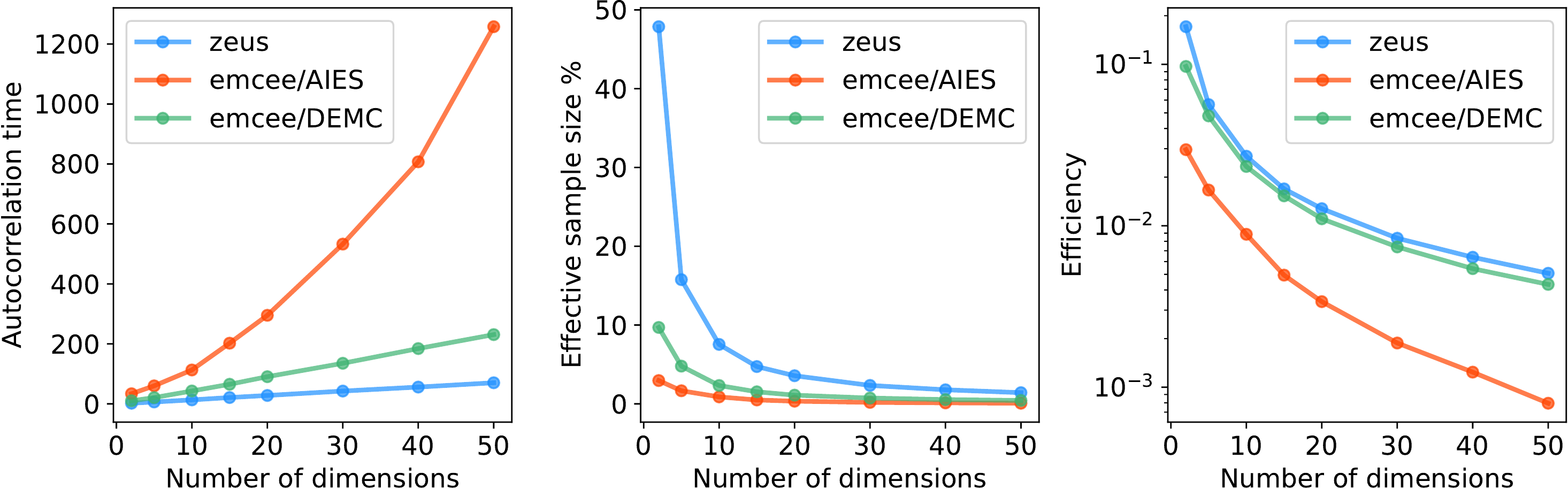}}
    \caption{The figure shows numerical estimates of the integrated autocorrelation time (number of steps along a chain required to obtain an independent sample; left panel), the effective sample size (percentage of effectively independent samples in a chain; middle panel), and the sampling efficiency (i.e. effective sample size per model evaluation; right panel) for a normal target distribution and varying number of dimensions. The number of walkers was set to $4\times D$ for \texttt{zeus} and $16\times D$ for \texttt{emcee}, this was the optimal choice (i.e. the one maximising the efficiency for the given dimensionality) for each sampler. \texttt{zeus} and \texttt{emcee/DEMC} exhibit linear scaling of the autocorrelation time with the number of dimensions whereas \texttt{emcee/AIES} scales exponentially.}
    \label{fig_zeus:act}
\end{figure}
%\end{landscape}
We can also draw some useful insights about the sampling efficiency of those samplers and their scaling with the number of dimensions by estimating the integrated autocorrelation time of the chains. Given the autocorrelation time, we can also estimate the effective sample size as the percentage of effectively independent samples in a chain. By dividing the effective sample size by the computational cost of each method we can then estimate the sampling efficiency. The results of such a comparison are shown in Figure \ref{fig_zeus:act}. We immediately notice here that the autocorrelation times of \texttt{zeus} and \texttt{emcee}/DEMC scale linearly with the number of dimensions, whereas the autocorrelation time of \texttt{emcee}/AIES scales exponentially. The computational cost of \texttt{zeus} per iteration per walker, although somewhat higher than that of \texttt{emcee}, does not vary with the number of dimensions. This means that in high dimensions, \texttt{zeus} dominates over \texttt{emcee}/AIES in terms of sampling efficiency.

\begin{table}[ht!]
    \centering
    \caption{The table shows a comparison of \texttt{emcee}/AIES, \texttt{emcee}\allowbreak/DEMC and \texttt{zeus} in terms of the expected squared jump distance (ESJD; higher is better) for the five toy examples i.e. $50$-$D$ normal distribution, $25$-$D$ ring distribution, $25$-$D$ Gaussian mixture, $25$-$D$ Student's $t$-distribution, and $25$-$D$ truncated normal distribution.}
    \def\arraystretch{1.1}
    \begin{tabular}{lcccc}
        \toprule[0.75pt]
         & \texttt{emcee}/AIES   & \texttt{emcee}/DEMC   & \textbf{zeus}  \\
        \midrule[0.5pt]
        Normal    &    $0.5288$    &    $1.1162$    &   $\mathbf{2.1354}$   \\
        Ring   &    $0.0043$   &    $0.0006$    & $\mathbf{0.1257}$  \\
        Mixture   &    $0.0037$    &    $0.0056$    &  $\mathbf{0.1015}$ \\
        Student   &    $12.9124$    &    $2.4137$    &  $\mathbf{23.5720}$ \\
        Truncated   &    $0.0940$    &    $0.3501$    &  $\mathbf{0.5882}$ \\
        \bottomrule[0.75pt]
        \end{tabular}
    \label{tab_zeus:table1}
\end{table}

The above discussion allows us to clearly state a crucial distinction between the three methods, which is their response to the curse of dimensionality. As the number of dimensions increases, the probability mass of a distribution is concentrated into a thin shell within the tails of the distribution (i.e. the typical set). To account for this and maintain its efficiency, a sampling method has to adjust its proposal scale -- otherwise, the proposals will not be located in the typical set and thus they will not be accepted. The three methods that we mentioned so far deal with this in different ways. \texttt{emcee}/AIES's proposal scale is not adjusted and thus its proposals become increasingly inefficient in high dimensions. \texttt{emcee}/DEMC's proposal scale is adjusted based on the theoretical expectation for the case of the normal target distribution. Although both \texttt{emcee} methods perform well in this example, their sub-optimal scaling will degrade their performance in non-Gaussian target distributions as we will demonstrate in the next toy example. Finally, \texttt{zeus}'s proposal scale is continuously adapted, as the slice expands and contracts in every iteration, thus guaranteeing optimal scaling. \textcite{huijser2017properties} found that the suboptimal scaling of \texttt{emcee}/AIES with the number of dimensions can introduce biases into the expectation values derived from the chains in high dimensions that are hard to diagnose. The locally adaptive nature of \texttt{zeus} allows it to avoid this problem by adjusting its proposals accordingly.

\begin{figure}[ht!]
    \centering
	\centerline{\includegraphics[scale=0.45]{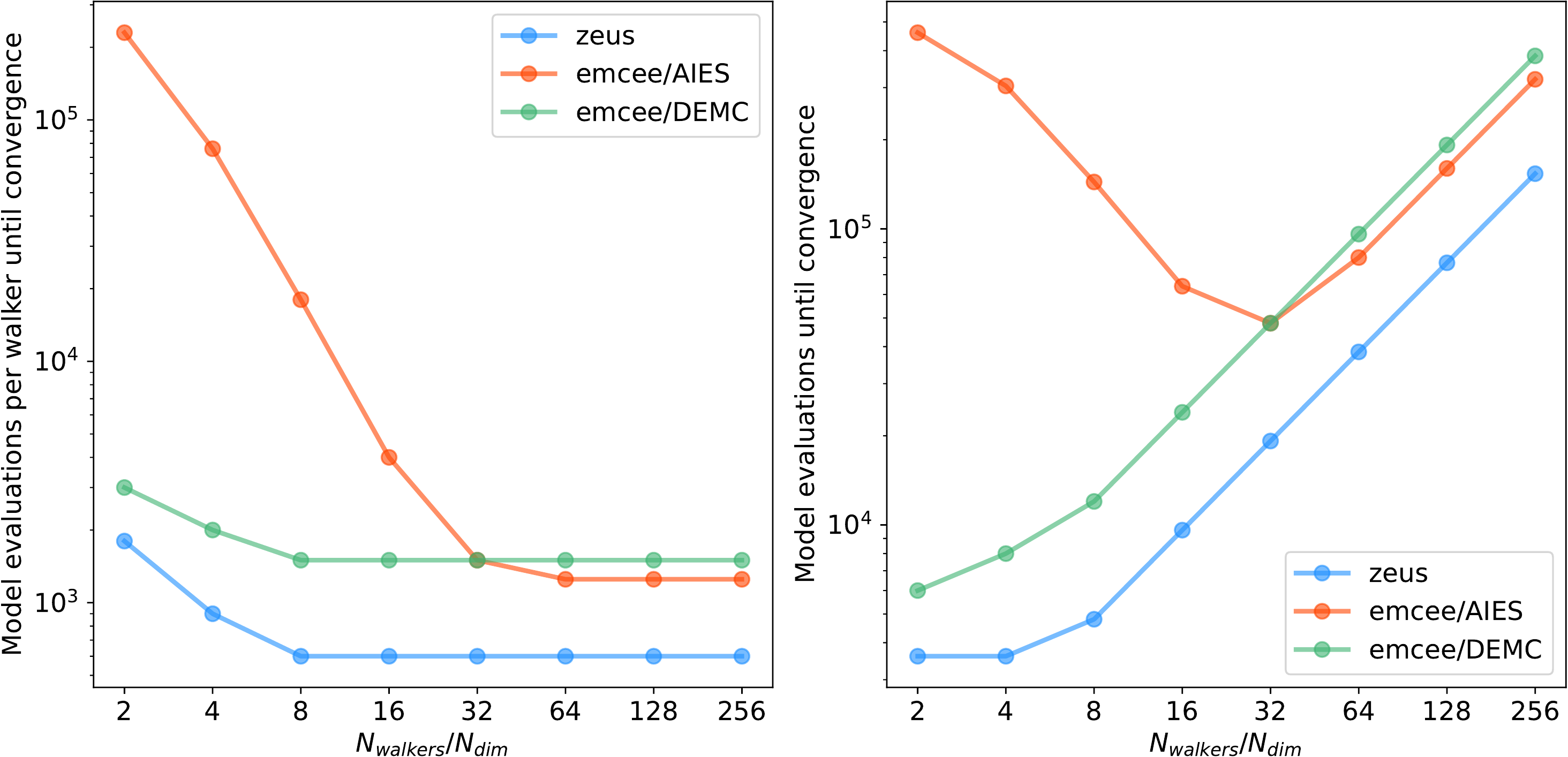}}
    \caption{The figure shows the computational cost until convergence is reached in terms of the number of model evaluations for the different ensemble samplers for a highly correlated 25--dimensional normal distribution. The left panel shows the computational cost for a single walker. From this we can see that the cost for a single walker decreases as we increase the number of walkers until it reaches a plateau. The high computational cost for low numbers of walkers can be attributed to the low variety or sparsity of possible proposals; this is significantly higher for \texttt{emcee}/AIES. The right panel takes into account the linear scaling of the total computational cost as we increase the number of walkers and shows the total computational cost for the whole ensemble until it converges.}
    \label{fig_zeus:convergence}
\end{figure}
Another kind of analysis we can perform is to use the highly correlated 25--dimensional normal distribution as the target distribution and estimate the convergence rate of the three samplers. Although simple, the normal distribution is a valid approximation of many realistic astronomical posterior distributions and as such we expect the results presented in this paragraph to be applicable to a wide range of other distributions that resemble the normal distribution to some extent. We acknowledge however that the \emph{no free lunch} theorem also applies to this case, and there are bound to be cases in which the results would be qualitatively different. That being said, we initialised the walkers from a compact normal distribution (i.e. standard deviation equal to $10^{-4}$ times that of the target distribution) centred around a point along the first axis of the parameter space at a distance of $100$ standard deviations from the mode. We then measured the number of model evaluations required until the samplers have converged to the target distribution. The results for varying number of walkers are presented in Figure \ref{fig_zeus:convergence}.

\begin{figure}[!ht]
    \centering
	\centerline{\includegraphics[scale=0.65]{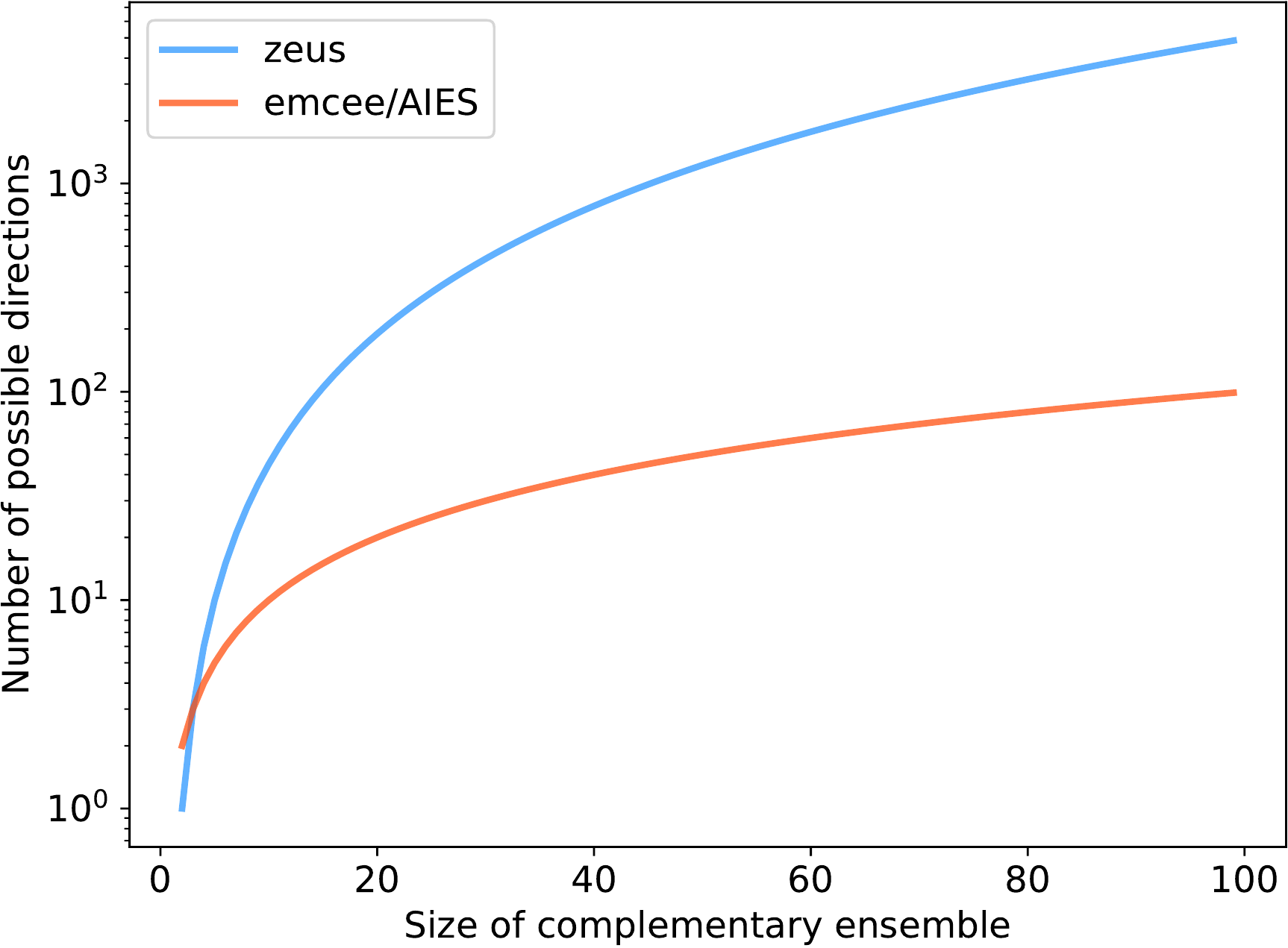}}
    \caption{The figure shows the number of possible directions along which \texttt{zeus} and \texttt{emcee}/AIES can propose new samples as a function of the number of walkers in the complementary ensemble. \texttt{emcee}/DEMC exhibits the same number of proposals as \texttt{zeus} and it is not plotted here. \texttt{zeus} has a much higher variety of possible directions compared to \texttt{emcee}/AIES for any given number of walkers, assuming that that number is greater than $2$.}
    \label{fig_zeus:proposals}
\end{figure}

In general, walkers move along directions defined by the walkers of the complementary ensemble. Thus, increasing the number of walkers offers a wider variety of available directions along which the walkers of \texttt{zeus} or \texttt{emcee} can move via slice sampling or Metropolis updates respectively. This is demonstrated in the left panel of Figure \ref{fig_zeus:convergence} in which the computational cost until convergence (i.e. number of model evaluations) for a single walker diminishes and then reaches a plateau as the number of walkers is increased. We notice however that, at the level of a single walker, the computational cost of \texttt{emcee}/AIES is significantly higher compared to that of either \texttt{zeus} or \texttt{emcee}/DEMC. This is due to the way that different samplers choose the directions along which walkers move. In particular, both \texttt{zeus} and \texttt{emcee}/DEMC define a direction vector as the difference between two walkers from the complementary ensemble, thus two walkers are required to define a direction. On the other hand, \texttt{emcee}/AIES requires only a single walker from the complementary ensemble as the direction is defined by the difference between the updated walker and the complementary one. This stark contrast between the way those samplers choose their direction vectors lies at the heart of the difference in the computational cost of \texttt{emcee}/AIES as compared to \texttt{zeus} and \texttt{emcee}/DEMC in the limit of low number of walkers. In order to dive a little deeper into this, we can compute the exact number of possible directions for all three methods. Since \texttt{emcee}/AIES requires only a single walker from the complementary ensemble the number of available directions is equal to the size of the complementary ensemble. On the other hand, \texttt{zeus}'s and \texttt{emcee}/DEMC's requirement for a pair of walkers means that the number of available directions is equal to $\binom{n}{2}$, meaning the 2--combination from a set of $n$ walkers that comprise the complementary ensemble. Clearly, as shown in Figure \ref{fig_zeus:proposals}, the latter increases faster with the size of the complementary ensemble, thus explaining the larger variety of possible directions available in the case of \texttt{zeus} and \texttt{emcee}/DEMC compared to \texttt{emcee}/AIES.

The discussion so far was about the computational cost of convergence in terms of the number of model evaluations for a single walker. Of course, the ensemble of walkers consists by definition of more than a single walker. Therefore, in order to compute the total number of model evaluations required until the ensemble converges we need to multiply the results of the single walker with the total number of walkers. Those results are presented in the right panel of Figure \ref{fig_zeus:convergence}. From this plot we can see that both \texttt{zeus} and \texttt{emcee}/DEMC converge faster when the number of walkers is close to its minimum value i.e. $2\times D$. \texttt{emcee}/AIES on the other hand prefers a higher number of walkers (i.e. $32\times D$) in order to overcome the sparsity of available directions in the limit of low number of walkers. This, however, means that even if we choose the optimal number of walkers for \texttt{emcee}/AIES it would still converge slower than either \texttt{zeus} or \texttt{emcee}/DEMC. Furthermore, we cannot know \emph{a priori} the optimal number of walkers for \texttt{emcee}/AIES unlike for \texttt{zeus} and \texttt{emcee}/DEMC in which the optimal size of the ensemble is close to $2\times D$. Finally, the faster convergence of \texttt{zeus} compared to \texttt{emcee}/DEMC can be attributed to the local adaptation that the former performs by extending the length of the slice and thus allowing larger steps in parameter space.

\subsubsection{The ring distribution}

\begin{figure}[!ht]
    \centering
	\centerline{\includegraphics[scale=0.45]{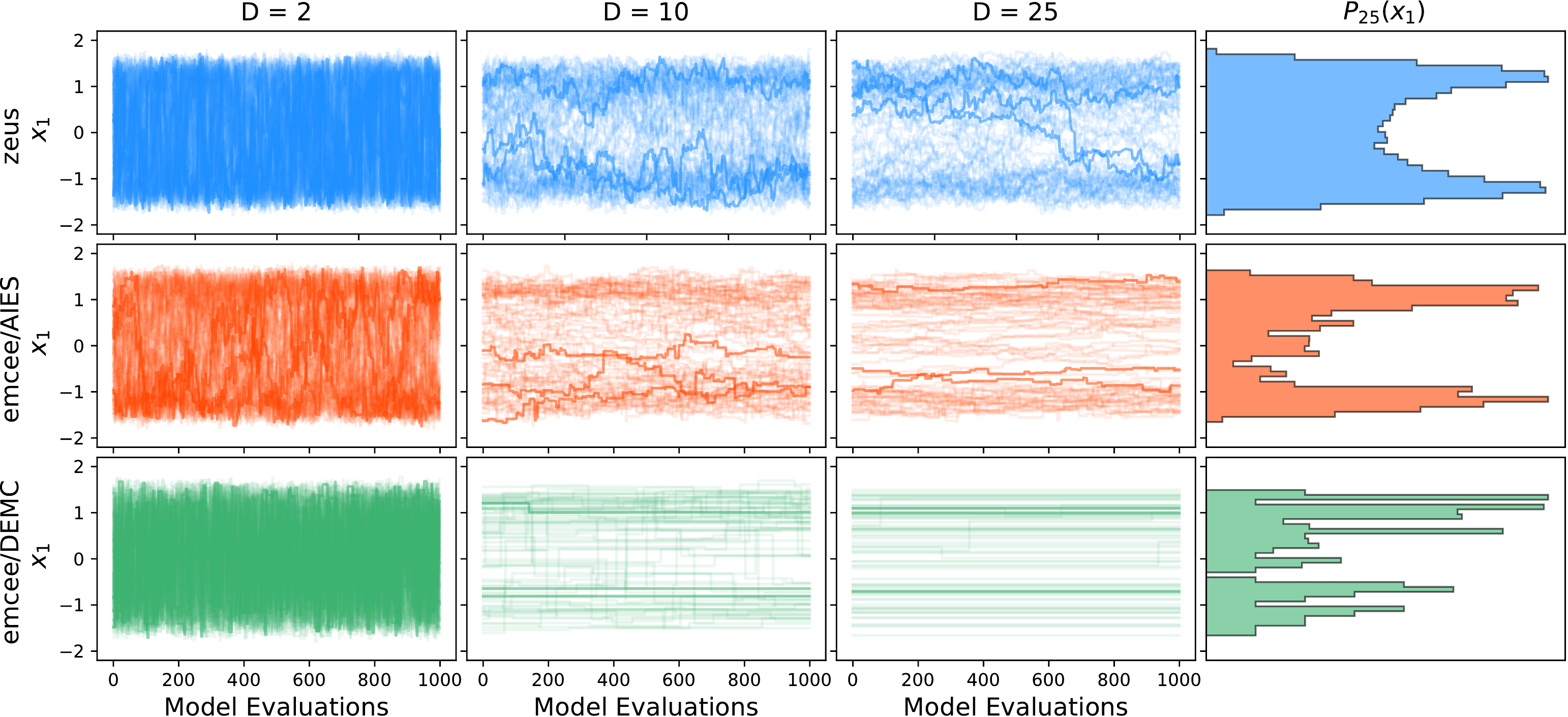}}
    \caption{The figure shows numerical results (i.e. walker trajectories/chains for the first parameter) demonstrating the performance of the three ensemble MCMC methods in the case of the ring target distribution in $2, 10$ and $25$ dimensions respectively. The last column illustrates the 1-D marginal posterior corresponding to the first parameter $x_{1}$ estimated directly from the samples for the 25-dimensional case. One can notice here that in $10$ and $25$ dimensions both \texttt{emcee} methods mix very slowly. In the 25-dimensional case almost all of \texttt{emcee}/DEMC's walkers are unable to move and the autocorrelation time is effectively infinite.}
    \label{fig_zeus:ring}
\end{figure}

The ring distribution defined as
\begin{equation}
\label{eq_zeus:ring}
    \ln P (x) = - \Bigg[ \frac{(x_{n}^{2} + x_{1}^{2} - a)^{2}}{b}\Bigg]^{2}  -\sum_{i=1}^{n-1} \Bigg[ \frac{(x_{i}^{2} + x_{i+1}^{2} - a)^{2}}{b}\Bigg]^{2} ,
\end{equation}
where $a=2$, $b=1$ and $n$ is the total number of parameters; this is an artificial target distribution that exhibits strong non-linear correlations between its parameters. This aspect of the ring distribution allows us to demonstrate the locally adaptive nature of \texttt{zeus}. Whereas \texttt{emcee}/AIES and \texttt{emcee}/DEMC use a single global proposal scale for all regions of the parameter space, \texttt{zeus} has the ability to adjust its proposal scale locally by expanding the slice appropriately. As expected, this will allow \texttt{zeus} to sample efficiently even in cases in which strong non-linear correlations are present. Looking at Figure \ref{fig_zeus:ring} one can see that \texttt{zeus} manages to generate multiple samples efficiently even in high dimensions. On the other hand, \texttt{emcee}/AIES and \texttt{emcee}/DEMC do not efficiently produce valid proposals: for \texttt{emcee}/AIES this leads to an inefficient random walk, characterised by small steps; for \texttt{emcee}/DEMC the acceptance rate almost vanishes beyond $D=2$. The expected squared jump distance of each method for the case of $D=25$ is shown in Table \ref{tab_zeus:table1}. It is important to note here that out of the three samplers only \texttt{zeus} manages to converge in all three cases (i.e. in 2, 10 and 25 dimensions). \texttt{emcee}/AIES and \texttt{emcee}/DEMC on the other hand converge successfully only in 2 dimensions.
%\begin{landscape}
%\end{landscape}

\begin{figure}[ht!]
    \centering
	\centerline{\includegraphics[scale=0.65]{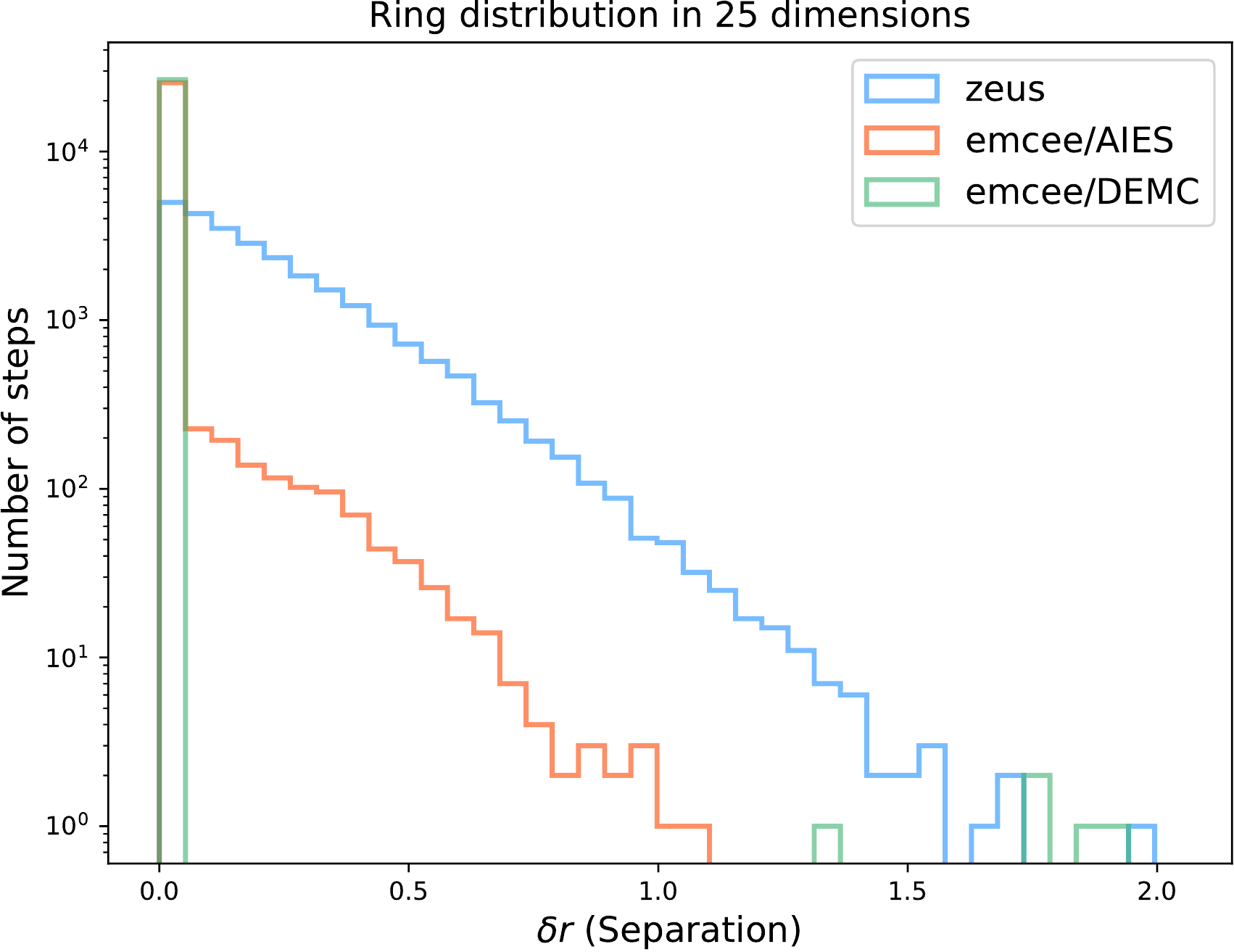}}
    \caption{This figure shows the distribution of step sizes of walkers for the three different samplers in the case of a ring target distribution in $D=25$. It is important to note here that both \texttt{emcee} algorithms exhibit a peak at zero separation; \texttt{zeus} on the other hand does not. The existence of the zero-peak in \texttt{emcee} is due to the high number of rejected proposals (i.e. low acceptance rate).}
    \label{fig_zeus:ring_sep}
\end{figure}

To explain this result one only has to look at the distribution of walker steps of the different methods at Figure \ref{fig_zeus:ring_sep}. \texttt{zeus}'s steps extend to large distances in parameter space whereas most of \texttt{emcee}/AIES's and \texttt{emcee}/DEMC's steps are rejected (i.e. shown as zero in the histogram). We can see that \texttt{emcee}/DEMC manages to perform some long distance steps but those are few and there is almost nothing in between. It is clear from this and the previous toy examples that the $\gamma = 2.38/\sqrt{D}$ scaling of \texttt{emcee}/DEMC's scale factor does not generalise well beyond the Gaussian case.

\subsubsection{The two-component Gaussian mixture distribution}

One other important aspect of astronomical posterior distributions is the fact that many of them exhibit multiple peaks. Multimodality can arise either from non-linear models or sparse and uninformative data. In either case, multimodal target distributions present a formidable challenge for most MCMC methods. Perhaps the simplest example of such a distribution is the two-component Gaussian mixture. In this example we will position the two, equal-mass, components at $\mathbf{-0.5}$ and $\mathbf{+0.5}$ respectively with standard deviation of $0.1$. Sampling from multimodal distributions requires two types of proposals, local proposals that sample different modes individually and global proposals that transfer walkers from one mode to the other. For this reason we will make use of \texttt{zeus}'s \texttt{GlobalMove} that uses a Dirichlet Process Gaussian Mixture model of the ensemble to efficiently propose between-mode and within-mode steps.

\begin{figure}[!ht]
    \centering
	\centerline{\includegraphics[scale=0.45]{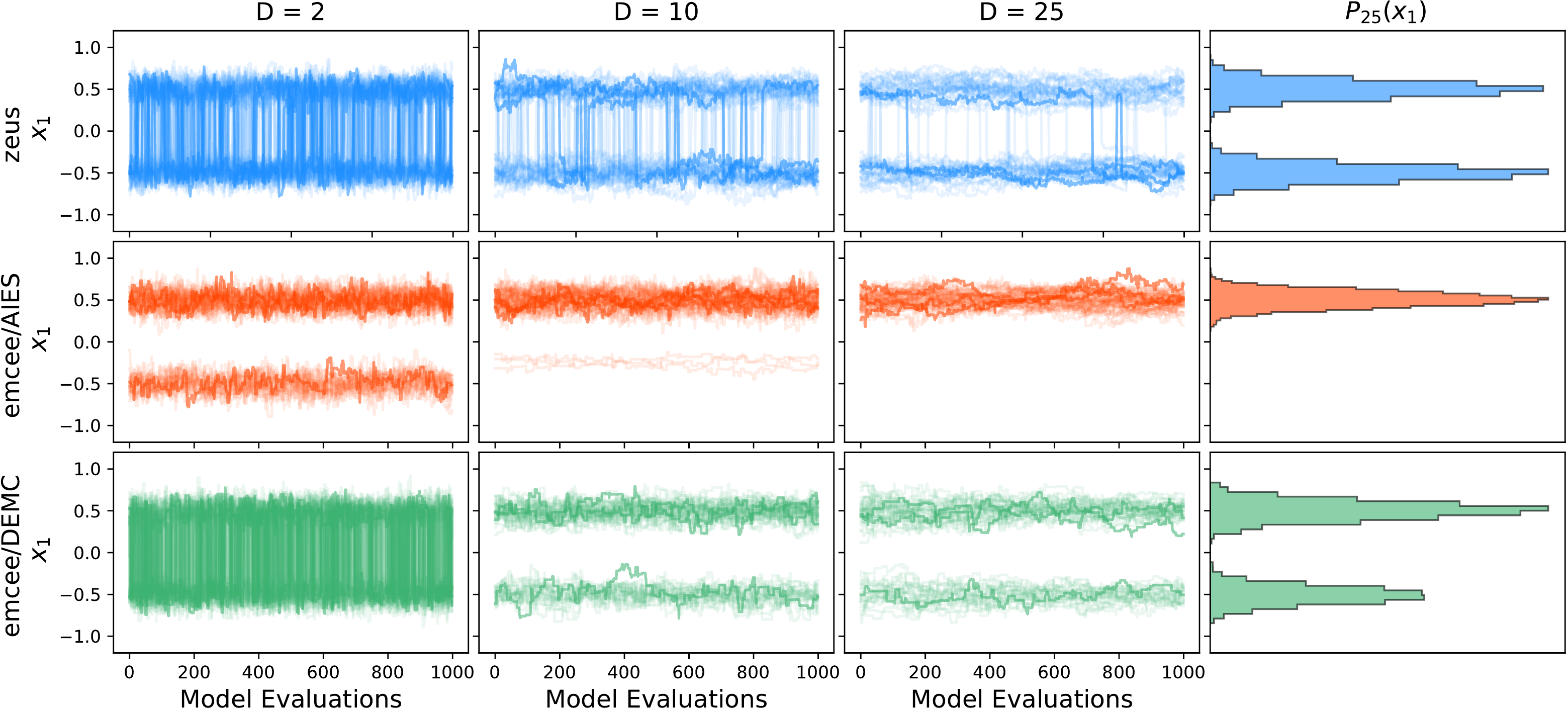}}
    \caption{The figure shows numerical results (i.e. walker trajectories/chains for the first parameter) demonstrating the performance of the three ensemble MCMC methods in the case of a two-component  Gaussian mixture target distribution in $2, 10$ and $25$ dimensions respectively. The last column illustrates the 1-D marginal posterior corresponding to the first parameter $x_{1}$ estimated directly from the samples for the 25-dimensional case. Whereas all three samplers make valid within-mode proposals, it is only \texttt{zeus} that manages to perform between-mode jumps and thus sample correctly from the target distribution in the 10 and 25-dimensional cases. Between-mode jumps are paramount in order to distribute the probability mass correctly between different modes.}
    \label{fig_zeus:bimodal}
\end{figure}

\begin{figure}[ht!]
    \centering
	\centerline{\includegraphics[scale=0.65]{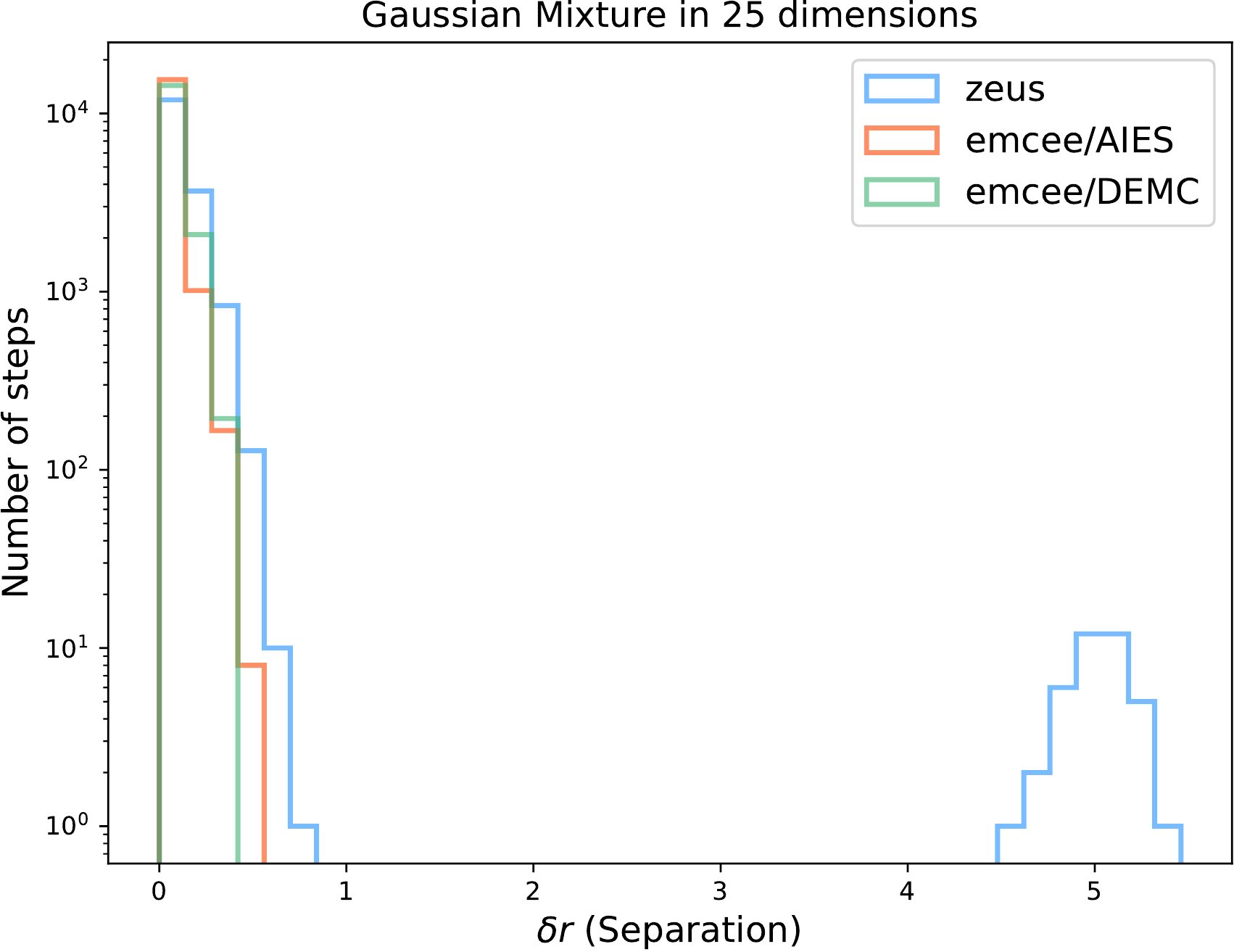}}
    \caption{This figure shows the distribution of step sizes of walkers for the three different samplers in the case of a two-component  Gaussian mixture target distribution in $D=25$. It is important to note here that both \texttt{emcee} algorithms exhibit a peak at zero separation; \texttt{zeus} on the other hand does not due to its non-rejection basis.}
    \label{fig_zeus:bimodal_sep}
\end{figure}

As seen in Figure \ref{fig_zeus:bimodal}, \texttt{zeus}'s walkers manage to move from one mode to the other frequently enough for mixing to be efficient even in the $D=25$ case. Out of \texttt{emcee}/AIES and \texttt{emcee}/DEMC, only the latter proposes valid steps from one mode to the other in the $D=2$ case. As for the $D=25$ case, one can see in Figure \ref{fig_zeus:bimodal_sep} that \texttt{zeus}'s walkers perform numerous jumps whereas \texttt{emcee}'s walkers are unable to do so. The ability of the walkers to jump from mode to mode is of paramount importance if we want to sample correctly from the target distribution. Lack of such proposals will lead to an improper probability mass ratio between the two modes and thus biased inference. The expected squared jump distance of each method for the case of $D=25$ is shown in Table \ref{tab_zeus:table1}.
%\begin{landscape}
%\end{landscape}

Clustering-based proposals have also been applied to MH-type ensemble MCMC methods but as shown in \textcite{karamanis2021ensemble}, they fail to generate valid proposals in problems with moderate number of dimensions. The reason is, as discussed in Section \ref{sec_zeus:tests}, that MH has to propose a valid point in the other mode. In other words, whereas Ensemble Slice Sampling only needs to determine the direction of the other mode relative to the chosen walker correctly, MH needs to guess both the direction and the distance, a task that rapidly becomes very hard as the number of dimensions rises.

\subsubsection{The Student's $t$-distribution}

The fourth toy example tests the case in which the target distribution is characterised by heavy-tails. In order to demonstrate \texttt{zeus}'s ability to sample efficiency is such cases we chose to use the multivariate Student's $t$-distribution with $2$ degrees of freedom. The aforementioned density exhibits heavier tails than a normal distribution which means that it is more likely to produce samples that are far away from the mean. The $t$-distribution arises when estimating the mean of a normally distributed sample with unknown standard deviation and small size. The probability density function of a $p$--dimensional Student's $t$-distribution with $\nu$ degrees of freedom is given by:
\begin{equation}
    \label{eq_zeus:student}
    P(x) = \frac{\Gamma[(\nu+p)/2]}{\Gamma(\nu/2)\nu^{p/2}\pi^{p/2}|\boldsymbol{\Sigma}|^{1/2}}\exp\bigg[ 1 + \frac{1}{\nu}(\mathbf{x}-\boldsymbol{\mu})^{T}\boldsymbol{\Sigma}^{-1}(\mathbf{x}-\boldsymbol{\mu})\bigg]^{-\frac{(\nu+p)}{2}}\,,
\end{equation}
where $\boldsymbol{\Sigma}$ is the $p\times p$ positive semi-definite shape matrix and $\boldsymbol{\mu}$ is the mean vector.
%\begin{landscape}
\begin{figure}[!ht]
    \centering
	\centerline{\includegraphics[scale=0.45]{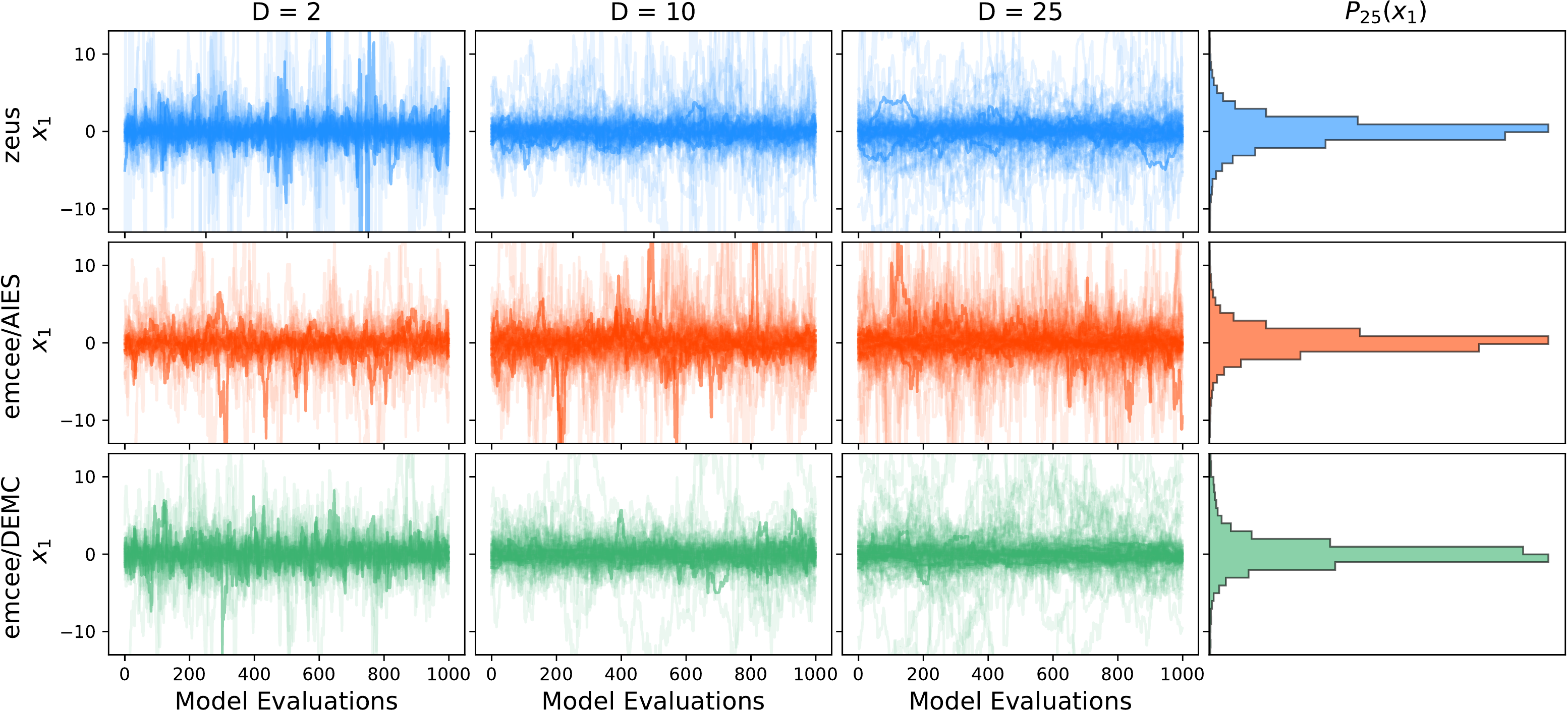}}
    \caption{The figure shows numerical results (i.e. walker trajectories/chains for the first parameter) demonstrating the performance of the three ensemble MCMC methods in the case of the Student's $t$-distribution with $2$ degrees of freedom in $2, 10$ and $25$ dimensions respectively. The last column illustrates the 1-D marginal posterior corresponding to the first parameter $x_{1}$ estimated directly from the samples for the 25-dimensional case.}
    \label{fig_zeus:student}
\end{figure}
%\end{landscape}

\begin{figure}[ht!]
    \centering
	\centerline{\includegraphics[scale=0.65]{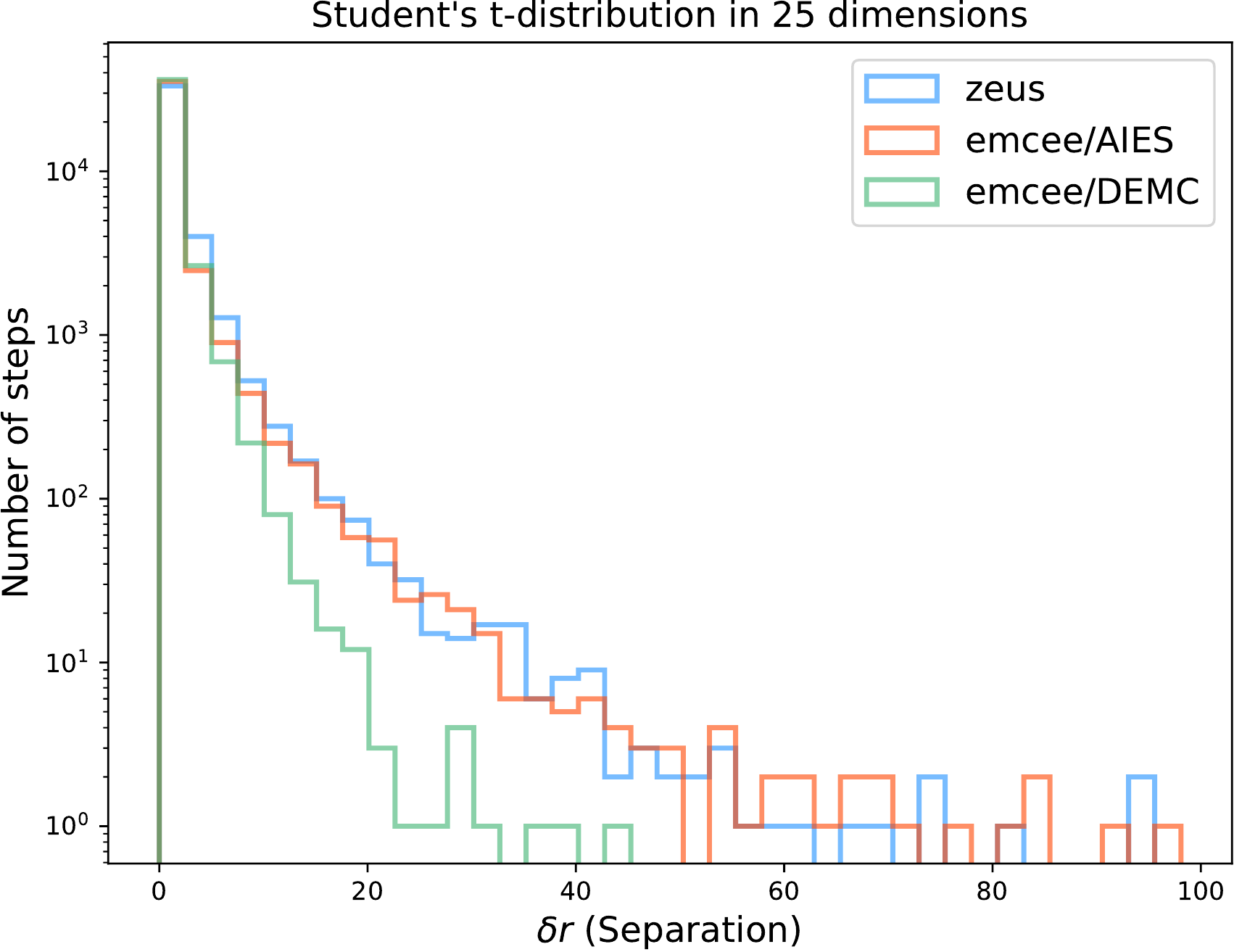}}
    \caption{This figure shows the distribution of step sizes of walkers for the three different samplers in the case of the Student's $t$-distribution with $2$ degrees of freedom in $D=25$. \texttt{zeus} and \texttt{emcee}/AIES exhibit similar distributions whereas \texttt{emcee}/DEMC performs shorter steps.}
    \label{fig_zeus:student_sep}
\end{figure}

We sampled the above distribution using the three samplers in $2$,  $10$ and $25$ dimensions respectively as shown in Figure \ref{fig_zeus:student}. The diagonal elements of shape matrix $\boldsymbol{\Sigma}$ were set to $1$ and the off-diagonal elements to $0.95$. The mean vector $\boldsymbol{\mu}$ was set to $\mathbf{0}$. All three samplers managed to sample efficiently in $2$,  $10$ and $25$ dimensions as shown in Figure~\ref{fig_zeus:student} and Table~\ref{tab_zeus:table1}. Overall, \texttt{zeus} was the most efficient method with \texttt{emcee}/AIES being second and \texttt{emcee}/DEMC last. One can see from Figure \ref{fig_zeus:student_sep} that the distributions of steps of \texttt{zeus} and \texttt{emcee}/AIES are very similar whereas that of \texttt{emcee}/DEMC is substantially shorter. Unlike the previous toy examples in which the proposal strategy of \texttt{emcee}/AIES was causing it to overshoot the bulk of posterior mass, in the case of the heavy-tailed $t$-distribution more proposals are accepted. On the other hand, \texttt{emcee}/DEMC's proposals which are optimised for Gaussian targets are more conservative in the case of the $t$-distribution and they do not extend far away. As also demonstrated in the previous toy examples, the locally adaptive nature of \texttt{zeus} allows it to perform efficient proposals that span large distances in parameter space.

\subsubsection{The truncated normal distribution}

The fifth and final toy example tests the case in which the target distribution is bounded from below or above. We chose to employ a truncated normal distribution similar to the one used in the first toy example, with the additional constraint being that $x > 0$. This effectively introduces a hard boundary along all dimensions. One of the reasons that we study this distribution is to assess the bias introduced by the presence of the hard boundary.
%\begin{landscape}
\begin{figure}[!ht]
    \centering
	\centerline{\includegraphics[scale=0.45]{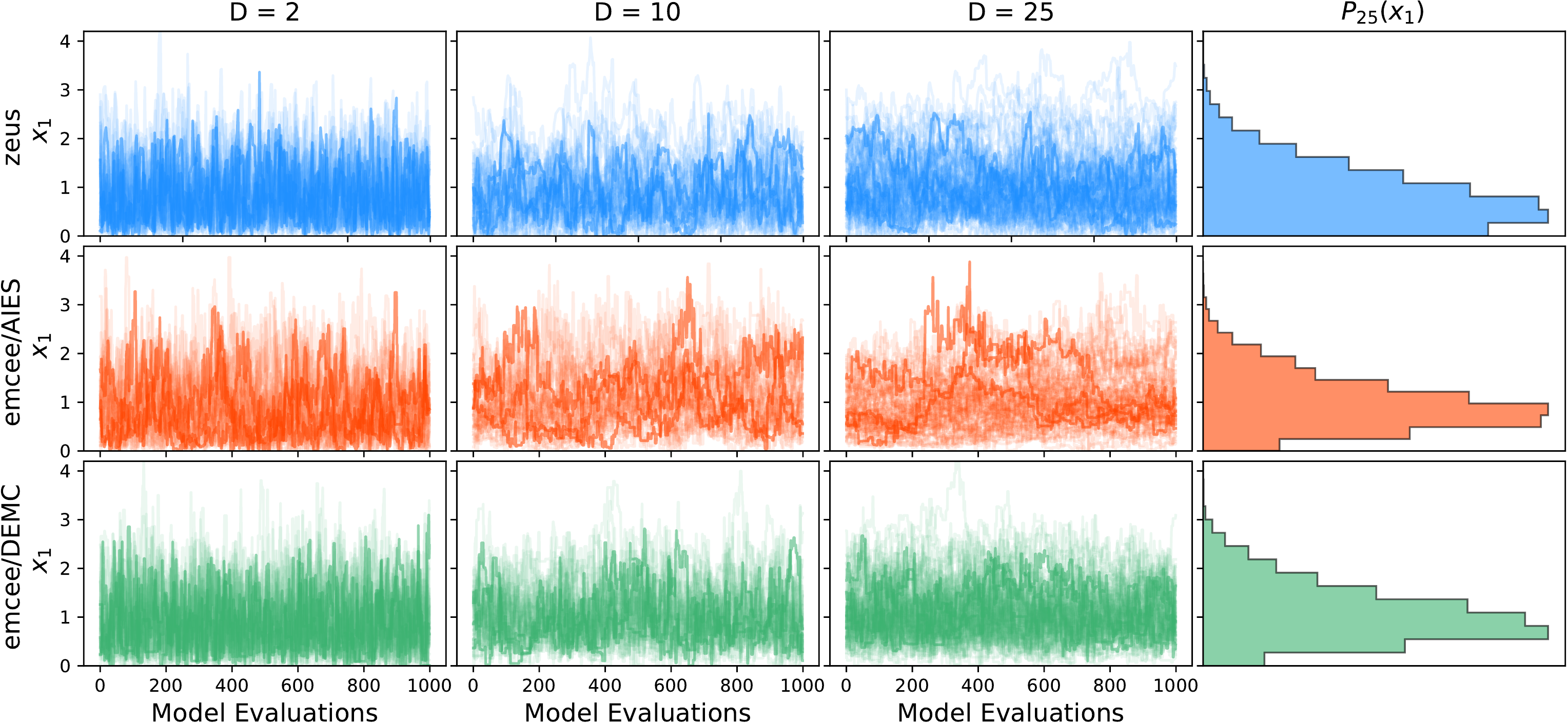}}
    \caption{The figure shows numerical results (i.e. walker trajectories/chains for the first parameter) demonstrating the performance of the three ensemble MCMC methods in the case of the truncated normal distribution in $2, 10$ and $25$ dimensions respectively. The last column illustrates the 1-D marginal posterior corresponding to the first parameter $x_{1}$ estimated directly from the samples for the 25-dimensional case. \texttt{zeus} exhibits the least amount of bias near the hard boundary at zero compared to \texttt{emcee}/AIES and \texttt{emcee}\allowbreak/DEMC.}
    \label{fig_zeus:truncated}
\end{figure}
%\end{landscape}

\begin{figure}[ht!]
    \centering
	\centerline{\includegraphics[scale=0.65]{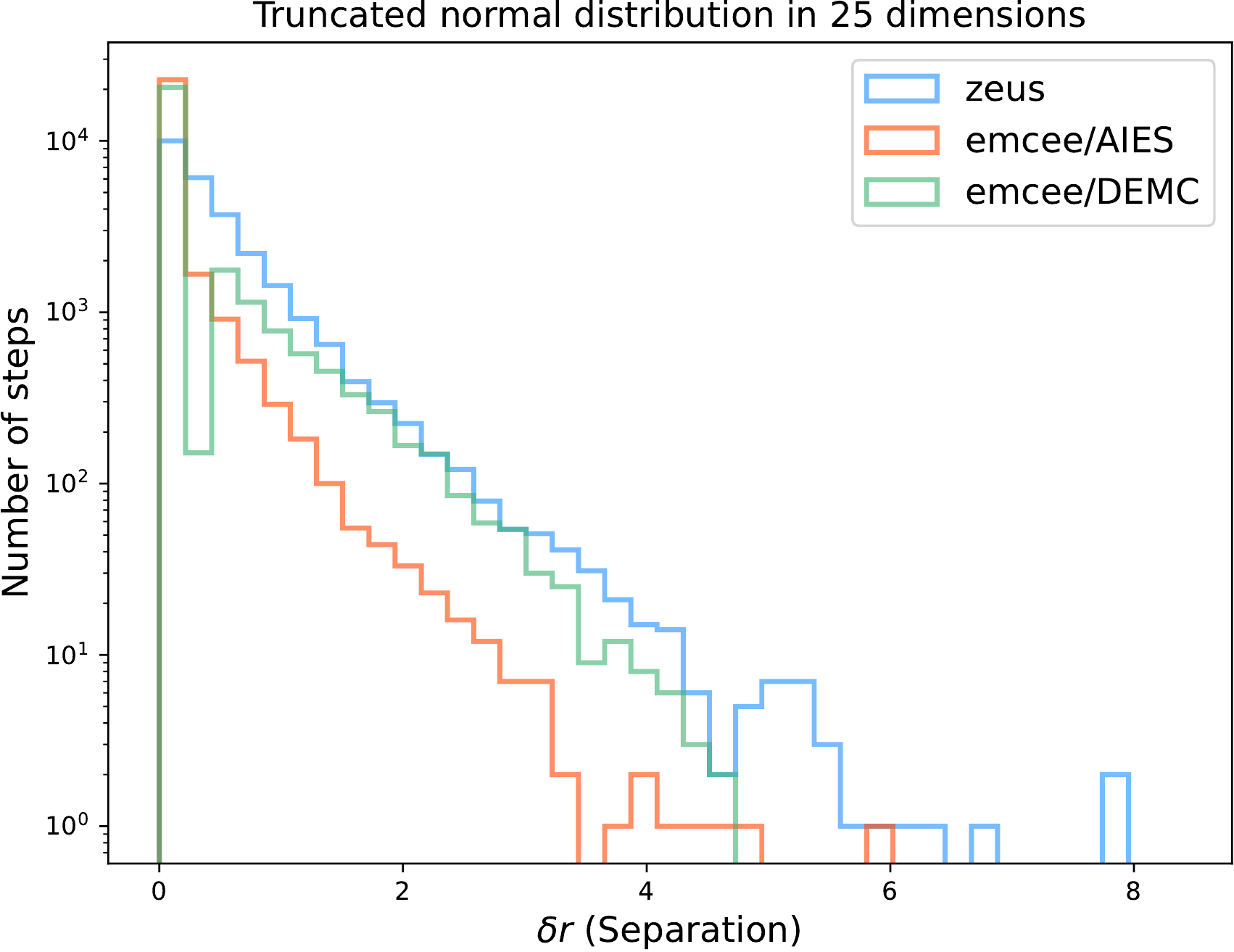}}
    \caption{This figure shows the distribution of step sizes of walkers for the three different samplers in the case of the truncated normal distribution in $D=25$. \texttt{zeus} and \texttt{emcee}/AIES exhibit similar distribution whereas \texttt{emcee}/DEMC performs shorter steps.}
    \label{fig_zeus:truncated_sep}
\end{figure}

We sampled the above distribution using the three samplers in $2$,  $10$ and $25$ dimensions respectively as shown in Figure \ref{fig_zeus:truncated}. The diagonal elements of the covariance matrix were set to $1$ and the off-diagonal to $0.95$. The mean vector $\boldsymbol{\mu}$ was set to $\mathbf{0}$. All three samplers managed to sample efficiently in $2$,  $10$ and $25$ dimensions as shown in Figure~\ref{fig_zeus:truncated} and Table~\ref{tab_zeus:table1}.  Overall, \texttt{zeus} was the most efficient method with \texttt{emcee}/DEMC being second and \texttt{emcee}/AIES last. One can see from Figure \ref{fig_zeus:truncated_sep} that the distributions of steps of \texttt{zeus} and \texttt{emcee}/AIES are very similar whereas that of \texttt{emcee}/AIES is slightly shorter. As shown in the right panels of Figure \ref{fig_zeus:truncated} \texttt{zeus} exhibits the least amount of bias compared to \texttt{emcee}/AIES and \texttt{emcee}/DEMC. In practical astronomical examples however, only one or two parameters would usually be bounded (e.g. the neutrino mass in galaxy clustering analyses) and thus unbiased sampling would be easier to perform by either of the three samplers.

\subsection{Real astronomical analyses}

The previous section employs toy examples in order to exhibit various scenarios that might emerge during sampling, and shows how \texttt{zeus} is better equipped to handle them. To demonstrate the efficiency of \texttt{zeus} compared to other samplers in realistic target distributions, we chose two common astronomical inference problems as the testing ground. Those are the cases of baryon acoustic oscillation (BAO) parameter inference and exoplanet parameter estimation.

We used the same three samplers in our comparison, namely \texttt{emcee} with AIES and DEMC, and of course \texttt{zeus}. We performed three distinct tests:

\begin{itemize}
    \item The first test was to estimate the efficiency for each sampler, defined as the number of independent samples produced per log-likelihood evaluation. To this end, we ran the MCMC procedure 5 times for each sampler and computed the mean efficiency using the estimated autocorrelation time of the chains. The autocorrelation time was estimated using the method presented in \textcite{karamanis2021ensemble}.
    \item The second test relates to the convergence rate of the three algorithms. As a measure of convergence rate, we adopt the inverse of the number of iterations required until all the convergence criteria specified below are met. In order to estimate the mean convergence rate we ran the sampling procedure 40 times for each sampler initialising the walkers close to the \textit{Maximum a Posteriori} (MAP) estimate. 
    \item Finally, we tested the sensitivity of the samplers to the initial conditions by running 40 realisations with the walkers initialised from a small sphere (of radius $10^{-4}$) around a randomly chosen point in the prior volume, counting how many of those attempts led to converged chains before a predetermined number of likelihood evaluations.
\end{itemize}

To determine whether a chain has converged we used four different metrics: the Gelman-Rubin split-$R$ statistic \parencite{gelman1992inference, gelman2013bayesian} using four independent ensembles of walkers; the Geweke test \parencite{geweke1992evaluating}; a minimum length of the chain as a multiple of the integrated autocorrelation time (IAT); as well as an upper bound on the rate of change of the IAT. Only the second half of the chains was used to evaluate the aforementioned criteria. The number of walkers used in both examples was close to the minimum value of $2\times D$ as specified below. As we will discuss in Section \ref{sec_zeus:discussion} this often leads to faster convergence.

\subsubsection{Cosmological inference}

The particular inference problem that we consider here is that of the anisotropic BAO parameter inference using estimates of the galaxy power spectrum. The data we used comes from the 12th data release (DR12) of the high-redshift North Galactic Cap (NGC) sample as observed by the Sloan Digital Sky Survey (SDSS) \parencite{SDSSIII} Baryon Oscillation Spectroscopic Survey (BOSS) \parencite{BOSS}. Our analysis follows closely that of \textcite{beutler2017clustering} with the difference that we chose not to fix any parameters and fit the hexadecapole multipole of the power spectrum as well as the monopole and quadrupole. Those choices were made solely to render the problem more challenging. Indeed the inclusion of the hexadecapole does not contribute any additional constraining power for the data that we used. However, such extended models will prove useful when analysing data from larger galaxy surveys such as DESI \parencite{DESI}. In terms of Bayesian inference, the problem has 22 free parameters. The results of our analysis are consistent with those of \textcite{beutler2017clustering}. We used weakly informative flat (uniform) priors for all parameters except for the two scaling parameters, $\alpha_{\parallel}$ and $\alpha_{\bot}$ for which we used normal (Gaussian) priors. We used $50$ walkers in total.

\begin{figure}[ht!]
    \centering
	\centerline{\includegraphics[scale=0.25]{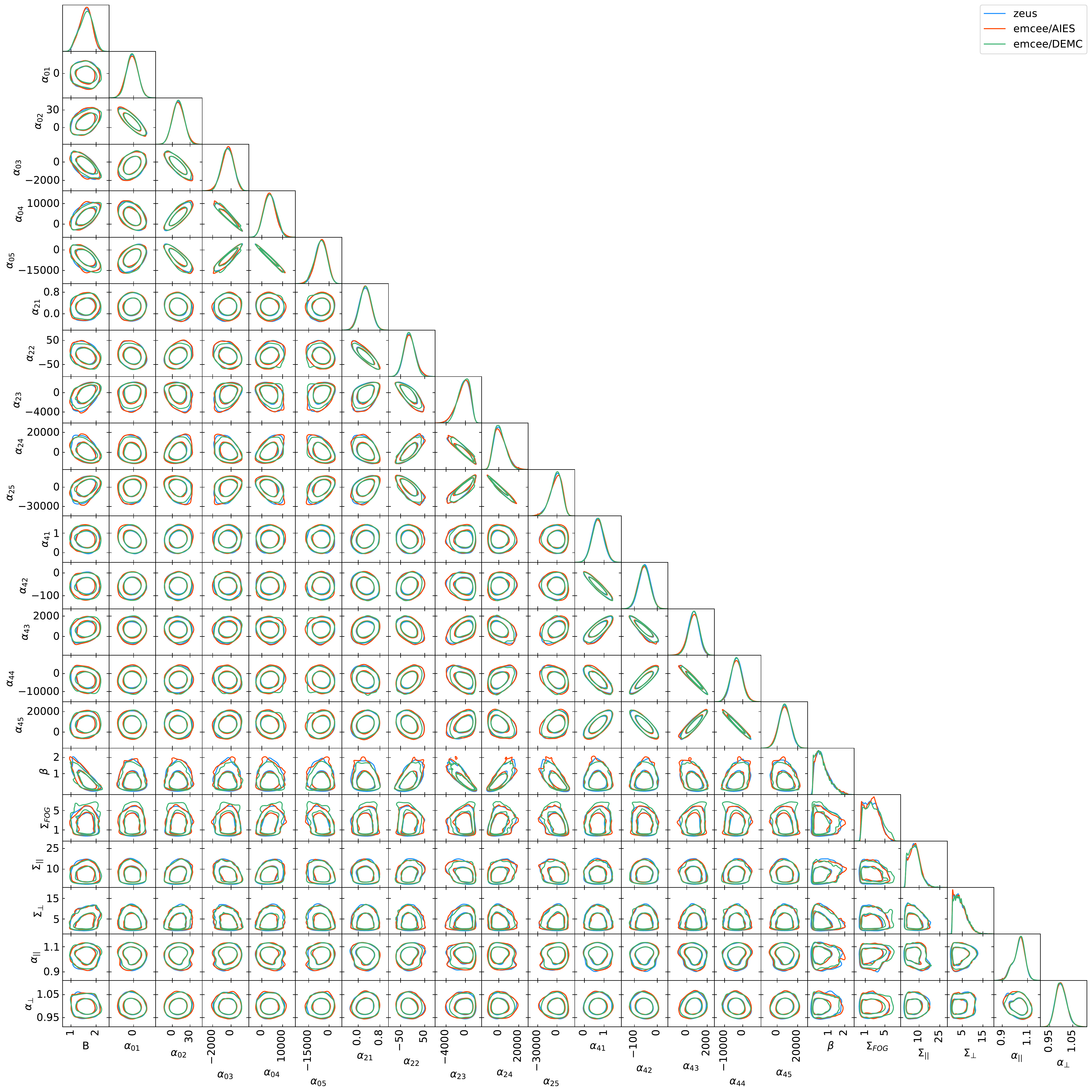}}
    \caption{A corner plot showing the 1-D and 2-D marginalised posteriors for the 22-parameter Baryon Acoustic Oscillation model as produced by the three different ensemble MCMC methods.}
    \label{fig_zeus:bao}
\end{figure}

In terms of efficiency, \texttt{zeus} generates at least 5 effectively independent samples for each one generated by \texttt{emcee}/DEMC and at least 9 for each one generated by \texttt{emcee}/AIES factoring in the different computational costs of the methods. As for the convergence rate, \texttt{zeus} converges more than 3 times faster than either \texttt{emcee} variant. Finally, we found that \texttt{zeus} is less sensitive to the initialisation than either of the other two methods. In particular, out of the 40 tests conducted with different initialisation, \texttt{zeus} converged 36 times, \texttt{emcee}/DEMC 14 times and \texttt{emcee}/AIES 7 times prior to the predetermined maximum number of likelihood evaluations (i.e. $5\times 10^6$ in this case). The aforementioned results are presented in detail in Table \ref{tab_zeus:table2}. The 1-D and 2-D marginal posterior distributions are shown in Figure \ref{fig_zeus:bao} demonstrating the agreement between the three methods\footnote{No upper limit on the number of likelihood evaluations or iterations was used for this run and convergence was diagnosed using all the metrics that we introduced.}.

\begin{table}[ht!]
    \centering
    \caption{The table shows a comparison of \texttt{emcee}/AIES, \texttt{emcee}\allowbreak/DEMC and \texttt{zeus} in terms of the inverse efficiency (i.e. reciprocal of the number of independent samples per model evaluation or the autocorrelation time estimate times the average number of model evaluations per iteration per walker), the convergence cost (i.e. number of model evaluations until convergence) and the convergence fraction (i.e. fraction of converged chains for given maximum number of model evaluations).}
    \def\arraystretch{1.1}
    \begin{tabular}{lcccc}
        \toprule[0.75pt]
         & \texttt{emcee}/AIES   & \texttt{emcee}/DEMC   & \textbf{zeus}  \\
        \midrule[0.5pt]
        \multicolumn{4}{l}{Cosmological inference} \\
        \midrule[0.5pt]
        efficiency$^{-1}$    &    12140    &    6750    &   $\mathbf{1320}$   \\
        convergence cost & $24\times10^{5}$ & $22\times10^{5}$ & $\mathbf{6.6\times10^{5}}$   \\
        convergence fraction   &    7/40   &    14/40    & $\mathbf{36/40}$  \\
        \midrule[0.5pt]
        \multicolumn{4}{l}{Exoplanet inference} \\
        \midrule[0.5pt]
        efficiency$^{-1}$    &    $1386$    &    $338$    &   $\mathbf{47}$   \\
        convergence cost & $36.0\times 10^{2}$ & $17.1\times 10^{2}$ & $\mathbf{4.8\times 10^{2}}$   \\
        convergence fraction   &    23/40    &    29/40    &  $\mathbf{38/40}$ \\
        \bottomrule[0.75pt]
        \end{tabular}
    \label{tab_zeus:table2}
\end{table}

\subsubsection{Exoplanet inference}

Another common application of MCMC methods in astronomy is the problem of exoplanet parameter inference through modelling of Keplerian orbits and radial velocity time series data. In this section we demonstrate the performance of \texttt{zeus} using a two-planet model with $14$ free parameters and real data from the K2-24 (EPIC-203771098) extrasolar system \parencite{k224} that is known to host two exoplanets. We used the popular \texttt{Python} package \texttt{RadVel} \parencite{radvel} for the Keplerian modelling of the planetary orbits. The results of our analysis are consistent with published constraints for the aforementioned extrasolar system \parencite{k224}. We used $30$ walkers in total for sampling.

\begin{figure}[ht!]
    \centering
	\centerline{\includegraphics[scale=0.40]{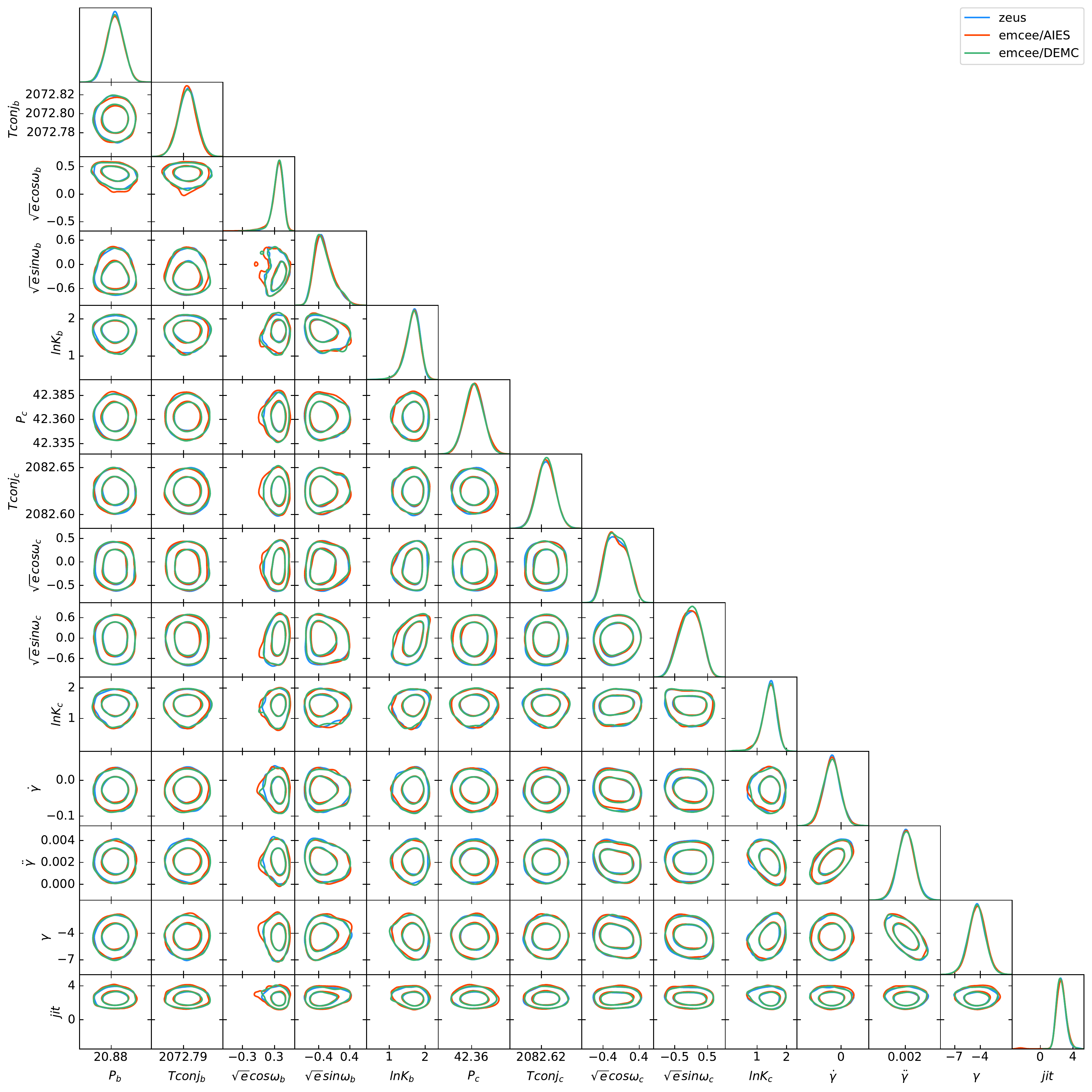}}
    \caption{A corner plot showing the 1-D and 2-D marginalised posteriors for the 14-parameter radial velocity model as produced by the three different ensemble MCMC methods.}
    \label{fig_zeus:exo}
\end{figure}

We performed the same suite of tests as in the cosmological inference case. In terms of efficiency, \texttt{zeus} generates more than $7$ independent samples per each one generated by \texttt{emcee}/DEMC and more than $29$ independent samples per each one generated by \texttt{emcee}/AIES. As for the convergence rate, \texttt{zeus} converges $7.5$ times faster than \texttt{emcee}/AIES and $3.5$ faster than \texttt{emcee}/DEMC on average. Finally, we found again that \texttt{zeus} is less sensitive to the specific initialisation of the walkers. In particular, out of the 40 tests conducted with different initialisation, \texttt{zeus} converged 38 times, \texttt{emcee}/DEMC 29 times and \texttt{emcee}/AIES 23 times prior to the predetermined maximum number of likelihood evaluations (i.e. $5\times 10^3$ in this case). Detailed results about the values of the used metrics are shown in Table \ref{tab_zeus:table2}. The 1-D and 2-D marginal posterior distributions are shown in Figure \ref{fig_zeus:exo}, demonstrating the agreement between the three methods.

\section{Discussion}
\label{sec_zeus:discussion}
Following the analysis we conducted in Section \ref{sec_zeus:tests} using the normal distribution there are two important questions that need to be answered about the initialisation of the walkers. First, how many walkers are necessary and, second, how to choose the initial positions of the walkers. Although there are many ways of answering those questions and there is no consistent solution that works for all target distributions, we will try to provide some general rules and heuristics to help ease the task of choosing the number and initial positions of the walkers for most cases.

Let us first discuss the effect of the number of the walkers on the general performance of \texttt{zeus}. Naively, one might expect that the minimum number of walkers should be $D+1$, where $D$ is the number of dimensions. However, the ensemble splitting technique, which was introduced in Section \ref{sec_zeus:tests} to render the algorithm parallelisable, requires at least $2\times D$ walkers in order to produce $2$ linearly independent samples. If a smaller number is chosen then the walkers can be trapped in a lower--dimensional hyper--plane of the parameter space, being unable to sample properly and leading to erroneous results. Although there is no upper bound on the number of walkers, we recommend to use between two to four times the number of dimensions. The reason is that increasing the number of dimensions can increase the cost of the burn-in period as we explained in detail in Section \ref{sec_zeus:tests}. Ideally, one wants to use the minimum number (or close to that) of walkers until the burn-in period is over and then increase the number of walkers to rapidly produce a great number of independent samples. It is also worth noting that in cases in which either non-linear correlations or multiple modes are present it is recommended to use more walkers (e.g. 4-8 times the number of parameters for a bimodal target distribution).

As for the initialisation of the walkers, there are many ways to choose their starting positions ranging from prior sampling to more localised initial positions. Empirical tests indicate that the latter often outperforms the former (i.e. leads to shorter burn-in periods). That is not surprising since the total probability of a prior-sampled initialisation can be very small when the number of parameters is high. In particular we found that initialising the walkers from a tight region in parameter space (i.e. normal distribution with small variance) consistently leads to good performance. For low to moderate dimensional problems initialising the walkers from a tight ball around the \textit{Maximum A Posteriori} (MAP) estimate can substantially reduce the burn-in period~\parencite{foreman2013emcee}.

Finally, while \texttt{emcee}/AIES and \texttt{emcee}/DEMC can sample acceptably from most target distributions with $D\lesssim 20$, the efficient scaling of \texttt{zeus} with the number of parameters allows us to extend this range and efficiently test more complicated models \parencite{karamanis2021ensemble}. Like most gradient-free methods, \texttt{zeus} will fail to sample efficiently in very high dimensional problems in which $D=\mathcal{O}(10^{2})$. In such cases, more sophisticated algorithms (e.g. tempering, block updating, Hamiltonian dynamics etc.) need to be used \parencite{2018arXiv180402719R}.

\section{Conclusions}
\label{sec_zeus:conclusions}

The aim of this project was to develop a tool that could facilitate Bayesian parameter inference in computationally demanding astronomical analyses and tackle the challenges posed by the complexity of the models and data that are often used by astronomers. To this end, we introduced \texttt{zeus}, a parallel, general-purpose and gradient-free \texttt{Python} implementation of Ensemble Slice Sampling.

After introducing the method in Section \ref{sec_zeus:ess}, we thoroughly demonstrated its performance compared to two popular alternatives (i.e. \texttt{emcee} with affine-invariant ensemble sampling and differential evolution Metropolis) using a variety of artificial and realistic target distributions in Section \ref{sec_zeus:tests}. The artificial toy examples helped to shed light on the general behaviour of the samplers in target distributions characterised by linear and non-linear correlations as well as multimodal densities. When compared to \texttt{emcee}/AIES and \texttt{emcee}/DEMC in the problems of Baryon Acoustic Oscillation parameter inference and exoplanet radial velocity fitting, \texttt{zeus} consistently converges faster (i.e. its burn-in is shorter by a factor of at least 3), it is less sensitive to the initialisation of the walkers and generates substantially more independent samples per likelihood evaluation (i.e. approximately $\times 9$ and  $\times 29$ speed-up compared to \texttt{emcee}/AIES in the cosmological and exoplanet examples, respectively). 

We have shown that \texttt{zeus} performs similarly or better than existing MCMC methods in a range of problems. We hope that \texttt{zeus} will prove useful to the astronomical and cosmological community by complementing existing approaches and facilitating the study of novel models and data over the coming years. \texttt{zeus} is publicly available at \url{https://github.com/minaskar/zeus} with detailed documentation and examples that can be found at \url{https://zeus-mcmc.readthedocs.io}.
% !TEX TS-program = pdflatex
% !TEX root = ../ArsClassica.tex

%************************************************
\chapter{Preconditioned Monte Carlo}
\label{chp:pmc}
%************************************************
 
\lstset{numbers=left,
    numberstyle=\scriptsize,
    stepnumber=1,
    numbersep=8pt
}    

This chapter presents \textit{Preconditioned Monte Carlo} which is the main contribution introduced in the paper titled \textit{Accelerating astronomical and cosmological inference with Preconditioned Monte Carlo} that was submitted for publication in the journal \textit{Monthly Notices of the Royal Astronomical Society} in July 2022~\parencite{karamanis2022pmc}. The content of the chapter is almost identical to that included in the aforementioned publication with the exception of minor text and figure formatting differences.

\begin{center}
\rule{0.5\textwidth}{.4pt}
\end{center}
\vspace{8pt}

\noindent We introduce \emph{Preconditioned Monte Carlo} (PMC), a novel Monte Carlo method for Bayesian inference that facilitates efficient sampling of probability distributions with non--trivial geometry. PMC utilises a \emph{Normalising Flow} (NF) in order to decorrelate the parameters of the distribution and then proceeds by sampling from the preconditioned target distribution using an adaptive \emph{Sequential Monte Carlo} (SMC) scheme. The results produced by PMC include samples from the posterior distribution and an estimate of the model evidence that can be used for parameter inference and model comparison respectively. The aforementioned framework has been thoroughly tested in a variety of challenging target distributions achieving state--of--the--art sampling performance. In the cases of \emph{primordial feature} analysis and \emph{gravitational wave} inference, PMC is approximately $50$ and $25$ times faster respectively than \emph{Nested Sampling} (NS). We found that in higher dimensional applications the acceleration is even greater. Finally, PMC is directly parallelisable, manifesting linear scaling up to thousands of CPUs. An open--source \texttt{Python} implementation of PMC, called \texttt{pocoMC}, is publicly available at \url{https://github.com/minaskar/pocomc}.

\section{Introduction}
\label{sec_pmc:intro}

Modern astronomical and cosmological analyses have largely adopted the framework of \emph{Bayesian probability} for tasks of parameter inference and model comparison. In the Bayesian context, the \emph{posterior probability distribution} $\mathcal{P}(\theta)=P(\theta |\mathcal{D}, \mathcal{M})$, meaning the probability distribution of the parameters $\theta$ of a model $\mathcal{M}$, given some data $\mathcal{D}$ and the model $\mathcal{M}$ is given by Bayes' theorem:
\begin{equation}
    \mathcal{P}(\theta) = \frac{\mathcal{L}(\theta)\pi(\theta)}{\mathcal{Z}}\,,
\end{equation}
where $\mathcal{L}(\theta) = P(\mathcal{D}|\theta, \mathcal{M})$ is the \emph{likelihood function}, $\pi(\theta)=P(\theta |\mathcal{M})$ is the \emph{prior} probability distribution, and $\mathcal{Z} = P(\mathcal{D}|\mathcal{M})$ is the \emph{model evidence} or \emph{marginal likelihood} that acts as a normalisation constant for the posterior probability distribution. For a detailed introduction to Bayesian probability theory we refer the reader to~\textcite{jaynes2003probability, gregory2005bayesian, mackay2003information} and the reviews~\textcite{trotta2017bayesian, sharma2017markov} for its use in astronomy and cosmology.

In tasks of parameter inference, the goal is to infer the values of physical and nuisance parameters from the data along with the respective uncertainties. Mathematically, this is formulated as the problem of estimating expectation values (e.g. mean values, standard deviations, 1--D and 2--D marginal posterior distributions, etc.) that correspond to high--dimensional integrals over the posterior probability density. During the past two decades, \emph{Markov chain Monte Carlo} (MCMC) has been established as the standard computational tool for the calculation of such integrals (see e.g. ~\parencite{speagle2019conceptual} for a review). 
MCMC methods generate a sequence of correlated samples, called a Markov chain, that are distributed according to the posterior probability distribution. Those samples can then be used in order to numerically estimate expectation values. Examples of MCMC software implementations in the astronomical and cosmological community are \textit{emcee}~\parencite{foreman2013emcee} and \textit{zeus}~\parencite{karamanis2021zeus}.

Most modern MCMC methods are based upon the \emph{Metropolis--Hastings} (MH) paradigm that consists of two steps~\parencite{metropolis1953equation, hastings1970}. In the first step, known as the \emph{proposal step}, a new sample is drawn from a known proposal distribution that depends only on the position of the current sample/state. The validity of the new sample, and thus the decision on whether to add it or not to the Markov chain, is determined in the second step, known as the \emph{acceptance step}, which takes into account the new sample, the old sample (i.e. current state) and the proposal distribution that was used in order to generate it. Arguably, the most important element of an efficient MCMC method is the choice of the proposal distribution. The degree to which the proposal distribution characterises the local geometry of the target distribution determines the sampling efficiency (i.e. rate of effectively independent samples) of the method. Unfortunately, choosing  or tuning the optimal proposal distribution for a given target distribution is not an easy task. However, certain optimal proposal distributions are known for specific classes of target distributions. For instance, in the case of a normal or Gaussian target distribution, using a normal proposal distribution of the form $\mathcal{N}(\theta,2.38^{2}\Sigma / D)$, where $\Sigma$ is the covariance matrix of the target density, $\theta$ is the current state of the chain, and $D$ is the number of dimensions yields the maximum sampling efficiency scheme with acceptance rate of $23.4\%$ in the acceptance step of MH~\parencite{gelman1997weak}. Alternatively, one can use a simpler proposal distribution of the form $\mathcal{N}(u,1)$ where $u=f(\theta)$ and $f$ is a suitable transformation. In this case, $f(\theta)$ is proportional to $L^{-1}\theta$ where $L$ is the lower triangular matrix of the \emph{Cholesky decomposition} of the covariance matrix $\Sigma = L L^{T}$. In other words, assuming that a suitable transformation can be found, one can increase the sampling efficiency of an MCMC method. This notion of preconditioning is central for the discussion that will follow in the next section.

In recent years, the need for higher sampling efficiency when the correlations between parameters are strong enough or the posterior exhibits multiple modes, as well as the required computation of the model evidence $\mathcal{Z}$ for model comparison tasks, motivated the development of more advanced sampling methodologies and algorithms. One very popular approach is the \emph{Sequential Monte Carlo} (SMC) algorithm~\parencite{del2006sequential}, which evolves a set of particles through a series of intermediate steps that bridge the gap between the prior distribution and the posterior distribution by geometrically interpolating between them. Another class of algorithms called \emph{Nested Sampling} (NS)~\parencite{skilling2004nested} attempts to approach the problem of Bayesian computation from a slightly different perspective. Instead of evolving a set of particles though a series of geometrically--interpolated steps between prior and posterior distribution, NS splits the posterior distribution into many slices and attempts to sample each slice individually with an appropriate weighting scheme. Many popular versions and implementations of NS exist in the astronomical literature~\parencite{speagle2020dynesty, buchner2021ultranest, handley2015polychord, feroz2009multinest}. Whereas both SMC and NS largely addressed the problem of multimodality, the performance of both methods is still very sensitive to the geometry of the target distribution, meaning the presence of strong non--linear correlations.

In this paper, we introduce \emph{Preconditioned Monte Carlo} (PMC), a novel Monte Carlo method for Bayesian inference that extends the range of applications of SMC to target distributions with non--trivial geometry, strong non--linear correlations between parameters, and severe multimodality. PMC achieves this by first preconditioning, or transforming the geometry of the target distribution into a more manageable one using a generative model known as a \emph{Normalising Flow} (NF)~\parencite{papamakarios2021normalizing}, before sampling using a SMC scheme. \textcite{hoffman2019neutra} used a NF to neutralise the bad geometry in Hamiltonian Monte Carlo (HMC)~\parencite{betancourt2017conceptual} achieving great results in terms of sampling speed but unreliable estimates for unknown target distributions. \textcite{moss2020accelerated} used a NF in order to parameterise efficient MCMC proposals and used it in the context of NS achieving a substantial speedup on several challenging distributions. Both of the aforementioned works used NFs as preconditioning transformations, the first in the context of HMC and the second in NS. In the context of NS and SMC, NFs have also been used as a sampling component of the algorithm~\parencite{albergo2019flow,williams2021nested, arbel2021annealed}, albeit not as a preconditioner but as a density from which new samples can be generated independently. The novelty of our work lies in the use of NFs as preconditioning transformations in the context of SMC, thus achieving both robustness and high sampling efficiency.

The structure of the rest of the paper is the following: Section \ref{sec_pmc:method} consists of a detailed presentation of the method, Section \ref{sec_pmc:tests} includes a wide range of empirical tests that act as a demonstration of PMC's sampling performance, and Section \ref{sec_pmc:conclusions} is reserved for the conclusions.

We also release a \texttt{Python} implementation of PMC, called \texttt{pocoMC}, which is publically available at \url{https://github.com/minaskar/pocomc} and detailed documentation with installation instructions and examples at \url{https://pocomc.readthedocs.io}. The code implementation is described in the accompanying paper~\parencite{karamanis2022pocomc}.

\section{Method}
\label{sec_pmc:method}

\subsection{Sequential Monte Carlo}

In this subsection, we will present a brief introduction to SMC algorithms. For a more detailed exposition, we refer the reader to~\textcite{naesseth2019elements}. We begin by first introducing the concept of \emph{importance sampling}, which is crucial for understanding the function of SMC. Assuming that we have a target probability density $p(\theta)$ that we are able to evaluate up to an unknown multiplicative constant, then if we define another density $\rho(\theta)$, called the \emph{importance sampling density}, such that $\rho(\theta) =0 \Rightarrow p(\theta) = 0$ then the following relation holds for any expectation value:
\begin{equation}
    \label{eq_pmc:importance_sampling}
    \begin{split}
        \mathrm{E}_{p}[f(\theta)] & = \int f(\theta) w(\theta) \rho(\theta) d\theta \Big/ \int w(\theta) \rho(\theta) d\theta \\
        & = \mathrm{E}_{\rho}[f(\theta)w(\theta)]/\mathrm{E}_{\rho}[w(\theta)]\, ,
    \end{split}
\end{equation}
for any function $f(\theta)$ where $w(\theta) = p(\theta) / \rho(\theta)$ are called importance weights. What is important here is that one can use samples from the importance density $\rho(\theta)$ in order to estimate the aforementioned expectation value without explicitly sampling from the target density $p(\theta)$.

A common measure of the quality of using the importance sampling density $\rho(\theta)$ to approximate $p(\theta)$ is the \emph{Effective Sample Size}, defined as:
\begin{equation}
    \label{eq_pmc:effective_sample_size}
    \text{ESS} = \mathrm{E}_{\rho}[w(\theta)]^{2} / \mathrm{E}_{\rho}[w(\theta)^{2}]\,.
\end{equation}
Unfortunately, in high--dimensional scenarios it is difficult to find an appropriate importance sampling density that ensures that the ESS is high enough for the variance of the expectation value to be low. This is exactly the problem that SMC methods address.

SMC samplers extend the importance sampling procedure from the setting of two densities (i.e. importance sampling density and target density) to a sequence of $T$ probability distribution densities $\lbrace p_{t}\rbrace _{t=1}^{T}$ in which each individual density $p_{t}$ acts as the importance density for the next one in the series. The method proceeds by pushing a collection of $N$ particles $\lbrace \theta_{t}^{k}\rbrace _{k=1}^{N}$ through this sequence of densities until the last one is reached. Each iteration of a SMC algorithm consists of three main steps:
\begin{enumerate}
    \item \textbf{Mutation} -- The population of particles is moved from $\lbrace \theta_{t-1}^{k}\rbrace _{k=1}^{N}$ to $\lbrace \theta_{t}^{k}\rbrace _{k=1}^{N}$ using a \emph{Markov transition kernel} $K_{t}(\theta '|\theta)$ that defines the next importance sampling density
    \begin{equation}
        \label{eq_pmc:Markov_kernel}
        p_{t}(\theta ') = \int p_{t-1}(\theta) K_{t}(\theta '|\theta) d\theta \, .
    \end{equation}
    In practice, this step consists of running multiple short MCMC chains (i.e. one for each particle) to get the new states $\theta '$ starting from the old ones $\theta$.
    
    \item \textbf{Correction} -- The particles are reweighted according to the next density in the sequence. This step consists of multiplying the current normalised weight $W_{t}^{k}$ of each particle by the appropriate importance weight:
    \begin{equation}
        \label{eq_pmc:importance_weight}
        w_{t}(\theta_{t}) = p_{t}(\theta_{t-1})/p_{t-1}(\theta_{t-1})\,.
    \end{equation}
    
    \item \textbf{Selection} -- The particles are resampled according to their normalised weights $W_{t}^{k}$ which are then set to $1/N$. This can be done using \emph{multinomial resampling} or more advanced schemes. The purpose of this step is to eliminate particles with low weight and multiply the ones with high weights.
\end{enumerate}

An important feature of the SMC method is that it allows for the unbiased estimation of the ratios of normalising constants
\begin{equation}
    \label{eq_pmc:normalising_ratios}
    \mathcal{Z}_{t}/\mathcal{Z}_{t-1} = \sum_{k=1}^{N}W_{t-1}^{k}w_{t}(\theta_{t-1}^{k})\,,
\end{equation}
between subsequent densities, where $W_{0}^{k}=1/N$. This is of paramount importance in cases in which the first density in the series corresponds to the prior distribution (i.e. with $\mathcal{Z}=1$) and the last to the posterior distribution. Then, SMC methods can be used in order to compute the model evidence $\mathcal{Z}$ for tasks of model comparison.

In principle, there are arbitrary many ways to construct the sequence of densities $\lbrace p_{t}\rbrace _{t=1}^{T}$.  A very common way to do so is to geometrically interpolate between two densities $\rho(\theta)$ and $p(\theta)$:
\begin{equation}
    \label{eq_pmc:temperature_annealing}
    p_{t}(\theta) \propto \rho(\theta)^{1-\beta_{t}}p(\theta)^{\beta_{t}},\quad t = 1, \dots, T\,
\end{equation}
parameterised by a \emph{temperature annealing} ladder:
\begin{equation}
    \label{eq_pmc:temperature_ladder}
    \beta_{1} = 0 < \beta_{2} < \dots < \beta_{T} = 1\,.
\end{equation}
In the Bayesian context, a natural choice of geometric interpolation is from the prior $\pi(\theta)$ to the posterior:
\begin{equation}
    \label{eq_pmc:temperature_annealing_bayesian}
    p_{t}(\theta) \propto \pi(\theta)\mathcal{L}(\theta)^{\beta_{t}},\quad t = 1, \dots, T\,
\end{equation}
where $\mathcal{L}(\theta)$ is the likelihood function. In practice, it can still be difficult to choose a good temperature schedule. However, this can be done adaptively by selecting the next value of $\beta_{t}$ such that the ESS is a constant $\alpha$ fraction of the number of particles $N$. Numerically, this can be done by solving
\begin{equation}
    \label{eq_pmc:bisection_method}
    \bigg( \sum_{k=1}^{N}w_{t+1}^{k}(\beta_{t+1})\bigg)^{2}\Big/ \sum_{k=1}^{N}w_{t+1}^{k}(\beta_{t+1})^{2}=\alpha N\,,
\end{equation}
the next $\beta_{t+1}$ such that $\beta_{t}<\beta_{t+1}\leq 1$ using, for instance, the \emph{bisection method}.

\begin{figure}
    \centering
	\centerline{\includegraphics[scale=0.90]{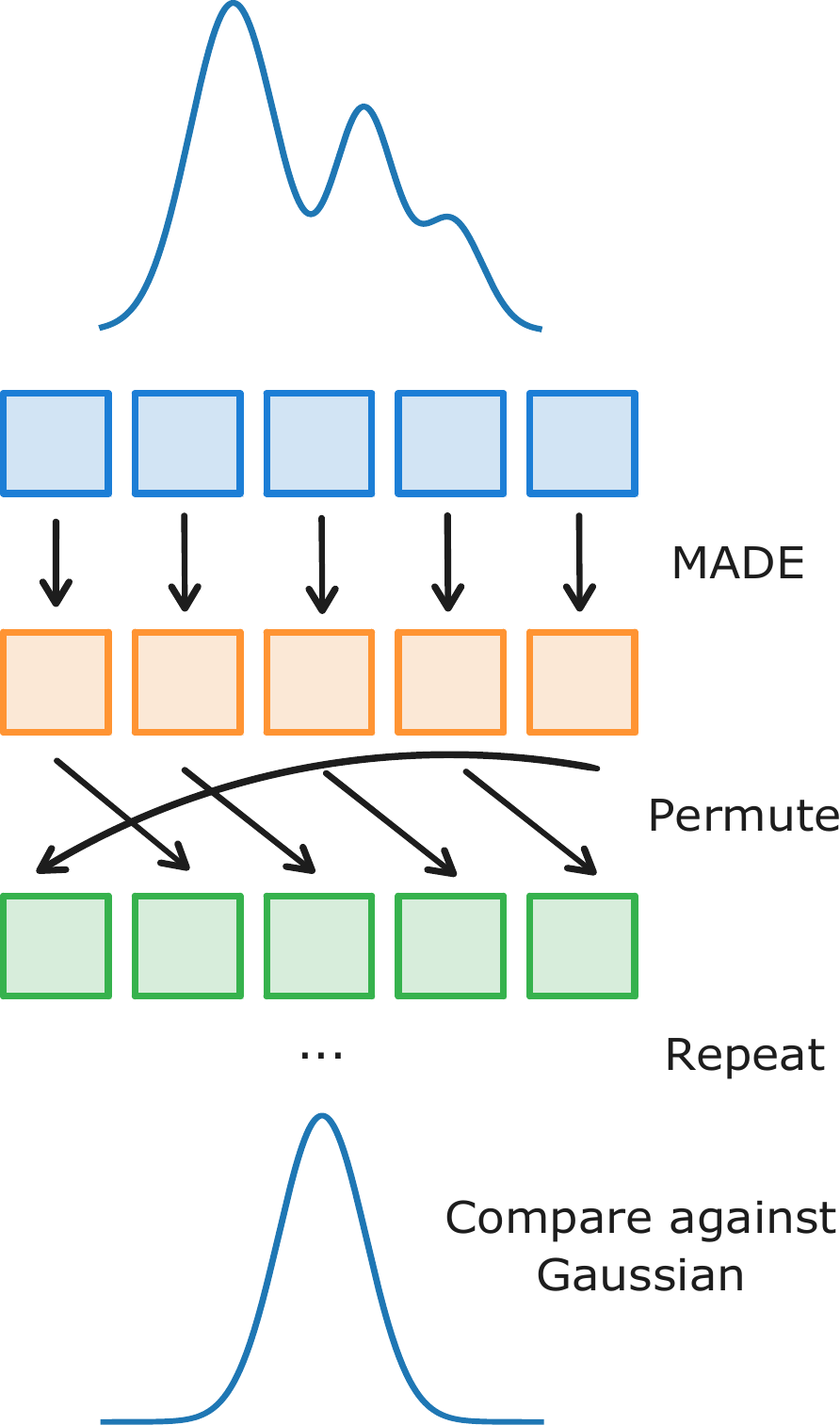}}
    \caption{Illustration of the inference scheme of a \emph{Masked Autoregressive Flow} (MAF). The arrows show the conditional dependence of the variables as well as the action of the \emph{Masked Autoregressive Density Estimation} (MADE) layer. The input target probability density (top) is mapped into a multivariate normal distribution (bottom). A sequence of MADE layers and permutations is repeated multiple times in order to increase the flexibility of the flow.}
    \label{fig_pmc:maf}
\end{figure}

\subsection{Normalising Flows}

Normalising flows (NF) are generative models, which can facilitate efficient and exact density estimation~\parencite{papamakarios2021normalizing}. They are based on the formula of change--of--variables $\theta = f(u)$ where $u$ is sampled from a base distribution $u~p_{u}(u)$ (i.e. usually a normal distribution). The NF is a bijective mapping between the base distribution $p_{u}(u)$ and the often more complex target distribution $p_{\theta}(\theta)$ that can be evaluated exactly using
\begin{equation}
    \label{eq_pmc:jacobian_transform}
    p_{\theta}(\theta) = p_{u}(f^{-1}(\theta))\bigg|\det\bigg( \frac{\partial f^{-1}}{\partial \theta}\bigg) \bigg| \,,
\end{equation}
where the Jacobian determinant is tractable.

NFs are usually parameterised by neural networks. However, neural networks are not in general invertible, and the Jacobian is not generally tractable. Thus special care needs to be taken when choosing the architecture of the neural network to ensure the invertability of the transformation and the tractability of the Jacobian. For instance, if the forward transformation is $\theta_{i} = u_{i}\exp(\alpha_{i})+\mu_{i}$ and inverse transformation is $u_{i} = (\theta_{i} -  \mu_{i})\exp(-\alpha_{i})$, where $\mu_{i}$ and $\alpha_{i}$ are constants, then it is straightforward to show that the Jacobian satisfies
\begin{equation}
    \label{eq_pmc:jacobian_example}
    \bigg|\det\bigg( \frac{\partial f^{-1}}{\partial \theta}\bigg) \bigg| = \exp \bigg( -\sum_{i}\alpha_{i}\bigg)\,.
\end{equation}

To this end, we chose to use the \emph{Masked Autoregressive Flow} (MAF), which has been used many times successfully for density estimation tasks due to its superior performance and high flexibility compared to alternative models~\parencite{papamakarios2017masked}. A MAF consists of many stacked layers of a simpler generative model, called \emph{Masked Autoregressive Density Estimator} (MADE)~\parencite{germain2015made}, with subsequent permutations of its outputs as shown in Figure \ref{fig_pmc:maf}. A MADE model decomposes a joint density $p(\theta)$ as a product of conditionals $p(\theta) = \prod_{i}p(\theta_{i}|\theta_{1:i-1})$ that ensures that any given value $\theta_{i}$ is only a function of the previous values thus maintaining the \emph{autoregressive property}. When the MADE is based on an \emph{autoencoder}, then \emph{masking} is required in order to remove connections between different units in different layers, so as to preserve the aforementioned autoregressive property.

\subsection{Preconditioning}

\begin{figure*}
    \centering
	\centerline{\includegraphics[scale=0.45]{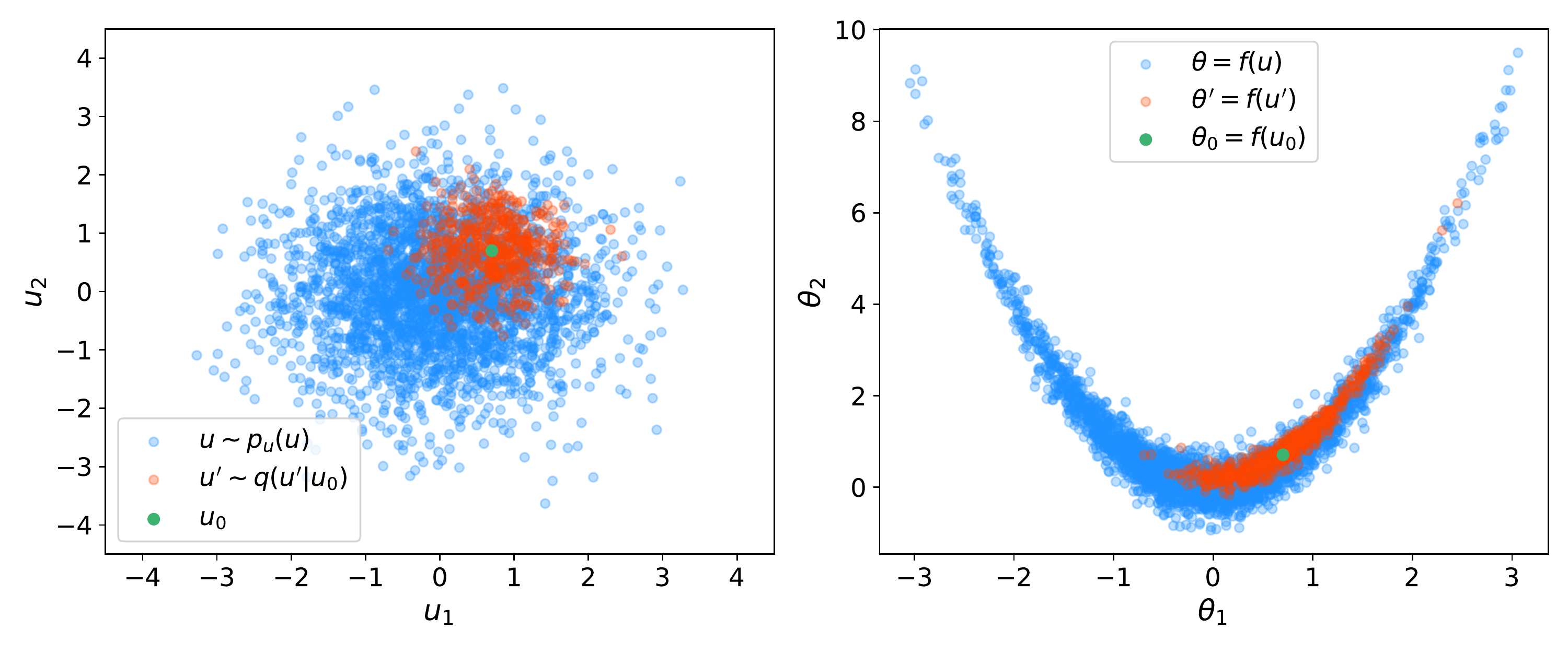}}
    \caption{The figure illustrates the effect of preconditioning on the \emph{Rosenbrock} distribution. The right panel shows samples (blue) from the true correlated distribution and the left panel shows samples (blue) from the preconditioned/transformed one. The orange samples in the left panel are drawn from a symmetric normal proposal distribution centred around the green point $u_{0}$ and they correspond to the respective orange points in the right panel. In other words, the transformed samples from the simple proposal in the left panel correspond to samples that capture the local geometry of the true target distribution in the right panel.}
    \label{fig_pmc:proposal}
\end{figure*}

Most \emph{Markov chain Monte Carlo} (MCMC) methods struggle to sample efficiently from highly correlated or skewed target distributions. Often, transforming the parameters of the distribution before sampling, a process also known as \emph{preconditioning}, using appropriate change--of--variable transformations, can help ameliorate this effect by disentangling the dependence between parameters. This is equivalent to choosing an appropriate proposal distribution in the context of \emph{Metropolis--Hastings} (MH) methods. However, finding a valid transformation and selecting an appropriate proposal distribution is often difficult a priori; and there is no obvious way of making this joint choice in an optimal way. For instance, a linear transformation $\theta \leftarrow L^{-1}\theta$ where $L$ is the lower triangular matrix of the \emph{Cholesky decomposition} of the \emph{sample covariance} matrix $\Sigma = LL^{T}$ can remove only linear correlations and is not effective against non--linear ones. More sophisticated transformations, such as the use of the \emph{chirp mass} and \emph{mass ratio} instead of the individual black--hole masses in gravitational wave astronomy requires expert knowledge that is problem--specific.

The \emph{Metropolis acceptance criterion} employed by MH methods in order to maintain detailed balance is
\begin{equation}
    \label{eq_pmc:Metropolis_criterion}
    \alpha = \min \bigg( 1, \frac{p_{\theta}(\theta ')q(\theta | \theta ')}{p_{\theta }(\theta)q(\theta' | \theta)}\bigg)\,,
\end{equation}
where $p_{\theta}(\theta)$ is the target distribution and $q(\theta' | \theta)$ is the proposal distribution. For a general transformation $\theta = f(u)$ and its inverse $u = f^{-1}(\theta)$ the modified \emph{Metropolis acceptance criterion} takes the following form
\begin{equation}
    \label{eq_pmc:Metropolis_criterion_transform}
    \alpha = \min \left( 1, \frac{p_{\theta}(f^{-1}(u'))q(u | u ')\Big|\det \frac{\partial f^{-1}(u')}{\partial u'}\Big|}{p_{\theta }(f^{-1}(u))q(u ' | u)\Big|\det \frac{\partial f^{-1}(u)}{\partial u}\Big|}\right)\,,
\end{equation}
where the Jacobian determinant also appears. In this formulation of MH, the sampler samples the distribution in the transformed space and then samples are pushed through the $\theta = f(u)$ transformation to the original space. Assuming that the transformation $\theta = f(u)$ induces a simpler geometry onto the transformed space, sampling using the above acceptance criterion can be substantially more efficient. 

Figure \ref{fig_pmc:proposal} shows one such transformation that transforms the banana--shaped \emph{Rosenbrock} distribution into a unit--variance normal distribution and \emph{vice versa}. The same figure also demonstrates the effectiveness of simple proposal distributions $q(u'|u)$ in the transformed/latent space. A symmetric normal proposal distribution $q(u'|u_{0})$ centred around a point $u_{0}$ corresponds to a highly effective proposal distribution in the original space, which captures the local geometry of the target distribution around that point.

\subsection{Preconditioned Monte Carlo}

\emph{Preconditioned Monte Carlo} (PMC) is the result of the amalgamation of SMC, NFs and preconditioning as they were introduced in the previous paragraphs. In particular, we suggest the use of the transformation $\theta = f(u)$ of a NF in order to precondition the \emph{Mutation} step of SMC. A pseudocode of the algorithm is presented at \Algo{pmc}. The \emph{Mutation} step in this case consists of $N$ \emph{Random--Walk Metropolis} (RWM) steps, meaning MH with an isotropic Gaussian proposal distribution centred around the current state of the Markov chain, in which the algorithm targets the preconditioned density. We fix the acceptance rate of MH to its optimal value $23.4\%$ between temperature steps by adapting the proposal scale~\parencite{gelman1997weak}. As the optimal proposal scale of MH for a Gaussian target distribution is
\begin{equation}
    \label{eq_pmc:optimal_scale}
    \sigma_{\textrm{opt}} = \frac{2.38}{\sqrt{D}}\,,
\end{equation}
where $D$ is the number of dimensions/parameters, we can assess the performance of the NF preconditioner by estimating the ratio of the true scale $\sigma$ to the optimal one $\sigma_{\textrm{opt}}$. Assuming that the NF preconditions perfectly the target density and maps it into a unit--variance Gaussian distribution, this ratio should be equal to one. In practice, this ratio can deviate slightly from the optimal value of unity, and one can utilise this ratio as a metric of the preconditioning quality. The number $N$ of the MCMC steps performed in each iteration is determined adaptively during the run. The process we used is based on the mean correlation coefficient between the initial positions of the particles in the beginning of an iteration and their current positions. In particular, the particles are updated, using MCMC, until their mean correlation coefficient drops below a prespecified threshold value. The lower the value of this threshold, the higher the number $N$ of MCMC steps. It is important to note that the correlation coefficient is computed in the preconditioned $u$ space.

\begin{algorithm}
\caption{Preconditioned Monte Carlo}
    \algolabel{pmc}
\begin{algorithmic}[1]
\STATE{\textbf{input} Number of particles $N$}
\STATE{$t\leftarrow 1$, $\beta_{1}\leftarrow 0$, $\mathcal{Z}\leftarrow 1$}
\STATE{\textbf{for} $k=1$ \textbf{to} N \textbf{do} sample $\theta_{1}^{k}\sim \pi(\theta)$ and set $W_{1}^{k}=1/N$}
\STATE{train $\theta = f(u)$ using $\lbrace \theta_{1}^{k}\rbrace _{k=1}^{N}$}
\WHILE{$\beta_{t}\neq 1$}
    \STATE{$t\leftarrow t + 1$}
    \STATE{$\beta_{t}\leftarrow$ solution to Eq. \ref{eq_pmc:bisection_method}}
    \STATE{\textbf{for} $k=1$ \textbf{to} N \textbf{do} $w_{t}^{k}\leftarrow W_{t-1}^{k}\mathcal{L}(\theta)^{\beta_{t}-\beta_{t-1}}$}
    \STATE{$\mathcal{Z}\leftarrow \mathcal{Z}N^{-1}\sum_{k=1}^{N}w_{t}^{k}$}
    \STATE{$\lbrace \Tilde{\theta}_{t-1}^{k}\rbrace _{k=1}^{N} \leftarrow$ resample $\lbrace \theta_{t-1}^{k}\rbrace _{k=1}^{N}$ according to $\lbrace W_{t}^{k}\rbrace _{k=1}^{N}$ where $W_{t}^{k}=w_{t}^{k}/\sum_{k'=1}^{N}w_{t}^{k'}$}
    \STATE{\textbf{for} $k=1$ \textbf{to} N \textbf{do} $W_{t}^{k}\leftarrow 1/N$}
    \STATE{$\lbrace \theta_{t}^{k}\rbrace _{k=1}^{N}\leftarrow$ move $\lbrace \Tilde{\theta}_{t-1}^{k}\rbrace _{k=1}^{N}$ according to $K_{t}\left( \lbrace \theta_{t}^{k}\rbrace _{k=1}^{N}\leftarrow \lbrace \Tilde{\theta}_{t-1}^{k}\rbrace _{k=1}^{N}\, ; f \right)$}
    \STATE{train $\theta = f(u)$ using $\lbrace \theta_{t}^{k}\rbrace _{k=1}^{N}$}
\ENDWHILE
\STATE{\textbf{return} samples $\lbrace \theta_{t}^{k}\rbrace _{k=1}^{N}$ and estimate of the marginal likelihood $\mathcal{Z}$}
\end{algorithmic}
\end{algorithm}

\subsection{Hyperparameters}

We can classify the hyperparameters of PMC into two groups, those that have to do with the normalising flow and those that have to do with the SMC algorithm. The first group consists of structure and training hyperparamaters for the NF. The NF structure parameters include the number of MADE layers (\texttt{blocks}), as well as the number of neurons per hidden layer (\texttt{neurons}). The NF training hyperparameters include the learning rate (\texttt{lr}) of the Adam optimiser~\parencite{kingma2014adam}, the maximum number of epochs (\texttt{epochs}), the training batch size (\texttt{batch}), the tolerance for early stopping (\texttt{tolerance}), and the scale of $L_{1}$ regularisation (\texttt{l1}). On the other hand, the SMC hyperparameters include the number of particles (\texttt{particles}), the desired effective sample size (\texttt{ESS}), and the correlation coefficient threshold (\texttt{threshold}). The default values for those hyperparameters are shown in Table \ref{tab_pmc:default}. We found that this configuration was robust and efficient for a wide range of applications and thus decided to recommend this as the default choice.

\begin{table}
    \centering
    \caption{The table shows the default values for the hyperparameters of PMC.}
    \def\arraystretch{1.1}
    \begin{tabular}{lc|lc}
        \toprule[0.75pt]
        \multicolumn{2}{l}{NF hyperparameters} & \multicolumn{2}{l}{SMC hyperparameters} \\
        \midrule[0.5pt]
        \texttt{blocks} & $6$ & \texttt{particles} & $1000 - 4000$ \\
        \texttt{neurons} & $3\times D$ & \texttt{ESS} & $95\%$ \\
        \texttt{batch} & $1000$ & \texttt{threshold} & $75\%$ \\
        \texttt{epochs} & $500$ &   &   \\
        \texttt{tolerance} & $30$ &   &   \\
        \texttt{lr} & $10^{-2} - 10^{-5}$ &  &  \\
        \texttt{l1} & $0.2$ &  &  \\
        \bottomrule[0.75pt]
    \end{tabular}
    \label{tab_pmc:default}
\end{table}

\subsection{Parallelisation}

\begin{figure}
    \centering
	\centerline{\includegraphics[scale=0.65]{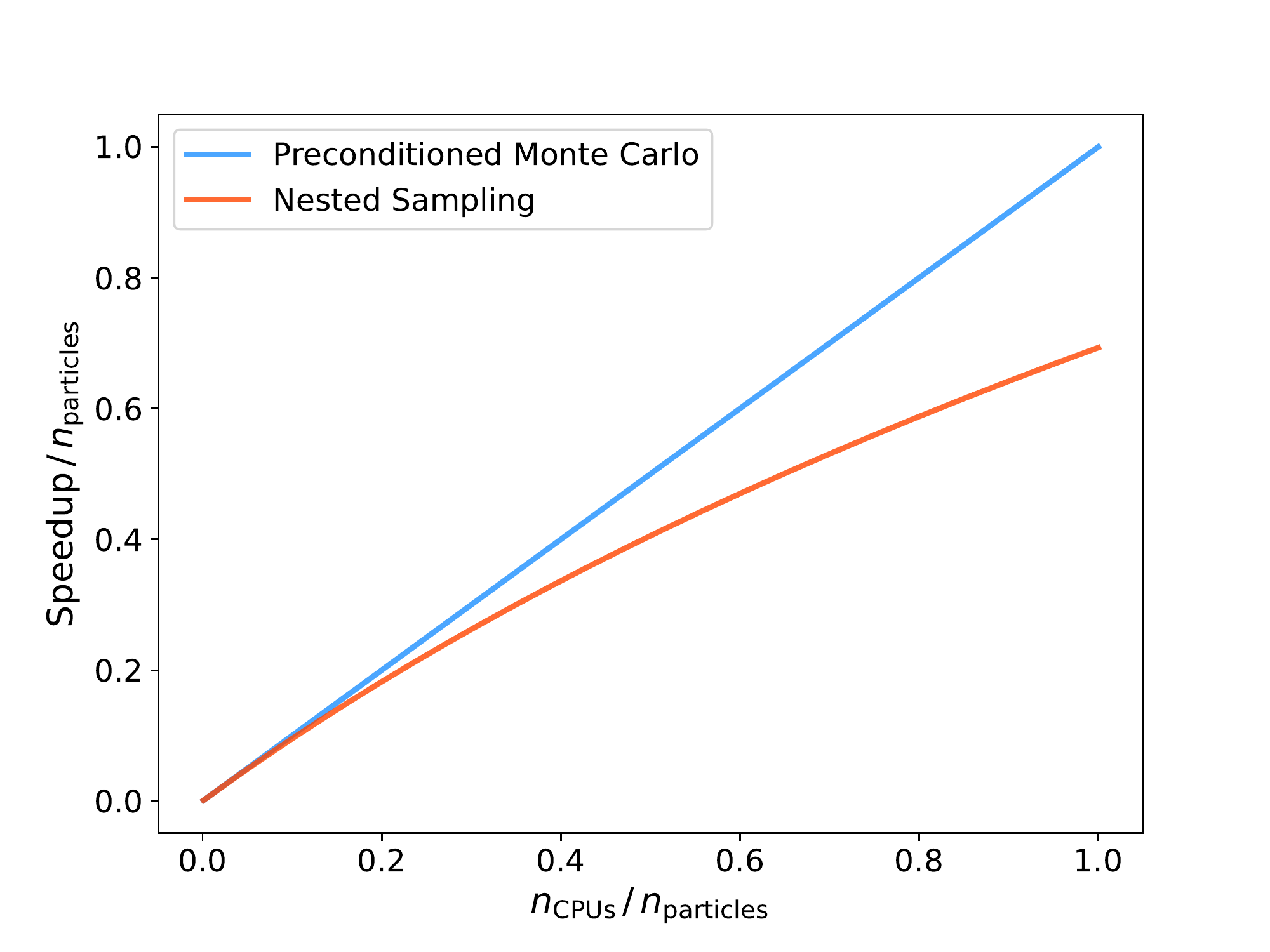}}
    \caption{Parallelization of PMC compared to nested sampling. PMC (blue) exhibits linear speedup compared to the sub--linear one achieved by NS (orange).}
    \label{fig_pmc:speedup}
\end{figure}

An important property of PMC, and indeed of any SMC algorithm, is its ideal scaling with the available number of CPUs. In particular, the mutation step of PMC is exactly parallelisable, meaning that that the speedup gained by using more than one CPU scales linearly with the number of CPUs as long as $n_{\mathrm{CPUs}} \leq n_{\mathrm{particles}}$. Similar methods that also use a large collection of particles scale less favourably. For instance, \emph{Nested Sampling} (NS) exhibits sub--linear scaling as shown in Figure \ref{fig_pmc:speedup} of \cite{handley2015polychord}. The aforementioned characteristic of PMC renders it ideal for computationally costly applications that are often encountered in astronomy and cosmology.

\section{Empirical Evaluation}
\label{sec_pmc:tests}

In this section we present two toy examples and two realistic parameter inference examples that reproduce common astronomical and cosmological analyses. In all cases, the hyperparameters of PMC were set to their default values as shown in Table \ref{tab_pmc:default}. In both analyses, the performance of PMC is compared to that of SMC without preconditioning but otherwise using the same settings (e.g. number of particles, ESS, etc.) as PMC, as well as \emph{Nested Sampling} (NS), a popular particle Monte Carlo alternative~\footnote{We used the popular \texttt{Python} implementation \texttt{dynesty}~\parencite{speagle2020dynesty} for NS.}. The metric that we use in order to evaluate the performance of each method is the total number of model evaluations performed until convergence. Convergence in all methods is well--defined: in PMC and SMC the algorithm converges when $\beta=1$, whereas in NS the run stops when less than $1\%$ of the model evidence is left unaccounted.  All other computational costs are negligible, including the training and evaluation of the normalising flow in the case of PMC that only required a few seconds for the whole inference procedure. All methods used $1000$ particles.

\subsection{Rosenbrock distribution}

\begin{figure}
    \centering
	\centerline{\includegraphics[scale=0.65]{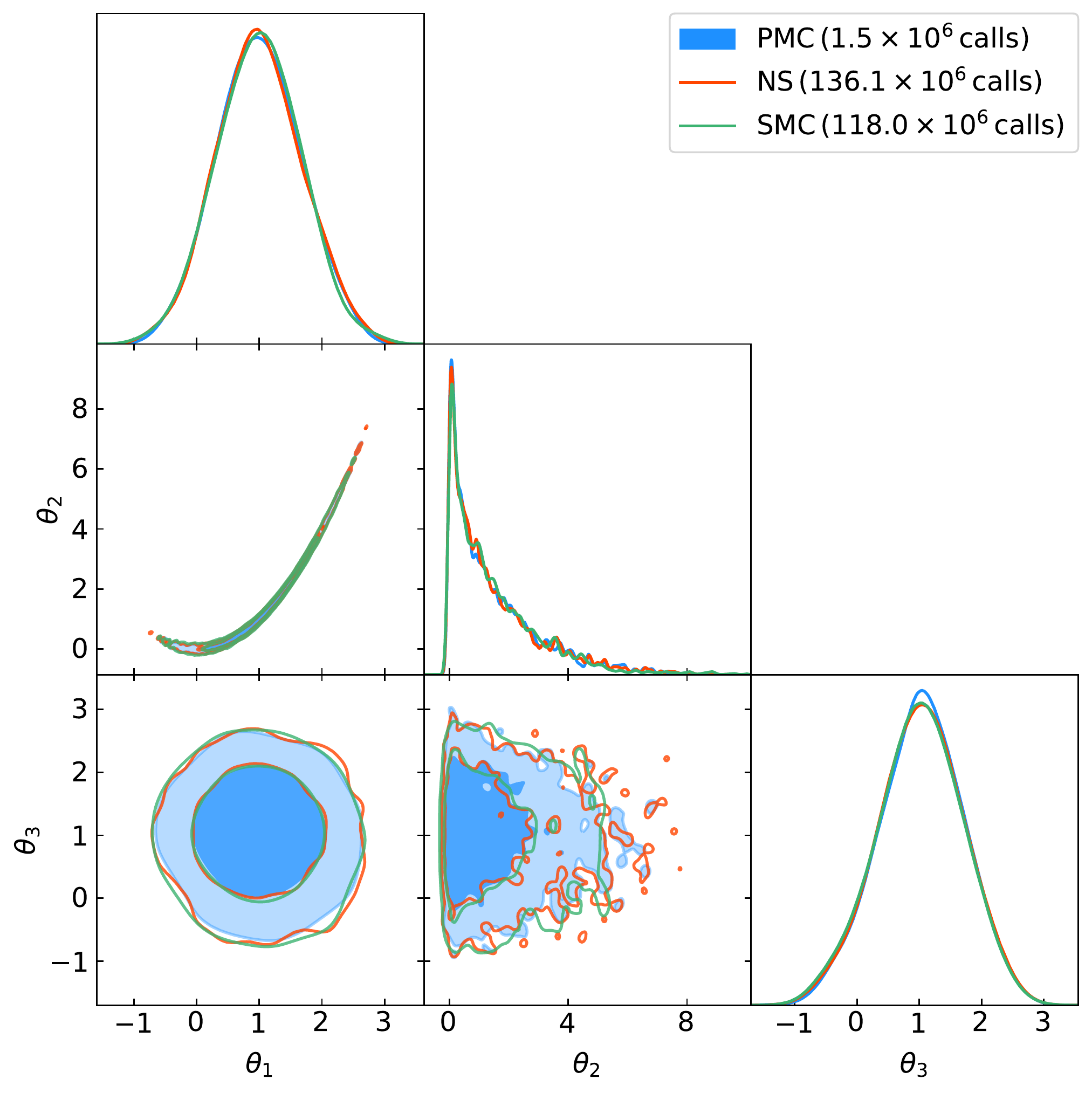}}
    \caption{Illustration of the 1--dimensional and 2--dimensional marginal posteriors for the first three out of $20$ parameters of the \emph{Rosenbrock} distribution. The figure shows the 1--$\sigma$ and 2--$\sigma$ contours generated by \emph{Preconditioned Monte Carlo} (PMC) in \emph{blue}, \emph{Nested Sampling} (NS) in \emph{orange}, and \emph{Sequential Monte Carlo} (SMC) in \emph{green}. The legend also shows the computational cost of each method in terms of the total number of required model evaluations until convergence is reached.}
    \label{fig_pmc:rosenbrock}
\end{figure}

The first toy example that we used is the \emph{Rosenbrock} distribution, which exhibits strong non--linear correlation between its parameters. For this reason, the \emph{Rosenbrock} distribution has often been used as a benchmark target for optimization and sampling tasks. Here we use a 20--dimensional generalisation of the distribution which is defined through the probability density function given by:
\begin{equation}
    \label{eq_pmc:rosenbrock}
    \log P(\theta) = -\sum_{i=1}^{N/2}\left[10\left(\theta_{2i-1}^{2}-\theta_{2i}\right)^{2}+\left(\theta_{2i-1}-1\right)^{2} \right]\,.
\end{equation}
Furthermore, we use flat priors $\mathcal{U}(-10,10)$ for all parameters. Figure \ref{fig_pmc:rosenbrock} shows the 2--dimensional marginal posterior for the first two parameters as generated by the three methods. The total computational cost of PMC, NS, and SMC is $1.5\times 10^{6}$, $136.1\times 10^{6}$, and $118.0\times 10^{6}$ model evaluations, respectively. PMC requires approximately $1/91$ of the number of model evaluations that NS does, and approximately $1/79$ of those that SMC does. 

\begin{table}
    \centering
    \caption{The table shows a comparison of PMC, NS, and SMC in terms of their computational cost (i.e. total number of model evaluations until convergence).}
    \def\arraystretch{1.1}
    \begin{tabular}{lcccc}
        \toprule[0.75pt]
        & & \multicolumn{3}{c}{Model evaluations $(\times 10^{6})$} \\
        \cmidrule(lr){3-5}
        Distribution & & \textbf{PMC} & NS & SMC \\
        \midrule[0.5pt]
        Rosenbrock & & $\mathbf{1.5}$ & $136.1$ & $118.0$  \\
        Gaussian Mixture & & $\mathbf{1.6}$ & $222.1$ & $9.6$  \\
        Primordial Features & & $\mathbf{0.4}$ & $21.3$ & $19.5$  \\
        Gravitational Waves &  & $\mathbf{0.4}$ & $10.2$ & $4.6$  \\
        \bottomrule[0.75pt]
    \end{tabular}
    \label{tab_pmc:comparison}
\end{table}

\subsection{Gaussian Mixture}

\begin{figure}
    \centering
	\centerline{\includegraphics[scale=0.65]{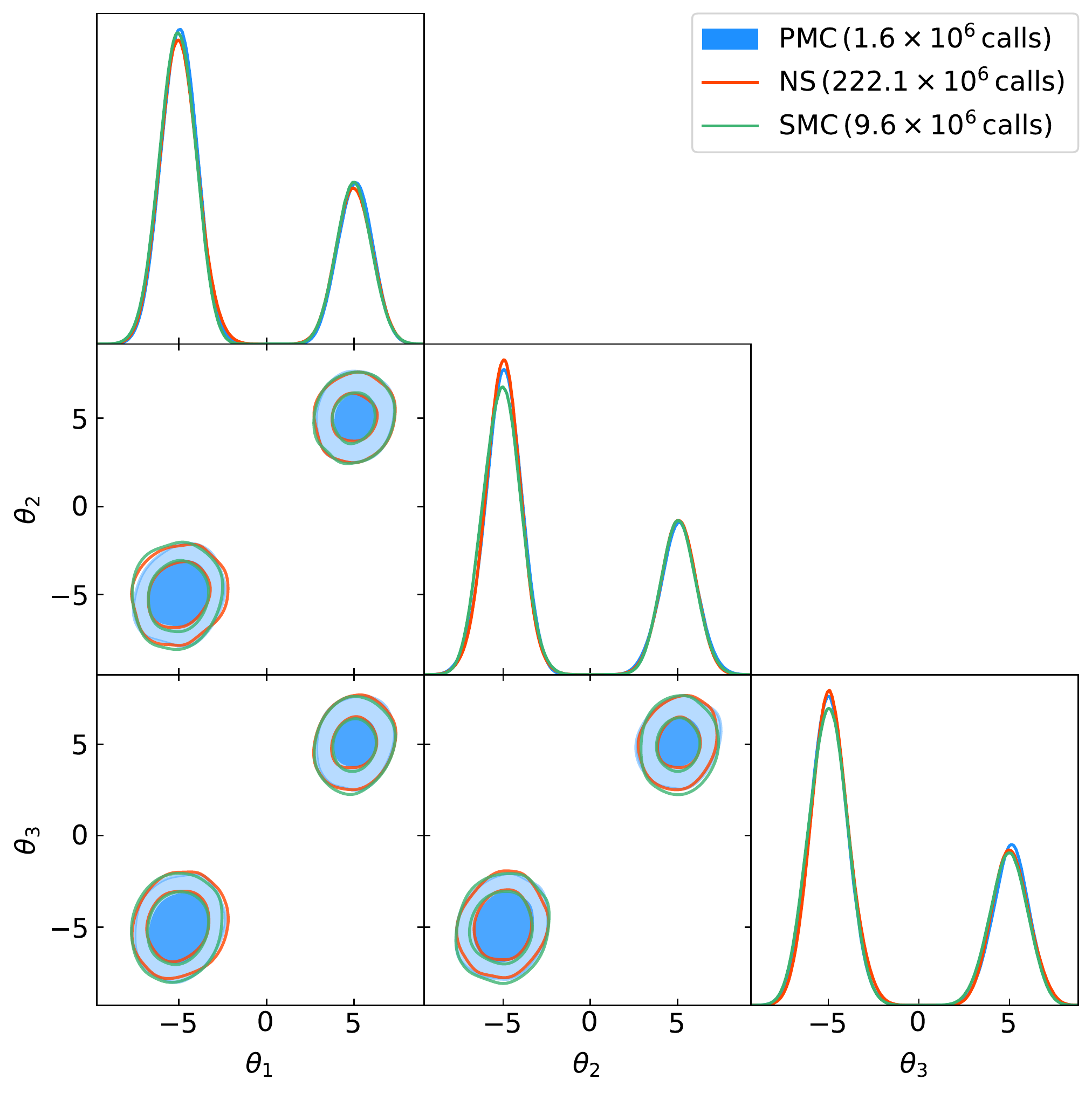}}
    \caption{Illustration of the 1--dimensional and 2--dimensional marginal posteriors for the first three out of $50$ parameters of the two--component Gaussian mixture distribution. The figure shows the 1--$\sigma$ and 2--$\sigma$ contours generated by \emph{Preconditioned Monte Carlo} (PMC) in \emph{blue}, \emph{Nested Sampling} (NS) in \emph{orange}, and \emph{Sequential Monte Carlo} (SMC) in \emph{green}. The legend also shows the computational cost of each method in terms of the total number of required model evaluations until convergence is reached.}
    \label{fig_pmc:mixture}
\end{figure}

The second toy example that we used is a 50--dimensional Gaussian Mixture with two components, one of them being twice as massive as the other. This is a highly multimodal problem as the target distribution exhibits two distinct modes that are well separated. Just as in the \emph{Rosenbrock} case, we use flat priors $\mathcal{U}(-10,10)$ for all parameters. Figure \ref{fig_pmc:rosenbrock} shows the 1--dimensional and 2--dimensional marginal posteriors for the first three parameters as generated by the three methods. The total computational cost of PMC, NS, and SMC is $1.6\times 10^{6}$, $222.1\times 10^{6}$, and $9.6\times 10^{6}$ model evaluations respectively. PMC requires approximately $1/139$ of the number of model evaluations that NS does, and $1/6$ of those that SMC does.

\subsection{Primordial Features}

\begin{figure*}
    \centering
	\centerline{\includegraphics[scale=0.11]{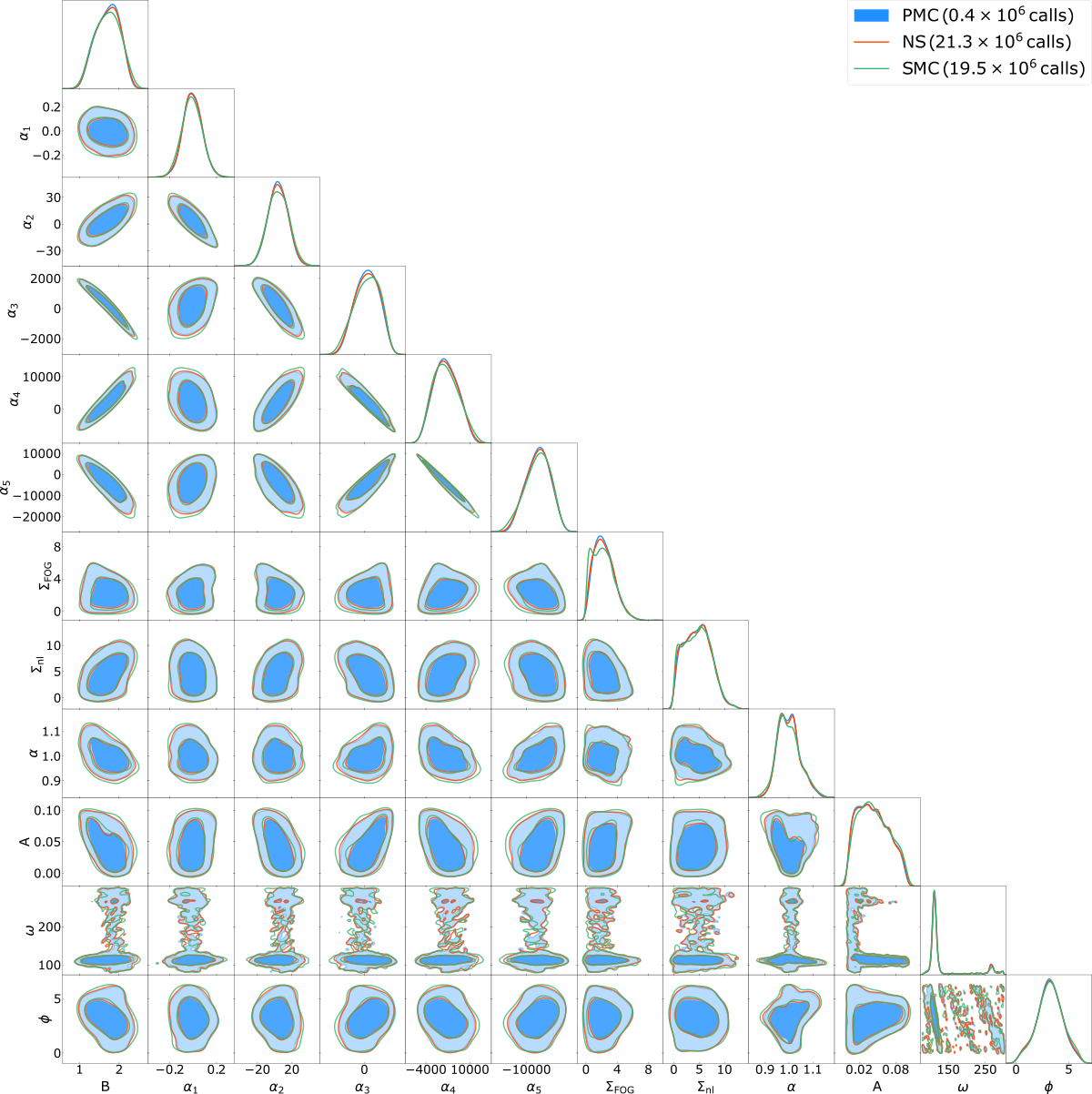}}
    \caption{Illustration of the 1--dimensional and 2--dimensional marginal posteriors for the $12$ parameters of the primordial features posterior. The figure shows the 1--$\sigma$ and 2--$\sigma$ contours generated by \emph{Preconditioned Monte Carlo} (PMC) in \emph{blue}, \emph{Nested Sampling} (NS) in \emph{orange}, and \emph{Sequential Monte Carlo} (SMC) in \emph{green}. The legend also shows the computational cost of each method in terms of the total number of required model evaluations until convergence is reached.}
    \label{fig_pmc:feature}
\end{figure*}

The first realistic application that we study is the the search for primordial features along the Baryon Accoustic Oscillation (BAO) signature in the distribution of galaxies observed by the Sloan Digital Sky Survey (SDSS)~\parencite{SDSSIII}. In particular, the data that we analysed come from the 12th data release (DR12) of the high--redshift North Galactic Cap (NGC) sample of the Baryon Oscillation Spectroscopic Survey (BOSS)~\parencite{BOSS}. Our analysis follows closely that of ~\textcite{beutler2019primordial} for the linear oscillation model. The inference problem includes $12$ free parameters with either flat/uniform or normal priors. Figure \ref{fig_pmc:feature} shows the 1--dimensional and 2--dimensional marginal posteriors of the aforementioned analysis. The posterior distribution exhibits a highly non--Gaussian geometry that can hinder the sampling performance of conventional methods. The total computational cost of PMC, NS, and SMC is $0.4\times 10^{6}$, $21.3\times 10^{6}$, and $19.5\times 10^{6}$ model evaluations respectively. PMC requires approximately $1/53$ of the number of model evaluations that NS does, and $1/49$ of those that SMC does.

\subsection{Gravitational Waves}

\begin{figure*}
    \centering
	\centerline{\includegraphics[scale=0.11]{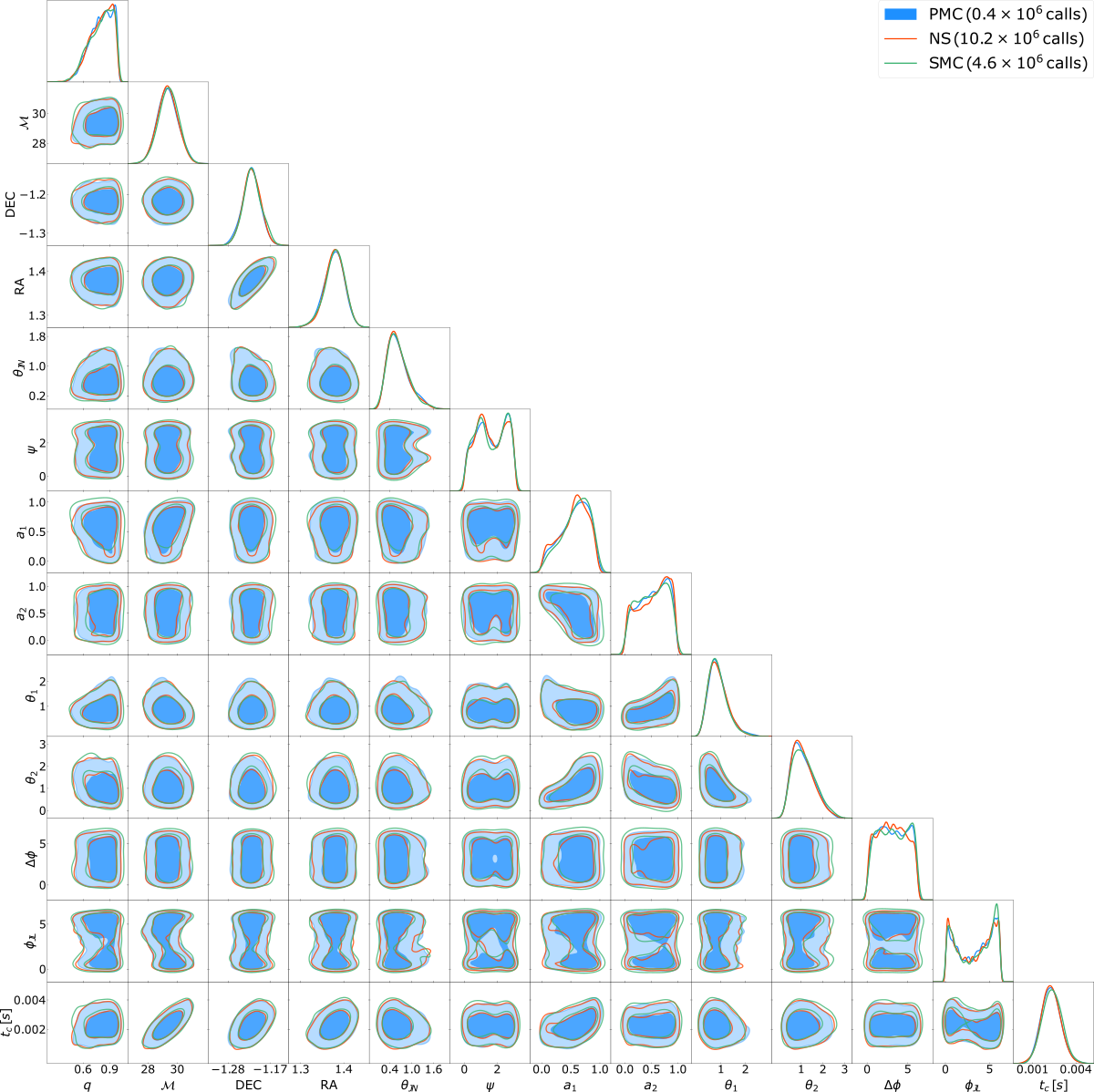}}
    \caption{Illustration of the 1--dimensional and 2--dimensional marginal posteriors for the $13$ parameters of the gravitational waves posterior. The figure shows the 1--$\sigma$ and 2--$\sigma$ contours generated by \emph{Preconditioned Monte Carlo} (PMC) in \emph{blue}, \emph{Nested Sampling} (NS) in \emph{orange}, and \emph{Sequential Monte Carlo} (SMC) in \emph{green}. The legend also shows the computational cost of each method in terms of the total number of required model evaluations until convergence is reached.}
    \label{fig_pmc:gw}
\end{figure*}

The second realistic application is the simulated gravitational wave analysis of an injected signal. For this, we used the standard CBC injected signal configuration provided by \texttt{BILBY}~\parencite{ashton2019bilby}. The inference problem includes $13$ free parameters with a variety of common priors. Figure \ref{fig_pmc:gw} shows the 1--dimensional and 2--dimensional marginal posteriors of the aforementioned analysis. The posterior distribution exhibits a highly non--Gaussian geometry that can hinder the sampling performance of conventional methods. The total computational cost of PMC, NS, and SMC is $0.4\times 10^{6}$, $10.2\times 10^{6}$, and $4.6\times 10^{6}$ model evaluations respectively. PMC requires approximately $1/25$ of the number of model evaluations that NS does, and $1/11$ of those that SMC does.

\section{Discussion}
\label{sec_pmc:discussion}

While we have demonstrated PMC's superior sampling performance for a number of target distributions, including two real--world applications, the real test is based on researchers applying the method to their analyses. Different applications pose different computational challenges and there is no one single sampler to rule them all. Sometimes, certain kinds of distributions will be better handled by other, perhaps simpler, approaches.

In general, we expect PMC to be a useful tool when dealing with computationally expensive likelihood functions and highly correlated or multimodal posteriors. There two main reasons for this. First, training of the normalising flow takes about $\mathcal{O}(1\, {\rm s})$ per iteration, whereas the actual vectorised evaluation of the bijective mapping takes almost $\mathcal{O}(10\, {\rm ms})$ per MCMC step for the whole population of particles. This means that if the cost of evaluating the likelihood is low enough to be comparable to that of the normalising flow, as discussed above, the chances are that there are simpler methods (e.g. MCMC) that can obtain the results more quickly. The second reason has to do with the geometry of the posterior distribution. If the latter is trivial enough, for instance, if the target is approximately Gaussian with no non--linear correlation or multiple modes, then the use of the normalising flow as a preconditioner would offer no benefit and instead only help delay the run.

On the other hand, if both of these conditions are met, that is, the likelihood function is computationally expensive, as it is often the case in cosmology, and the posterior is non--Gaussian, then PMC can be a valuable asset in the astronomer's toolkit. Furthermore, when the cost of evaluating the likelihood function is large enough to dominate both the normalising flow evaluation and any potential \textit{MPI} communication overhead, one can capitalise on the availability of multiple CPUs in order to accelerate PMC. In particular, if the evaluation of the likelihood function takes $\mathcal{O}(1\, {\rm s})$, one should be able to use up to thousands of CPUs, potentially parallelising all or a substantial fraction of the particles simultaneously.

\section{Conclusions}
\label{sec_pmc:conclusions}

The goal of this work was to develop a novel sampling method that can accelerate Bayesian parameter inference and model comparison in computationally challenging astronomical and cosmological analyses. To this end, we introduced PMC, a preconditioned generalisation of the standard SMC algorithm.

After introducing the method in Section \ref{sec_pmc:method}, we presented a thorough demonstration of \textit{Preconditioned Monte Carlo}'s sampling capabilities by comparing its sampling performance to that of \textit{Nested Sampling} and \textit{Sequential Monte Carlo} in a range of  target distributions characterised by non--trivial geometry. The results are presented in Table \ref{tab_pmc:comparison}. In general, we found that \textit{Preconditioned Monte Carlo} is one to two orders of magnitude faster than either \textit{Nested Sampling} or \textit{Sequential Monte Carlo}, both of which performed similarly to each other. Furthermore, in the realistic analyses of primordial features and gravitational waves, \textit{Preconditioned Monte Carlo} required approximately $50$ and $25$ times fewer model evaluations compared to NS in order to converge. The reduced computational cost, combined with the superior parallisation scaling, renders \textit{Preconditioned Monte Carlo} ideal for astronomical and cosmological Bayesian analyses with computationally expensive, strongly correlated, multimodal and high--dimensional posteriors. 

We hope that \textit{Preconditioned Monte Carlo} will prove useful to the astronomical community by facilitating challenging Bayesian data analyses and enabling the investigation of complex models and sparse datasets. We also release a \texttt{Python} implementation of \textit{Preconditioned Monte Carlo}, called \texttt{pocoMC}, which is publically available at \url{https://github.com/minaskar/pocomc} and detailed documentation with installation instructions and examples at \url{https://pocomc.readthedocs.io}.

\section{Appendix: Comparison to Independent Metropolis--Hastings Sequential Monte Carlo}

Recent practice in the literature \parencite{albergo2019flow,williams2021nested, arbel2021annealed} is to use normalising flows as auxiliary densities for \textit{Importance Sampling (IS)} and \textit{Independent Metropolis--Hastings (IMH)} estimators. The latter approach can also be accommodated in the context of \textit{Sequential Monte Carlo (SMC)} as an alternative to PMC. For this reason, we will offer an experimental comparison of PMC to IMH--SMC.

\begin{figure}[!ht]
    \centering
	\centerline{\includegraphics[scale=0.65]{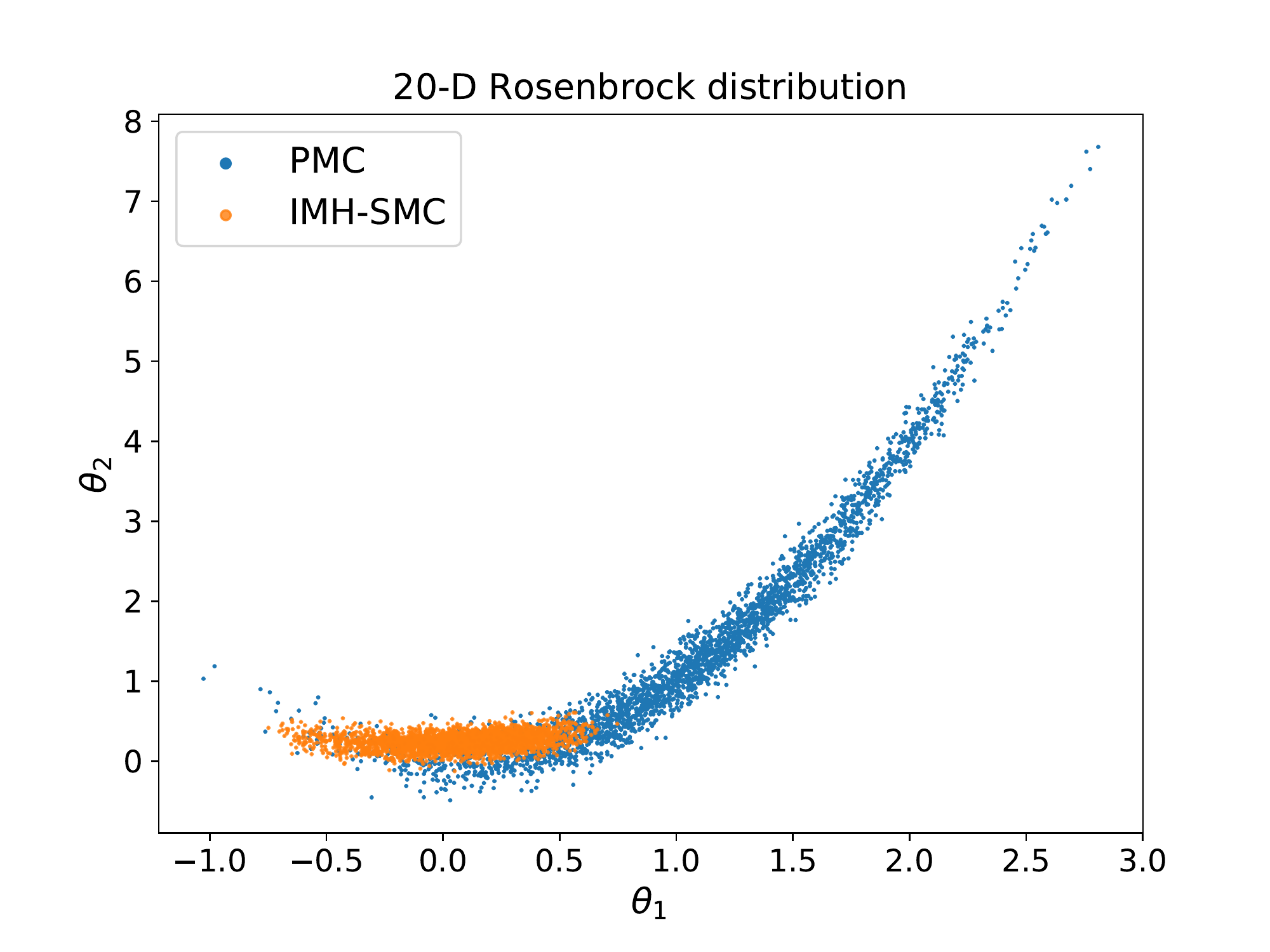}}
    \caption{Comparison of the first two parameters of samples generated using PMC (\textit{blue}) and IMH--SMC (\textit{orange}) for the $20$--D Rosenbrock target distribution. PMC produces representative samples, whereas IMH--SMC does not.}
    \label{fig_pmc:rosenbrock_comparison}
\end{figure}
\begin{figure}[!ht]
    \centering
	\centerline{\includegraphics[scale=0.65]{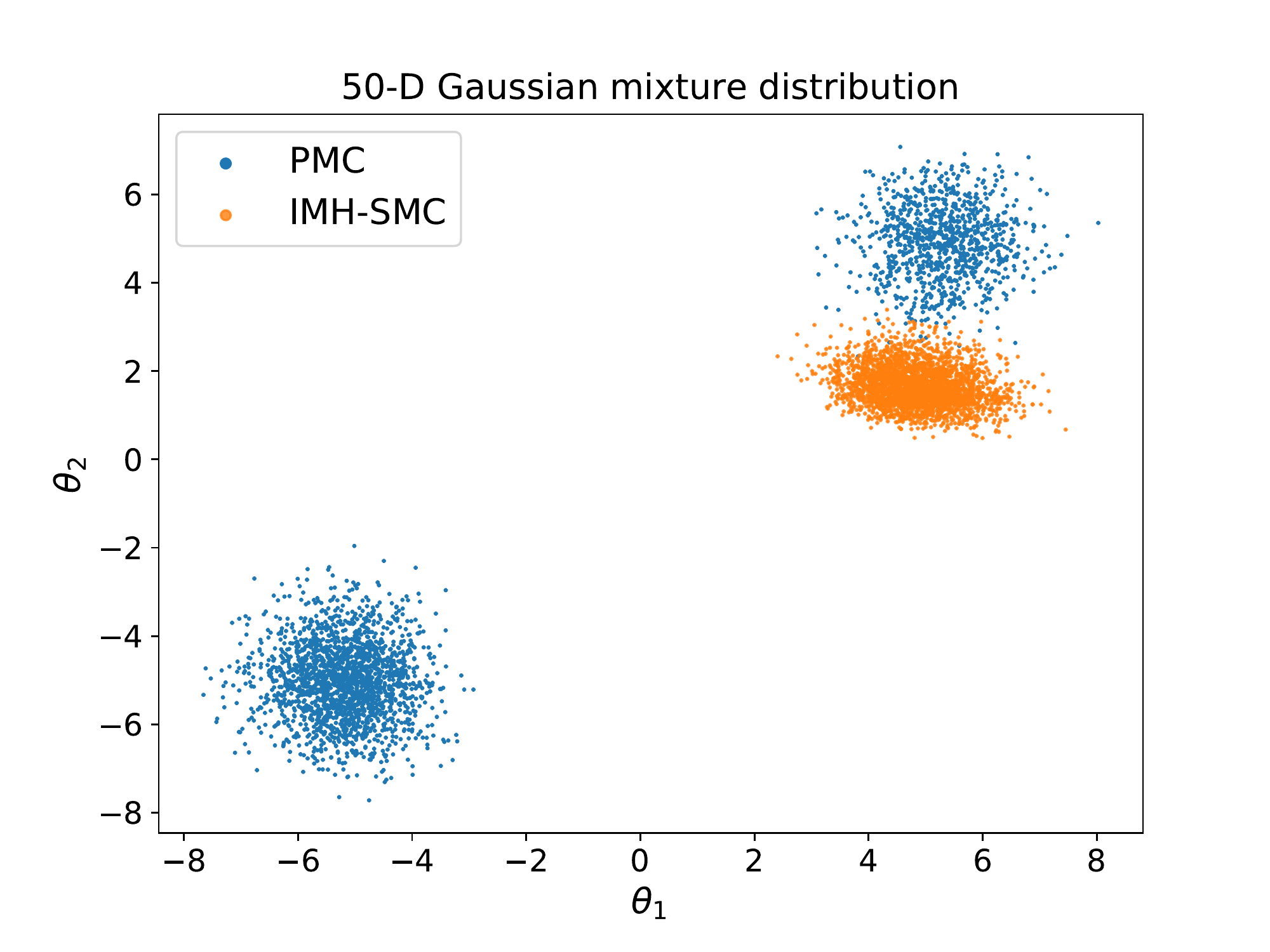}}
    \caption{Comparison of the first two parameters of samples generated using PMC (\textit{blue}) and IMH--SMC (\textit{orange}) for the $50$--D two--component Gaussian mixture target distribution. PMC produces representative samples, whereas IMH--SMC does not.}
    \label{fig_pmc:gaussian_mixture_comparison}
\end{figure}
The IMH--SMC allgorithm is identical to \Algo{pmc} with the exception that the mutation step of line $12$ takes place using the modified \textit{Metropolis acceptance criterion}
\begin{equation}
    \label{eq_pmc:Metropolis_criterion_transform_independent}
    \alpha = \min \left( 1, \frac{p_{\theta}(f^{-1}(u'))q(u)\Big|\det \frac{\partial f^{-1}(u')}{\partial u'}\Big|}{p_{\theta }(f^{-1}(u))q(u ')\Big|\det \frac{\partial f^{-1}(u)}{\partial u}\Big|}\right)\,,
\end{equation}
instead of that of equation \ref{eq_pmc:Metropolis_criterion_transform}. The difference between the two criteria is that the proposal distribution $q(u)=\mathcal{N}(u\vert 0,1)$ is no longer conditional on the previous state of the Markov chain.

The number $M$ of IMH steps performed in each iteration of IMH--SMC is determined adaptively during the run, based on the observed acceptance rate $\alpha$, as
\begin{equation}
    \label{eq_pmc:number_of_imh_steps}
    M = \frac{\log (1 - p)}{\log (1 - \alpha)}\,,
\end{equation}
where $p$ is the target probability of generating a new independent sample. In our examples below, the value of $p$ is chosen such that the computational cost of IMH--SMC is similar to that of PMC for the same example. This results in $p>0.99$ which corresponds to very conservative sampling. 

Despite this, as shown in Figures \ref{fig_pmc:rosenbrock_comparison} and \ref{fig_pmc:gaussian_mixture_comparison}, for the $20$--dimensional Rosenbrock and the $50$--dimensional two--component Gaussian mixture studied in the main text respectively, IMH--SMC does not manage to produce typical samples from the posterior distribution. It is important to note here that the acceptance rate of IMH--SMC was high throughout both runs, and as such offered no indication on its own that NF is not correct.

The origin of this discrepancy between IMH--SMC and PMC in both cases, and the ultimate inability of IMH--SMC to compete with PMC, originates in the substantial mismatch between the importance/NF distribution and target distribution in high dimensions and the subsequent over--fitting of the NF to the particle distribution leading to a narrower distribution. The high acceptance rate does not imply the high quality of NF solution, and other tests of the quality of solution are needed, such as comparing expectation of $\log p$ between samples from NF and true MCMC samples. On the other hand, PMC does not suffer from this pathology as the local exploration offered by MCMC helps diversify the particles in order to avoid over--fitting. Furthermore, local MCMC methods generally scale better with the number of dimensions compared to IMH and IS.
% !TEX TS-program = pdflatex
% !TEX root = ../ArsClassica.tex

%************************************************
\chapter{pocoMC}
\label{chp:pocomc}
%************************************************
 
\lstset{numbers=left,
    numberstyle=\scriptsize,
    stepnumber=1,
    numbersep=8pt
}    

This chapter presents \textit{pocoMC} which is the main contribution introduced in the paper titled \textit{pocoMC: A Python package for accelerated Bayesian inference in astronomy and cosmology} that was submitted for publication in the \textit{Journal of Open Source Software} in July 2022~\parencite{karamanis2022pocomc}. The content of the chapter is almost identical to that included in the aforementioned publication with the exception of minor text and figure formatting differences.

\begin{center}
\rule{0.5\textwidth}{.4pt}
\end{center}
\vspace{8pt}

\section{Summary}

\texttt{pocoMC} is a Python package for accelerated Bayesian inference in astronomy and cosmology. The code is designed to sample efficiently from posterior distributions with non--trivial geometry, including strong multimodality and non--linearity. To this end, \texttt{pocoMC} relies on the Preconditioned Monte Carlo algorithm which utilises a Normalising Flow in order to decorrelate the parameters of the posterior. It facilitates both tasks of parameter estimation and model comparison, focusing especially on computationally expensive applications. It allows fitting arbitrary models defined as a log--likelihood function and a log--prior probability density function in Python. Compared to popular alternatives (e.g. nested sampling) \texttt{pocoMC} can speed up the sampling procedure by orders of magnitude, cutting down the computational cost substantially. Finally, parallelisation to computing clusters manifests linear scaling.

\section{Statement of need}

Over the past few decades the volume of astronomical and cosmological data has increased substantially. At the same time, theoretical and phenomenological models in these fields have grown even more complex. As a response to that, a number of methods aiming at efficient Bayesian computation have been developed with the sole task of comparing those models to the available data \parencite{trotta2017bayesian, sharma2017markov}. 
In the Bayesian context, scientific inference proceeds though the use of Bayes' theorem:
\begin{equation}\label{eq_pocomc:bayes}
\mathcal{P}(\theta) = \frac{\mathcal{L}(\theta)\pi(\theta)}{\mathcal{Z}}
\end{equation}
where the posterior $\mathcal{P}(\theta)\equiv p(\theta\vert d,\mathcal{M})$ is the probability of the parameters $\theta$ given the data $d$ and the model $\mathcal{M}$. The other components of this equation are: the likelihood function $\mathcal{L}(\theta)\equiv p(d\vert \theta,\mathcal{M})$, the prior $\pi(\theta) \equiv p(\theta\vert \mathcal{M})$, and the model evidence $\mathcal{Z}=p(d\vert \mathcal{M})$. The prior and the likelihood are usually provided as input in this equation and one seeks to estimate the posterior and the evidence. Knowledge of the posterior, in the form of samples, is paramount for the task of parameter estimation whereas the ratio of model evidences yields the Bayes factor which is the cornerstone of Bayesian model comparison.

Markov chain Monte Carlo (MCMC) has been established as the standard tool for Bayesian computation in astronomy and cosmology, either as a standalone algorithm or as part of another method (e.g. nested sampling \parencite{skilling2006nested}). However, as MCMC relies on the local exploration of the posterior, the presence of non-linear correlation between parameters and multimodality can at best hinder its performance and at worst violate its theoretical guarantees of convergence (i.e. ergodicity). Usually those challenges are partially addressed by reparameterising the model using a common change--of--variables parameter transformation. However, guessing the right kind of reparameterisation a priori is not trivial as it often requires a deep knowledge of the physical model and its symmetries. These problems are usually complicated further by the substantial computational cost of evaluating astronomical and cosmological models. \texttt{pocoMC} is designed to tackle exactly these kinds of difficulties by automatically reparameterising the model such that the parameters of the model are approximately uncorrelated and standard techniques can be applied. As a result, \texttt{pocoMC} produces both samples from the posterior distribution and an unbiased estimate of the model evidence thus facilitating both scientific tasks with excellent efficiency and robustness. Compared to popular alternatives such as nested sampling, \texttt{pocoMC} can reduce the computational cost, and thus, the total run time of the analysis by orders of magnitude, in both artificial and realistic applications \parencite{karamanis2022pmc}. Finally, the code is well-tested and is currently used for research work in the field of gravitational wave parameter estimation \parencite{vretinaris2022postmerger}.

\begin{figure}[H]
    \centering
	\centerline{\includegraphics[scale=0.175]{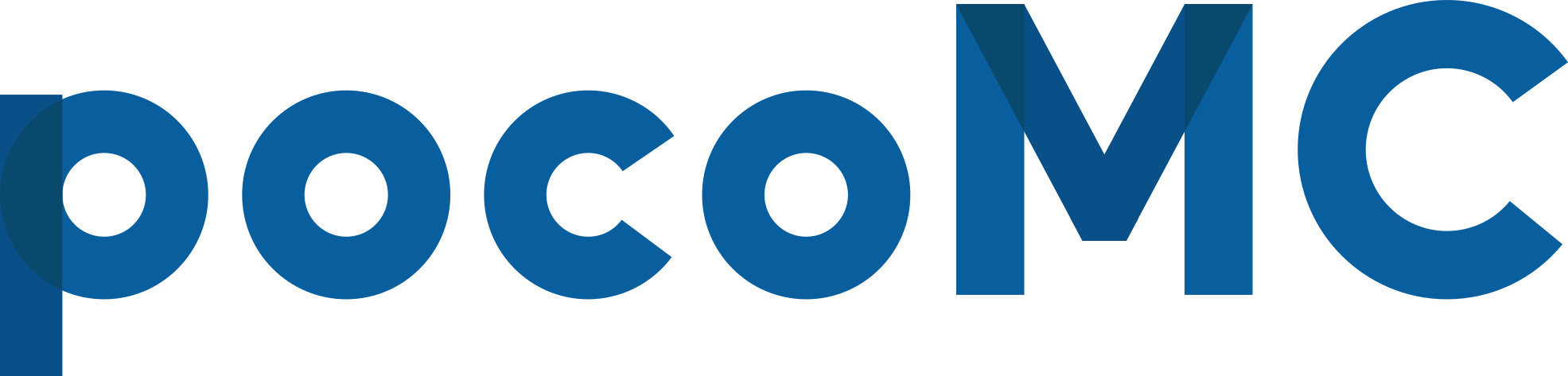}}
    \caption{Logo of \texttt{pocoMC}.}
    \label{fig_pocomc::logo}
\end{figure}

\section{Method}

\texttt{pocoMC} implements the Preconditioned Monte Carlo (PMC) algorithm. PMC combines the popular Sequential Monte Carlo (SMC) \parencite{del2006sequential} method with a Normalising Flow (NF) \parencite{papamakarios2021normalizing}. The latter works as a preconditioner for the target distribution of the former. As SMC evolves a population of particles, starting from the prior distribution and gradually approaching the posterior distribution, the NF transforms the parameters of the target distribution such that any correlation between parameters or presence of multimodality is removed. The effect of this bijective transformation is the substantial rise in the sampling efficiency of the algorithm as the particles are allowed to sample freely from the target without being hindered by its locally--curved geometry. The method is explained in detail in the accompanying publication \parencite{karamanis2022pmc} and we provide only a short summary here.

\subsection{Sequential Monte Carlo}

The basic idea of basic SMC is to sample from the posterior distribution $\mathcal{P}(\theta)$ by first defining a path of intermediate distributions starting from the prior $\pi(\theta)$. In the
case of \texttt{pocoMC} the path has the form:
\begin{equation}\label{eq_pocomc:path}
p_{t}(\theta) = \pi(\theta)^{1-\beta_{t}} \mathcal{P}(\theta)^{\beta_{t}}
\end{equation}
where $0=\beta_{1}<\beta_{2}<\dots<\beta_{T}=1$. Starting from the prior, each distribution with density $p_{t}(\theta)$ is sampled in turn using a collection of particles propagated by a number of MCMC steps. Prior to MCMC sampling, the particles are re-weighted using importance sampling and then re-sampled to account for the transition from $p_{t}(\theta)$ to $p_{t+1}(\theta)$. \texttt{pocoMC} utilises the importance weights of this step to define an estimator for the effective sample size (ESS) of the population of particles. Maintaining a fixed value of ESS during the run allows \texttt{pocoMC} to adaptively specify the $\beta_{t}$ schedule.

\subsection{Preconditioned Monte Carlo}

In vanilla SMC, standard MCMC methods (e.g. Metropolis-Hastings) are used to update the positions of the particles during each iteration. This however can become highly inefficient if the distribution $p_{t}(\theta)$ is characterised by a non--trivial geometry. \texttt{pocoMC}, which is based on PMC, utilises a NF to learn an invertible transformation that simplifies the geometry of the distribution by mapping $p_{t}(\theta)$ into a zero-mean unit-variance normal distribution. Sampling then proceeds in the latent space in which correlations are substantially reduced. The positions of the particles are transformed back to the original parameter space at the end of each iteration. This way, PMC and \texttt{pocoMC} are able to sample from very challenging posteriors very efficiently using simple Metropolis-Hastings updates in the preconditioned/uncorrelated latent space.

\section{Features}

\begin{itemize}
    \item User--friendly black-box API (only the log-likelihood, log-prior and some prior samples required from the user)
    \item Default configuration sufficient for most applications (no tuning is required but is possible for experienced users)
    \item Posterior corner, trace, and run plotting tools
    \item Support for both MAF and RealNVP normalising flows with added regularisation \parencite{papamakarios2017masked, dinh2016density}
    \item Straightforward parallelisation using MPI or multiprocessing
    \item Continuous integration, unit tests and wide range of examples available
    \item Extensive documentation available  online \url{http://pocomc.readthedocs.io}
\end{itemize}

% !TEX TS-program = pdflatex
% !TEX root = ../ArsClassica.tex

%************************************************
\chapter{Conclusions}
\label{chp:conclusions}
%************************************************

\begin{flushright}
\itshape
So long, and thanks for all the fish. \\
\medskip
--- Douglas Adams, The Hitchhiker's Guide to the Galaxy
\end{flushright}

\lstset{numbers=left,
    numberstyle=\scriptsize,
    stepnumber=1,
    numbersep=8pt
}    

% Intro paragraphs
Over the past couple of decades, Bayesian inference has been established as the standard mathematical framework for conducting scientific inference in the physical sciences. This progress has been largely facilitated by the recent advances in computer technology and probabilistic computational methods. However, the specific characteristics of the mathematical models and available data used in astronomy and cosmology still pose significant challenges for existing computational tools.

From the perspective of theoretical modelling, many astrophysical models involve computationally expensive operations which are almost always non--differentiable. This limits the potential range of application of a plethora of MCMC methods, particularly those that rely on the use of the gradient of the posterior density function or are unable to scale to a large number of parallel CPUs. On the other hand, the commonly sparse nature of the available data often induces a level of multimodality in the studied posterior distributions. The existence of multiple modes in the posterior distribution can hinder the sampling procedure of most computational tools and in the case of most MCMC methods, make the results unreliable. This thesis has introduced two methods and their software implementations that were specifically designed with this kind of challenge in mind.

% Ensemble slice sampling
In Chapter \ref{chp:ess} we introduced \textit{Ensemble Slice Sampling (ESS)}, a method that extends the applicability of the univariate slice sampler to multivariate target distributions, by utilising an ensemble of parallel walkers. The method requires minimal tuning and no gradient information, demonstrates affine--invariant sampling performance, and is trivially parallelisable to a large number of CPUs. Chapter \ref{chp:zeus} presents \texttt{zeus}, an open--source \texttt{Python} implementation of ESS. Compared to the popular MCMC sampler \texttt{emcee}, the sampling efficiency of \texttt{zeus} scales more favourably with the total number of dimensions. Furthermore, the generated Markov chains exhibit substantially lower autocorrelation levels for a wide range of target distributions and the method generally requires significantly fewer walkers than \texttt{emcee}. Finally, in the problems of BAO and exoplanet parameter estimation, \texttt{zeus} is $9$ and $29$ times more efficient than the competition, respectively.

% Preconditioned Monte Carlo
Chapter \ref{chp:pmc} is devoted to \textit{Preconditioned Monte Carlo (PMC)}, a novel Monte Carlo method for sampling from posteriors with non--trivial geometry (i.e. non--linear correlations, multimodality). PMC utilises a Normalising Flow (NF) transformation in order to precondition the target distribution by approximately removing the correlations between its parameters. PMC then relies on a \textit{Sequential Monte Carlo (SMC)} in order to produce posterior samples and an estimate of the model evidence. Empirical tests validate the high sampling efficiency of PMC. In the cases of primordial feature analysis and gravitational wave inference, PMC is approximately $50$ and $25$ times faster respectively than nested sampling. Finally, Chapter \ref{chp:pocomc} offers a short overview of \texttt{pocoMC}, an open--source \texttt{Python} implementation of PMC. The basic principles of PMC are presented along with the various options and features provided in the package. In terms of parallelisation, \texttt{pocoMC} manifests linear scaling up to thousands of CPUs.

% Future outlook
The methods introduced in the aforementioned chapters aim to address the various computational challenges currently presented by modern astrophysical models and data. Despite their empirical success, as demonstrated by the provided tests and their adoption by the astronomical community, their application in higher dimensions (e.g. $D>100$) is still hindered by the \textit{curse of dimensionality}. In the future, in order to accommodate for subtle effects present in the data, astrophysical models will necessarily become increasingly complicated. As a response, sampling methods such as the ones presented in this thesis will have to evolve in order to cope with the additional computational challenges. A possible avenue of future research could be the self--supervised construction of surrogate models (e.g. emulators) for either the likelihood function, posterior density, or model, thus enabling the use of gradient--based MCMC methods in the context of advanced schemes such as PMC. We sincerely hope that, in the meantime, methods and packages such as ESS $\&$ PMC, and \texttt{zeus} $\&$ \texttt{pocoMC} will prove useful to the astronomical community by facilitating the next generation of Bayesian data analyses.

%\clearpage
\cleardoublepage
%\bibliographystyle{mnras}
% !TEX TS-program = pdflatex
% !TEX root = ../ArsClassica.tex

%*******************************************************
% Bibliography
%*******************************************************
\nocite{*}
\printbibliography
\end{document}